\UseRawInputEncoding

\documentclass[a4paper,aps,prl,amsmath,amssymb,floatfix,preprint,citeautoscript,noeprint,superscriptaddress,12pt]{revtex4-2}

\usepackage{dsfont}
\usepackage{lipsum} 
\usepackage{bibentry}
\usepackage[english]{babel}
\usepackage[dvipsnames]{xcolor}
\usepackage{graphicx}
\usepackage[caption=false]{subfig} 
\usepackage{amsmath,amssymb,bm}
\usepackage[version=3]{mhchem}
\usepackage{verbatim}
\usepackage{multirow}
\usepackage{dcolumn}
\usepackage{float}
\usepackage{nicefrac}
\usepackage{siunitx}
\usepackage{booktabs}
\usepackage{chemformula}
\usepackage{wrapfig}
\usepackage{enumitem}  
\usepackage{transparent}
\usepackage[colorlinks,allcolors=black,citecolor=blue,urlcolor=blue]{hyperref}
\usepackage{tcolorbox}
\usepackage{relsize}
\hypersetup{
    citecolor=purple,
    colorlinks=true,
    linkcolor=black,    
    urlcolor=purple,
    }
\usepackage{etoolbox}

\usepackage{sansmathfonts}
\usepackage[justification=centerlast, font={sf}, labelfont={bf}]{caption}
\makeatletter
\def\@fnsymbol#1{%
  \ifcase#1\or a\or *\or *\or *\or *\else\@ctrerr\fi}
\makeatother

\makeatletter
\input{si-labels.aux}
\makeatother

\begin{document}

\author{Benjamin X. Shi}
\affiliation{Initiative for Computational Catalysis, Flatiron Institute, 160 5th Avenue, New York, NY 10010}

\author{Kristina M. Herman}
\altaffiliation{Present address: Department of Chemistry, University of Chicago, Chicago, IL 60637, United States}
\affiliation{Department of Chemistry, University of Washington, Seattle, Washington 98195, United States}

\author{Flaviano Della Pia}
\affiliation{Yusuf Hamied Department of Chemistry, University of Cambridge, Lensfield Road, Cambridge CB2 1EW, United Kingdom}

\author{Venkat Kapil}
\affiliation{Department of Physics and Astronomy, University College London, 7-19 Gordon St, London WC1H 0AH, UK}
\affiliation{Thomas Young Centre and London Centre for Nanotechnology, 9 Gordon St, London WC1H 0AH}

\author{Andrea Zen}
\affiliation{Dipartimento di Fisica Ettore Pancini, Universit\`{a} di Napoli Federico II, Monte S. Angelo, I-80126 Napoli, Italy}
\affiliation{Department of Earth Sciences, University College London, Gower Street, London WC1E 6BT, United Kingdom}

\author{P\'eter R. Nagy}
\affiliation{Department of Physical Chemistry and Materials Science, Faculty of Chemical Technology and Biotechnology, Budapest University of Technology and Economics, M\H uegyetem rkp. 3., H-1111 Budapest, Hungary}
\affiliation{HUN-REN-BME Quantum Chemistry Research Group, M\H uegyetem rkp. 3., H-1111 Budapest, Hungary}
\affiliation{MTA-BME Lend\"ulet Quantum Chemistry Research Group, M\H uegyetem rkp. 3., H-1111 Budapest, Hungary}

\author{Sotiris Xantheas}
\email[Electronic address: ]{sotiris.xantheas@pnnl.gov}
\affiliation{Department of Chemistry, University of Washington, Seattle, Washington 98195, United States}
\affiliation{Advanced Computing, Mathematics and Data Division, Pacific Northwest National Laboratory, 902 Battelle Boulevard, MS J7-10, Richland, Washington 99352, United States}
\affiliation{Computational and Theoretical Chemistry Institute (CTCI), Pacific Northwest National Laboratory, Richland, Washington 99352, United States}

\author{Angelos Michaelides}
\email[Electronic address: ]{am452@cam.ac.uk}
\affiliation{Yusuf Hamied Department of Chemistry, University of Cambridge, Lensfield Road, Cambridge CB2 1EW, United Kingdom}%

\title{Efficient first-principles modeling of complex\\ molecular crystals at sub-chemical accuracy}

\date{\today}

\begin{abstract}
Molecules can form myriad crystalline polymorphs, each with distinct properties affecting their performance across diverse applications, from pharmaceuticals to functional materials and more.
Predicting the thermodynamically most stable polymorph from first principles remains a formidable challenge.
It requires methods that scale to large, technologically-relevant molecules while achieving very high accuracy (below $1\,$kJ/mol) on relative lattice energies.
Such accuracy, often termed sub-chemical accuracy, is generally beyond the reach of the workhorse density functional theory (DFT).
In this work, we introduce a framework, combining advances in correlated wavefunction theory (cWFT) and the many-body expansion, to deliver accurate, cost-effective predictions of complex molecular crystals.
For 23 organic molecules and 13 ice polymorphs, we predict crystal lattice energies to within experimental uncertainties at costs comparable to hybrid DFT, while being several orders of magnitude more efficient than previous cWFT approaches.
We extend this approach to a set of large, drug-like molecules including  axitinib and ROY, previously inaccessible to cWFT and where DFT is insufficient, achieving sub-chemical accuracy on the relative energies between challenging polymorphs.
With the reference data generated throughout this work, we have been able to further parametrize a DFT functional with unprecedented accuracy aligning with our predictions.
This cWFT framework as well as DFT functional are made openly available, providing new ranking tools to facilitate efficient high-throughput screening of molecular crystal polymorphs.
\end{abstract}

\maketitle

\section{Introduction}

The constituent molecules in a molecular crystal can adopt numerous possible crystallographic arrangements~\cite{maddoxCrystalsFirstPrinciples1988,woodleyCrystalStructurePrediction2008}, with each polymorph characterized by distinct properties.
This diversity offers exciting opportunities to design molecular crystals tailored for its widespread applications across pharmaceuticals~\cite{neumannCombinedCrystalStructure2015,bhardwajProlificSolvateFormer2019}, agrochemicals~\cite{lamberthCurrentChallengesTrends2013}, gas storage materials~\cite{jonesModularPredictableAssembly2011}, functional devices~\cite{baldoHighlyEfficientPhosphorescent1998,boltonActivatingEfficientPhosphorescence2011}, and the food industry.
More fundamentally, molecular crystals underpin many natural phenomena, from biological sugars such as glucose and sucrose~\cite{brownADGlucosePreciseDetermination1965} to heat and mass transport in high-pressure planetary ices~\cite{sunPhaseDiagramHighpressure2015,hernandezMeltingCurveSuperionic2023}.
Mapping the relative stability of these polymorphs~\cite{pricePredictingCrystalStructures2014} and identifying the thermodynamically most stable form is essential towards these applications.
However, experimental determination is often hindered by inaccessible thermodynamic conditions or high synthesis costs.
Consequently, substantial effort has focused on addressing this challenge through computational modeling, giving rise to the field of organic crystal structure prediction~\cite{beranModelingPolymorphicMolecular2016b,pulidoFunctionalMaterialsDiscovery2017,marzariElectronicstructureMethodsMaterials2021}.

\begin{figure}[t]
    \includegraphics[width=\textwidth]{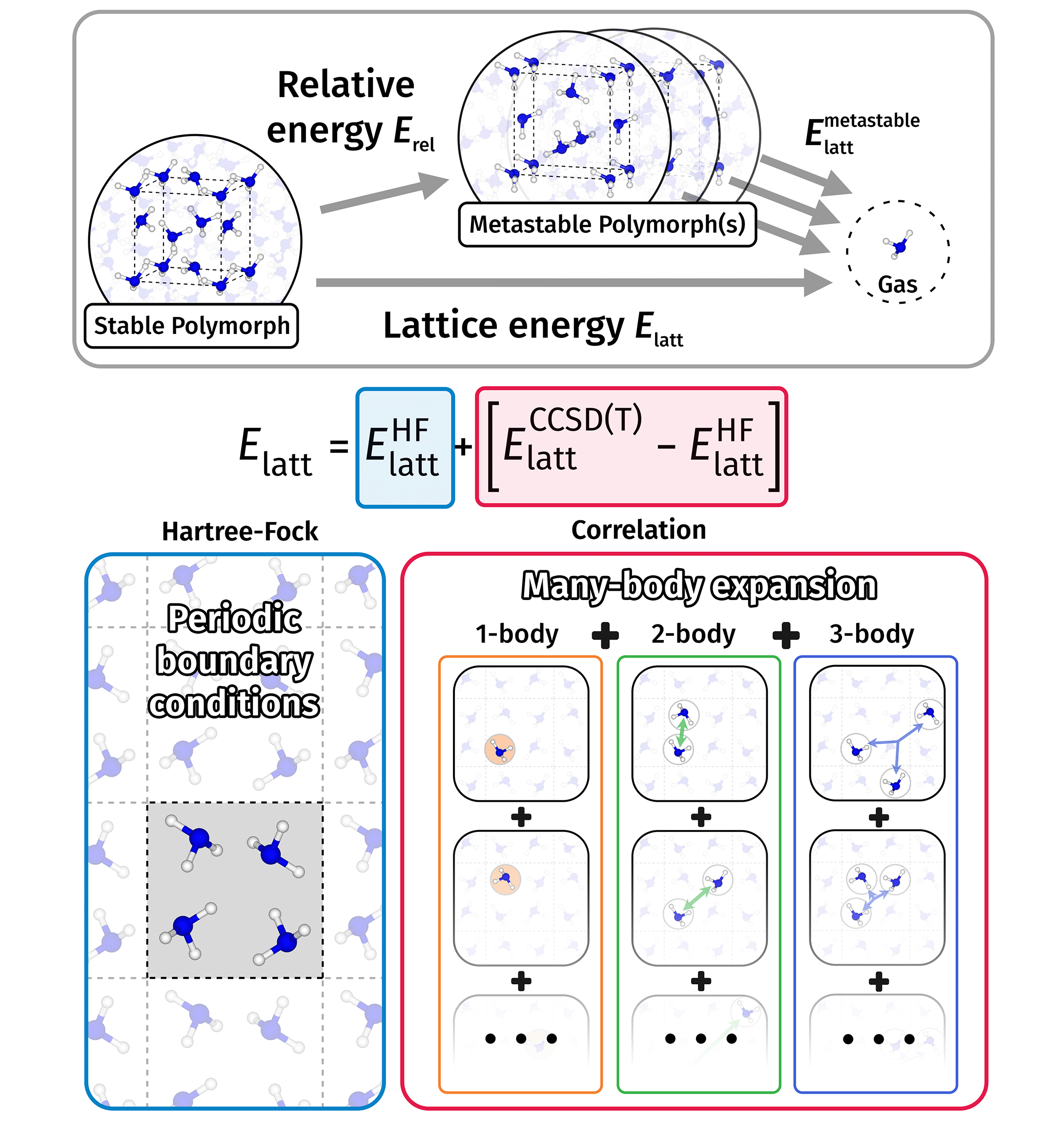}
    \caption{\label{fig:schematic}\textbf{Reliable and efficient procedure for molecular crystal energetics.} 
    We employ a divide-and-conquer approach to reach `gold-standard' CCSD(T) estimates of the lattice $E_\text{latt}$ and relative $E_\text{rel}$ energies (shown schematically in the top panel and defined in the Methods).
    As a post-Hartree-Fock (HF) method, the CCSD(T) $E_\text{latt}$ is composed of a (mean-field) HF and an (electron-electron) correlation contribution.
    We treat HF under periodic boundary conditions (PBC) while the correlation energy is treated with the many-body expansion (MBE).
    This division provides a natural description for each component, allowing us to leverage new developments in solid-state modeling under PBCs together with molecular quantum chemistry for the MBE.
    }
\end{figure}

Predicting polymorph stability from first principles remains a longstanding challenge~\cite{beranHowManyMore2022}.
The key computational quantities (Figure~\ref{fig:schematic}), within a zeroth-order approximation~\cite{priceZerothOrderCrystal2018} neglecting thermal~\cite{hojaFirstprinciplesModelingMolecular2017a} and quantum nuclear effects~\cite{kapilCompleteDescriptionThermodynamic2022a}, are the lattice ($E_\text{latt}$) and relative ($E_\text{rel}$) energies.
The former compares the energy per molecule of a crystalline polymorph to the gas phase, $E_\text{latt} = E^\text{crys} - E^\text{gas}$, while the latter compares the energy between two polymorphs (e.g., A and B), $E_\text{rel} = E^\text{crys,A} - E^\text{crys,B}$.
The value of $E_\text{rel}$ is often less than $2\,$kJ/mol, observed within more than half of previously examined polymorph pairs~\cite{nymanStaticLatticeVibrational2015}.
Thus, reliably distinguishing polymorph stability requires methods that achieve sub-chemical accuracy~\cite{yangInitioDeterminationCrystalline2014a} ($1\,$kJ/mol).
In addition to this exceptionally stringent accuracy, such a method must also scale efficiently to large, complex molecular crystals of technological importance.
While challenging, if these demands can be met, the resulting methods will offer major technological, economic, and public health benefits~\cite{kaiserShortagesCancerDrugs2011,bucarDisappearingPolymorphsRevisited2015a}.
Beyond enabling materials discovery and exploration of experimentally inaccessible phases~\cite{johalExploringOrganicChemical2025}, such methods can also be used to de-risk~\cite{kazantsevSuccessfulPredictionModel2011} existing applications, especially pharmaceuticals, where late-emerging stable polymorphs (notably rotigotine) have led to costly recalls~\cite{bucarDisappearingPolymorphsRevisited2015a,taylorMinimizingPolymorphicRisk2020}.

Density functional theory (DFT) has become the primary tool for modeling molecular crystals~\cite{firahaPredictingCrystalForm2023a}.
Recently, its scope has been greatly expanded by machine-learning interatomic potentials (MLIPs) that aim to reproduce its accuracy at a fraction of the cost~\cite{kalitaMachineLearningInteratomic2025,gharakhanyanOpenMolecularCrystals2025}.
The development of advanced dispersion corrections now allow DFT to attain chemical accuracy for a relatively wide range of properties~\cite{beranPredictingOrganicCrystal2010,hojaReliablePracticalComputational2019,priceAccurateEfficientPolymorph2023a,a.priceXDMcorrectedHybridDFT2023} ($4\,$kJ/mol).
However, achieving sub-chemical accuracy across broad sets of molecular crystals remains difficult.
Results can vary significantly~\cite{dellapiaAccurateEfficientMachine2025} with the choice of density functional approximation (DFA) to the exchange-correlation functional~\cite{perdewJacobLadderDensity2001b} and dispersion correction~\cite{grimmeDispersionCorrectedMeanFieldElectronic2016}.
Notably, systems prone to delocalization error~\cite{bryentonDelocalizationErrorGreatest2023} pose challenges for DFT, such as ionic co-crystals~\cite{leblancPervasiveDelocalisationError2018}, flexible molecular crystals~\cite{bernsteinConformationalPolymorphismInfluence1978,chattopadhyayLatticeEnergyPartitions2025} and tautomeric polymorphs~\cite{perryTamingTautomerismOrganic2025}.
These challenges are exemplified by the prototypical ROY molecule - named for its red, orange, and yellow polymorphs~\cite{chenNewPolymorphsROY2005,levesqueROYReclaimsIts2020,weatherstonPolymorphicROYalty14th2025} - with DFT consistently overstabilizing the red forms over stable yellow ones~\cite{nymanAccuracyReproducibilityCrystal2019,greenwellOvercomingDifficultiesPredicting2020b,beranHowManyMore2022}.
Developing methods that overcome these limitations of DFT represents a major frontier in computational modeling~\cite{o.beranFrontiersMolecularCrystal2023}.

Methods from correlated wavefunction theory (cWFT), notably coupled cluster with single, double and perturbative triple excitations~\cite{bartlettCoupledclusterTheoryQuantum2007a} [CCSD(T)] and quantum diffusion Monte Carlo~\cite{foulkesQuantumMonteCarlo2001} (DMC), can achieve sub-chemical accuracy on molecular crystal predictions.
Both methods have been shown to agree with experimental $E_\text{latt}$ measurements for several organic molecules~\cite{otero-de-la-rozaBenchmarkNoncovalentInteractions2012,reillyUnderstandingRoleVibrations2013b,dolgonosRevisedValuesX232019,dellapiaHowAccurateAre2024} and polymorphs of ice~\cite{zenFastAccurateQuantum2018a,dellapiaDMCICE13AmbientHigh2022b}.
However, their computational cost has limited these methods to small systems, where we show in this work that a single polymorph treated with DMC can require hundreds of thousands of CPU core-hours.
This cost is comparable to a full DFT search over thousands of polymorphs for a large molecule~\cite{hunnisettSeventhBlindTest2024}.
Conventional CCSD(T) features a steeper computational scaling than DMC and is typically applied to crystals through a many-body expansion (MBE)\cite{gordonFragmentationMethodsRoute2012}.
Here, the total energy of the full periodic system is decomposed into many smaller calculations of monomer (one-), dimer (two-), trimer (three-) and higher-body contributions.
Although this expansion is exact when all body-order terms are included, its practicality comes from the fact that high-order (beyond tetramer) and long-range contributions (i.e., dimers or trimers that are far apart) can be neglected or approximated with cheaper methods.
Nonetheless, attaining sub-chemical accuracy in the MBE remains challenging, requiring \textit{tour-de-force} efforts only achievable for model systems such as benzene~\cite{ringerFirstPrinciplesComputation2008,yangInitioDeterminationCrystalline2014a,borcaBenchmarkCoupledclusterLattice2023a}.
In fact, recent studies~\cite{sargentBenchmarkingTwobodyContributions2023a,sytyMultiLevelCoupledClusterDescription2025} show that dimers of moderate-sized molecules like anthracene (C$_{14}$H$_{10}$) already exceed practical limits for conventional CCSD(T).

In this work, we introduce LNO-MBE-CCSD(T), a framework that combines recent methodological developments in CCSD(T) with modeling advances in the MBE to enable accurate and efficient treatment of molecular crystals.
Notably, this framework extends CCSD(T) to the regime of large drug-like molecules that are more than two times the size of anthracene.
We employ a divide-and-conquer approach (discussed in the Methods) that treats the Hartree-Fock (HF) contribution [to the CCSD(T) total energy] under periodic boundary conditions (PBCs) - made highly efficient by recent GPU implementations and sparsity-based algorithms~\cite{linAdaptivelyCompressedExchange2016} - and the remaining CCSD(T) correlation energy with the MBE.
The main advance employed in this work is an improved local natural orbital (LNO) approximation for CCSD(T)~\cite{nagyApproachingBasisSet2019,nagyStateoftheartLocalCorrelation2024} that makes it practical to use within MBE calculations for large molecular crystals.
As described in Supplementary Section~\ref{si-sec:opt_lno}, our modifications to LNO-CCSD(T) prevent the accumulation of local-approximation errors from the numerous two- and three-body terms in the MBE, enabling reliable and affordable convergence.
In addition, we show that the number of MBE calculations can be greatly reduced by combining new formulations that exploit translational symmetry~\cite{hermanFormulationManyBodyExpansion2023b} with machine-learning descriptors~\cite{ruppFastAccurateModeling2012} to eliminate all redundant terms.
Together, the resulting LNO-MBE-CCSD(T) framework can make $E_\text{latt}$ predictions at orders of magnitude lower cost than DMC, comparable to hybrid DFT with PBCs.

We apply LNO-MBE-CCSD(T) to study an unprecedented set of molecular crystals to high accuracy.
We predict $E_\text{latt}$ for 23 molecular crystals and 13 ice phases to an accuracy that surpasses experimental uncertainties.
Furthermore, we demonstrate agreement with experiments on $E_\text{rel}$ to sub-chemical accuracy for competing polymorphs of ROY (C$_{12}$H$_9$N$_3$O$_2$S), axitinib (C$_{22}$H$_{18}$N$_4$OS) and rotigotine (C$_{19}$H$_{25}$NOS), with the latter two being currently marketed pharmaceuticals.
Using the reference-quality datasets generated here, we benchmark and parametrize a new functional, B86bPBE50-revXDM, that outperforms existing DFAs, and use it to train cost-efficient MLIPs to capture thermal and nuclear quantum effects missing from static $E_\text{latt}$ and $E_\text{rel}$ estimates.
As a final outlook, this work delivers an efficient toolkit - all freely available and well-documented (see Methods) - for high-throughput screening and ranking of molecular crystal polymorphs, ranging from high-accuracy LNO-MBE-CCSD(T) for final-stage refinement to a cost-effective B86bPBE50-revXDM functional suitable for intermediate stages.

\section{Results and discussion}

\begin{figure}[h]
    \includegraphics[width=\textwidth]{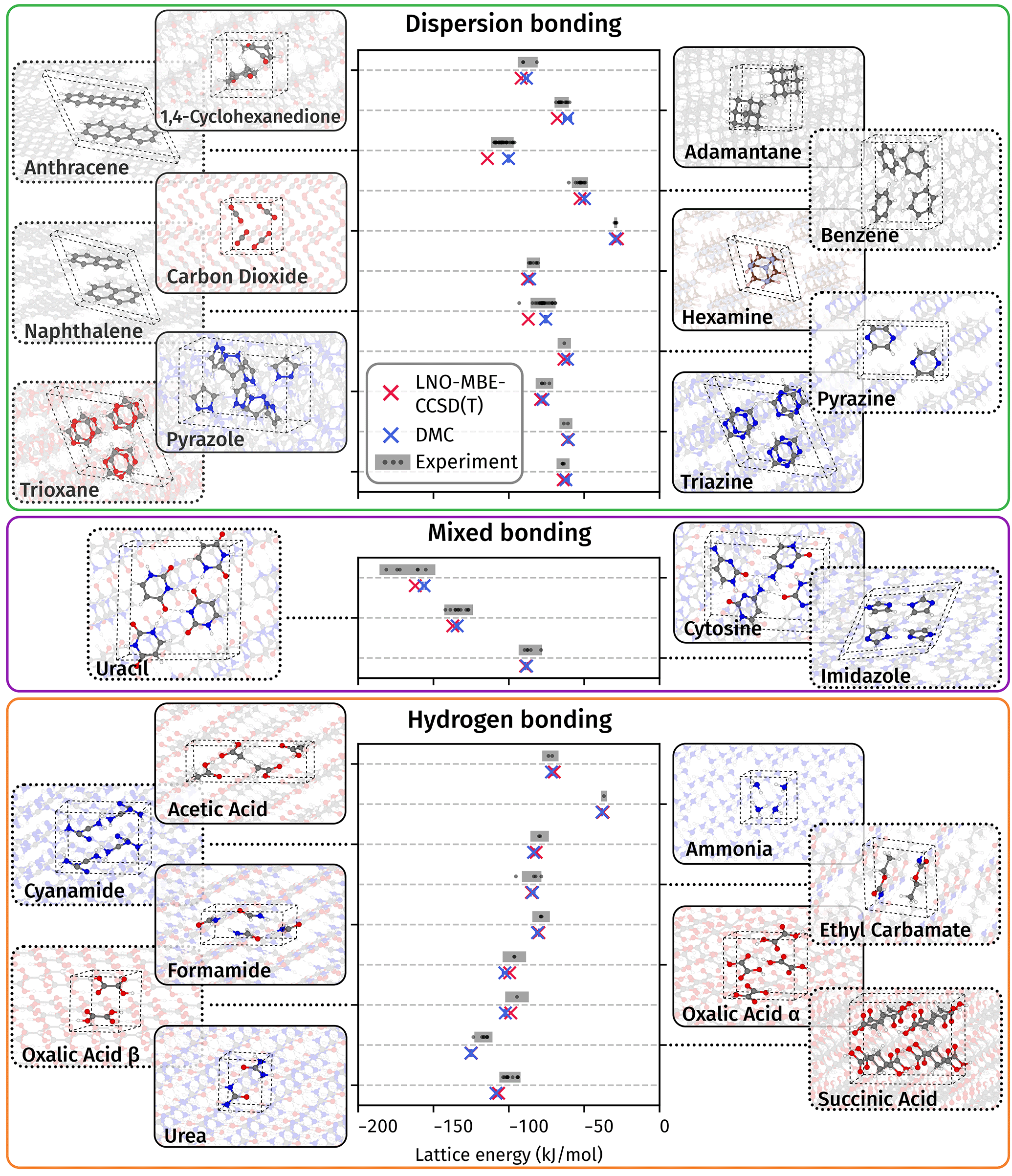}
    \caption{\label{fig:x23_comparison}\textbf{Agreement across diverse molecular crystals.} Comparison of the lattice energy predicted by LNO-MBE-CCSD(T) against previous DMC estimates~\cite{dellapiaHowAccurateAre2024} as well as experimental measurements for the 23 organic molecular crystals in the X23 dataset. The full dataset is split into crystals held primarily by dispersion, hydrogen-bonding, or a mixture in the top, bottom and middle panels, respectively. The experimental sublimation enthalpies were converted to lattice energies using corrections computed with DFT-trained machine-learning interatomic potentials in Ref.~\citenum{dellapiaAccurateEfficientMachine2025} and expressed as a range based on the set of previous measurements (see Supplementary Section~\ref{si-sec:x23_exp_analysis}).}
\end{figure}

\subsection{Agreement with experiments across diverse datasets}

We first assess LNO-MBE-CCSD(T) at predicting $E_\text{latt}$ for the X23 dataset of small organic molecular crystals.
This dataset has been a fundamental benchmark for new methods, whether semi-empirical~\cite{mortazaviStructureStabilityMolecular2018,thomasAccurateLatticeEnergies2018}, DFT~\cite{a.priceXDMcorrectedHybridDFT2023,hojaMultimerEmbeddingApproach2024} or machine-learning~\cite{kovacsMACEOFFShortRangeTransferable2025,dellapiaAccurateEfficientMachine2025} based.
As shown in Figure~\ref{fig:x23_comparison}, X23 consists of 23 systems spanning a wide range of lattice energies ($-25$ to $-170\,$kJ/mol) from weak dispersion to strong hydrogen-bonding.
They are well-characterized by experimental~\cite{otero-de-la-rozaBenchmarkNoncovalentInteractions2012,reillyUnderstandingRoleVibrations2013b,dolgonosRevisedValuesX232019} references, with LNO-MBE-CCSD(T) yielding mean absolute deviations (MADs) of $3.4\,$kJ/mol.
As experimental uncertainties exceed $4\,$kJ/mol for 21 systems in the X23 dataset~\cite{dellapiaHowAccurateAre2024} (see Supplementary Section~\ref{si-sec:x23_exp_analysis}), we conclude that both LNO-MBE-CCSD(T) and recent DMC estimates~\cite{dellapiaHowAccurateAre2024} are accurate to within experimental uncertainty, with DMC also having a MAD of 3.4 kJ/mol relative to experiment.
On the other hand, the differences between DMC and LNO-MBE-CCSD(T) are below $2\,$kJ/mol for most systems (15 of 23), yielding a MAD of $3.1\,$kJ/mol relative to DMC.
Consistent with recent observations on dimers~\cite{al-hamdaniInteractionsLargeMolecules2021,shiSystematicDiscrepanciesReference2025}, CCSD(T) predicts slightly stronger binding for van der Waals systems and weaker binding for H-bonded systems.
Anthracene and naphthalene are notable exceptions, where the source of their discrepancies remains presently debated~\cite{al-hamdaniInteractionsLargeMolecules2021,fishmanAnotherAngleBenchmarking2025} and we show in Supplementary Section~\ref{si-sec:ct_effects} that higher-order correlation effects beyond CCSD(T)~\cite{schaferUnderstandingDiscrepanciesNoncovalent2025} may not be negligible.

We briefly highlight the capabilities enabled by our methodological developments introduced in the Methods.
The systems from the X23 dataset have been the focus of numerous MBE-based studies, summarized and compared in Supplementary Section~\ref{si-sec:x23_mbe_lit}.
The high-cost of going beyond DFT [to CCSD(T)] has typically limited these studies to only subsets of the X23.
In addition, multilevel approaches are often required, where CCSD(T) is applied only to two-body terms (sometimes to short cutoff distances) while using lower-level methods for higher-body terms.
Examples include the random phase approximation (RPA)~\cite{sytyMultiLevelCoupledClusterDescription2025}, second-order M{\o}ller-Plesset perturbation theory (MP2)~\cite{beranPredictingOrganicCrystal2010,xieAssessmentThreebodyDispersion2023a}, HF~\cite{hermannGroundStatePropertiesCrystalline2008}, DFT~\cite{boeseEmbeddedDFTCalculations2017a} or force fields~\cite{bukowskiPredictionsPropertiesWater2007,wenAccurateMolecularCrystal2011a}. 
These neglected terms can introduce uncontrolled errors which can be significant~\cite{xieAssessmentThreebodyDispersion2023a,nelsonConvergenceManybodyExpansion2024} (exceeding $4\,$kJ/mol on $E_\text{latt}$) in some cases.
Our optimizations to the LNO approximation within this work enable all dimers to be treated accurately with LNO-CCSD(T), while also allowing three-body terms to be computed directly at the same level of theory.
As a result, this study is the most comprehensive to date.
It is the only work to apply (LNO-)CCSD(T) up to the trimers for the entire X23 dataset, needed for an accurate CCSD(T)-quality $E_\text{latt}$ estimate,  while ensuring a cost competitive with DFT (discussed later).
In Supplementary Section~\ref{si-sec:x23_mbe_lit}, we have used LNO-MBE-CCSD(T) to benchmark the previous MBE estimates, showing that some multi-level approximations~\cite{sytyMultiLevelCoupledClusterDescription2025} can reach MADs as low as $2\,$kJ/mol.

\begin{figure}[p]
    \includegraphics[width=\textwidth]{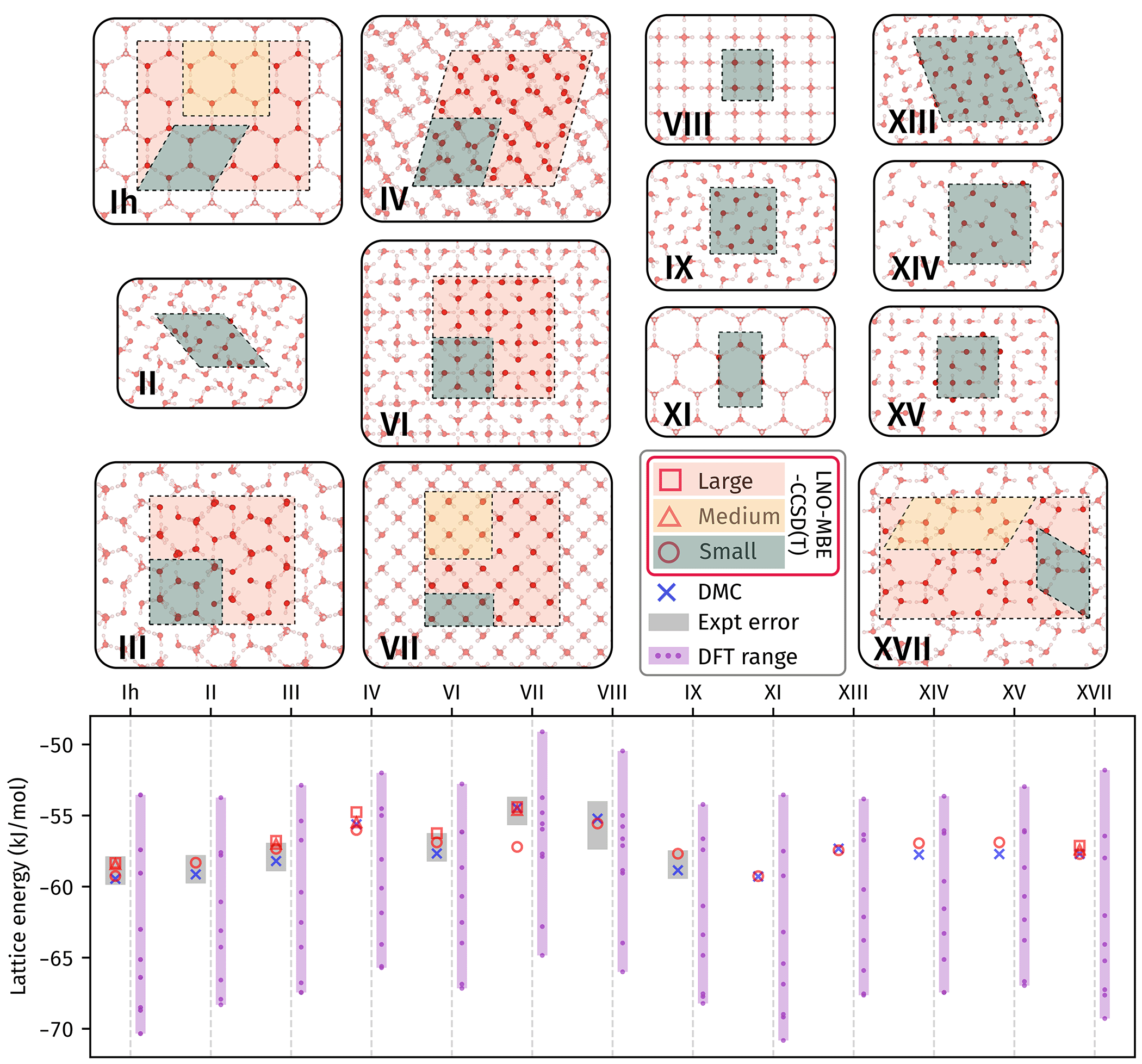}
    \caption{\label{fig:ice13_elatt}\textbf{Agreement across 13 polymorphs of ice.} Comparison of the lattice energy predicted by LNO-MBE-CCSD(T) against previous DMC estimates~\cite{dellapiaDMCICE13AmbientHigh2022b} and experiments~\cite{whalleyEnergiesPhasesIce1984} (when available) for the 13 ice polymorphs in the ICE13 dataset.
    The lattice energy for a range of widely-used density functional approximations [B3LYP-D4, revPBE0-D3, PBE0-MBD, SCAN+rVV10, optB86b-vdW, rev-vdW-DF2, vdW-DF, revPBE-D3, PBE-D3(BJ)] from Ref.~\citenum{dellapiaDMCICE13AmbientHigh2022b} is illustrated in purple.
    We extend beyond standard small ferroelectric ordered cells ($6{-}16$ water molecules) to increasingly hydrogen-disordered antiferroelectric cells of medium ($12{-}16$ molecules) and large (${>}80$ molecules) sizes.}
\end{figure}

We next examine 13 ice polymorphs (out of more than 20 polymorphs experimentally characterized to date) which comprise the ICE13 dataset~\cite{dellapiaDMCICE13AmbientHigh2022b} in Figure~\ref{fig:ice13_elatt}.
These systems are challenging due to the small relative energy differences (${<}5\,$kJ/mol) and the need for large unit cells to capture hydrogen disorder~\cite{bernalTheoryWaterIonic1933,paulingStructureEntropyIce1935} present in several anti-ferroelectric polymorphs, including Ih, III, VI, VII, and XVII~\cite{salzmannPolymorphismIceFive2011}.
LNO-MBE-CCSD(T) resolves these small energy differences, yielding an MAD of $0.5\,$kJ/mol on $E_\text{latt}$ (and $0.6\,$kJ/mol on $E_\text{rel}$; see Supplementary Section~\ref{si-sec:ice13_dataset}) relative to DMC for the ICE13 dataset, within experimental uncertainties where available~\cite{whalleyEnergiesPhasesIce1984}.
This agreement surpasses that of any DFA~\cite{dellapiaDMCICE13AmbientHigh2022b}, which can show deviations up to $10\,$kJ/mol.
For hydrogen-disordered polymorphs in Figure~\ref{fig:ice13_elatt}, we move beyond the small ferroelectric unit cells (typically ${\leq}12$ water molecules) used in most high-level studies~\cite{santraHydrogenBondsVan2011,macherRandomPhaseApproximation2014,dellapiaDMCICE13AmbientHigh2022b}, toward experimentally observed anti-ferroelectric, hydrogen-disordered structures of medium (${\sim}$16 molecules) and large (${>}$80 molecules) sizes.
The linear scaling of LNO-MBE-CCSD(T) enables efficient treatment of these larger systems while retaining high accuracy, owing to our optimizations that prevent error accumulation across the many three-body contributions.
The importance of modeling hydrogen disorder is evident for ice VII: the small unit cell yields poor agreement with experiment due to its large dipole moment (see Supplementary Section~\ref{si-sec:geom_details}), whereas the medium and large cells - whose dipole moments are near zero - restore agreement.

\subsection{Efficient benchmarks for lower-level theories}

\begin{figure}[h]
    \includegraphics[width=\textwidth]{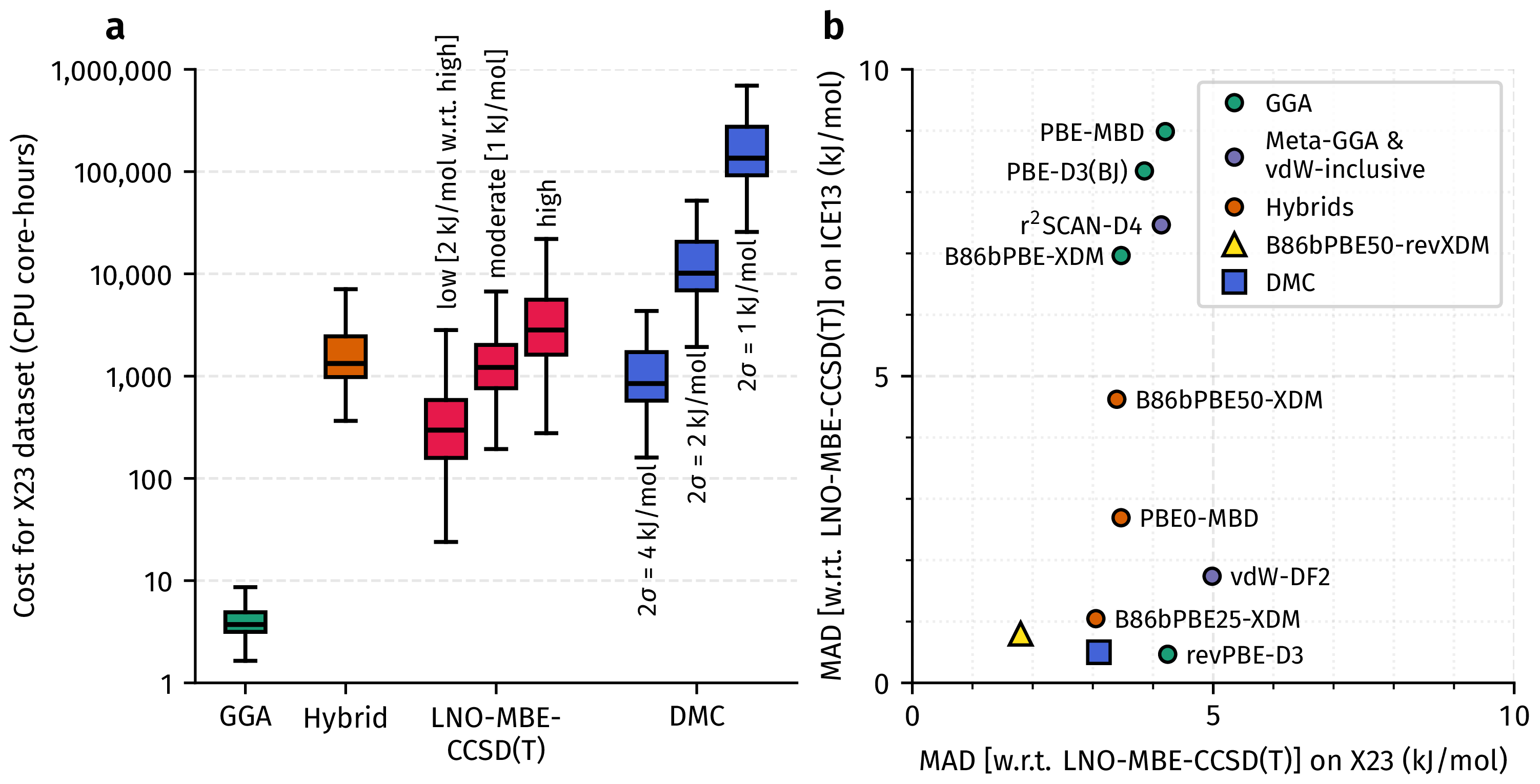}
    \caption{\label{fig:cost_accuracy_cc}\textbf{High accuracy benchmarks at affordable costs.} \textbf{a} Comparison of the cost of LNO-MBE-CCSD(T) against periodic DFT, for both the generalized gradient approximation (GGA) and hybrids, as well as DMC.
    We report LNO-MBE-CCSD(T) results using low, moderate, and high settings defined by different two- and three-body cutoff distances, where the low and moderate settings have a mean absolute deviation (MAD) of 1 and $2\,$kJ/mol to the high settings, respectively.
    DMC costs are also reported for different levels of statistical sampling corresponding to 95\% confidence intervals (2$\sigma$) of 1, 2, and $4\,$kJ/mol.
    The DMC costs neglect the costs to generate the trial wave-function and optimize the Jastrows while the CCSD(T) cost is only for the correlation energy.
    \textbf{b} We have computed the mean absolute deviation (MAD) to our LNO-MBE-CCSD(T) estimates for a wide set of density functional approximations to benchmark their performance on the ICE13 and X23 datasets.
    These benchmarks have been further used to reparameterize the XDM coefficients of the B86bPBE50 functional, yielding B86bPBE50-revXDM with unprecedented accuracy across both X23 and ICE13.}
\end{figure}

We now assess the computational cost of LNO-MBE-CCSD(T) on the X23 dataset.
Developing methods that improve the accuracy-cost Pareto front is essential for high-throughput applications~\cite{wengertDataefficientMachineLearning2021,firahaPredictingCrystalForm2023a,hunnisettSeventhBlindTest2024} such as materials screening and training machine-learning interatomic potentials. 
Figure~\ref{fig:cost_accuracy_cc}a compares the cost of LNO-MBE-CCSD(T) with periodic DFT using GGAs and hybrid DFAs (i.e., functionals), as well as with DMC (tabulated in Supplementary Section 5).
Remarkably, LNO-MBE-CCSD(T) is comparable in cost to periodic hybrid DFT.
Specifically, using stringent (dubbed ``high'') distance cutoffs (to converge 2B and 3B terms to within $0.5\,$kJ/mol each) leads to costs that are on average $4{,}762\,$CPU core hours (CPUh) compared to $2{,}412$ for periodic hybrid DFT.
It is two orders of magnitude lower than DMC, where ${\sim}200{,}000\,$CPUh is needed on average to achieve a 95\% confidence interval of $1\,$kJ/mol on statistical uncertainties in the X23 dataset, reaching up to $700{,}000\,$CPUh for some systems.

As discussed in Supplementary Section~\ref{si-sec:comp_cost}, the cost of LNO-MBE-CCSD(T) can be lowered substantially by relaxing 2B and 3B distance cutoffs.
We have defined ``moderate'' and ``low'' settings that yield MADs of 1.0 and $1.8\,$kJ/mol relative to the high setting, with average costs of 1,906 and $502\,$CPUh, respectively.
Moreover, our method runs efficiently (and if needed massively parallel) on commodity hardware, requiring as little as $46\,$GB of memory and 8 CPU cores per calculation - significantly less than the demands of a recent periodic MP2 study~\cite{liangCanSpinComponentScaled2023} on the X23 dataset.
We have not included the cost of periodic HF calculations within our analysis of CPU times as the HF part is greatly accelerated via new GPU implementations, discussed in Supplementary Section~\ref{si-sec:x23_hf}; this can also be used to accelerate periodic hybrid DFT.

The extensive reference data (from X23 and ICE13) computed within this work provide a reliable benchmark for lower-cost methods, like DFAs or MLIP foundation models~\cite{yuanFoundationModelsAtomistic2026}, free from thermal, quantum nuclear and geometric errors.
On the other hand, using experiments as benchmarks imposes fundamental limits on precision of the obtained MAD due to the wide range of values~\cite{dellapiaHowAccurateAre2024}.
For example, experimental uncertainties exceed $4\,$kJ/mol for 21 systems within the X23 dataset (see Supplementary Section~\ref{si-sec:x23_exp_analysis}).
Figure~\ref{fig:cost_accuracy_cc}b compares commonly used DFAs against our CCSD(T) references, ranging from PBE with various flavors of dispersion corrections to the B86bPBE-based GGA and hybrid functionals, along with trusted DFAs such as r$^2$SCAN-D4 and PBE0-MBD.
As in earlier studies~\cite{a.priceXDMcorrectedHybridDFT2023}, balancing performance across X23 and ICE13 is difficult for DFT.
For example, while revPBE-D3 performs best for ICE13 (MAD $=0.5\,$kJ/mol), its performance is worse for X23 ($4.2\,$kJ/mol).
The good overall performance of PBE0-MBD~\cite{tkatchenkoAccurateEfficientMethod2012} and B86bPBE25-XDM~\cite{a.priceXDMcorrectedHybridDFT2023} on both sets is also notable.

Until now, the lack of condensed phase references have meant that dispersion corrections are parametrized only to binding energies of gas-phase dimers~\cite{rezacBenchmarkCalculationsInteraction2016a} or trimers.
Our LNO-MBE-CCSD(T) references here allow for optimization towards molecular crystal $E_\text{latt}$ and in Supplementary Section~\ref{si-sec:opt_xdm}, we have revised the XDM coefficients for B86bPBE50 to our X23 benchmarks.
As shown in Figure~\ref{fig:cost_accuracy_cc}, this new DFA, dubbed B86bPBE50-revXDM, reduces the MAD for X23 from $3.4$ to $1.8\,$kJ/mol and on the out-of-training ICE13 set from $1.8$ to $0.8\,$kJ/mol.
Its high fraction (50\%) of exact-exchange makes it more robust~\cite{broderickDelocalizationErrorPoisons2024} towards treating delocalization error-prone systems, which we will discuss for ROY and axitinib in Figure~\ref{fig:big_pharma_crystals}.
Furthermore, we have optimized these coefficients for the small, cost-effective (`lightdense') basis sets, allowing for the X23 dataset to be computed by this hybrid DFT within $10\,$CPUh on average - comparable to a GGA.
These qualities of B86bPBE50-revXDM - enabled through optimization against the LNO-MBE-CCSD(T) references - makes the functional a robust and effective tool, e.g., for initial screening in high-throughput searches.
In Supplementary Section~\ref{si-sec:mlip_thermal}, we show that B86bPBE50-revXDM can be integrated into recent MLIP frameworks~\cite{dellapiaAccurateEfficientMachine2025}, requiring ${\sim}900$ small-cell geometries (under $40{,}000\,$CPUh) to train robust models for axitinib and ROY.
These models capture the full thermodynamics of the polymorph pairs~\cite{hojaReliablePracticalComputational2019}, including anharmonic and nuclear quantum effects via path-integral molecular dynamics.
As a result, these contributions can be removed from experimental relative enthalpies (as we have done for ROY and axitinib in Figure~\ref{fig:big_pharma_crystals}), enabling direct comparison with $0\,$K estimates.

\subsection{Resolving complex and challenging systems}

\begin{figure}[h]
    \includegraphics[width=\textwidth]{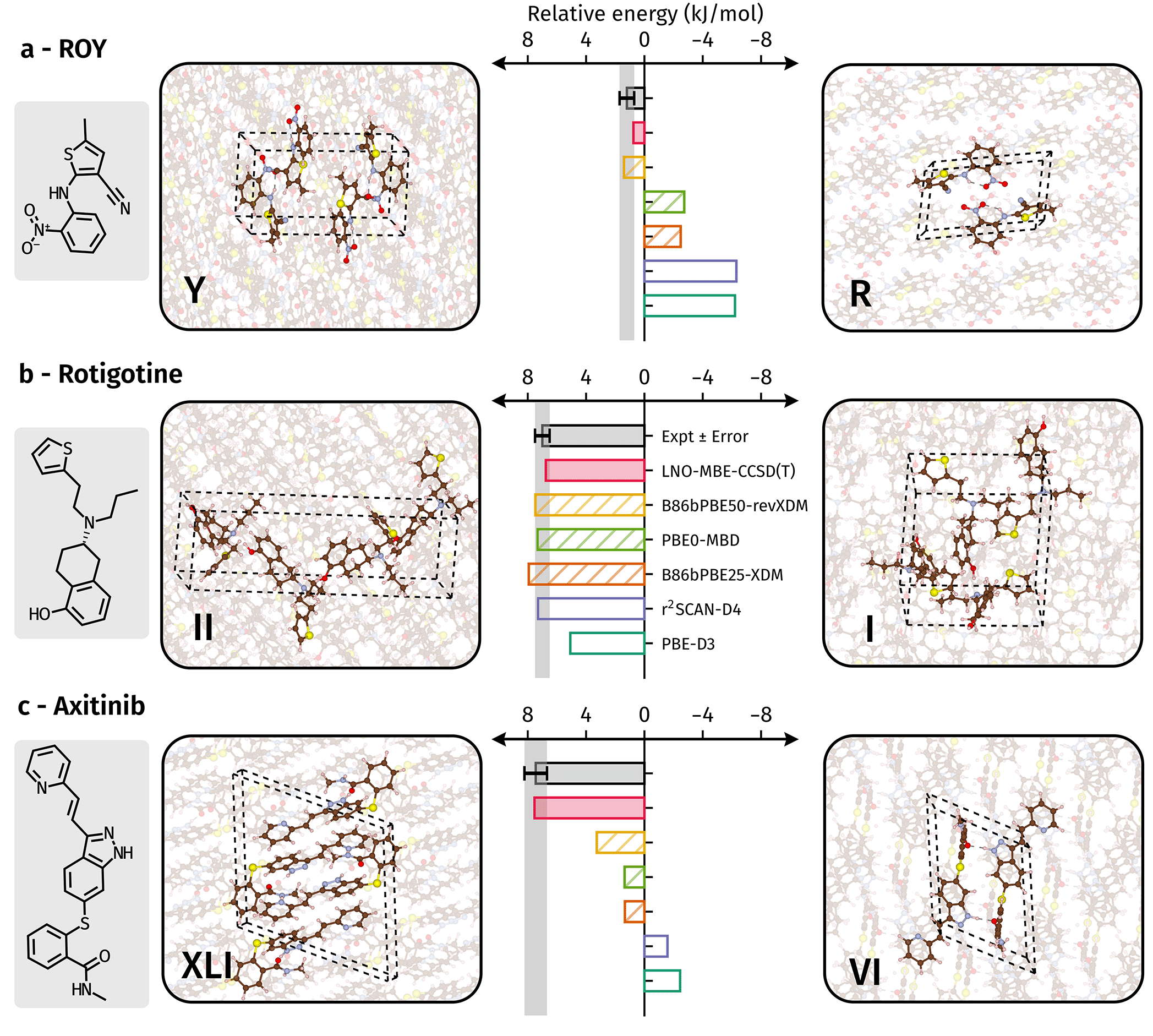}
    \caption{\label{fig:big_pharma_crystals}\textbf{Sub-chemical accuracy on pharmaceutical-scale polymorph pairs.} Comparison of the relative energy predicted by LNO-MBE-CCSD(T) against experiments (where the gray bar indicates uncertainties) for competing polymorphs of \textbf{a} ROY~\cite{yuPolymorphismMolecularSolids2010b} (Y and R), \textbf{b} rotigotine~\cite{mortazaviComputationalPolymorphScreening2019} (II and I) and \textbf{c} axitinib~\cite{campetaDevelopmentTargetedPolymorph2010a} (XLI and VI).
    The experimentally observed form is given to the left, with a negative relative energy favoring its formation while a positive value favoring the (incorrect) metastable form.
    We also report the predictions for a set of commonly-employed density functional approximations and also the B86PBE50-revXDM parametrized in this work.}
\end{figure}

We now show that the LNO-MBE-CCSD(T) framework can be used to predict $E_\text{rel}$ for large drug-like molecules.
Compared to $E_\text{latt}$, converging $E_\text{rel}$ requires looser LNO-MBE-CCSD(T) parameters (demonstrated in Supplementary Section~\ref{si-sec:rel_ene_details}) - including smaller basis sets, looser local thresholds, and shorter distance cutoffs in the MBE - enabling further cost reductions.
The specific systems examined include competing polymorphs pairs of ROY~\cite{beranHowManyMore2022} (R and Y) as well as axitinib~\cite{vasileiadisPredictionCrystalStructures2015,beranPolymorphsCocrystalsSalts2025} (VI and XLI) and rotigotine~\cite{mortazaviComputationalPolymorphScreening2019} (I and II).
While ROY represents a moderate-sized molecule with 27 atoms, axitinib and rotigotine represent large drug-like molecules with 46 and 47 atoms, respectively, and are currently marketed drugs used for treating cancer and Parkinson's disease, respectively.
In Figure~\ref{fig:big_pharma_crystals}, we compare LNO-MBE-CCSD(T) $E_\text{rel}$ predictions against experimental measurements (with thermal and vibrational contributions removed, as described in Supplementary Section~\ref{si-sec:mlip_thermal}).
The values of $E_\text{rel}$ range from $1.2\,$kJ/mol for the ROY polymorph pair up to ${\sim}8\,$kJ/mol for the axitinib and rotigotine polymorph pairs.
In all cases, LNO-MBE-CCSD(T) reproduces these differences to within experimental uncertainty (all less than $1\,$kJ/mol).

The level of agreement on $E_\text{rel}$ achieved with LNO-MBE-CCSD(T) is not trivial, with the polymorphs of ROY and axitinib representing long-standing challenges for DFT~\cite{greenwellInaccurateConformationalEnergies2020}.
Both systems feature conformations with varying $\pi$ conjugation, with DFT tending to overstabilize planar conformations due to delocalization error~\cite{beranInterplayIntraIntermolecular2022a,bryentonDelocalizationErrorGreatest2023}.
Figure~\ref{fig:big_pharma_crystals} illustrates this challenge across widely used functionals/DFAs for the competing polymorph pairs of these two systems.
We include representative DFAs from the generalized gradient approximation (GGA), given by PBE-D3(BJ) and r$^2$SCAN-D4, to hybrids (B86bPBE25-XDM and PBE0-MBD), with a more comprehensive list tabulated in Supplementary Section~\ref{si-sec:pharma_dfa}.
The GGAs predict incorrect polymorphs for both systems, with errors exceeding $7\,$kJ/mol on the $E_\text{rel}$, and while hybrids reduce these errors, they still deviate by more than $3\,$kJ/mol from experimental and LNO-MBE-CCSD(T) values.
Here, the higher exact exchange in B86bPBE50-revXDM allow for the correct polymorph to be predicted in all cases, albeit with remaining quantitative errors on $E_\text{rel}$.
Similarly, quantitative agreement to experiment cannot be achieved by lower-level cWFT methods such as MP2 and CCSD (see Supplementary Sections~\ref{si-sec:x23_mp2_ccsd},~\ref{si-sec:ice13_final_latt}, and~\ref{si-sec:big_pharma_final_rel}) nor more approximate MBE-based methods which only correct up to CCSD(T) for one-body contributions~\cite{greenwellInaccurateConformationalEnergies2020,beranPolymorphsCocrystalsSalts2025}.

\section{Conclusion}

To conclude, we have introduced the LNO-MBE-CCSD(T) framework within this work, delivering sub-chemical accuracy on relative energies for molecular crystals at costs competitive with periodic hybrid DFT and two orders of magnitude lower than other high-level methods.
For 23 organic molecular crystals and 13 ice polymorphs, comprising the X23 and ICE13 datasets, respectively, we reproduce lattice energies to within the uncertainties of experiments.
Its low cost significantly expands the scope and complexity of systems that can now be tackled with high accuracy, enabling sub-chemical-accuracy predictions on the relative energies of drug-like molecules such as ROY (forms R and Y), axitinib (forms VI and XLI), and rotigotine (forms I and II).
The first two systems challenge existing density functional approximations~\cite{beranHowManyMore2022,greenwellInaccurateConformationalEnergies2020}, with agreement to experimental relative energies achieved only by LNO-MBE-CCSD(T).

Our ability to address technologically relevant molecular crystals with the LNO-MBE-CCSD(T) framework builds on a combination of advances developed over the past decade, both within our groups and by others.
This represents a natural progression from a landmark study~\cite{yangInitioDeterminationCrystalline2014a} in 2014, which achieved sub-chemical accuracy for the lattice energy of benzene using advances from the preceding decade.
Since then, efficient linear-scaling local approximations~\cite{nagyApproachingBasisSet2019,riplingerSparseMapsSystematic2016,jiangAccurateEfficientOpensource2024a} have significantly expanded the range of accessible systems.
However, their application within the MBE has been limited by the accumulation of small errors across many-body terms, an issue we address here.
We also significantly decrease the large number of MBE calculations by combining a recently developed formulation exploiting translational-symmetry~\cite{hermanFormulationManyBodyExpansion2023b} with machine-learning descriptors~\cite{borcaCrystaLattEAutomatedComputation2019b,himanenDScribeLibraryDescriptors2020}.
Finally, we reduce MBE distance cutoffs, leading to a further reduction of calculations, through subtractive embedding with modern dispersion corrections~\cite{caldeweyherGenerallyApplicableAtomiccharge2019} fitted to CCSD(T) data.

The relatively low computational cost of the LNO-MBE-CCSD(T) framework - requiring only tens of thousands of CPU core-hours for drug-like molecular crystals - makes it well suited for high-throughput workflows such as crystal structure prediction (CSP) and computational benchmarking~\cite{zhouRobustCrystalStructure2025,priceOneSizeFits2025}.
The systems studied here, including rotigotine and axitinib, are comparable to those from the latest CSP Blind Test~\cite{hunnisettSeventhBlindTest2024}, where some DFT-based protocols required millions of CPU hours to screen thousands of polymorphs.
We anticipate that LNO-MBE-CCSD(T) can be an economical choice for studying a final shortlist of candidate structures.
Moreover, the B86bPBE50-revXDM functional parametrized from our LNO-MBE-CCSD(T) data can be used to replace existing functionals within the initial stages to filter generated polymorphs~\cite{habgoodEfficientHandlingMolecular2015,nikharReliableCrystalStructure2022,galanakisRapidPredictionMolecular2024}.
It achieves unprecedented accuracy on molecular crystals for a DFA, while reliably predicting the qualitative ordering of difficult systems such as ROY and axitinib.
Together, we anticipate that these advances will enable reliable and practical identification of late-emerging stable polymorphs, as exemplified by rotigotine and axitinib.
In the case of rotigotine, a new polymorph discovered decades after clinical use led to mass recalls with major economic and public-health consequences~\cite{rietveldRotigotineUnexpectedPolymorphism2015}, while a late-emerging polymorph similarly delayed the development of axitinib~\cite{campetaDevelopmentTargetedPolymorph2010a}.

To support its broad use in molecular crystal modeling, the LNO-MBE-CCSD(T) framework will be released as an open-source GitHub package upon publication.
It will include worked examples from this study, enabling immediate adoption by users.
More broadly, these tools are expected to facilitate applications beyond organic molecular crystals to a wide range of systems in computational chemistry and materials science, including inorganic materials.
With routine CCSD(T) static calculations becoming feasible with LNO-MBE-CCSD(T), the next frontier is to develop CCSD(T)-level machine-learning interatomic potentials that capture finite-temperature~\cite{olehnovicsAccurateLatticeFree2025} and anharmonic effects, to understand thermodynamic stability~\cite{priceZerothOrderCrystal2018,cervinkaInitioPredictionPolymorph2018,correaAssessingPolymorphStability2025} as a function of pressure and temperature.

\section{Acknowledgements}

The Flatiron Institute is a division of the Simons Foundation.
A.M. and B.X.S. acknowledges support from the European Union under the ``n-AQUA'' European Research Council project (Grant No.\ 101071937).
P.R.N. acknowledges support from ERC Starting Grant No. 101076972, ``aCCuracy'', the  National Research, Development, and Innovation Office (NKFIH, Grant No. FK142489), and the J\'anos Bolyai Research Scholarship of the Hungarian Academy of Sciences.
A.Z. acknowledges support from the European Union under the Next Generation EU (Project Nos. 20222FXZ33 and P2022MC742).
K.M.H. and S.S.X. acknowledge support from the Center for Scalable Predictive Methods for Excitations and Correlated Phenomena (SPEC), which is funded by the U.S. Department of Energy (DOE), Office of Science, Basic Energy Sciences (BES), Chemical Sciences, Geosciences, and Biosciences Division (CSGB), as part of the Computational Chemical Sciences (CCS) program under FWP 70942 at Pacific Northwest National Laboratory (PNNL), a multi-program national laboratory operated for DOE by Battelle. 
The authors are grateful for resources provided by the Cirrus UK National Tier-2 HPC Service at EPCC (\href{http://www.cirrus.ac.uk}{http://www.cirrus.ac.uk}) funded by the University of Edinburgh and EPSRC (EP/P020267/1) and computational support from the UK national high performance computing service, ARCHER 2, as obtained via the UKCP consortium and funded by EPSRC grant ref EP/P022561/1
We also acknowledge the EuroHPC Joint Undertaking for awarding this project access to the EuroHPC supercomputer LEONARDO, hosted by CINECA (Italy) and the LEONARDO consortium through an EuroHPC Regular Access call.
This research also used resources of the National Energy Research Scientific Computing Center, which is supported by the Office of Science of the U.S. Department of Energy under Contract No. DE-AC02-05CH11231.

\section{Competing Interests Statement}
The authors declare no competing interests.

\section{Methods}

\subsection{The divide-and-conquer approach}

We employ a divide-and-conquer approach to reach a CCSD(T)-level of accuracy in an efficient manner.
As a post-Hartree-Fock (HF) method, CCSD(T) builds on the HF approximation to the Schr\"{o}dinger equation by incorporating electron-electron interactions neglected in the mean-field treatment of HF.
The resulting total CCSD(T) energy consists of the HF and (electron) correlation energy: the former (qualitatively) captures electrostatic, exchange, and induction effects, while the latter accounts for van der Waals dispersion, charge transfer and other correlation effects.
It is known that HF leads to slow and erratic convergence with cutoff distance in the MBE~\cite{bygraveEmbeddedManybodyExpansion2012,hofierkaBindingEnergiesMolecular2021} (see Supplementary Section~\ref{si-sec:ice_conv}), necessitating higher body orders (up to five molecules).
In contrast, the correlation energy converges smoothly within the MBE, requiring only up to three-body terms and shorter cutoff distances, even for the challenging high-pressure ice VIII phase.
Given these considerations, we treat HF under periodic boundary conditions (PBC) while only the correlation energy is treated with the MBE.
This division further allows us to leverage methodological advances from the past decade to enhance these methods.

In (periodic) solid-state electronic structure modeling, major progress has been achieved in reducing the costs of calculations with hybrid DFAs.
Here, the HF-like exchange term dominates, where most development has been focused on.
Specifically, these developments include GPU implementations~\cite{haceneAcceleratingVASPElectronic2012,giannozziQuantumESPRESSOExascale2020a} and efficient sparsity-based~\cite{kokottEfficientAllelectronHybrid2024} or low-rank decomposition methods such as the adaptively compressed exchange (ACE) operator~\cite{linAdaptivelyCompressedExchange2016}, included in multiple popular solid state DFT packages~\cite{giannozziQuantumESPRESSOExascale2020a}.
These advances can be readily utilized in periodic HF calculations, enabling accurate computation of large molecular crystals within tens of GPU hours, as shown in Supplementary Section~\ref{si-sec:x23_hf}.

For CCSD(T), local correlation approximations~\cite{riplingerSparseMapsSystematic2016,niemeyerSubsystemQuantumChemistry2023,jiangAccurateEfficientOpensource2024a,maExplicitlyCorrelatedLocal2018,nagyOptimizationLinearScalingLocal2018,nagyApproachingBasisSet2019} pioneered within the molecular quantum chemistry community have greatly expanded the size of molecules that are now accessible~\cite{nagyStateoftheartLocalCorrelation2024}.
For example, the local natural orbital (LNO)\cite{nagyOptimizationLinearScalingLocal2018,nagyApproachingBasisSet2019} approximation used in this work has been used to evaluate the binding energy of 
a protein-ligand complex of 1023 atoms~\cite{nagyApproachingBasisSet2019,nagyStateoftheartLocalCorrelation2024}
However, the application of local CCSD(T) to the MBE was limited previously by the accumulation of individually 
small errors due to the large number of three-body terms.
As described in Supplementary Section~\ref{si-sec:opt_lno}, we have optimized the LNO approximation for the MBE, eliminating this error accumulation while enabling a more affordable overall treatment.

There have also been new advances to the efficiency of the MBE.
Symmetrically-equivalent body terms can now be identified using new formulations exploiting crystal translational symmetry~\cite{hermanFormulationManyBodyExpansion2023b}.
In addition, recently-developed machine-learning descriptors~\cite{borcaCrystaLattEAutomatedComputation2019b} allow for a further reduction. 
We also introduce a subtractive embedding of the correlation energy with Grimme's D4 dispersion~\cite{caldeweyherGenerallyApplicableAtomiccharge2019}, a negligible-cost addition that reduces cutoff distances.
Together, these developments lower the number of required CCSD(T) calculations by orders of magnitude compared to a full enumeration and significantly expand the scope of molecular crystals that can be tackled by CCSD(T) with the MBE.

We combine all these developments within several packages that will be released upon publication.
Our MBE implementation - combining translation symmetry with machine-learning descriptors - is provided as a separate standalone Python dependency, pMBE.
It generates and manages the two- and three-body (or beyond) terms, storing them in a database file.
The LNO-MBE-CCSD(T) framework interfaces with pMBE to generate an efficient set of MBE fragments, while using the QuAcc workflow library~\cite{rosenQuaccQuantumAccelerator2024} to manage the full set of calculations to compute their final contributions to the lattice energy.
Our optimizations to the LNO-CCSD(T) approximation is available in the latest 2025 version of {\sc Mrcc}~\cite{kallayMRCCProgramSystem2020,mesterOverviewDevelopmentsMRCC2025}, with the necessary options needed for the MBE turned on within the LNO-MBE-CCSD(T) package by default.
The use of the QuAcc workflow library also means that in principle, it is modular and can be interfaced with other
(local) CCSD(T) implementations~\cite{neeseORCAQuantumChemistry2020,jiangAccurateEfficientOpensource2024a}  as well as 
MBE codes~\cite{borcaCrystaLattEAutomatedComputation2019b,burnsQCManyBodyFlexibleImplementation2024,hojaMultimerEmbeddingApproach2024,sytyMultiLevelCoupledClusterDescription2025}.
In addition, the inputs to perform B86bPBE50-revXDM are provided and we have also made it available as a custom calculator for the atomic simulation environment (ASE).

\subsection{Coupled cluster theory}

All CCSD(T) calculations were performed in {\sc Mrcc}~\cite{kallayMRCCProgramSystem2020,mesterOverviewDevelopmentsMRCC2025}, accelerated by the local natural orbital (LNO)~\cite{nagyOptimizationLinearscalingLocal2017,nagyOptimizationLinearScalingLocal2018,nagyApproachingBasisSet2019,nagyStateoftheartLocalCorrelation2024} approximation.
The aug-cc-pV$X$Z basis set family~\cite{petersonAccurateCorrelationConsistent2002} was used to compute $E_\text{latt}$ while the jul-cc-pV$X$Z basis set family was used for $E_\text{rel}$.
The complete basis set (CBS) limit was reached using a two point extrapolation~\cite{neeseRevisitingAtomicNatural2011} of the triple (TZ) and quadruple-zeta (QZ) pair for the two-body contributions, while the double-zeta (DZ) and TZ pair was used for three-body contributions.
We use a composite scheme whereby CCSD(T) was calculated with the smaller basis set in the CBS pair, and a correction to the CBS limit was performed with (canonical) MP2.
For $E_\text{latt}$ calculations, we performed a two-point extrapolation to the local approximation free (i.e., conventional) CCSD(T)
limit from {\tt Tight} and {\tt vTight} LNO settings, while we used {\tt Tight} LNO settings only for $E_\text{rel}$.
Counterpoise corrections were employed when calculating all terms.

\subsection{Many-body expansion}

The many-body expansion fragmentation was performed using the periodic implementation of Herman and Xantheas~\cite{hermanFormulationManyBodyExpansion2023b}.
We truncated the MBE to the three-body level, with distance cutoffs (as well as convergence plots) described in Supplementary Section~\ref{si-sec:mbe_details}.
The Coulomb Matrix (CM) global descriptor, computed with DScribe~\cite{himanenDScribeLibraryDescriptors2020,laaksoUpdatesDScribeLibrary2023}, was further used to identify redundant terms.
As described in Supplementary Section~\ref{si-sec:coulomb_matrix}, we have made additional modifications to enable permutational and rotational symmetry of the CM descriptor, together with a regularizing term based on the distance cutoff.

\subsection{Density functional theory}
The periodic HF calculations were performed in the GPU implementation of the Vienna \textit{ab-initio} simulation package~\cite{kresseInitioMolecularDynamics1993a,kresseEfficiencyAbinitioTotal1996b,kresseEfficientIterativeSchemes1996c} (VASP) version 6.4.2.
We used the hard (denoted by \texttt{X\_h}, where \texttt{X} is the element) PBE projector augmented wave (PAW) potentials and an energy cutoff of $1000\,$eV.
A $\Gamma$-centered $k$-point mesh was used with the $k$-point density (\texttt{KSPACING}) of $0.2\,$\AA{}\textsuperscript{-1}.
We used the truncated Coulomb potential (\texttt{HFRCUT=-1}) and the ACE~\cite{linAdaptivelyCompressedExchange2016} operator.
For the DFT benchmarks in Figure~\ref{fig:cost_accuracy_cc}, the calculations for PBE-D3(BJ), revPBE-D3(0), r$^2$SCAN-D4, optB86b-vdW and vdW-DF2 were performed in VASP, while some of the DFT functionals (B86bPBE-XDM, B86bPBE-XDM-25, B86bPBE-XDM-50, PBE0-MBD, PBE+TS) were performed in the Fritz Haber Institute \textit{ab initio} molecular simulations (FHI-aims) package~\cite{blumInitioMolecularSimulations2009}.
In the latter, we aim to reach the `tight' basis set level, using a composite process described in Supplementary Section~\ref{si-sec:periodic_dft_settings} for hybrid functionals.
The CPU costs for both the GGA and hybrid DFT in Figure~\ref{fig:cost_accuracy_cc}a were performed in Quantum Espresso~\cite{giannozziQUANTUMESPRESSOModular2009a} as described in Supplementary Section~\ref{si-sec:periodic_dft_settings}.

\section{Data Availability}

See the supplementary information for a detailed compilation of the obtained results as well as further data and analysis to support the points made throughout the text. The input and output files associated with this work and all analysis will be made available on Github and Zeonodo upon publication.

\section{Code Availability}
The LNO-MBE-CCSD(T) framework and the standalone pMBE software will be made freely available on Github upon publication, with documentation containing instructions and examples on how to run the two codes.

\bibliography{references.bib}

\end{document}


\author{Benjamin X. Shi}
\affiliation{Initiative for Computational Catalysis, Flatiron Institute, 160 5th Avenue, New York, NY 10010}

\author{Kristina M. Herman}
\affiliation{Department of Chemistry, University of Washington, Seattle, Washington 98195, United States}

\author{Flaviano Della Pia}
\affiliation{Yusuf Hamied Department of Chemistry, University of Cambridge, Lensfield Road, Cambridge CB2 1EW, United Kingdom}

\author{Venkat Kapil}
\affiliation{Department of Physics and Astronomy, University College London, 7-19 Gordon St, London WC1H 0AH, UK}
\affiliation{Thomas Young Centre and London Centre for Nanotechnology, 9 Gordon St, London WC1H 0AH}

\author{Andrea Zen}
\affiliation{Dipartimento di Fisica Ettore Pancini, Universit\`{a} di Napoli Federico II, Monte S. Angelo, I-80126 Napoli, Italy}
\affiliation{Department of Earth Sciences, University College London, Gower Street, London WC1E 6BT, United Kingdom}

\author{P\'eter R. Nagy}
\affiliation{Department of Physical Chemistry and Materials Science, Faculty of Chemical Technology and Biotechnology, Budapest University of Technology and Economics, M\H uegyetem rkp. 3., H-1111 Budapest, Hungary}
\affiliation{HUN-REN-BME Quantum Chemistry Research Group, M\H uegyetem rkp. 3., H-1111 Budapest, Hungary}
\affiliation{MTA-BME Lend\"ulet Quantum Chemistry Research Group, M\H uegyetem rkp. 3., H-1111 Budapest, Hungary}

\author{Sotiris Xantheas}
\email[Electronic address: ]{sotiris.xantheas@pnnl.gov}
\affiliation{Department of Chemistry, University of Washington, Seattle, Washington 98195, United States}
\affiliation{Advanced Computing, Mathematics and Data Division, Pacific Northwest National Laboratory, 902 Battelle Boulevard, MS J7-10, Richland, Washington 99352, United States}
\affiliation{Computational and Theoretical Chemistry Institute (CTCI), Pacific Northwest National Laboratory, Richland, Washington 99352, United States}

\author{Angelos Michaelides}
\email[Electronic address: ]{am452@cam.ac.uk}
\affiliation{Yusuf Hamied Department of Chemistry, University of Cambridge, Lensfield Road, Cambridge CB2 1EW, United Kingdom}%

\title{Supplementary information: Efficient first-principles predictions of complex molecular crystals at sub-chemical accuracy}

\date{\today}

\maketitle
\tableofcontents

\newpage 
\section{Computational details}
\subsection{\label{sec:geom_details}Geometries}

\begin{figure}[b]
    \includegraphics[width=\textwidth]{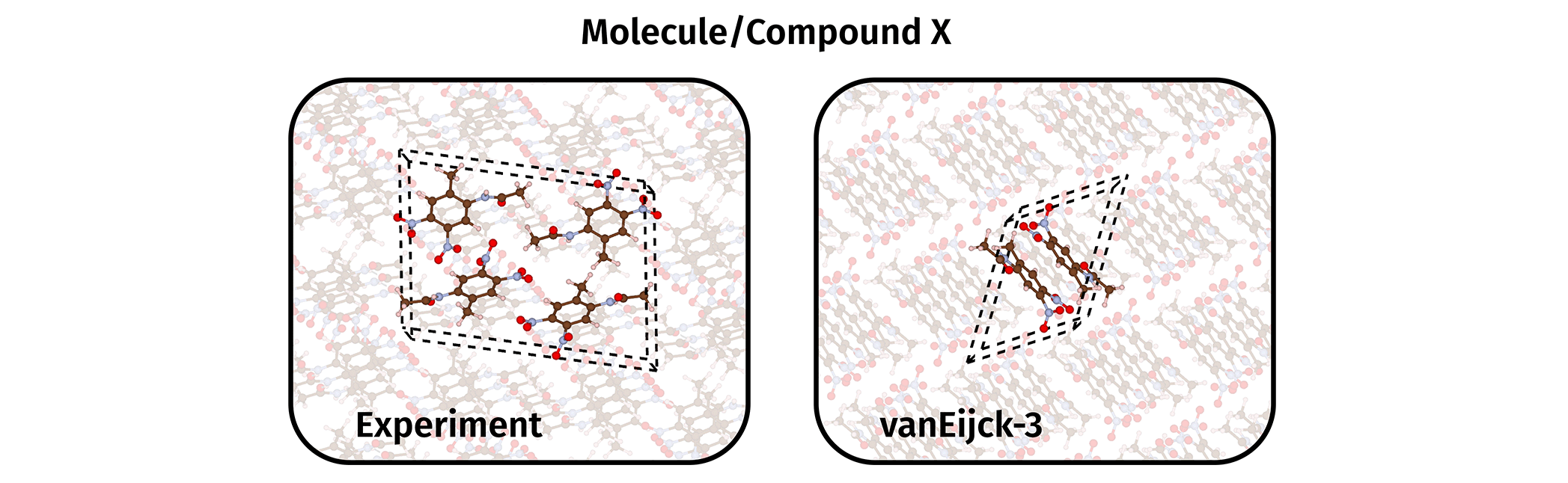}
    \caption{\label{fig:molecule_x_viz}Visualization of the vanEijck-3 and experimental forms of molecule X.}
\end{figure}

The geometries used for all of the molecular crystals studied in this work were taken from previous work.
For ease of reproduction, we have provided these geometries in our GitHub data repository (available upon publication).
The geometries for the X23 dataset were compiled by~\citet{dellapiaHowAccurateAre2024}, with most geometries taken from~\citet{klimesLatticeEnergiesMolecular2016} for the optB88-vdW~\cite{klimesChemicalAccuracyVan2010} density functional approximation (DFA) besides hexamine and succinic acid, which come from~\citet{reillyUnderstandingRoleVibrations2013b} optimised using the PBE+TS~\cite{perdewGeneralizedGradientApproximation1996d,tkatchenkoAccurateMolecularVan2009} DFA.
The ICE13 geometries in Figure~3 of the main text were also taken from~\citet{dellapiaDMCICE13AmbientHigh2022b} for the `small' unit cells.
The `medium' and `large' unit cells were generated with GenIce~\cite{matsumotoGenIceHydrogendisorderedIce2018,matsumotoNovelAlgorithmGenerate2021} then optimized within this work using the PBE~\cite{perdewGeneralizedGradientApproximation1996d} DFA, using the same input parameters as Ref.~\citenum{dellapiaDMCICE13AmbientHigh2022b}.
The large unit cells exhibit greater proton disorder, more closely approximating antiferroelectric unit cells with zero net dipole, as shown in Table~\ref{tab:ice13_dipole_moments}.
The geometries for forms I and II of rotigotine were taken from~\citet{mortazaviComputationalPolymorphScreening2019}.
Within experiments, this system has a disordered crystal structure, where its thiophene rings randomly takes on two alternative orientations that are related by a 180$^\circ$ rotation, as visualized in Supplementary Figure~2 of Ref.~\citenum{mortazaviComputationalPolymorphScreening2019}.
We have modeled this system using a 50\%:50\% average of the two respective configurations, meaning we compute the two configurations for each form, following the procedure of \citet{mortazaviComputationalPolymorphScreening2019}.
We have considered the experimental and vanEijck-3 forms of molecule X from the Third CSP Blind Test~\cite{dayThirdBlindTest2005}, with the geometries provided by Prof. Erin Johnson from Ref.~\citenum{whittletonExchangeHoleDipoleDispersion2017}.
This system was used for testing purposes and the only one not visualized in the main text, add we display it in Figure~\ref{fig:molecule_x_viz} here.
These geometries were optimized by the B86bPBE-XDM~\cite{beckeExchangeholeDipoleMoment2007,otero-de-la-rozaVanWaalsInteractions2012} DFA.
The geometries for forms VI and XLI of axitinib geometries were provided by Prof.\ Greg O.\ Beran from Ref.~\citenum{greenwellInaccurateConformationalEnergies2020}, optimized using the B86bPBE-XDM DFA.
The geometries for forms Y and R of ROY were taken from~\citet{beranHowManyMore2022}, optimized using the B86bPBE-XDM DFA.

\begin{table}[h]

\caption{\label{tab:ice13_dipole_moments}Estimated dipole moment per monomer in the unit cell (in Debye) for the 13 ice systems in the ICE13 dataset for the large, medium and small sizes. We used the TTM2.1-F~\cite{fanourgakisFlexiblePolarizableTholetype2006} polarizable water model to compute the dipole moments. We generated $3\times3\times3$ supercell in a box with large vacuum and computed the net diple of the inner unit cell.}
\begin{tabular}{lrrr}
\toprule
 & Large & Medium & Small \\ 
\midrule
Ih & 0.00 & 0.00 & 1.39 \\
II & 0.01 & 0.02 & 1.21 \\
III & 0.05 & 0.54 & 0.96 \\
IV & 0.02 & - & 0.00 \\
VI & 0.00 & 0.02 & 1.54 \\
VII & 0.00 & 0.02 & 0.80 \\
VIII & - & - & 0.00 \\
IX & - & - & 0.01 \\
XI & - & - & 0.00 \\
XIII & - & - & 1.46 \\
XIV & - & - & 0.00 \\
XV & - & - & 0.02 \\
XVII & - & - & 0.00 \\
\bottomrule
\end{tabular}

\end{table}

\subsection{\label{sec:periodic_dft_settings}Periodic HF and DFT}
\subsubsection{\label{sec:vasp_details}VASP}
The periodic HF calculations were all performed in the GPU implementation of the Vienna \textit{ab-initio} simulation package~\cite{kresseInitioMolecularDynamics1993a,kresseEfficiencyAbinitioTotal1996b,kresseEfficientIterativeSchemes1996c} (VASP) 6.4.2.
We used version 54 of the hard (denoted by \texttt{X\_h}, where \texttt{X} is the element) PBE projector augemented wave (PAW) potentials, together with an energy cutoff of $1000\,$eV.
A $\Gamma$-centered $k$-point mesh was used in all the systems, with the $k$-point density set (via the \texttt{KSPACING} parameter) to $0.2\,$\AA{}\textsuperscript{-1} and the truncated Coulomb potential (\texttt{HFRCUT=-1}).
We also utilized the Adaptively Compressed Exchange~\cite{linAdaptivelyCompressedExchange2016} (ACE) operator to accelerate the HF calculations, enabled by setting \texttt{ALGO=Normal}.
In Section~\ref{sec:dft_benchmark}, the benchmarks for the PBE-D3(BJ)~\cite{perdewGeneralizedGradientApproximation1996d,grimmeConsistentAccurateInitio2010}, revPBE-D3(0)~\cite{zhangCommentGeneralizedGradient1998,grimmeConsistentAccurateInitio2010}, r$^2$SCAN-D4~\cite{furnessAccurateNumericallyEfficient2020a,ningWorkhorseMinimallyEmpirical2022}, optB86b-vdW~\cite{klimesVanWaalsDensity2011} and vdW-DF2~\cite{hamadaVanWaalsDensity2014,leeHigheraccuracyVanWaals2010} DFAs were performed in VASP, utilizing the same parameters as the periodic HF calculations.
%

\subsubsection{\label{sec:fhiaims_parameters}FHI-AIMs}

The benchmarking for a subset of DFAs (B86bPBE-XDM, B86bPBE-XDM-25~\cite{a.priceXDMcorrectedHybridDFT2023}, B86bPBE-XDM-50~\cite{a.priceXDMcorrectedHybridDFT2023}, PBE0-MBD~\cite{adamoReliableDensityFunctional1999d,tkatchenkoAccurateEfficientMethod2012} and  PBE+TS~\cite{perdewGeneralizedGradientApproximation1996d,tkatchenkoAccurateMolecularVan2009} in Section~\ref{sec:dft_benchmark} was performed in the Fritz Haber Institute \textit{ab initio} molecular simulations (FHI-aims) package~\cite{blumInitioMolecularSimulations2009}.
We used slightly different procedures to reach the basis set limit depending on the type (i.e., cost) of the underlying DFA.
B86bPBE-XDM was performed with the \texttt{tight} basis sets, while we approximated the \texttt{tight} limit for B86bPBE-XDM-25 and B86bPBE-XDM-25 using the procedure of~\citet{a.priceXDMcorrectedHybridDFT2023}:
\begin{equation}
    E_{\text{hybrid}}^{\text{tight}} \approx E_{\text{hybrid}}^{\text{lightdense}} + \left( E_{\text{GGA}}^{\text{tight}} - E_{\text{GGA}}^{\text{lightdense}} \right),
\end{equation}
where the GGA is B86bPBE-XDM and the hybrid is either B86bPBE-XDM-25 or B86bPBE-XDM-50, which were performed with the \texttt{lightdense} basis sets.
We make use of a similar composite process for PBE0-MBD, which were also both performed with the \texttt{lightdense} basis set and subsequently corrected to the \texttt{tight} basis set using PBE-MBD and PBE-D3(BJ) DFAs, respectively.
PBE-TS was performed directly at the \texttt{tight} basis set level.
The DFT calculations were performed with scalar relativistic effects and a $\Gamma$-centred k-point density (using the parameter \texttt{k\_grid\_density} of 4, increased to 6 for MBD-based DFAs).

\subsubsection{Quantum Espresso}

We used Quantum Espresso~\cite{giannozziQUANTUMESPRESSOModular2009a} to benchmark the cost of hybrid DFT on the CPU with the ACE operator in Section~\ref{sec:comp_cost}.
We used the \texttt{Precision} Standard Solid State Pseudopotentials~\cite{prandiniPrecisionEfficiencySolidstate2018}, together with an energy cutoff of $80\,$Ry and charge density cutoff of $320\,$Ry.
We used the same $k-$point grids as used for the corresponding VASP calculations.

\subsection{\label{sec:lno_comp_details}Correlated wavefunction theory}

We leverage developments in the local natural orbital (LNO)~\cite{nagyOptimizationLinearscalingLocal2017,nagyOptimizationLinearScalingLocal2018,nagyApproachingBasisSet2019,nagyStateoftheartLocalCorrelation2024} approximation to coupled cluster theory with single, double, and perturbative triple excitations [CCSD(T)]~\cite{raghavachariFifthorderPerturbationComparison1989} in {\sc Mrcc}~\cite{kallayMRCCProgramSystem2020,mesterOverviewDevelopmentsMRCC2025,mrcceng3}.
In addition, we also performed density-fitted second-order M{\o}ller-Plesset perturbation theory (DF-MP2)~\cite{mollerNoteApproximationTreatment1934} to correct for local correlation and basis set errors as described in Section~\ref{sec:cheaper_settings}. 
%
%
LNO-CCSD(T) was performed with a patched 2023 version of the {\sc Mrcc}~\cite{kallayMRCCProgramSystem2020,mesterOverviewDevelopmentsMRCC2025,mrcceng3} program, described in Section~\ref{sec:opt_lno}, that has been optimized for many-body expansion calculations.
The improvements described in Section~\ref{sec:opt_lno} are available and made to be the default setting for LNO methods since the 2025 version of {\sc Mrcc}.
%
%
To reach the local approximation free (LAF) limit, we use the following formula from Refs.~\citenum{nagyApproachingBasisSet2019,nagyStateoftheartLocalCorrelation2024}:
\begin{equation}
E^\text{LAF N-T} = E^\text{Tight} + 0.5\left(E^\text{Tight} - E^\text{Normal}\right).
\end{equation}

We used the Dunning family~\cite{petersonAccurateCorrelationConsistent2002} of correlation consistent basis sets, specifically the aug-cc-pV$X$Z basis set family for lattice energies and jul-cc-pV$X$Z basis set family for relative energies.
The latter coresponds to only using cc-pV$X$Z on the H atoms, while using the aug-cc-pV$X$Z basis set on the other atoms.
We have tested the CBS(DZ/TZ) and CBS(TZ/QZ) two-point complete basis set (CBS) extrapolations for the double-zeta (DZ), triple-zeta (TZ) and quadruple-zeta (QZ) basis sets.
We employed the following two-point-extrapolation formulae:
\begin{equation}
    E^\text{CBS}_\text{corr} = \dfrac{X^\beta E^X_\text{corr} - Y^\beta E^Y_\text{corr}}{X^\beta - Y^\beta},
\end{equation}
\begin{equation}
E^\text{CBS}_\text{HF} = E^X_\text{HF} - \dfrac{E^Y_\text{HF} - E^X_\text{HF}}{\exp  (-\alpha \sqrt{Y}  ) -  \exp(-\alpha \sqrt{X})} \exp (-\alpha \sqrt{X} ),
\end{equation}
for the correlation and HF components of the energy, with $X$ and $Y=X+1$ denoting the (zeta) size of the basis set.
We use $\alpha = 5.79$ and $\beta = 3.05$ when using the aug-cc-pV$X$Z and $\alpha = 5.625$ and $\beta = 3.05$ when using the jul-cc-pV$X$Z, as adapted from Ref.~\citenum{neeseRevisitingAtomicNatural2011}.
We use the def2-QZVPP-RI-JK auxiliary basis function for density-fitting/resolution-of-identity  Hartree--Fock (HF) computations, and the RI auxiliary basis sets from Weigend~\cite{weigendRIMP2OptimizedAuxiliary1998,hellwegOptimizedAccurateAuxiliary2007} corresponding to the atomic orbital basis sets for subsequent cWFT calculations.
Counterpoise corrections were employed when calculating all terms.

\subsection{\label{sec:mbe_details}Many-body expansion}

The cWFT calculations described in the preceding section were performed on molecular clusters generated by the many-body expansion~\cite{heindelManyBodyExpansionAqueous2020,heindelManyBodyExpansionAqueous2021,hermanManybodyExpansionAqueous2021} (MBE).
The MBE allows for the interaction or total energy of a system formed of (molecular) monomers to be expressed as a sum of 1-body, 2-body, 3-body, and higher-order terms:
\begin{equation}
E = \sum_i E_i + \sum_{i>j} \Delta^2 E_{ij} + \sum_{i>j>k} \Delta^3 E_{ijk} + \cdots
\end{equation}
Here, the first term is a sum over the total energy $E_i$ minus a reference energy $E_\text{ref}$ (typically taken as the optimized gas-phase energy) for each molecule which makes up the unit cell.
The second term sums over the 2-body (2B) interaction energies for each enumerated pair of molecules $i$ and $j$:
\begin{equation}
\Delta^2 E_{ij} = E_{ij} - E_i - E_j.
\end{equation}
The third term gives the 3-body contributions involving molecules $i$, $j$, and $k$ which are nonadditive (i.e., beyond the sum of the individual pairwise interactions):
\begin{equation}
\Delta^3 E_{ijk} = E_{ijk} - \Delta^2 E_{ij} - \Delta^2 E_{ik} - \Delta^2 E_{jk} - E_i - E_j - E_k.
\end{equation}

\begin{figure}[h]
    \includegraphics[width=\textwidth]{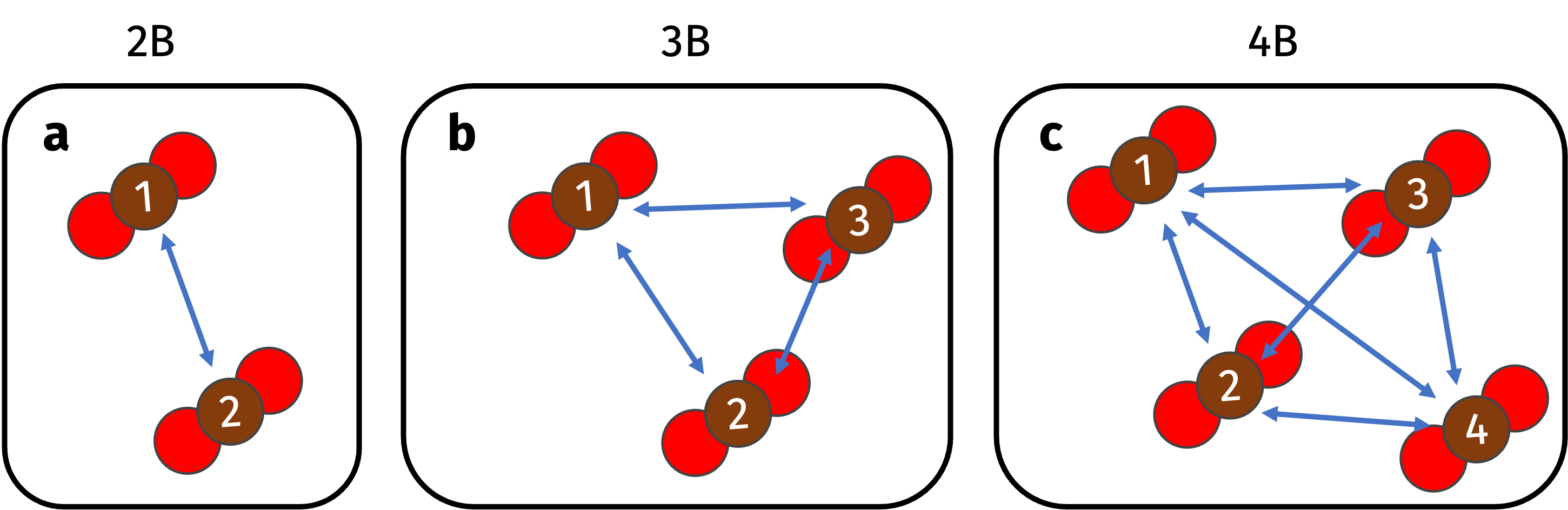}
    \caption{\label{fig:mbe_schematic}Schematic describing the distance metrics used for the (a) two-body (2B), (b) three-body (3B) and (c) four-body (4B) terms.}
\end{figure}

The main advantage of this approach is that it restricts the (expensive) cWFT calculations to smaller fragments as opposed to a full calculation of the whole unit cell.
For example, for ice, it has been shown that the total lattice energy converges at the 4-body term~\cite{hermanFormulationManyBodyExpansion2023b}.
Furthermore, it is possible to truncate the distance required to compute each of the these contributions, lowering the cost more.
While the distance metric between two monomers is straightforward -- being the distance between their center of masses or a specific atom within each molecule -- it is less clear for three or four bodies.
We follow the convention of~\citet{hermanFormulationManyBodyExpansion2023b}, where the distance metric is formed as the product of pairwise distances between all monomers.
We have described the standard MBE here, but we explore further methods to lower its cost in Section~\ref{sec:lower_cost}.

We report the final 2B and 3B distance cutoffs for the system studied within this work in Table~\ref{tab:system_cutoffs}.
These cutoffs were selected based upon convergence of these terms for the various systems reported throughout, given in the table.

\begin{table}[h]
\caption{\label{tab:system_cutoffs}Two (2B) and three (3B) body cutoffs for the systems studied within this work.}
\begin{tabular}{@{}lrrrr@{}}
\toprule
System     & 2B cutoff (\AA{}) & 3B cutoff (\AA$^3$) & Convergence & Reference \\ \midrule
X23        & 12        & 300       & Section~\ref{sec:d4_corr_ene} &  Table~\ref{tab:x23_final_compare}       \\
ICE13      & 9         & 150       &  Section~\ref{sec:ice_conv} &  Table~\ref{tab:ice13_lattice_energies}      \\
ROY        & 16        & 800       &  Section~\ref{sec:pharma_dist_conv} & Table~\ref{tab:final_relative_energies}       \\
Axitinib   & 20        & 720       &  Section~\ref{sec:pharma_dist_conv} &  Table~\ref{tab:final_relative_energies}    \\
Rotigotine & 19        & 750       &  Section~\ref{sec:pharma_dist_conv} &   Table~\ref{tab:final_relative_energies}       \\
Molecule X & 16        & 450       &  Section~\ref{sec:pharma_dist_conv} &   Table~\ref{tab:final_relative_energies}       \\ \bottomrule
\end{tabular}
\end{table}

\clearpage

\section{\label{sec:lower_cost}Lowering computational cost of MBE}

In this section, we describe the techniques employed within this work to lower the cost of standard MBE -- dubbed `full enumeration'.
In particular, these approaches enhance the efficiency of the MBE by decreasing the number of terms and maximum order contribution required in the MBE.
We emphasize here that several of these approaches have been explored within previous work (and we will endeavour to highlight the prior work).
The key development here is that we have combined all these approaches to make the most efficient MBE strategy to date for studying molecular crystals that can reach sub-chemical accuracy on $E_\text{rel}$.
When combined with an LNO-CCSD(T) approach that has been further optimized for the MBE within this work, it has allowed for complex (pharmaceutical-sized) molecular crystals to be studied.
We have dubbed this framework LNO-MBE-CCSD(T) in this paper.
The final comparison of the cost to perform LNO-MBE-CCSD(T) is tabulated and discussed in Section~\ref{sec:comp_cost}.

\subsection{\label{sec:ice_conv}Removing HF contribution to handle only electron correlation}

As a post-HF method, the CCSD(T) total energy is made up of HF energy contribution as well as the CCSD(T) correlation energy contribution.
These two terms account for different types of physical interactions~\cite{szalewiczSymmetryadaptedPerturbationTheory2012}.
The HF contribution primarily accounts for the electrostatic, exchange, and induction components of the lattice energy. The CCSD(T) correlation energy includes van der Waals dispersion, charge-transfer effects and the electron correlation contribution to the higher order coupling of all the above.
As such, they have differing behaviors in their separate MBE.
In fact, the MBE for HF is significantly more challenging than the MBE for the CCSD(T) correlation energy as it (1) requires longer distance cutoffs to converge any of the terms and (2) higher-order many-body terms.

\begin{figure}[h]
    \includegraphics[width=\textwidth]{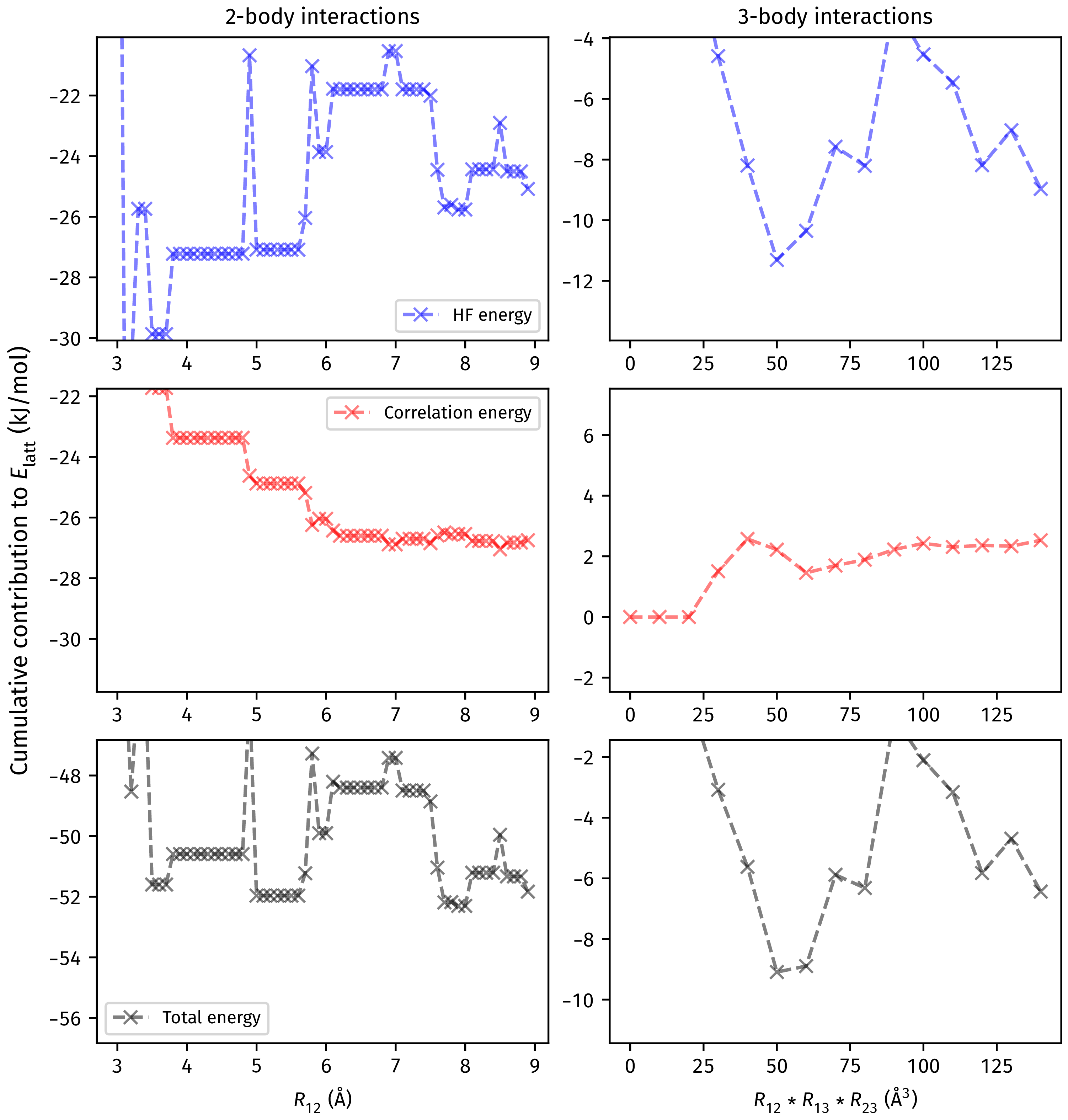}
    \caption{\label{fig:total_corr_comparison}Comparison of the convergence behaviour of the HF, CCSD(T) correlation and total energy for the two-body and three-body contributions to the lattice energy of ice VIII. The aug-cc-pVTZ and aug-cc-pVDZ basis sets were used for the 2B and 3B contributions, respectively. CCSD(T) calculations are accelerated with the LNO approximation using {\tt vTight} thresholds.}
\end{figure}

We illustrate the differing behavior of HF and CCSD(T) in Figure~\ref{fig:total_corr_comparison} for the two-body (2B) and three-body (3B) contribution to the total energy of ice VIII.
As seen for both the 2B and 3B contributions, the HF energy converges non-monotonically as a function of distance while the correlation energy components converge rapidly with distance cutoffs.
For example, the CCSD(T) correlation energy contribution to the two-body terms has converged to sub ${\sim}0.1\,$kJ/mol by a $7\,$\AA{} while there are still $4\,$kJ/mol jumps for the HF energy, which manifest in the total energy as well.
Beyond just a slower convergence with the distance cutoff, there is also a worse convergence with number of body terms for the total energy.
In Table~\ref{tab:total_corr_4b}, we show for ice VIII that the total energy contribution of four-body terms -- computed with MBpol~\cite{meddersDevelopmentFirstPrinciplesWater2014,zhuMBpol2023SubchemicalAccuracy2023} from~\citet{hermanFormulationManyBodyExpansion2023b} -- still has a significant contribution of ${\sim}0.8\,$kJ/mol while the correlation energy we have computed with CCSD(T) contributes less than $0.1\,$kJ/mol. 

\begin{table}[h]
\caption{\label{tab:total_corr_4b}MBE contributions up to four-body (4B) for the total and correlation energy contribution to ice VIII. The correlation energy contribution comes from the `Canonical' approach described in Section~\ref{sec:cheaper_settings}.}
\begin{tabular}{@{}lrr@{}}
\toprule
Contribution (kJ/mol) & Correlation & Total (MBpol~\cite{hermanFormulationManyBodyExpansion2023b}) \\ \midrule
1B                    & -5.3        & 1.2           \\
2B                    & -31.0       & -53.9         \\
3B                    & 2.5         & -6.7          \\
4B                    & -0.1        & 3.2           \\ \bottomrule
\end{tabular}
\end{table}

These observations thus highlight the advantage of separating the treatment of both HF and CCSD(T).
Furthermore, as we highlight in Section~\ref{sec:comp_cost}, advancements in the treatment of HF under periodic boundary conditions now make it economical to treat this component directly without resorting to the MBE.

It should be emphasized here that the idea to separate the treatment of HF and CCSD(T) correlation has been explored across several different studies for various systems.
For ice Ih, it was shown by~\citet{hermannGroundStatePropertiesCrystalline2008b} that correlation energies up to only 2B terms are needed to compute the total energy of this system.
This has also been further demonstrated for both water clusters and ice clusters by~\citet{gillanEnergyBenchmarksWater2013a} as well as for driving molecular dynamics simulations of water clusters by~\citet{heindelMolecularDynamicsDriven2021}.
For molecular crystals, this has been recently used by~\citet{sytyMultiLevelCoupledClusterDescription2025} for the X23 dataset, with 2B contributions computed by LNO-CCSD(T) and 3B contributions from the random phase approximation (RPA).
Similarly, such subtractive embedding approaches have been shown earlier for diverse molecular crystal systems by~\citet{mullerIncrementallyCorrectedPeriodic2013} as well as Beran and co-workers~\cite{beranPredictingOrganicCrystal2010,wenPracticalQuantumMechanicsbased2012a,cervinkaInitioPredictionPolymorph2018}.
Such an approach has also been employed to study hydrates~\cite{kosataStabilityHydrogenHydrates2018}.
Separate from the goal to reach cWFT accuracy, Boese and co-workers~\cite{lobodaHybridDensityFunctional2018,hojaMultimerEmbeddingApproach2024} have demonstrated its usefulness to bring GGA-level DFT calculations up to the hybrid-DFT level from 2B contributions.

\subsection{\label{sec:d4_corr_ene}Enhancing distance convergence using empirical dispersion corrections}

The CCSD(T) correlation energy is largely dominated by van der Waals dispersion interaction.
There exist widely-used force-fields~\cite{neumannTailorMadeForceFields2008,beranApproximatingQuantumManybody2009,priceModellingOrganicCrystal2010,nymanAccurateForceFields2016,bowskillCrystalStructurePrediction2021} that approximate this contribution, most relevantly to the present work through a family of dispersion corrections developed by Grimme~\cite{grimmeSemiempiricalGGAtypeDensity2006,grimmeConsistentAccurateInitio2010,grimmeDispersionCorrectedMeanFieldElectronic2016,caldeweyherGenerallyApplicableAtomiccharge2019} for DFT.
The D4 dispersion correction has also been parametrized for HF and we used this parametrization as it corresponds to a direct approximation of the dispersion contributions from the correlation energy.
This family of dispersion corrections are trivial and cost-effective to compute as they only depend on atomic numbers and positions.
The latest is the D4 correction~\cite{caldeweyherGenerallyApplicableAtomiccharge2019}, which improves upon the previous D3 correction by incorporating an atomic-charge dependence (based on atomic charges computed using classical electronegativity equilibration partial charges).
In In addition, three-body dispersion effects are described using an Axilrod-Teller-Muto term~\cite{j.pms.jpn-17-6-629,axilrodInteractionVanWaals1943}.
We remove the D4 term from the CCSD(T) correlation energy and instead compute the difference between the CCSD(T) and D4 energy ($\Delta_\text{D4}^\text{CCSD(T)}$), which we can converge instead.
The full D4 term with periodic boundary conditions can be trivially computed and added onto the periodic HF calculation.

\begin{figure}[h!]
\includegraphics[width=\textwidth]{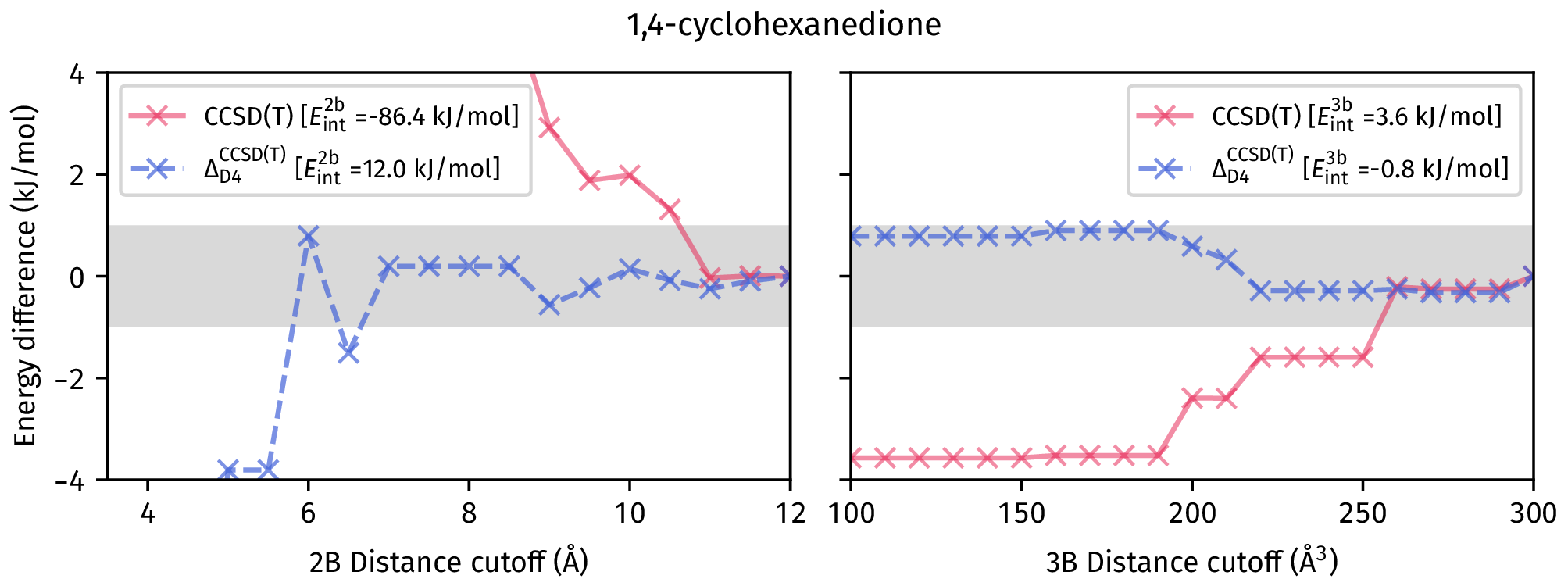}
\caption{\label{fig:x23_d4_conv_01} The convergence with distance cutoff of the two-body (2B) and three-body (3B) components of the many-body expansion (MBE) for the lattice energy $E_\text{latt}$ of 1,4-cyclohexanedione from the X23 dataset. The red line shows the CCSD(T) result, while the blue line shows the difference to the D4 result. The shaded area indicates the range of energies between -1 and 1 kJ/mol. We use the pairwise distance $R_{12}$ between the two monomers in the 2B case, and the product of the pairwise distances $R_{12}*R_{13}*R_{23}$ in the 3B case.}
\end{figure}

\begin{figure}[h!]
\includegraphics[width=\textwidth]{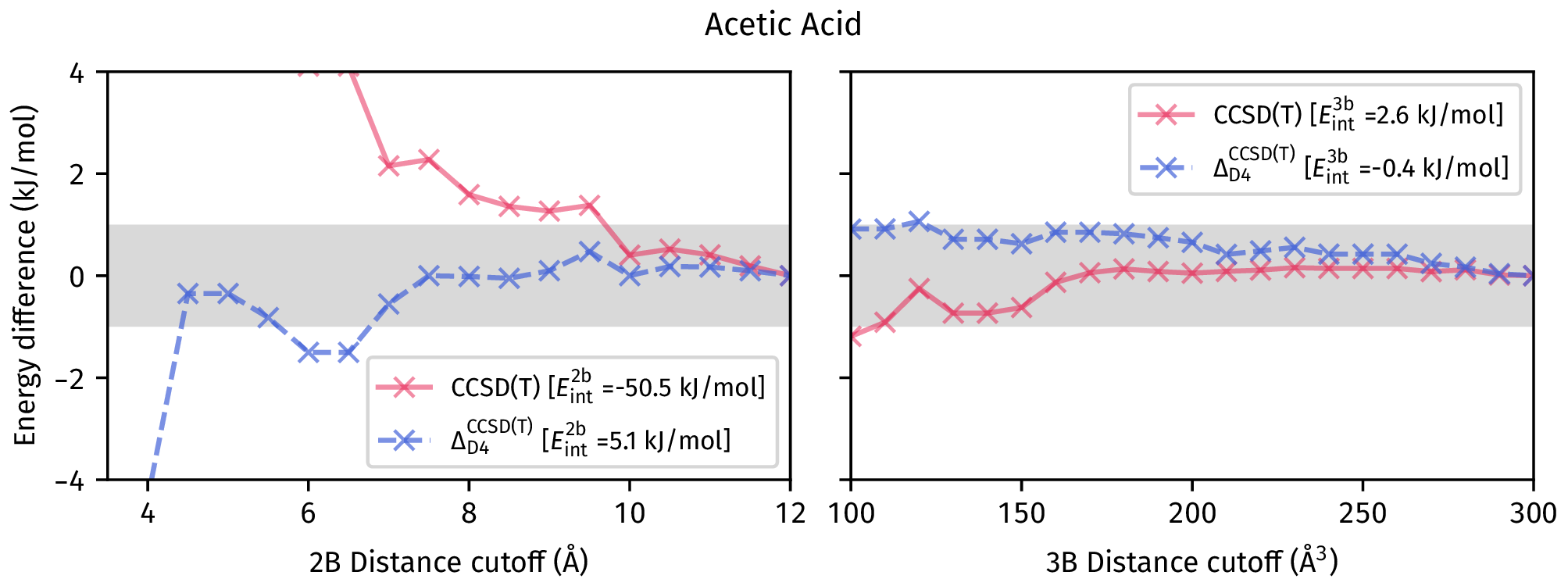}
\caption{\label{fig:x23_d4_conv_02} The convergence of the 2B and 3B MBE components to the $E_\text{latt}$ of acetic acid.}
\end{figure}

\begin{figure}[h!]
\includegraphics[width=\textwidth]{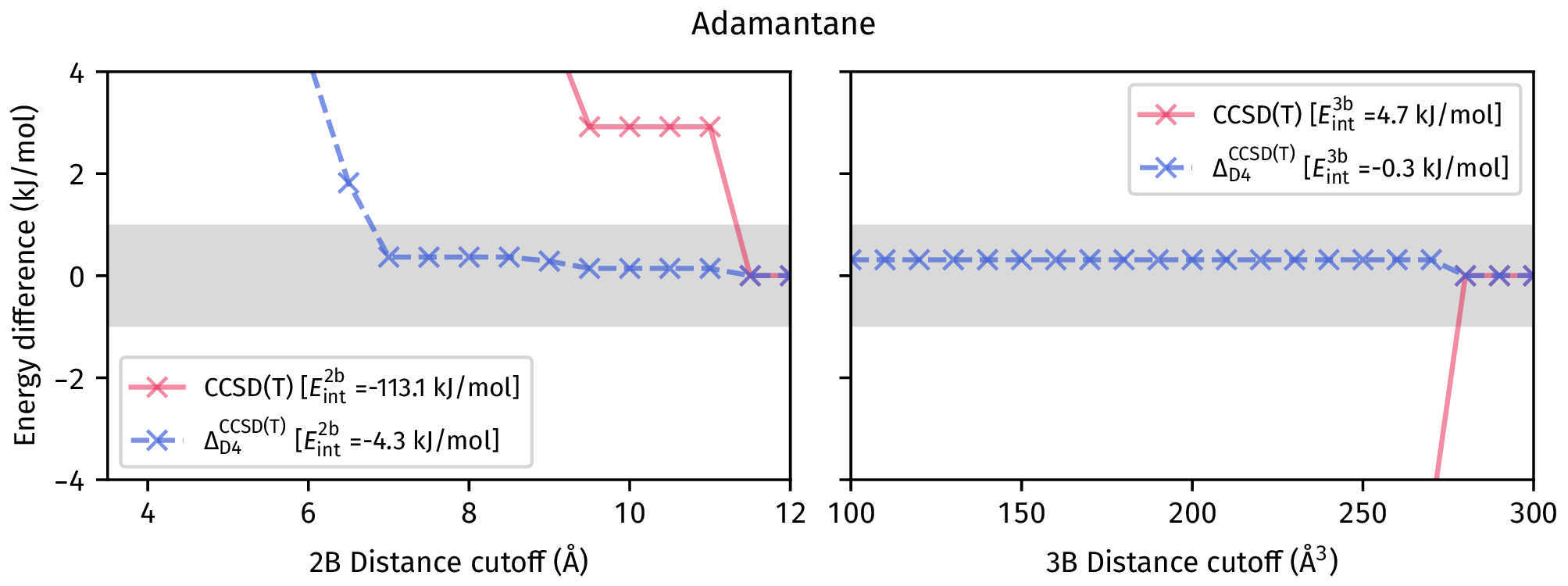}
\caption{\label{fig:x23_d4_conv_03} The convergence of the 2B and 3B MBE components to the $E_\text{latt}$ of adamantane.}
\end{figure}

\begin{figure}[h!]
\includegraphics[width=\textwidth]{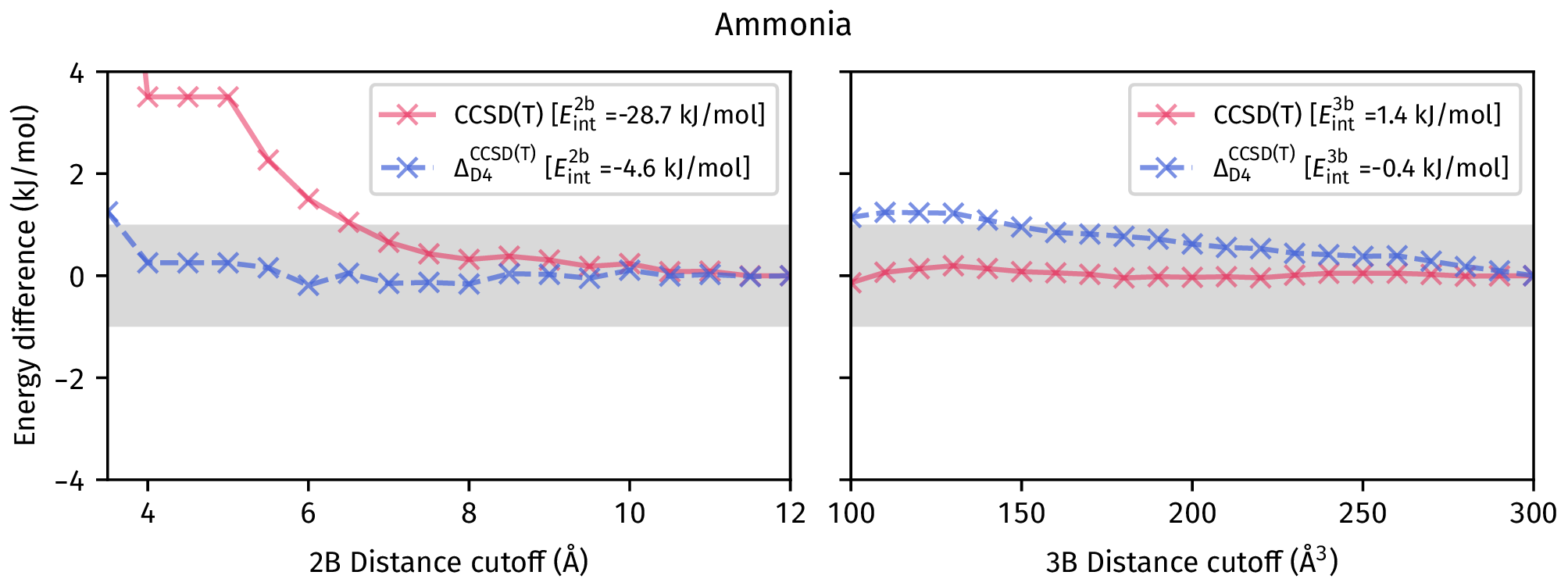}
\caption{\label{fig:x23_d4_conv_04} The convergence of the 2B and 3B MBE components to the $E_\text{latt}$ of ammonia.}
\end{figure}

\begin{figure}[h!]
\includegraphics[width=\textwidth]{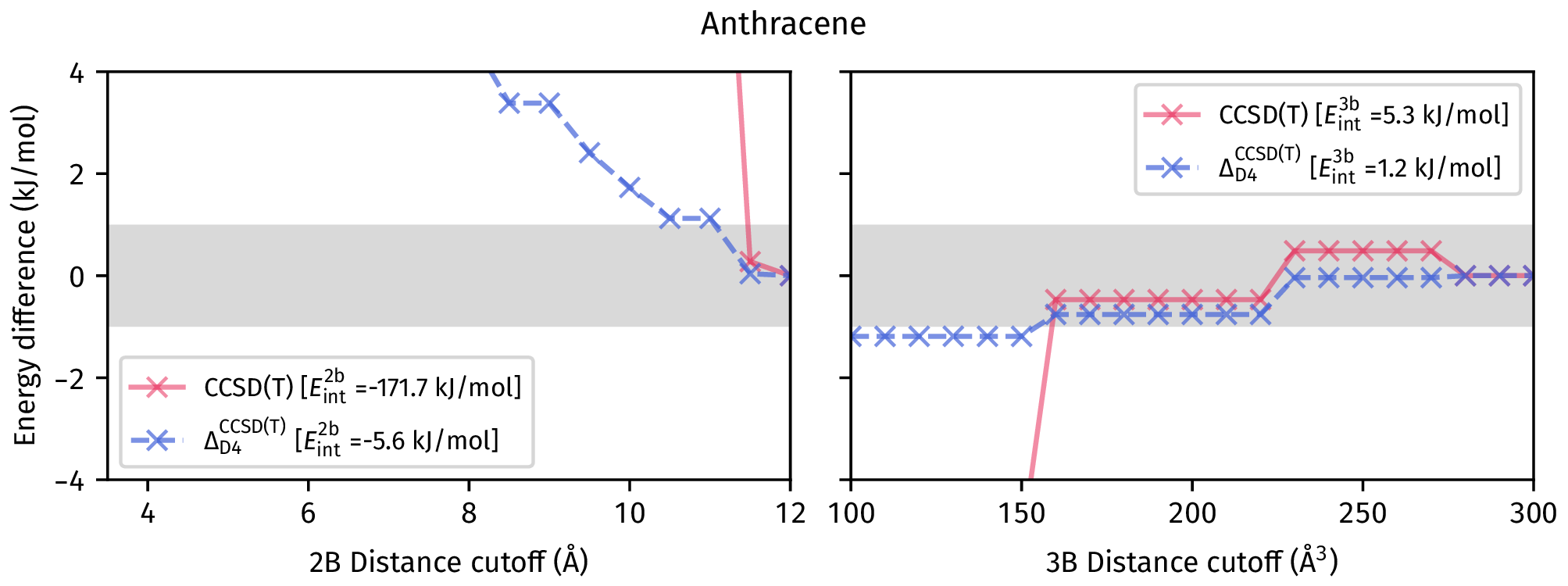}
\caption{\label{fig:x23_d4_conv_05} The convergence of the 2B and 3B MBE components to the $E_\text{latt}$ of anthracene.}
\end{figure}

\begin{figure}[h!]
\includegraphics[width=\textwidth]{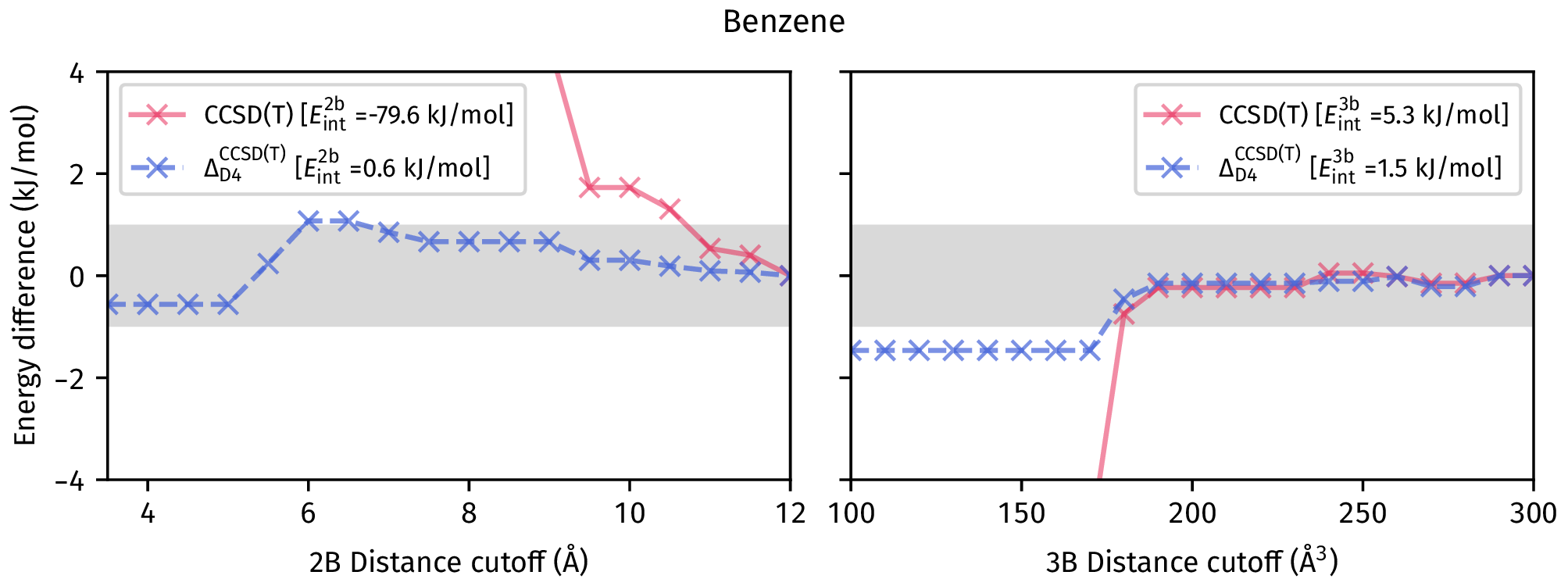}
\caption{\label{fig:x23_d4_conv_06} The convergence of the 2B and 3B MBE components to the $E_\text{latt}$ of benzene.}
\end{figure}

\begin{figure}[h!]
\includegraphics[width=\textwidth]{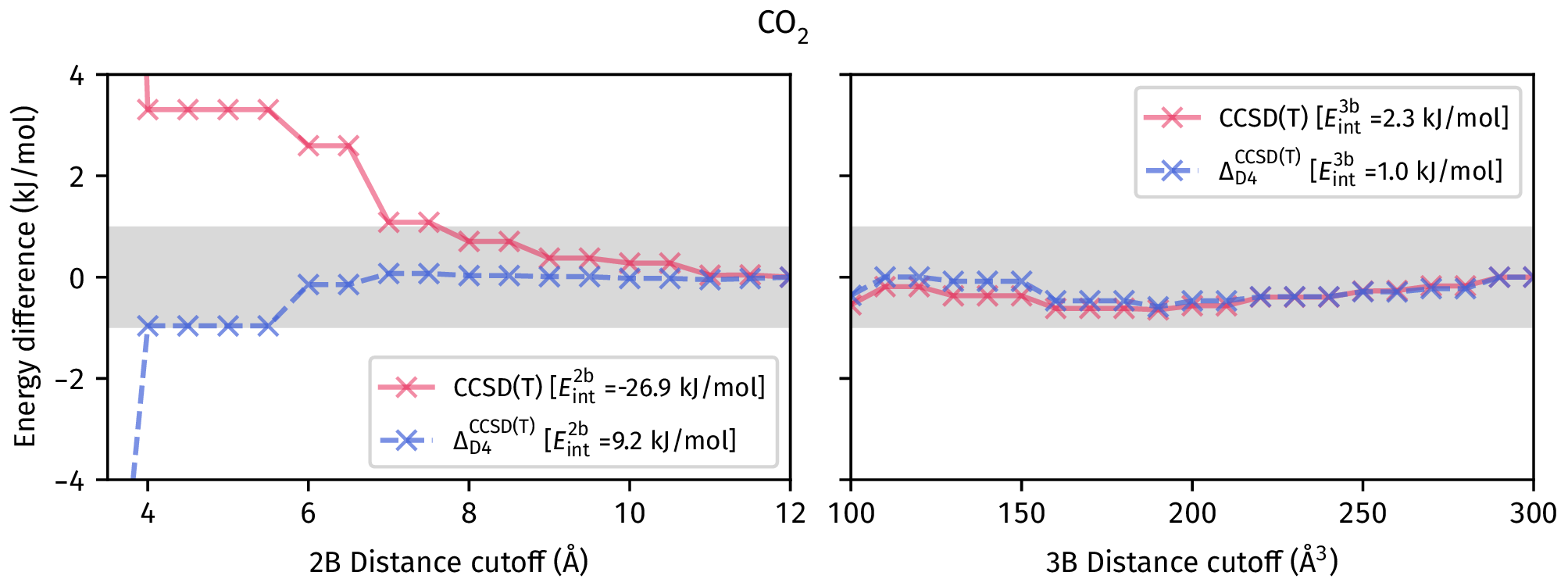}
\caption{\label{fig:x23_d4_conv_07} The convergence of the 2B and 3B MBE components to the $E_\text{latt}$ of CO$_2$.}
\end{figure}

\begin{figure}[h!]
\includegraphics[width=\textwidth]{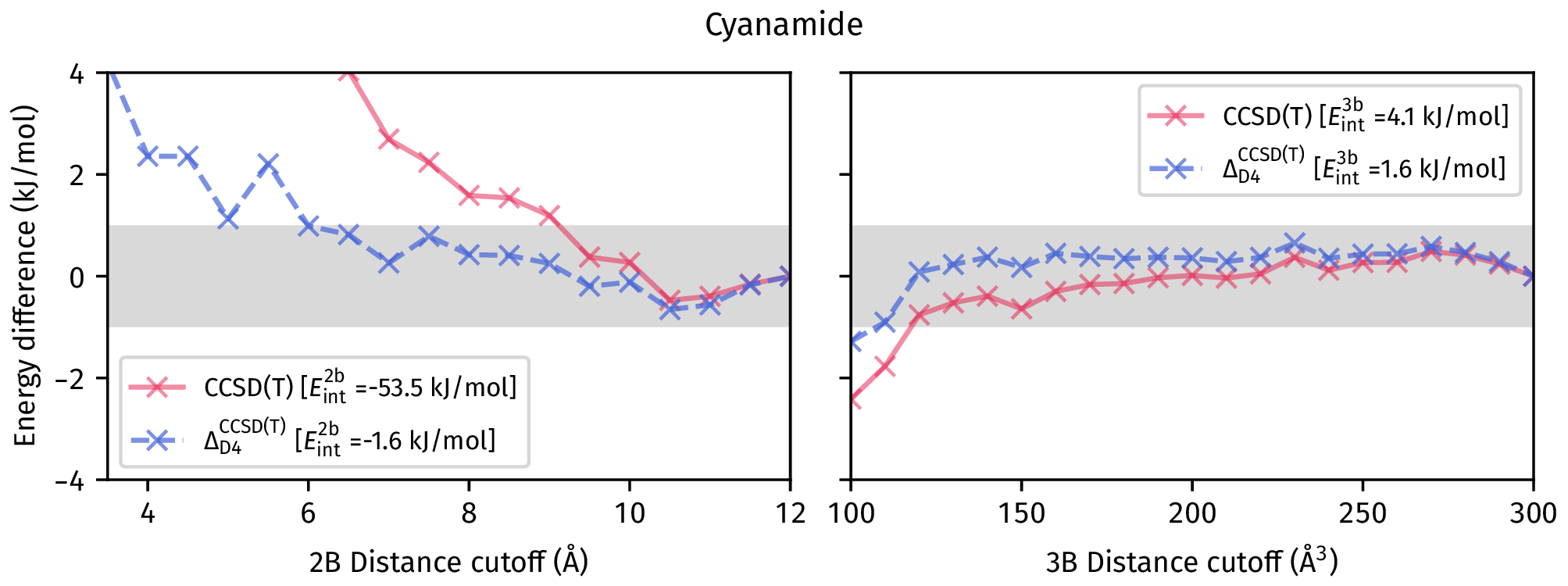}
\caption{\label{fig:x23_d4_conv_08} The convergence of the 2B and 3B MBE components to the $E_\text{latt}$ of cyanamide.}
\end{figure}

\begin{figure}[h!]
\includegraphics[width=\textwidth]{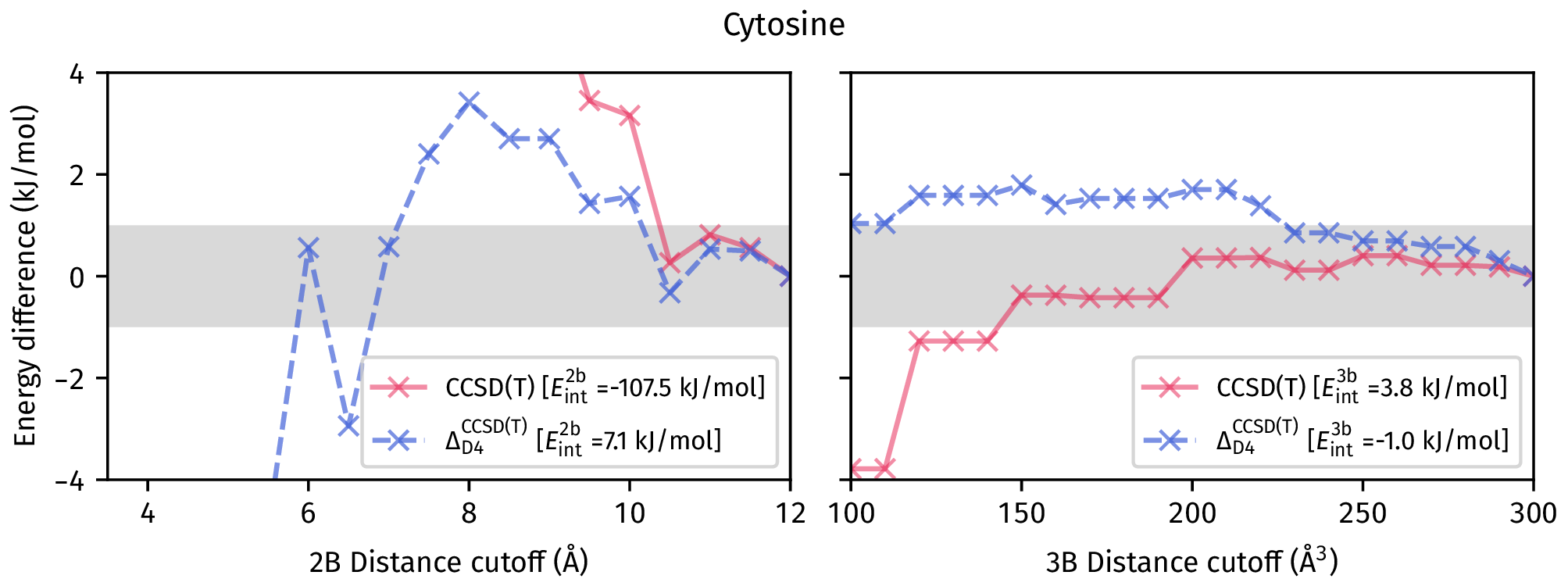}
\caption{\label{fig:x23_d4_conv_09} The convergence of the 2B and 3B MBE components to the $E_\text{latt}$ of cytosine.}
\end{figure}

\begin{figure}[h!]
\includegraphics[width=\textwidth]{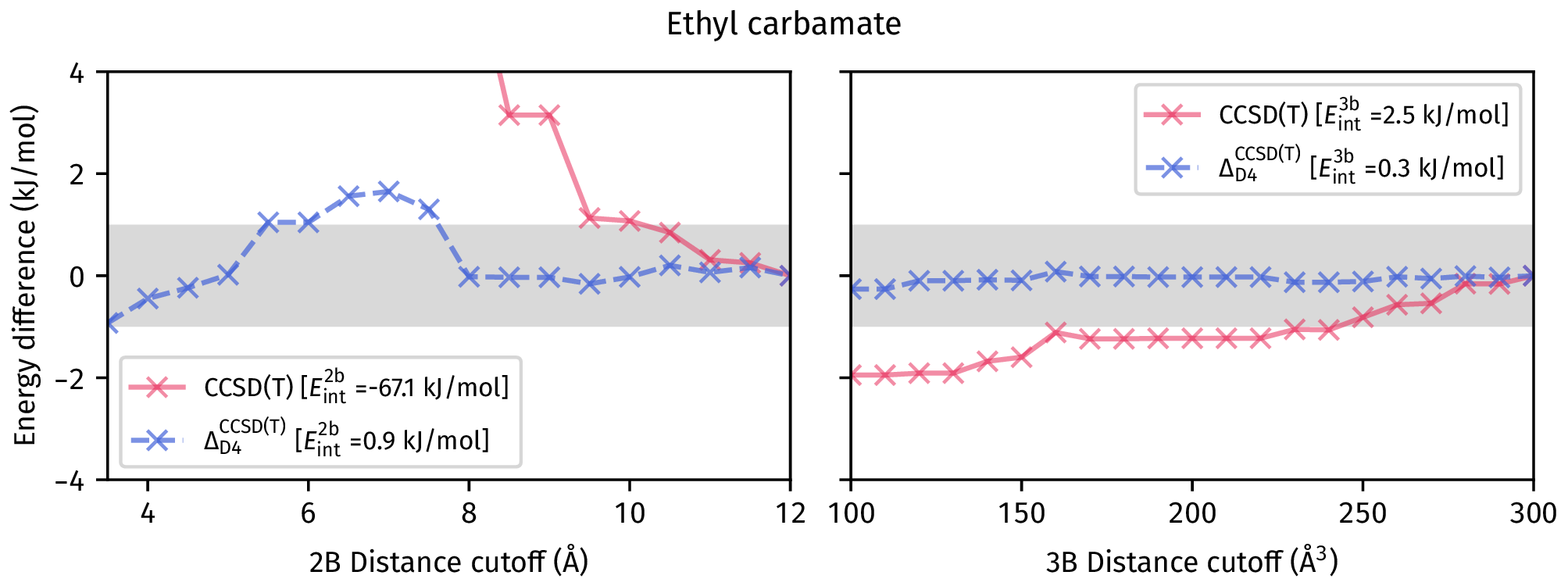}
\caption{\label{fig:x23_d4_conv_10} The convergence of the 2B and 3B MBE components to the $E_\text{latt}$ of ethyl carbamate.}
\end{figure}

\begin{figure}[h!]
\includegraphics[width=\textwidth]{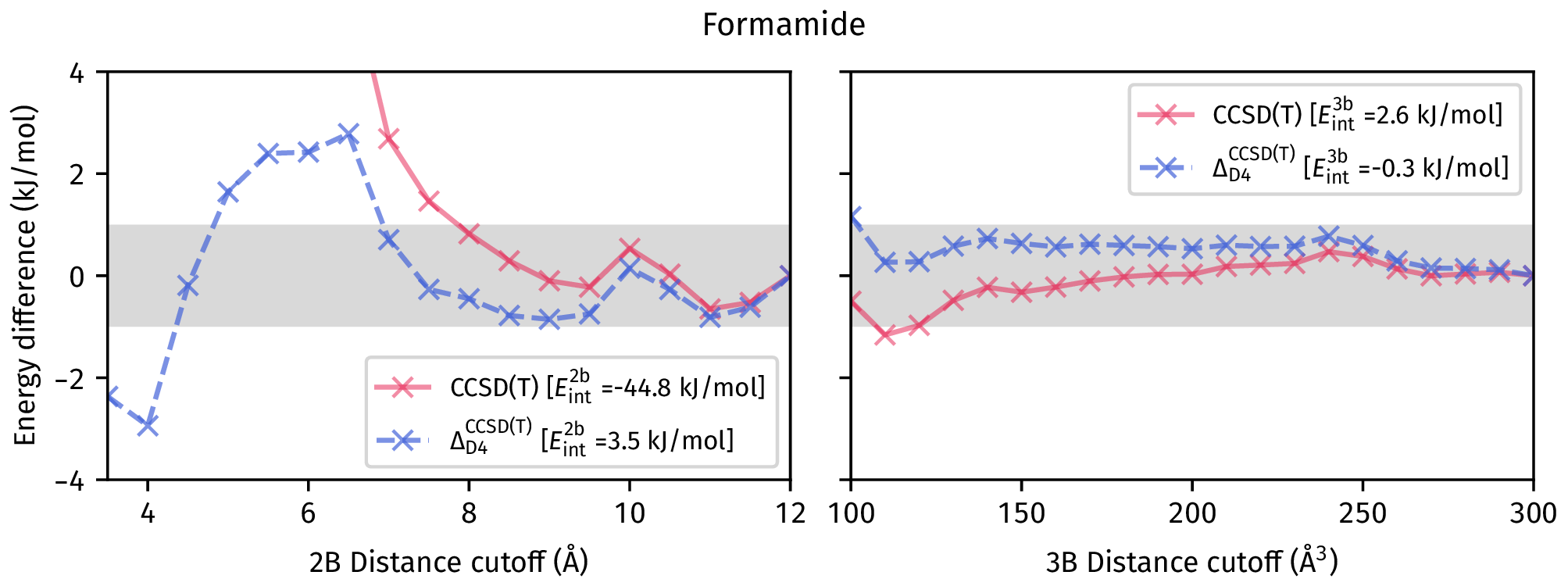}
\caption{\label{fig:x23_d4_conv_11} The convergence of the 2B and 3B MBE components to the $E_\text{latt}$ of formamide.}
\end{figure}

\begin{figure}[h!]
\includegraphics[width=\textwidth]{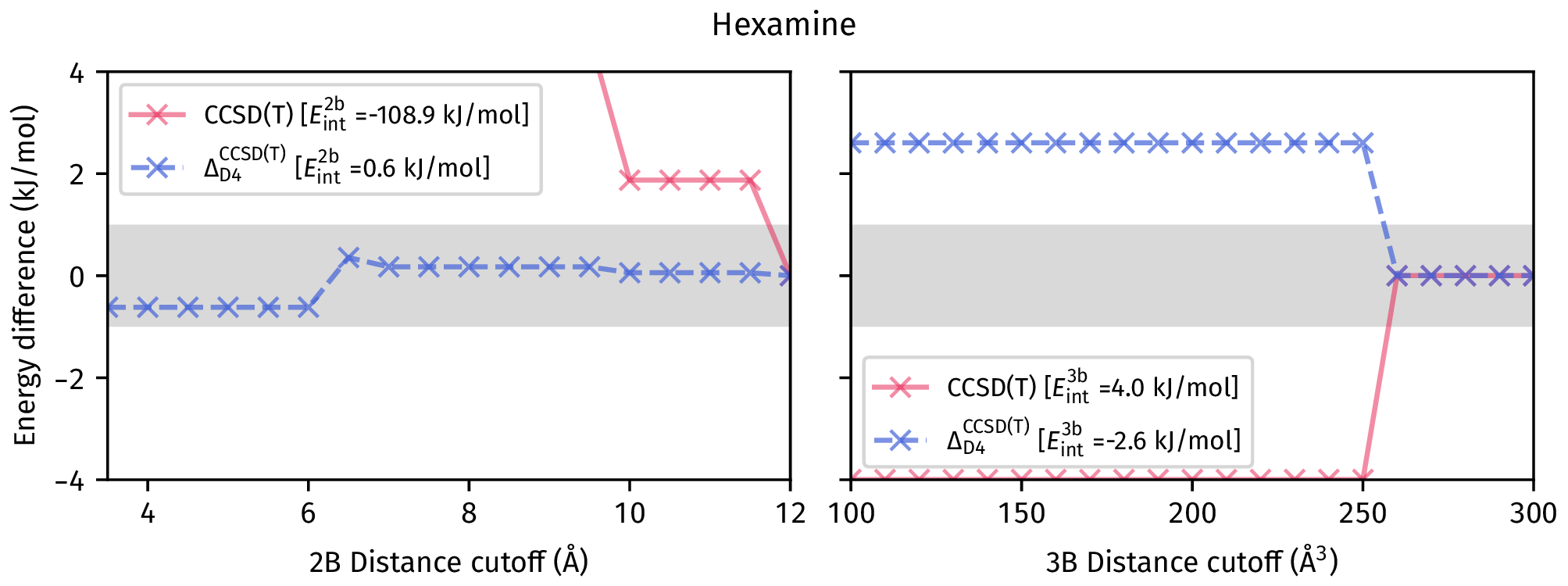}
\caption{\label{fig:x23_d4_conv_12} The convergence of the 2B and 3B MBE components to the $E_\text{latt}$ of hexamine.}
\end{figure}

\begin{figure}[h!]
\includegraphics[width=\textwidth]{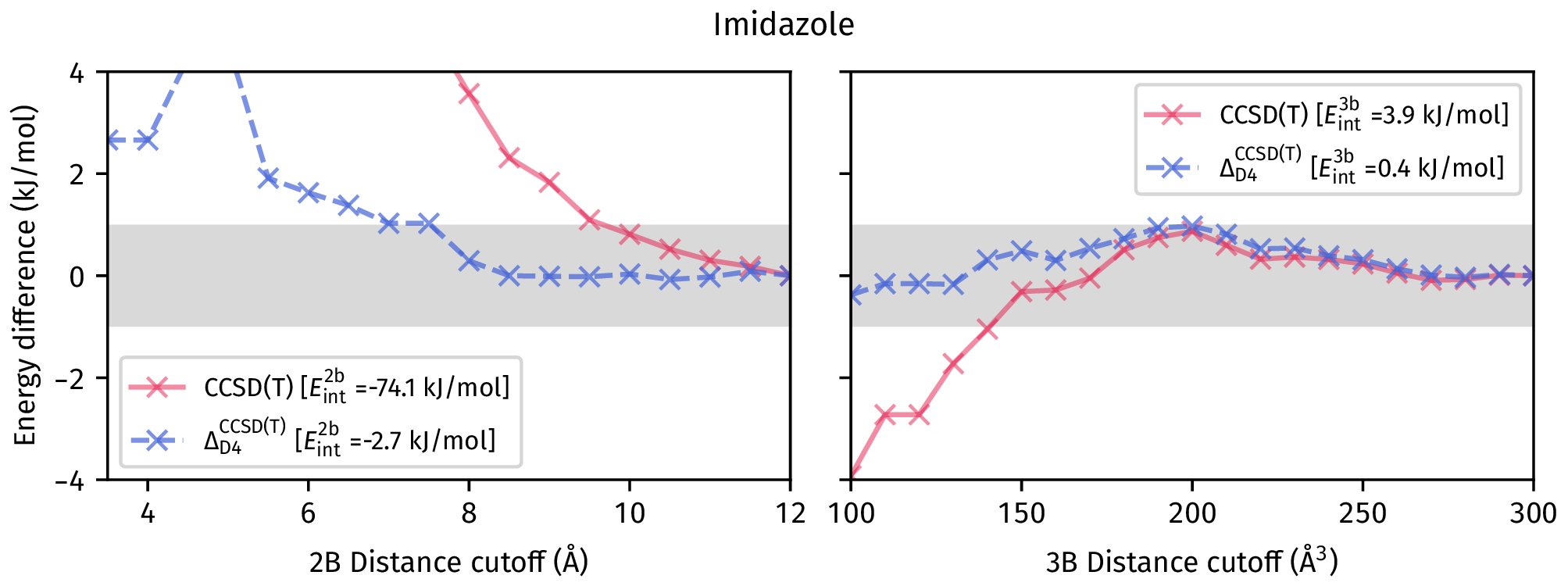}
\caption{\label{fig:x23_d4_conv_13} The convergence of the 2B and 3B MBE components to the $E_\text{latt}$ of imidazole.}
\end{figure}

\begin{figure}[h!]
\includegraphics[width=\textwidth]{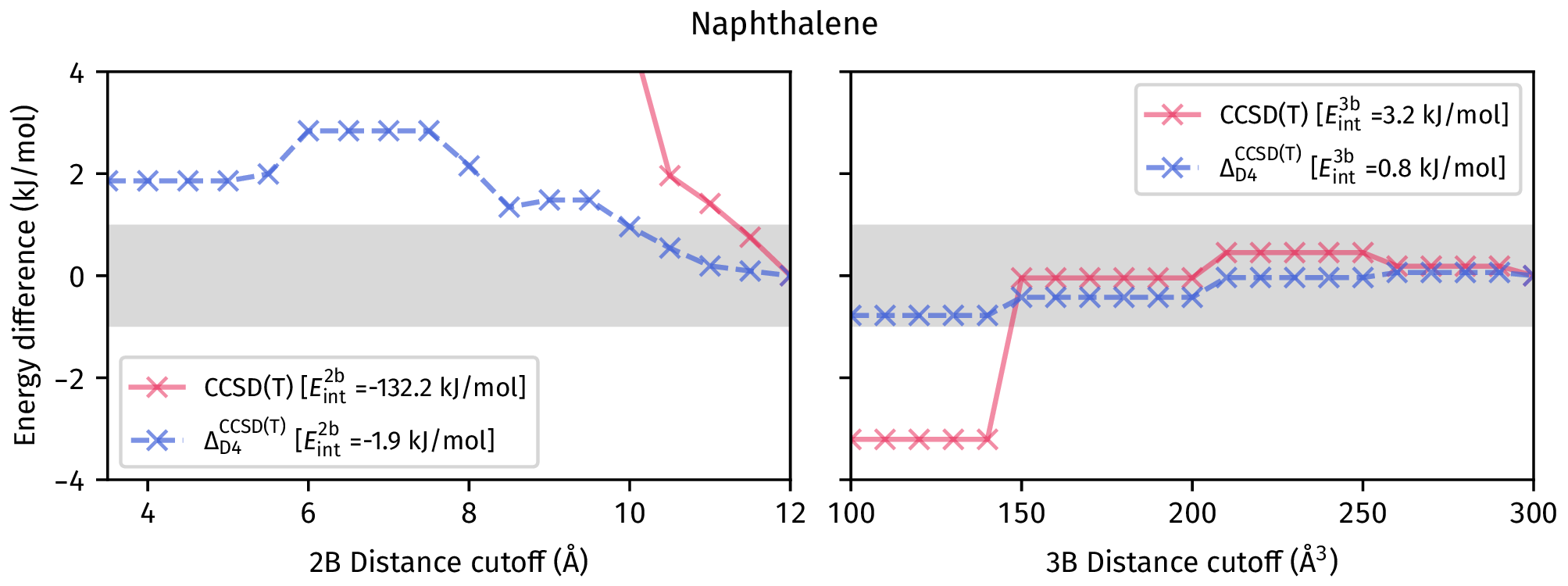}
\caption{\label{fig:x23_d4_conv_14} The convergence of the 2B and 3B MBE components to the $E_\text{latt}$ of naphthalene.}
\end{figure}

\begin{figure}[h!]
\includegraphics[width=\textwidth]{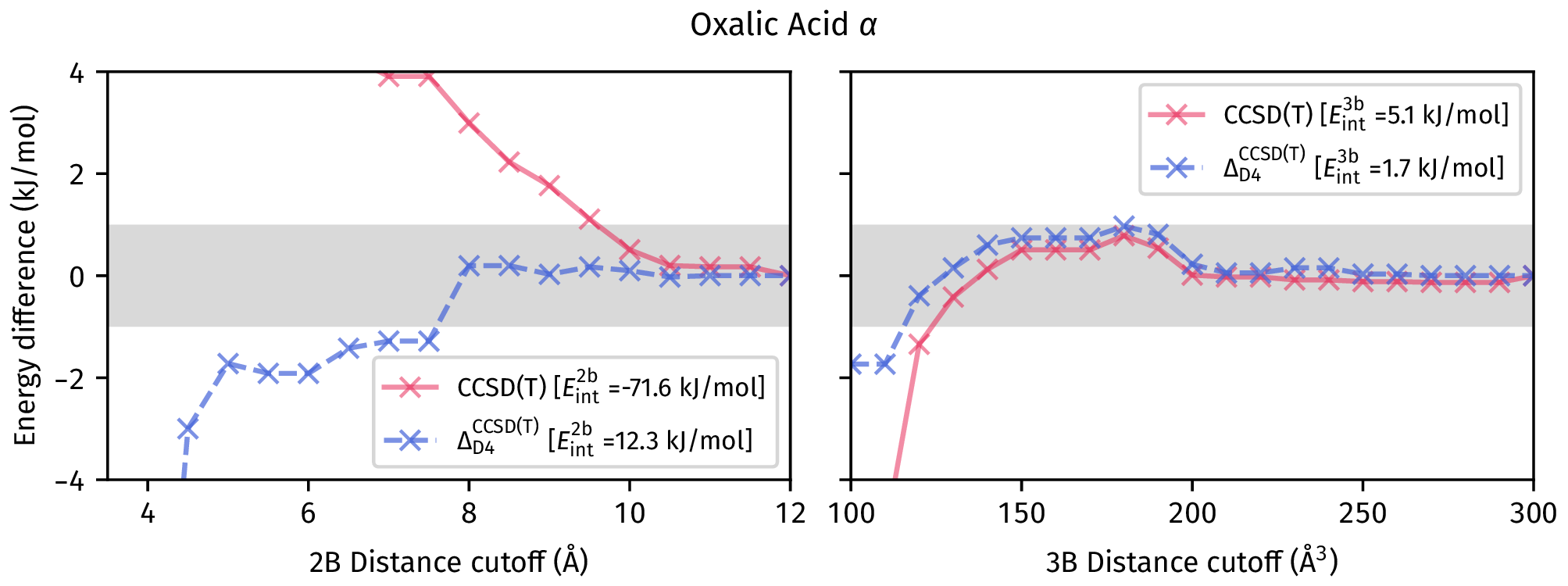}
\caption{\label{fig:x23_d4_conv_15} The convergence of the 2B and 3B MBE components to the $E_\text{latt}$ of oxalic acid $\alpha$.}
\end{figure}

\begin{figure}[h!]
\includegraphics[width=\textwidth]{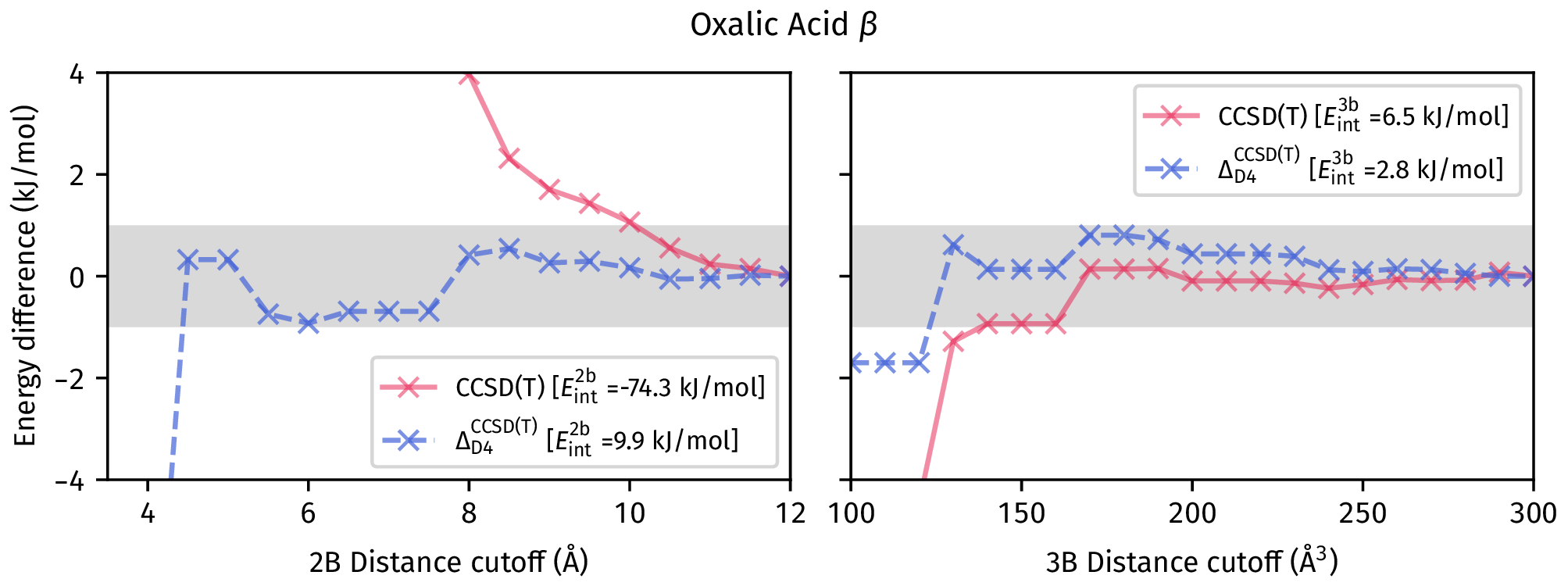}
\caption{\label{fig:x23_d4_conv_16} The convergence of the 2B and 3B MBE components to the $E_\text{latt}$ of oxalic acid $\beta$.}
\end{figure}

\begin{figure}[h!]
\includegraphics[width=\textwidth]{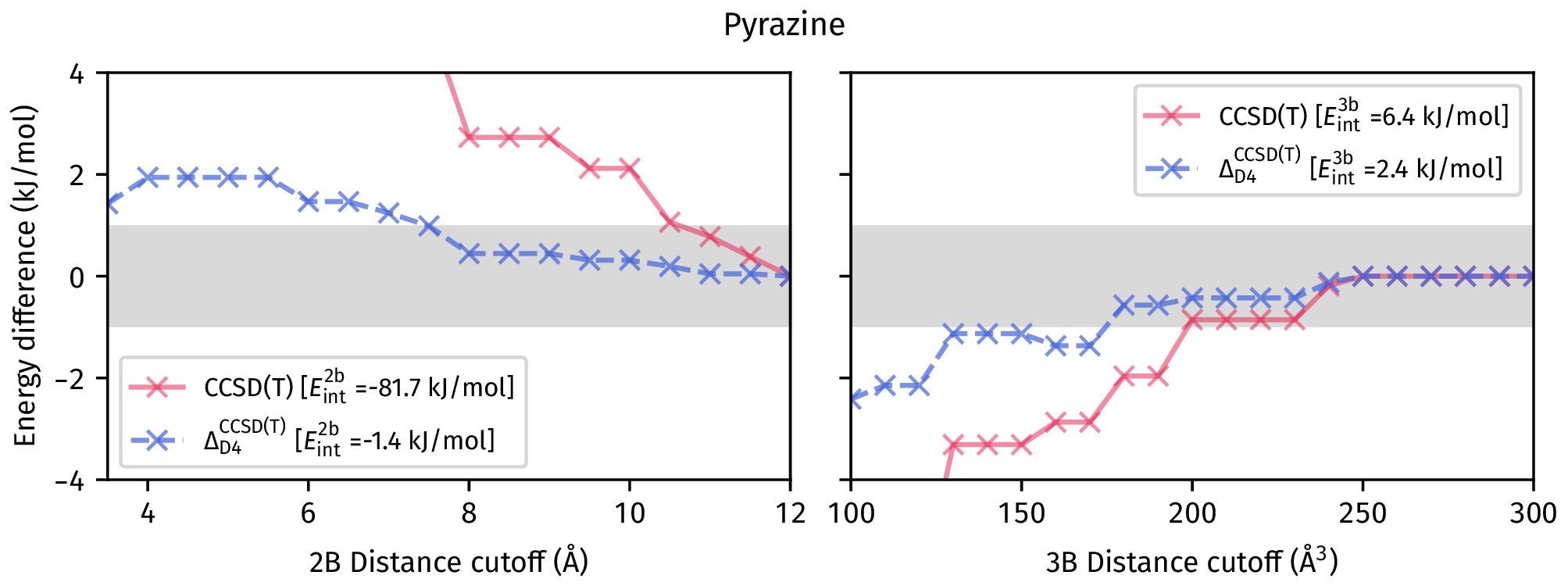}
\caption{\label{fig:x23_d4_conv_17} The convergence of the 2B and 3B MBE components to the $E_\text{latt}$ of pyrazine.}
\end{figure}

\begin{figure}[h!]
\includegraphics[width=\textwidth]{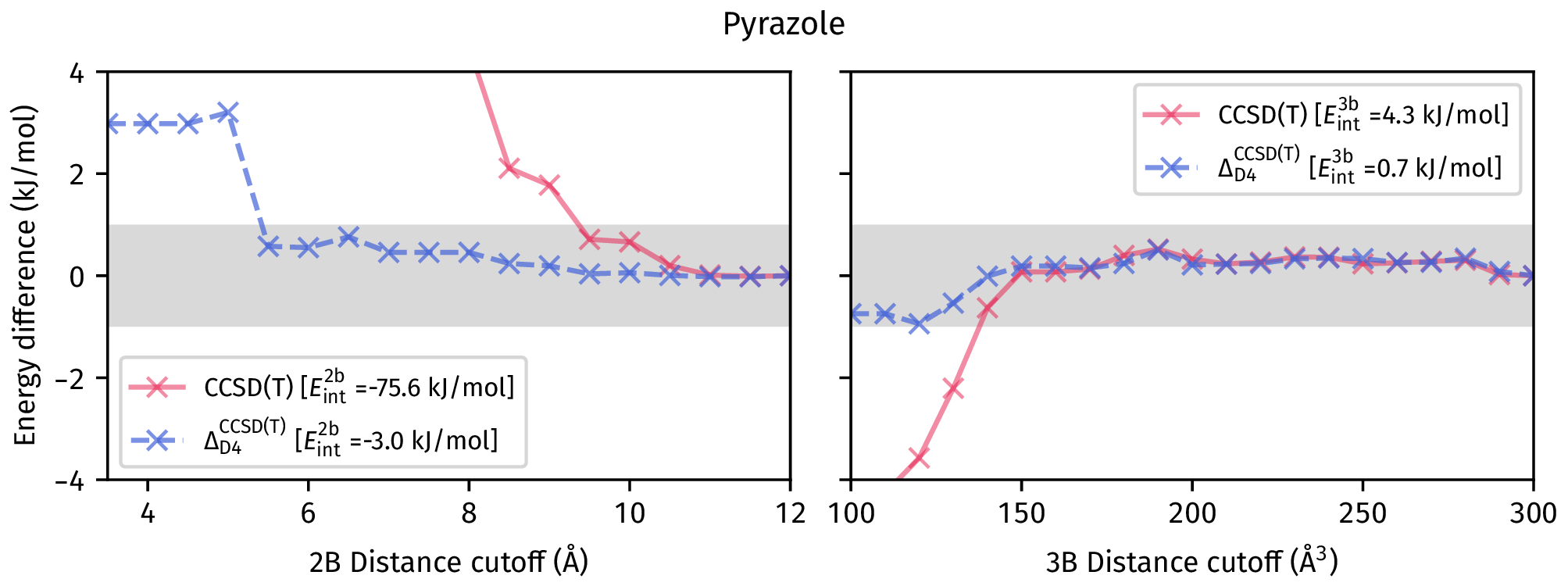}
\caption{\label{fig:x23_d4_conv_18} The convergence of the 2B and 3B MBE components to the $E_\text{latt}$ of pyrazole.}
\end{figure}

\begin{figure}[h!]
\includegraphics[width=\textwidth]{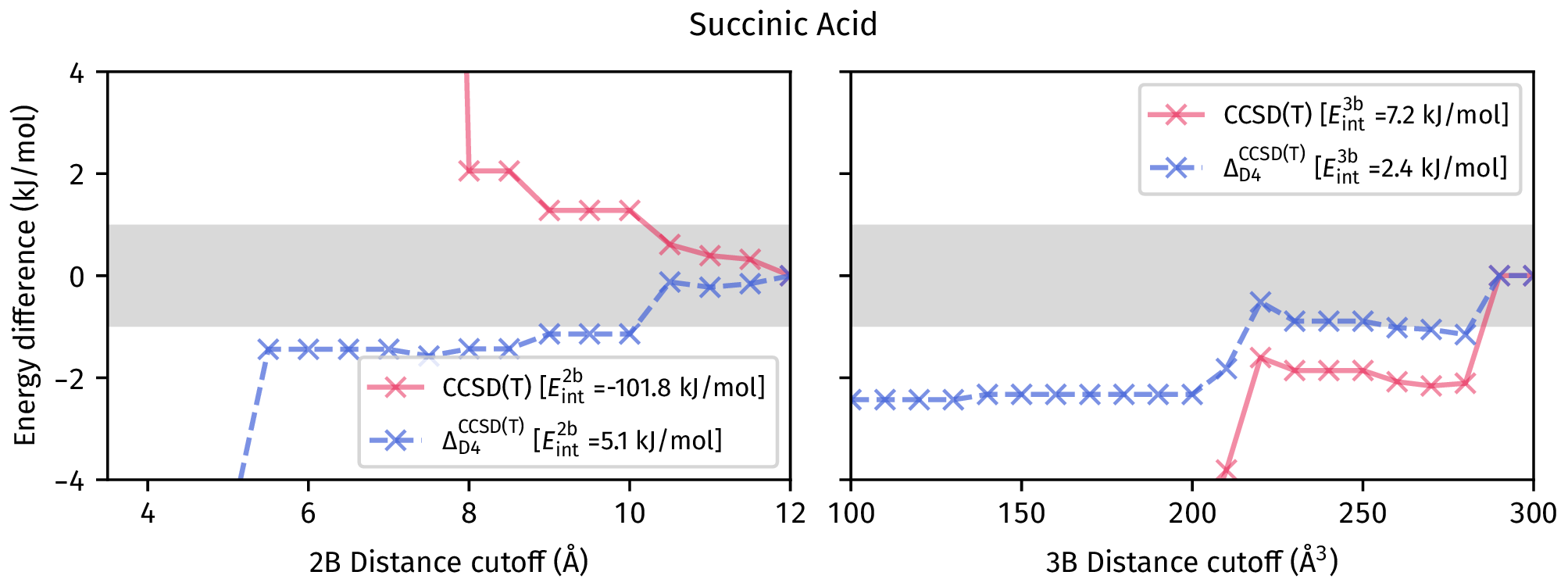}
\caption{\label{fig:x23_d4_conv_19} The convergence of the 2B and 3B MBE components to the $E_\text{latt}$ of succinic acid.}
\end{figure}

\begin{figure}[h!]
\includegraphics[width=\textwidth]{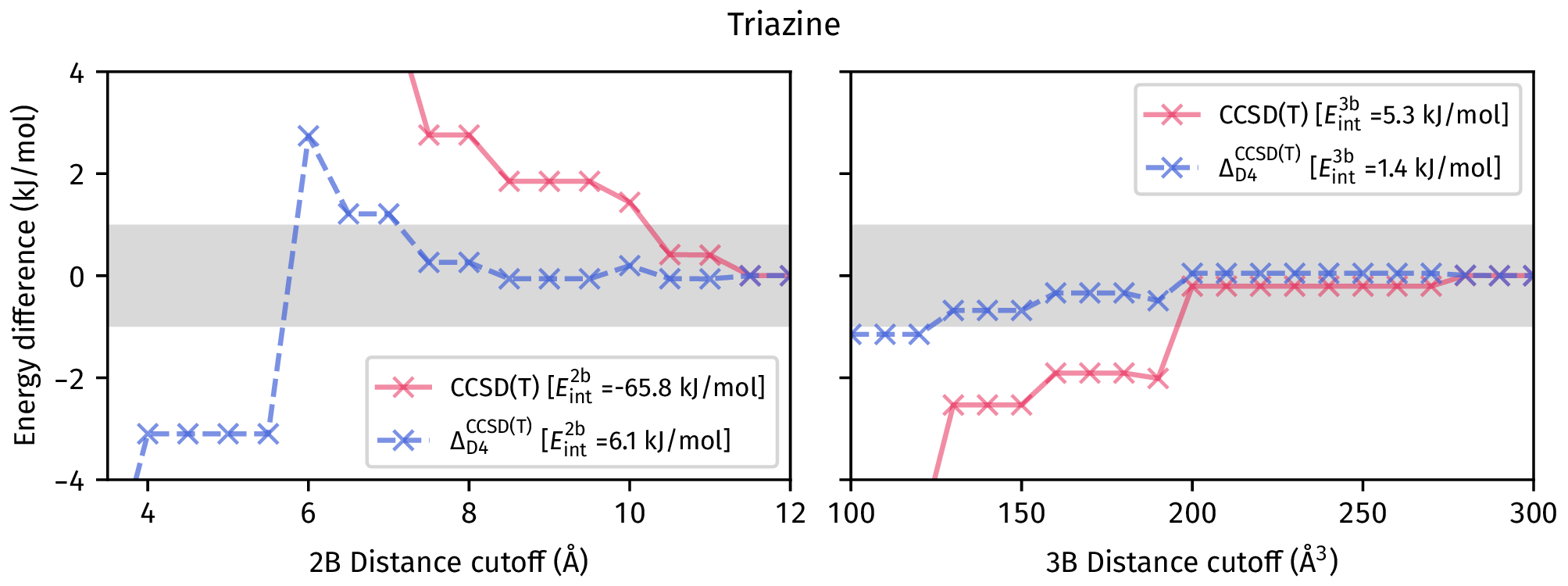}
\caption{\label{fig:x23_d4_conv_20} The convergence of the 2B and 3B MBE components to the $E_\text{latt}$ of triazine.}
\end{figure}

\begin{figure}[h!]
\includegraphics[width=\textwidth]{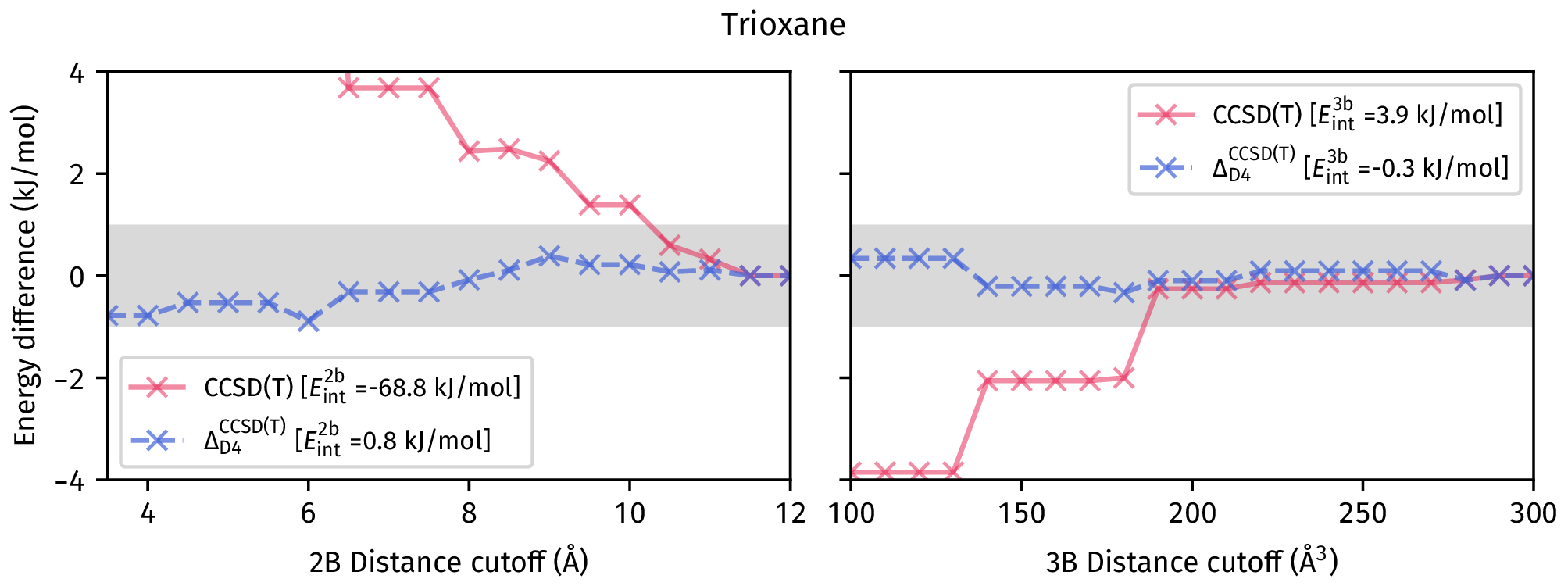}
\caption{\label{fig:x23_d4_conv_21} The convergence of the 2B and 3B MBE components to the $E_\text{latt}$ of trioxane.}
\end{figure}

\begin{figure}[h!]
\includegraphics[width=\textwidth]{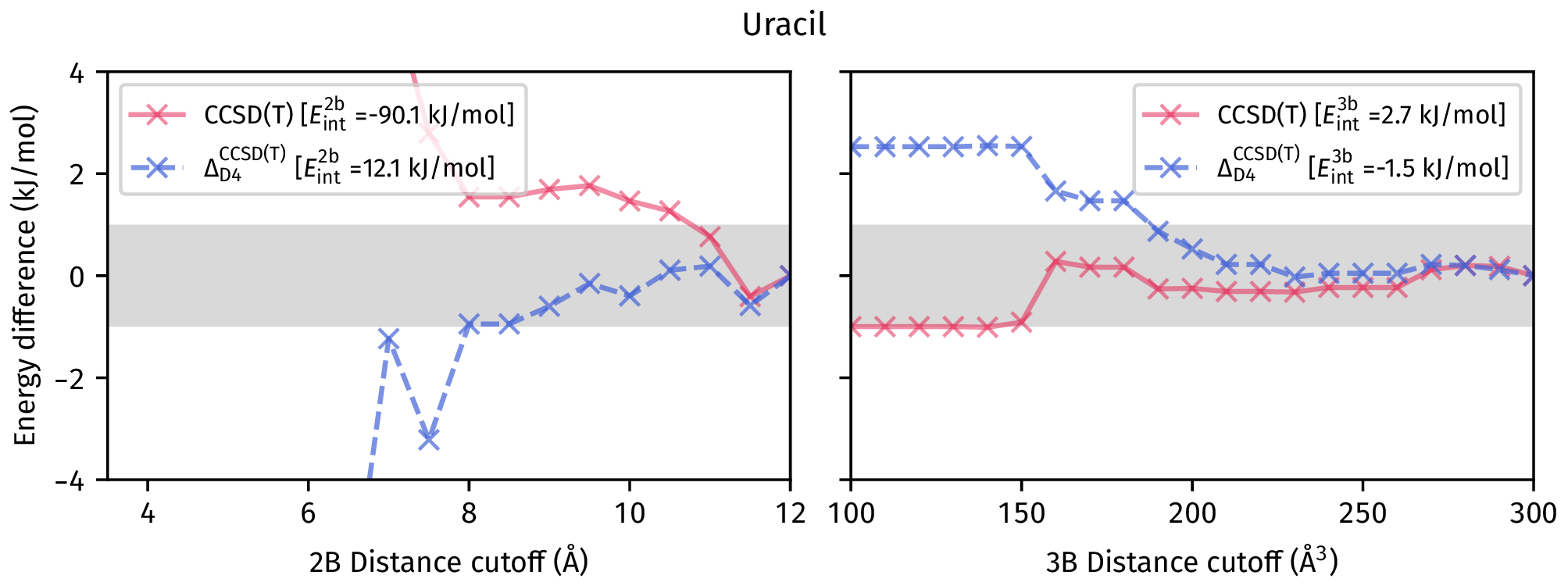}
\caption{\label{fig:x23_d4_conv_22} The convergence of the 2B and 3B MBE components to the $E_\text{latt}$ of uracil.}
\end{figure}

\begin{figure}[h!]
\includegraphics[width=\textwidth]{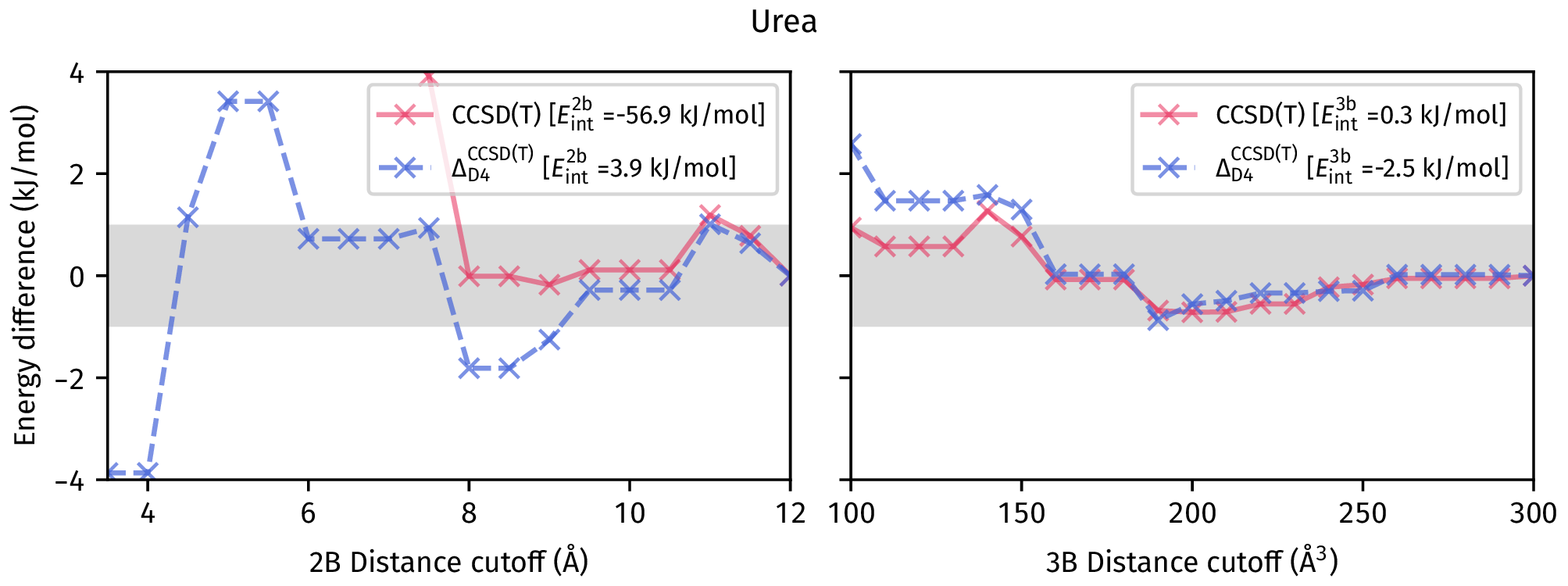}
\caption{\label{fig:x23_d4_conv_23} The convergence of the 2B and 3B MBE components to the $E_\text{latt}$ of urea.}
\end{figure}

In Figs.~\ref{fig:x23_d4_conv_01} to~\ref{fig:x23_d4_conv_23}, we illustrate the convergence of the 2B and 3B terms as a function of distance cutoffs relative to the converged value of both the CCSD(T) correlation energy and $\Delta_\text{D4}^\text{CCSD(T)}$ for all 23 molecules in the X23 dataset.
Here, we have used $R_{12} = 12\,$\AA{} and $R_{12}*R_{23}*R_{13} = 300\,$\AA{}$^3$ distance cutoffs as the converged value for the 2B and 3B terms, respectively.
It can be seen that $\Delta_\text{D4}^\text{CCSD(T)}$ converges faster with distance cutoff than the CCSD(T) correlation energy, allowing shorter truncation distances and fewer overall calculations.
For the latter, by a $9\,$\AA{} cutoff on the 2B terms, 18 out of 23 were converged to within $1\,$kJ/mol for $\Delta_\text{D4}^\text{CCSD(T)}$ while only 4 were converged for the CCSD(T) correlation energy.
Similarly, by a $200\,$\AA$^3$ cutoff on the 3B terms, 20 out of 23 were converged to within $1\,$kJ/mol with $\Delta_\text{D4}^\text{CCSD(T)}$ while 18 were converged for the CCSD(T) correlation energy.

Part of the faster convergence of $\Delta_\text{D4}^\text{CCSD(T)}$ arises from the fact that it has a smaller value than the CCSD(T) correlation energy.
For example, its lies in the range between -5.6 to $12.3\,$kJ/mol for the 2B terms, an order of magnitude smaller range than the correlation energy, which lies between -171.7 to -26.9 kJ/mol for the 2B terms.
The resulting decrease in required cutoff distance allows us to decrease the number of terms required.
For example, the number of 2B terms at $9\,$\AA{} is half that at $12\,$\AA{} for most systems in Table~\ref{tab:x23_2b_num_terms}, and there is a similar decrease in number of 3B terms between $250\,$\AA{}$^3$ and $300\,$\AA{}$^3$ distance cutoff in Table~\ref{tab:x23_3b_num_terms}.

A final outcome is that we now have two approaches that converge systematically towards the final result with distance cutoff, given a means to validate our results.
In Table~\ref{tab:x23_d4_compare} of Section~\ref{sec:x23}, we show that there is a small difference in the resulting lattice energy $E_\text{latt}$ at for 2B and 3B cutoffs of $12\,$\AA{} and $300\,$\AA{}$^3$, respectively, when using either the $\Delta_\text{D4}^\text{CCSD(T)}$ or CCSD(T) correlation energy in the MBE treatment.
Specifically, we find a mean-absolute-deviation (MAD) of $0.42\,$kJ/mol and root-mean-square-deviation of $0.79\,$kJ/mol, confirming the sub-kJ/mol convergence of the chosen distance cutoffs.

It is instructive to highlight again that this type of subtractive embedding between two levels of theory in the MBE has been explored by other authors.
For example, Sherrill and co-workers~\cite{ringerFirstPrinciplesComputation2008,sargentBenchmarkingTwobodyContributions2023a,xieAssessmentThreebodyDispersion2023a,nelsonConvergenceManybodyExpansion2024} have considered several variations of second-order M{\o}ller-Plesset perturbation theory.
Similarly, this has been explored by Beran and co-workers~\cite{beranApproximatingQuantumManybody2009,beranPredictingOrganicCrystal2010,wenAccurateMolecularCrystal2011a,wenPracticalQuantumMechanicsbased2012a} through their Hybrid Many-body Interaction (HMBI) approach.
Klimes and Modrzejewski~\cite{hofierkaBindingEnergiesMolecular2021,modrzejewskiRandomPhaseApproximation2020,modrzejewskiRandomPhaseApproximationManyBody2021,sytyMultiLevelCoupledClusterDescription2025} have explored subtractive embedding on top of variants of the random phase approximation (RPA).
Finally, the partitioning of the CCSD(T) total energy into the HF and correlation energy described in the preceding section is itself a type of subtractive embedding.
To our knowledge, no study has used the D4 dispersion (on top of HF) proposed within this work.

\clearpage

\subsection{Identifying symmetrically and translationally redundant terms}

\begin{figure}[h]
    \includegraphics[width=\textwidth]{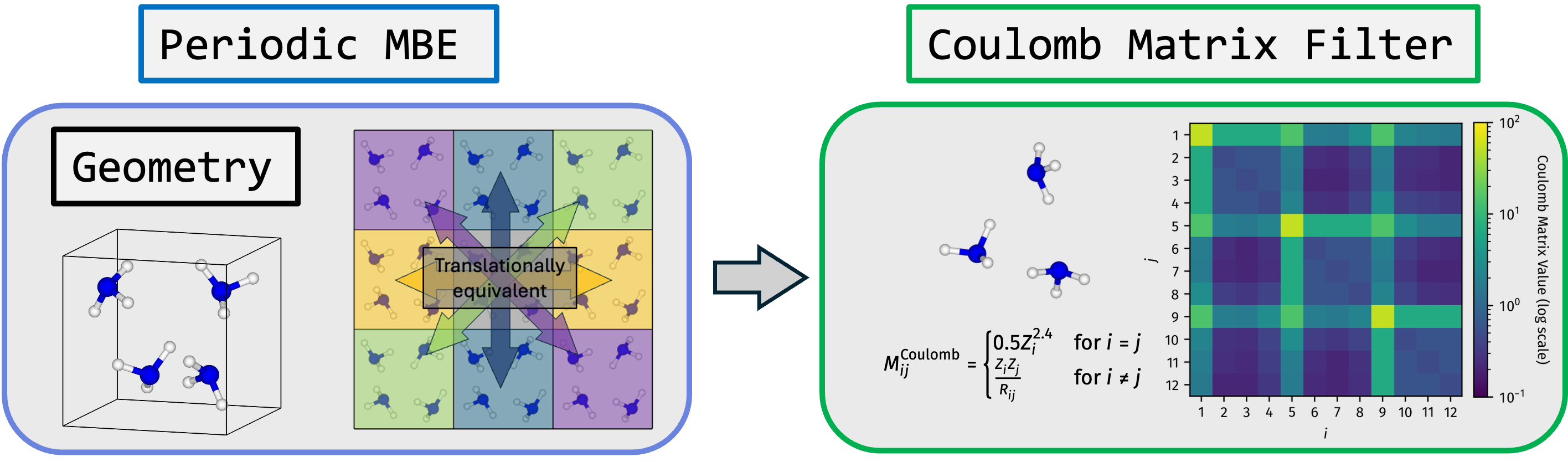}
    \caption{\label{fig:workflow}A schematic of the workflow to generate the dimers and trimers in the MBE. This entire workflow is implemented and automated in the pMBE Github code (available upon publication)}
\end{figure}

So far, we have dealt with physical approaches to remove or decrease the number of calculations required to reach convergence of the MBE.
Here, we will consider mathematical or machine-learning means to identify equivalent subsystems due to symmetry so that there are no redundant calculations.
As illustrated in Figure~\ref{fig:workflow}, within this work, we use the periodic MBE approach developed by~\citet{hermanFormulationManyBodyExpansion2023b} to remove terms which are translationally equivalent relative to a full enumeration of the MBE, followed by a Coulomb-matrix descriptor to further filter the remaining terms.
This entire process is automated within the pMBE code on Github that will be available upon publication.
In Tables~\ref{tab:x23_2b_num_terms} and~\ref{tab:x23_3b_num_terms}, we tabulate the cumulative number of 2B and 3B terms, respectively, as a function of distance cutoff.
For the former, we use the distance between the center of mass of the two molecules while in the latter, we use the product of the (three) pairwise distances between the three molecules.

\subsubsection{Periodic MBE}

The periodic MBE (shortened to pMBE) leverages the inherent translational symmetry of the periodic system. That said, the maximum reduction in the number of subsystems (in comparison to a full enumeration) is by a factor of $n$ where $n$ is the order of the expansion. For a unit cell of 1 molecule (i.e., hexamine), the reduction in the 2- and 3-body terms is exactly a factor of 2 and 3, respectively. When the $n$-body terms are enumerated for each monomer in the unit cell (as is necessary to account for the differing molecular environments in the unit cells of several ice polymorphs), the factor that the terms are reduced are slightly less than $n$ due to the dimers, trimers, etc. that exist within the same unit cell. Naturally, for the case where the $n$-body terms are enumerated for a single molecule in the unit cell, the savings are dependent on the size of the unit cell. For larger unit cells, the translational symmetry cannot be leveraged as effectively, leading to a smaller reduction of subsystems in comparison to smaller unit cells.

As input, the code requires a unit cell structure, unit cell vectors, and the cutoff distance for the many-body term of interest. It typically takes a fraction of a second on a laptop to enumerate the subsystems necessary to compute the $n$-body component of the lattice energy. The output of the pMBE code is an SQLite3 database which, for each body term, provides: (1) the Atomic Simulation Environment \texttt{Atoms} object (containing the atomic elements and positions) (2) cutoff distance and (3) weight contribution to the lattice energy.

\subsubsection{\label{sec:coulomb_matrix}Coulomb-matrix filter}

The Coulomb-matrix filter can be used to further decrease the number of subsystems taken from the pMBE, decreasing the number of 2B terms by half and in some highly symmetric systems, the number of terms are decreased significantly further.
For example, in hexamine at a 2B cutoff of $12\,$\AA{}, the pMBE approach decreases the number of 2B terms from 50 to 25, with the Coulomb-matrix filter further decreasing this down to 4.
The cost savings are even more prominent for 3B terms, where with respect to a full enumeration, the Coulomb matrix filter decreases the number of terms 3B terms within a $300\,$\AA{}$^3$ cutoff for CO$_2$ from 1050 down to 56.
We note that this approach has also been explored by the CrystaLattE~\cite{borcaCrystaLattEAutomatedComputation2019b} package from Sherrill and co-workers.

The Coulomb matrix for the combined 2B, 3B (or more) system is defined as:
\begin{equation}
    M^{\text{Coulomb}}_{ij} = 
\begin{cases} 
0.5 Z_i^{2.4} & \text{for } i = j \\
\frac{Z_i Z_j}{R_{ij}} & \text{for } i \ne j
\end{cases},
\end{equation}
where $i$ and $j$ are the indices of two atoms.
We sort the row and columns of the matrix by its $l^2$-norm to ensure permutation symmetry, before flattening the matrix into a one-dimensional array (called the Coulomb array).
We use the DScribe Python package~\cite{himanenDScribeLibraryDescriptors2020,laaksoUpdatesDScribeLibrary2023} to generate these Coulomb arrays.
We further regularize the Coulomb matrix by adding the distance cutoff ($R_{12}$ and $R_{12}*R_{23}*R_{13}$ for 2B and 3B systems, respectively, where $R_{12}$ is the distance between the centers of masses of monomers 1 and 2) to the front of this array.
Finally, we compare and group the Coulomb arrays using the Density-Based Spatial Clustering of Applications with Noise (DBSCAN) approach~\cite{schubertDBSCANRevisitedRevisited2017}, as provided by the \texttt{scikit-learn} package in Python.
Below, we provide an example of the Python code to apply the Coulomb Matrix Filter to filter out the ASE database generated by the pMBE code:
\newpage

\LTcapwidth=\textwidth

\begin{longtable}{lrrrrrrrrr}
\caption{\label{tab:x23_2b_num_terms}The number of two-body (2B) terms as a function of distance between the center of mass of the two molecules that make up the dimer subsystem.} \\

\toprule
2B distance (\AA) &  & 5 & 6 & 7 & 8 & 9 & 10 & 11 & 12 \\
\midrule
\endfirsthead

\caption[]{(continued)} \\
\endhead

\multicolumn{10}{r}{{Continued on next page}} \\
\endfoot

\bottomrule
\endlastfoot

\multirow[c]{3}{*}{1,4-cyclohexanedione} & Full Enumeration & 2 & 6 & 14 & 14 & 16 & 26 & 46 & 50 \\
 & pMBE & 2 & 6 & 11 & 11 & 12 & 18 & 35 & 39 \\
 & pMBE + Coulomb & 1 & 3 & 7 & 7 & 8 & 13 & 23 & 25 \\
\cline{1-10}
\multirow[c]{3}{*}{Acetic Acid} & Full Enumeration & 6 & 12 & 22 & 30 & 36 & 62 & 76 & 100 \\
 & pMBE & 5 & 10 & 18 & 25 & 31 & 55 & 69 & 91 \\
 & pMBE + Coulomb & 3 & 6 & 11 & 15 & 18 & 31 & 38 & 49 \\
\cline{1-10}
\multirow[c]{3}{*}{Adamantane} & Full Enumeration & 0 & 0 & 12 & 12 & 14 & 18 & 18 & 42 \\
 & pMBE & 0 & 0 & 10 & 10 & 11 & 13 & 13 & 33 \\
 & pMBE + Coulomb & 0 & 0 & 2 & 2 & 3 & 4 & 4 & 7 \\
\cline{1-10}
\multirow[c]{3}{*}{Ammonia} & Full Enumeration & 12 & 24 & 42 & 60 & 86 & 128 & 158 & 212 \\
 & pMBE & 12 & 21 & 39 & 51 & 73 & 115 & 142 & 184 \\
 & pMBE + Coulomb & 2 & 4 & 7 & 9 & 13 & 18 & 22 & 27 \\
\cline{1-10}
\multirow[c]{3}{*}{Anthracene} & Full Enumeration & 0 & 6 & 6 & 6 & 8 & 18 & 22 & 34 \\
 & pMBE & 0 & 5 & 5 & 5 & 6 & 15 & 17 & 25 \\
 & pMBE + Coulomb & 0 & 2 & 2 & 2 & 3 & 6 & 8 & 13 \\
\cline{1-10}
\multirow[c]{3}{*}{Benzene} & Full Enumeration & 0 & 12 & 14 & 16 & 16 & 34 & 50 & 66 \\
 & pMBE & 0 & 12 & 13 & 14 & 14 & 31 & 45 & 57 \\
 & pMBE + Coulomb & 0 & 3 & 4 & 5 & 5 & 10 & 15 & 21 \\
\cline{1-10}
\multirow[c]{3}{*}{CO$_2$} & Full Enumeration & 12 & 18 & 42 & 54 & 78 & 86 & 134 & 176 \\
 & pMBE & 12 & 15 & 39 & 45 & 69 & 73 & 121 & 160 \\
 & pMBE + Coulomb & 1 & 2 & 4 & 6 & 8 & 10 & 14 & 17 \\
\cline{1-10}
\multirow[c]{3}{*}{Cyanamide} & Full Enumeration & 9 & 16 & 29 & 42 & 53 & 81 & 104 & 136 \\
 & pMBE & 9 & 16 & 27 & 40 & 51 & 76 & 99 & 127 \\
 & pMBE + Coulomb & 5 & 9 & 16 & 23 & 29 & 42 & 55 & 70 \\
\cline{1-10}
\multirow[c]{3}{*}{Cytosine} & Full Enumeration & 2 & 6 & 10 & 20 & 24 & 32 & 48 & 58 \\
 & pMBE & 1 & 5 & 9 & 18 & 22 & 29 & 43 & 52 \\
 & pMBE + Coulomb & 1 & 3 & 5 & 10 & 12 & 16 & 24 & 29 \\
\cline{1-10}
\multirow[c]{3}{*}{Ethyl carbamate} & Full Enumeration & 3 & 5 & 8 & 20 & 22 & 30 & 48 & 60 \\
 & pMBE & 3 & 4 & 7 & 15 & 17 & 23 & 34 & 45 \\
 & pMBE + Coulomb & 3 & 4 & 7 & 15 & 17 & 23 & 34 & 45 \\
\cline{1-10}
\multirow[c]{3}{*}{Formamide} & Full Enumeration & 7 & 16 & 27 & 39 & 53 & 73 & 100 & 134 \\
 & pMBE & 6 & 15 & 25 & 35 & 48 & 64 & 89 & 116 \\
 & pMBE + Coulomb & 5 & 11 & 17 & 23 & 32 & 45 & 62 & 85 \\
\cline{1-10}
\multirow[c]{3}{*}{Hexamine} & Full Enumeration & 0 & 0 & 14 & 14 & 14 & 26 & 26 & 50 \\
 & pMBE & 0 & 0 & 7 & 7 & 7 & 13 & 13 & 25 \\
 & pMBE + Coulomb & 0 & 0 & 2 & 2 & 2 & 3 & 3 & 4 \\
\cline{1-10}
\multirow[c]{3}{*}{Imidazole} & Full Enumeration & 5 & 11 & 17 & 22 & 31 & 48 & 66 & 85 \\
 & pMBE & 5 & 10 & 16 & 20 & 29 & 42 & 57 & 74 \\
 & pMBE + Coulomb & 3 & 7 & 11 & 14 & 20 & 30 & 41 & 51 \\
\cline{1-10}
\multirow[c]{3}{*}{Naphthalene} & Full Enumeration & 0 & 6 & 6 & 12 & 16 & 24 & 36 & 46 \\
 & pMBE & 0 & 5 & 5 & 10 & 12 & 18 & 26 & 35 \\
 & pMBE + Coulomb & 0 & 2 & 2 & 4 & 6 & 9 & 14 & 17 \\
\cline{1-10}
\multirow[c]{3}{*}{Oxalic Acid $\alpha$} & Full Enumeration & 8 & 12 & 16 & 26 & 38 & 58 & 70 & 86 \\
 & pMBE & 8 & 12 & 14 & 23 & 33 & 51 & 61 & 73 \\
 & pMBE + Coulomb & 2 & 3 & 5 & 8 & 12 & 18 & 22 & 28 \\
\cline{1-10}
\multirow[c]{3}{*}{Oxalic Acid $\beta$} & Full Enumeration & 4 & 14 & 16 & 20 & 40 & 50 & 70 & 92 \\
 & pMBE & 4 & 11 & 12 & 16 & 30 & 37 & 53 & 70 \\
 & pMBE + Coulomb & 1 & 5 & 6 & 7 & 15 & 19 & 26 & 34 \\
\cline{1-10}
\multirow[c]{3}{*}{Pyrazine} & Full Enumeration & 2 & 12 & 16 & 26 & 26 & 32 & 52 & 80 \\
 & pMBE & 1 & 10 & 12 & 21 & 21 & 24 & 42 & 60 \\
 & pMBE + Coulomb & 1 & 3 & 5 & 7 & 7 & 10 & 13 & 20 \\
\cline{1-10}
\multirow[c]{3}{*}{Pyrazole} & Full Enumeration & 5 & 12 & 16 & 16 & 34 & 48 & 68 & 72 \\
 & pMBE & 5 & 12 & 15 & 15 & 32 & 46 & 64 & 68 \\
 & pMBE + Coulomb & 4 & 10 & 13 & 13 & 24 & 36 & 51 & 54 \\
\cline{1-10}
\multirow[c]{3}{*}{Succinic Acid} & Full Enumeration & 0 & 8 & 8 & 20 & 26 & 26 & 46 & 64 \\
 & pMBE & 0 & 8 & 8 & 20 & 25 & 25 & 43 & 61 \\
 & pMBE + Coulomb & 0 & 3 & 3 & 7 & 9 & 9 & 17 & 25 \\
\cline{1-10}
\multirow[c]{3}{*}{Triazine} & Full Enumeration & 2 & 8 & 14 & 22 & 28 & 34 & 54 & 72 \\
 & pMBE & 2 & 8 & 14 & 21 & 27 & 30 & 50 & 68 \\
 & pMBE + Coulomb & 1 & 2 & 3 & 5 & 6 & 7 & 10 & 13 \\
\cline{1-10}
\multirow[c]{3}{*}{Trioxane} & Full Enumeration & 2 & 8 & 14 & 20 & 28 & 34 & 58 & 64 \\
 & pMBE & 2 & 8 & 14 & 20 & 27 & 30 & 54 & 60 \\
 & pMBE + Coulomb & 1 & 2 & 3 & 4 & 6 & 7 & 11 & 12 \\
\cline{1-10}
\multirow[c]{3}{*}{Uracil} & Full Enumeration & 2 & 4 & 13 & 23 & 25 & 30 & 47 & 63 \\
 & pMBE & 1 & 3 & 12 & 21 & 23 & 28 & 42 & 56 \\
 & pMBE + Coulomb & 1 & 3 & 8 & 15 & 16 & 19 & 30 & 40 \\
\cline{1-10}
\multirow[c]{3}{*}{Urea} & Full Enumeration & 10 & 14 & 14 & 30 & 42 & 60 & 76 & 96 \\
 & pMBE & 9 & 11 & 11 & 21 & 33 & 46 & 58 & 76 \\
 & pMBE + Coulomb & 3 & 4 & 4 & 8 & 10 & 14 & 16 & 20 \\
\cline{1-10}
\end{longtable}
\LTcapwidth=\textwidth

\begin{longtable}{lrrrrrrrr}
\caption{\label{tab:x23_3b_num_terms}The number of three-body (3B) terms as a function of the product of the pairwise distances between the center of mass of the three molecules that make up the trimer subsystem.} \\

\toprule
3B distance (\AA\textsuperscript{3}) &  & 120 & 150 & 180 & 210 & 240 & 270 & 300 \\
\midrule
\endfirsthead

\caption[]{(continued)} \\
\endhead

\multicolumn{9}{r}{{Continued on next page}} \\
\endfoot

\bottomrule
\endlastfoot

\multirow[c]{3}{*}{1,4-cyclohexanedione} & Full Enumeration & 0 & 0 & 3 & 15 & 27 & 48 & 60 \\
 & pMBE & 0 & 0 & 2 & 10 & 18 & 32 & 40 \\
 & pMBE + Coulomb & 0 & 0 & 1 & 5 & 9 & 16 & 20 \\
\cline{1-9}
\multirow[c]{3}{*}{Acetic Acid} & Full Enumeration & 30 & 45 & 87 & 135 & 177 & 204 & 279 \\
 & pMBE & 20 & 31 & 59 & 97 & 131 & 150 & 205 \\
 & pMBE + Coulomb & 10 & 15 & 30 & 47 & 64 & 76 & 106 \\
\cline{1-9}
\multirow[c]{3}{*}{Adamantane} & Full Enumeration & 0 & 0 & 0 & 0 & 0 & 0 & 24 \\
 & pMBE & 0 & 0 & 0 & 0 & 0 & 0 & 16 \\
 & pMBE + Coulomb & 0 & 0 & 0 & 0 & 0 & 0 & 2 \\
\cline{1-9}
\multirow[c]{3}{*}{Ammonia} & Full Enumeration & 195 & 321 & 519 & 846 & 1125 & 1314 & 1683 \\
 & pMBE & 165 & 261 & 414 & 666 & 897 & 1062 & 1353 \\
 & pMBE + Coulomb & 19 & 28 & 43 & 69 & 91 & 108 & 136 \\
\cline{1-9}
\multirow[c]{3}{*}{Anthracene} & Full Enumeration & 0 & 0 & 6 & 6 & 12 & 12 & 18 \\
 & pMBE & 0 & 0 & 4 & 4 & 8 & 8 & 12 \\
 & pMBE + Coulomb & 0 & 0 & 1 & 1 & 2 & 2 & 4 \\
\cline{1-9}
\multirow[c]{3}{*}{Benzene} & Full Enumeration & 0 & 0 & 30 & 36 & 42 & 78 & 102 \\
 & pMBE & 0 & 0 & 28 & 32 & 36 & 68 & 92 \\
 & pMBE + Coulomb & 0 & 0 & 5 & 6 & 7 & 14 & 18 \\
\cline{1-9}
\multirow[c]{3}{*}{CO$_2$} & Full Enumeration & 132 & 150 & 222 & 366 & 510 & 834 & 1050 \\
 & pMBE & 120 & 132 & 180 & 300 & 396 & 648 & 840 \\
 & pMBE + Coulomb & 8 & 10 & 14 & 22 & 29 & 45 & 56 \\
\cline{1-9}
\multirow[c]{3}{*}{Cyanamide} & Full Enumeration & 87 & 123 & 198 & 261 & 393 & 540 & 696 \\
 & pMBE & 84 & 120 & 187 & 249 & 365 & 495 & 639 \\
 & pMBE + Coulomb & 26 & 38 & 54 & 71 & 110 & 148 & 188 \\
\cline{1-9}
\multirow[c]{3}{*}{Cytosine} & Full Enumeration & 9 & 12 & 24 & 42 & 48 & 60 & 93 \\
 & pMBE & 5 & 7 & 15 & 27 & 33 & 43 & 67 \\
 & pMBE + Coulomb & 3 & 4 & 8 & 14 & 16 & 20 & 31 \\
\cline{1-9}
\multirow[c]{3}{*}{Ethyl carbamate} & Full Enumeration & 6 & 15 & 21 & 24 & 39 & 66 & 84 \\
 & pMBE & 4 & 10 & 14 & 16 & 26 & 43 & 53 \\
 & pMBE + Coulomb & 2 & 5 & 7 & 8 & 13 & 22 & 28 \\
\cline{1-9}
\multirow[c]{3}{*}{Formamide} & Full Enumeration & 63 & 102 & 165 & 237 & 333 & 435 & 555 \\
 & pMBE & 48 & 80 & 135 & 194 & 268 & 349 & 442 \\
 & pMBE + Coulomb & 21 & 34 & 55 & 78 & 108 & 142 & 179 \\
\cline{1-9}
\multirow[c]{3}{*}{Hexamine} & Full Enumeration & 0 & 0 & 0 & 0 & 0 & 36 & 36 \\
 & pMBE & 0 & 0 & 0 & 0 & 0 & 12 & 12 \\
 & pMBE + Coulomb & 0 & 0 & 0 & 0 & 0 & 1 & 1 \\
\cline{1-9}
\multirow[c]{3}{*}{Imidazole} & Full Enumeration & 6 & 30 & 57 & 84 & 117 & 150 & 168 \\
 & pMBE & 5 & 27 & 49 & 72 & 99 & 129 & 141 \\
 & pMBE + Coulomb & 2 & 10 & 19 & 28 & 38 & 49 & 55 \\
\cline{1-9}
\multirow[c]{3}{*}{Naphthalene} & Full Enumeration & 0 & 6 & 6 & 12 & 12 & 18 & 30 \\
 & pMBE & 0 & 4 & 4 & 8 & 8 & 12 & 20 \\
 & pMBE + Coulomb & 0 & 1 & 1 & 2 & 2 & 4 & 6 \\
\cline{1-9}
\multirow[c]{3}{*}{Oxalic Acid $\alpha$} & Full Enumeration & 24 & 42 & 78 & 114 & 138 & 210 & 234 \\
 & pMBE & 24 & 36 & 68 & 100 & 124 & 172 & 196 \\
 & pMBE + Coulomb & 4 & 7 & 14 & 20 & 24 & 38 & 42 \\
\cline{1-9}
\multirow[c]{3}{*}{Oxalic Acid $\beta$} & Full Enumeration & 12 & 42 & 60 & 114 & 138 & 192 & 246 \\
 & pMBE & 8 & 28 & 38 & 74 & 90 & 116 & 146 \\
 & pMBE + Coulomb & 2 & 8 & 11 & 20 & 24 & 32 & 41 \\
\cline{1-9}
\multirow[c]{3}{*}{Pyrazine} & Full Enumeration & 3 & 15 & 51 & 63 & 75 & 87 & 87 \\
 & pMBE & 1 & 9 & 29 & 37 & 45 & 49 & 49 \\
 & pMBE + Coulomb & 1 & 2 & 6 & 7 & 9 & 11 & 11 \\
\cline{1-9}
\multirow[c]{3}{*}{Pyrazole} & Full Enumeration & 6 & 36 & 42 & 93 & 109 & 138 & 184 \\
 & pMBE & 6 & 34 & 40 & 88 & 102 & 130 & 172 \\
 & pMBE + Coulomb & 4 & 23 & 27 & 57 & 66 & 86 & 115 \\
\cline{1-9}
\multirow[c]{3}{*}{Succinic Acid} & Full Enumeration & 0 & 6 & 6 & 36 & 60 & 69 & 105 \\
 & pMBE & 0 & 6 & 6 & 36 & 58 & 66 & 102 \\
 & pMBE + Coulomb & 0 & 1 & 1 & 6 & 10 & 13 & 19 \\
\cline{1-9}
\multirow[c]{3}{*}{Triazine} & Full Enumeration & 3 & 21 & 39 & 75 & 75 & 75 & 117 \\
 & pMBE & 2 & 20 & 38 & 74 & 74 & 74 & 114 \\
 & pMBE + Coulomb & 1 & 2 & 3 & 6 & 6 & 6 & 10 \\
\cline{1-9}
\multirow[c]{3}{*}{Trioxane} & Full Enumeration & 0 & 21 & 39 & 57 & 75 & 75 & 111 \\
 & pMBE & 0 & 20 & 38 & 56 & 74 & 74 & 104 \\
 & pMBE + Coulomb & 0 & 3 & 5 & 9 & 10 & 10 & 16 \\
\cline{1-9}
\multirow[c]{3}{*}{Uracil} & Full Enumeration & 6 & 12 & 30 & 45 & 63 & 70 & 106 \\
 & pMBE & 3 & 7 & 19 & 30 & 44 & 52 & 75 \\
 & pMBE + Coulomb & 2 & 4 & 16 & 23 & 31 & 35 & 50 \\
\cline{1-9}
\multirow[c]{3}{*}{Urea} & Full Enumeration & 24 & 42 & 78 & 135 & 183 & 219 & 255 \\
 & pMBE & 16 & 28 & 52 & 81 & 113 & 133 & 157 \\
 & pMBE + Coulomb & 3 & 6 & 10 & 19 & 23 & 27 & 31 \\
\cline{1-9}
\end{longtable}

\subsection{\label{sec:opt_lno}Optimizing the local natural orbital approximation for the MBE}

A key development within this work is the optimization of some of the approximations in the local natural orbital methods to further improve its application to the three-body terms in the many-body expansion.
The correlation energy contribution of the three-body terms and the LNO errors in them  are individually very small.
However, adding a large number of these terms and the overall very high accuracy goals of this project motivated improvements targeting the 3B contributions.

\begin{figure}[h]
    \includegraphics{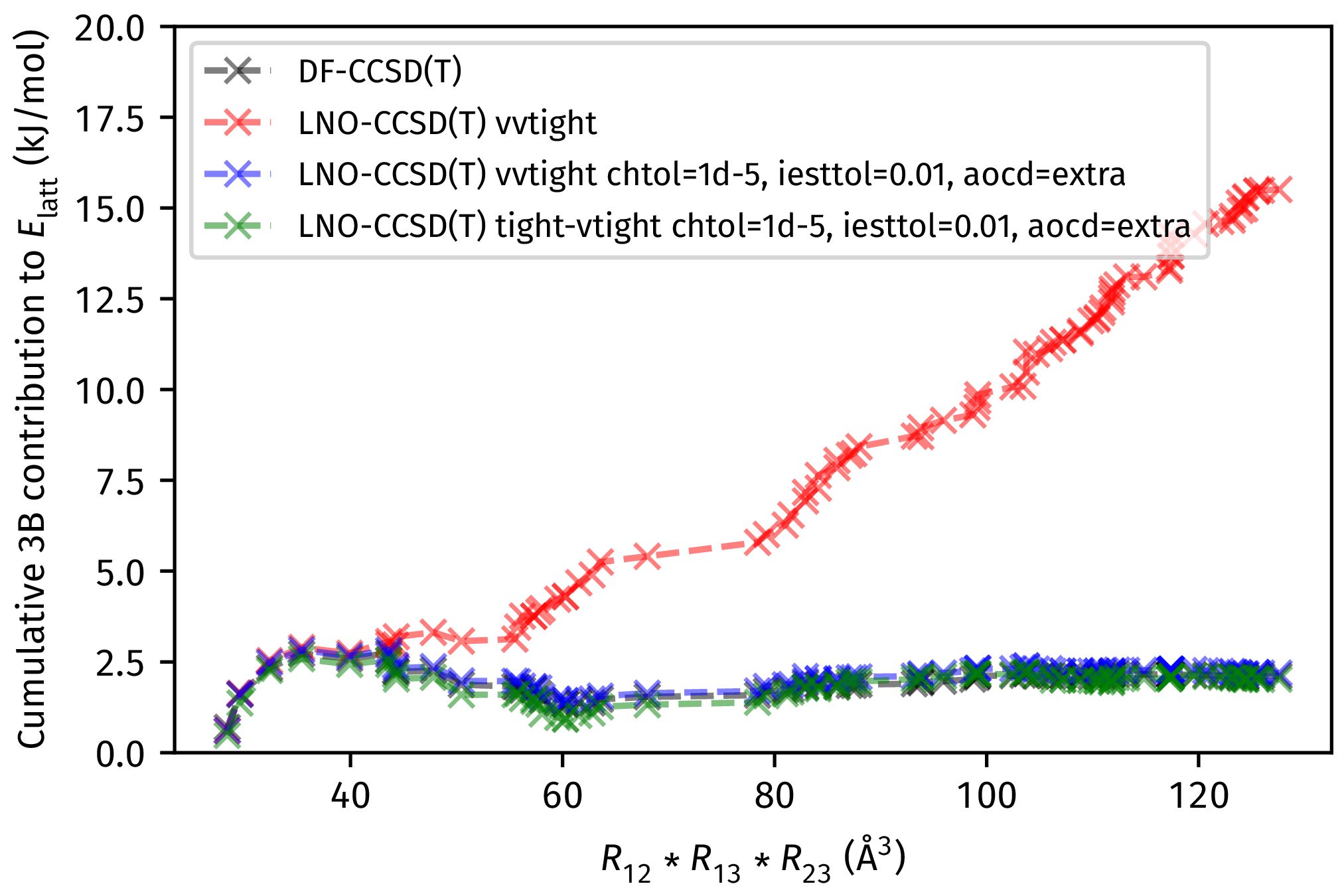}
    \caption{\label{fig:3b_err_buildup} LNO error in the three-body term to ice VIII for the aug-cc-pVDZ basis set compared to the canonical density-fitted (DF-)CCSD(T) reference.}
\end{figure}

As shown in Figure~\ref{fig:3b_err_buildup}, up to medium range of 40--50\AA$^3$, previous LNO-CCSD(T) results also provided accurate cumulative 3B contributions compared to approximation free DF-CCSD(T) references. 
At this point, the cumulative 3B contributions are also converged well within 1~kJ/mol due to the rapid decay of these terms beyond 40--50~\AA$^3$.
Going beyond that, the previous LNO-CCSD(T) results exhibited increasing errors originating from the buildup of small numerical noise from the rapidly increasing number of 3B contributions.
Detailed investigation revealed that this behavior originates from two usually very reliable approximations. 

First, the accuracy of the Cholesky-decomposition factorization of MP2 denominations has an effect on the the local MP2 part of the correlation energies, that is usually very small and cancels excellently upon formation of energy differences.\cite{nagyIntegralDirectLinearScalingSecondOrder2016a,szaboLinearScalingOpenShellMP22021} 
However, for the high relative accuracy requirement posed by the small 3B terms, this approximation was tightened via setting {\tt chtol=1d-5}.
In the updated LNO-CCSD(T) implementation, this setting is automatically adjusted, as it is tied to the {\tt lcorthr} keyword governing the local correlation threshold settings.\cite{mrcceng3} 

The second adjustment is related to the use of counterpoise corrections to increase the basis set convergence. For example for water 3B terms, ghost AOs are put on the position of some of the water atom centers when computing dimer and monomer contributions.
In this ghost positions the density of electrons becomes particularly small, especially with increasing trimer distances. 
The composition of the AO basis set used for expanding the localized orbitals are designed to introduce cost saving approximations, when we recognize the very small population of the localized orbitals on the some of the ghost AO centers. 
Previously, we extended this approach by considering the expected size of also the AO integrals on these ghost AO centers as an additional criteria.\cite{nagyPursuingBasisSet2023a}
This criteria is adjusted to be more precise via the {\tt iesttol=0.01} and {\tt aocd=extra} settings in the updated LNO-CCSD(T) version.~\cite{mesterOverviewDevelopmentsMRCC2025,mrcceng3}

An additional benefit of these developments is that the LNO thresholds ({\tt lcorthr}) required for convergence can be loosened.
In particular, we find that {\tt Tight} or {\tt vTight} settings are sufficient for converging two- and three-body terms, in contrast to the {\tt vvTight} settings used in previous work~\cite{sytyMultiLevelCoupledClusterDescription2025}.
For lattice energies in the X23 and ICE13 datasets, we extrapolate from {\tt Tight} to {\tt vTight}, while relative energies of pharmaceutical polymorphs require only {\tt Tight}, discussed in the next section.
As shown in Table~\ref{tab:anthracene_lno_costs}, these choices lead to substantial cost savings.
For example, {\tt Tight} calculations are roughly an order of magnitude faster and require $5{-}6\,\times$less memory than {\tt vvTight} for anthracene dimer and trimer calculations, making them far more affordable and suitable for commodity hardware.

\begin{table}[h]

\caption{\label{tab:anthracene_lno_costs}Computational walltime and memory requirements for LNO-CCSD(T) calculations on a dimer and trimer of anthracene with various LNO thresholds. These estimates are given for a calculation involving 32 cores on an Intel Icelake node with 400 GB of available memory.}
\begin{tabular}{lrrrrrrr}
\toprule
 &  & Loose & Normal & Tight & vTight & vvTight \\ 
System & Metric &  &  &  &  &  \\
\midrule
\multirow[c]{2}{*}{Dimer} & Walltime (min) & 121 & 233 & 763 & 2473 & 7005 \\
 & Memory (Mb) & 1328 & 3517 & 10696 & 26351 & 52202 \\
\cline{1-7}
\multirow[c]{2}{*}{Trimer} & Walltime (min) & 465 & 840 & 2595 & 9493 & 33822 \\
 & Memory (Mb) & 1524 & 4652 & 15618 & 42048 & 90182 \\
\cline{1-7}
\bottomrule
\end{tabular}

\end{table}

\newpage
\section{Cost-efficient electronic structure parameters}

It is important to ensure that the electronic structure parameters in the final LNO-CCSD(T) calculations are sufficiently accurate to reach e.g., $1\,$kJ/mol accuracy while being efficient enough to study large pharmaceutical-sized molecular crystals.
In this section, we will study the effect of the basis set and local correlation thresholds.
The required settings depend on the type of calculations, where the higher body-order terms in the MBE generally require smaller basis sets and the calculation of relative energies, as opposed to lattice energies, also allows for smaller basis sets and local thresholds.

\subsection{Lattice energy}

\subsubsection{Basis set dependence}

We show the convergence of the 2B contribution to $E_\text{latt}$ for the aug-cc-pV$X$Z set of basis sets for $X=$ D, T and Q, and also consider two-point complete basis set (CBS) extrapolations for the D/T and T/Q combinations in Table~\ref{tab:basis_effect}.
This was all performed with the Boys and Bernardi counterpoise corrections~\cite{boysCalculationSmallMolecular1970}. Taking the CBS(aug-cc-pVDZ/aug-cc-pVTZ) as the (most converged) reference, there is a strong basis set dependence of the 2B terms, where aug-cc-pVDZ is more than 10 kJ/mol away, while CBS(aug-cc-pVDZ/aug-cc-pVTZ) is sufficient to be within 0.6 kJ/mol of the CBS(aug-cc-pVTZ/aug-cc-pVQZ).

\begin{table}[h]

\caption{\label{tab:basis_effect}Effect of basis set to the two-body contribution of the $E_\text{latt}$ of ice VIII.}
\begin{tabular}{lr}
\toprule
 & 2B contribution to $E_\text{latt}$ (kJ/mol) \\ 
\midrule
CCSD(T) aug-cc-pVDZ & -20.7 \\
CCSD(T) aug-cc-pVTZ & -27.6 \\
CCSD(T) aug-cc-pVQZ & -29.6 \\
CCSD(T) CBS(aug-cc-pVDZ/aug-cc-pVTZ) & -31.6 \\
CCSD(T) CBS(aug-cc-pVTZ/aug-cc-pVQZ) & -31.0 \\
\bottomrule
\end{tabular}

\end{table}

In Table~\ref{tab:2b_3b_basis_conv}, we show the basis set convergence for 12 ice phases.
It confirms the weak basis set dependence of 3B terms, with aug-cc-pVDZ being only $0.1\,$kJ/mol away from the CBS estimates.
It also highlights the necessity of performing basis set extrapolations for 2B terms, as even the QZ basis set is off by $1.5\,$kJ/mol on average.

\begin{table}[h]

\caption{\label{tab:2b_3b_basis_conv}Three-body contributions have a significantly smaller basis set dependence than two-body contributions to the $E_\text{latt}$ of ice phases. These tests are for the aug-cc-pVXZ (X = D, T, Q) basis sets and their CBS, T/Q for 2B and D/T for 3B terms, extrapolations.}
\begin{tabular}{lrrrrrr}
\toprule
 & \multicolumn{3}{c}{2B} & \multicolumn{3}{c}{3B} \\ \cmidrule(lr){2-4} \cmidrule(lr){5-7} 
 & TZ & QZ & CBS & DZ & TZ & CBS \\
\midrule
Ih & -20.3 & -22.5 & -24.0 & 0.1 & 0.1 & 0.1 \\
II & -23.1 & -25.2 & -26.7 & 1.6 & 1.7 & 1.7 \\
III & -22.2 & -24.4 & -26.0 & 0.6 & 0.7 & 0.7 \\
IV & -24.1 & -26.1 & -27.6 & 1.3 & 1.4 & 1.5 \\
VI & -25.3 & -27.5 & -29.0 & 1.8 & 1.9 & 1.9 \\
VIII & -27.6 & -29.6 & -31.0 & 2.3 & 2.4 & 2.5 \\
IX & -21.4 & -23.6 & -25.1 & 1.0 & 1.0 & 1.1 \\
XI & -20.7 & -23.0 & -24.6 & 0.0 & 0.0 & 0.0 \\
XIII & -23.3 & -25.5 & -27.0 & 1.3 & 1.4 & 1.5 \\
XIV & -24.4 & -26.6 & -28.1 & 1.7 & 1.8 & 1.9 \\
XV & -24.8 & -26.9 & -28.4 & 1.7 & 1.8 & 1.9 \\
XVII & -19.7 & -21.9 & -23.5 & 0.2 & 0.3 & 0.3 \\
\midrule MAD & 3.7 & 1.5 & 0.0 & 0.1 & 0.0 & 0.0 \\
\bottomrule
\end{tabular}

\end{table}

In Table~\ref{tab:benzene_basis_conv}, we show the basis set convergence for the benzene $E_\text{latt}$ within the X23 dataset.
It can be seen that the basis set dependence is much weaker, where the aug-cc-pVTZ basis set is within $1\,$kJ/mol away from the CBS limit, with a similar weak basis set dependence for the 3B terms, showing no change to less than $0.1\,$kJ/mol for aug-cc-pVDZ.

\begin{table}[h]

\caption{\label{tab:benzene_basis_conv}Convergence of 2B with basis set: aug-cc-pVTZ, aug-cc-pVQZ and CBS(aug-cc-pVTZ/aug-cc-pVQZ) as well as 3B with basis set: aug-cc-pVDZ, aug-cc-pVTZ and CBS(aug-cc-pVDZ/aug-cc-pVTZ) for the MP2 level of theory.}
\begin{tabular}{lrr}
\toprule
 &  & MP2 (kJ/mol) \\ 
\midrule
\multirow[c]{3}{*}{2B} & TZ & -22.5 \\
 & QZ & -22.9 \\
 & CBS & -23.2 \\
\cline{1-3}
\multirow[c]{3}{*}{3B} & DZ & 0.5 \\
 & TZ & 0.5 \\
 & CBS & 0.5 \\
\cline{1-3}
\bottomrule
\end{tabular}

\end{table}

To summarize, we conclude from these tests on the ice phases and benzene that the CBS(DZ/TZ) or CBS(TZ/QZ) with the aug-cc-pV$X$Z family of basis sets is needed to consistently achieve sub-kJ/mol accuracy on 2B terms.
On the other hand, even aug-cc-pVDZ is sufficient for the 3B terms.
These conclusions reflect observations from several previous studies~\cite{richardAimingBenchmarkAccuracy2014a,richardUnderstandingManyBodyBasis2018,heindelManyBodyExpansionAqueous2020}

\subsubsection{Effect of core electrons}

For Ice VIII, we have also investigated the effect of core electrons on the 2B contribution to the lattice energy.
We expect core-electron effects to be strongest out of all the systems we study because this system is a high-pressure dense phase of ice.
This means incorporating the 1s electrons on the O atoms in the correlation treatment together with corresponding polarized weighted core valence basis sets at the CBS(aug-cc-pwCVTZ/aug-cc-pwCVQZ) level.
As shown in Table~\ref{tab:core_effect}, we find that differences w.r.t.\ a standard valence electron-only treatment are smaller than $0.3\,$kJ/mol, indicating that these effects are small for ice phases.
As such, we do not expect the core electrons to have a significant contribution to the lattice energy and can be safely neglected when aiming for sub-kJ/mol accuracy.

\begin{table}[h]

\caption{\label{tab:core_effect}Effect of core correlation to the two-body contribution of the $E_\text{latt}$ of ice VIII.}
\begin{tabular}{lr}
\toprule
 & 2B contribution to $E_\text{latt}$ (kJ/mol) \\ 
\midrule
CCSD(T) CBS(aug-cc-pVTZ/aug-cc-pVQZ) & -31.0 \\
CCSD(T) smallcore CBS(aug-cc-pwCVTZ/aug-cc-pwCVQZ) & -31.3 \\
\bottomrule
\end{tabular}

\end{table}

\subsubsection{Beyond-CCSD(T) contributions}

We further consider the effect of higher order correlations beyond CCSD(T) for ice VIII, performing calculations with CCSDT(Q) using the aug-cc-pVDZ basis set in Table~\ref{tab:core_effect}.
These effects are found to be small for ice VIII, at less than $0.2\,$kJ/mol.
Thus, we expect that these effects can be mostly neglected when aiming for sub-kJ/mol accuracy.

\begin{table}[h]

\caption{\label{tab:tq_effect}Effect of higher order correlation to the two-body contribution of the $E_\text{latt}$ of ice VIII.}
\begin{tabular}{lr}
\toprule
 & 2B contribution to $E_\text{latt}$ (kJ/mol) \\ 
\midrule
CCSD(T) aug-cc-pVDZ & -20.7 \\
CCSDT(Q) aug-cc-pVDZ & -20.9 \\
\bottomrule
\end{tabular}

\end{table}

\subsubsection{Convergence with many-body terms in the MBE}

\begin{figure}[h]
    \includegraphics[width=\textwidth]{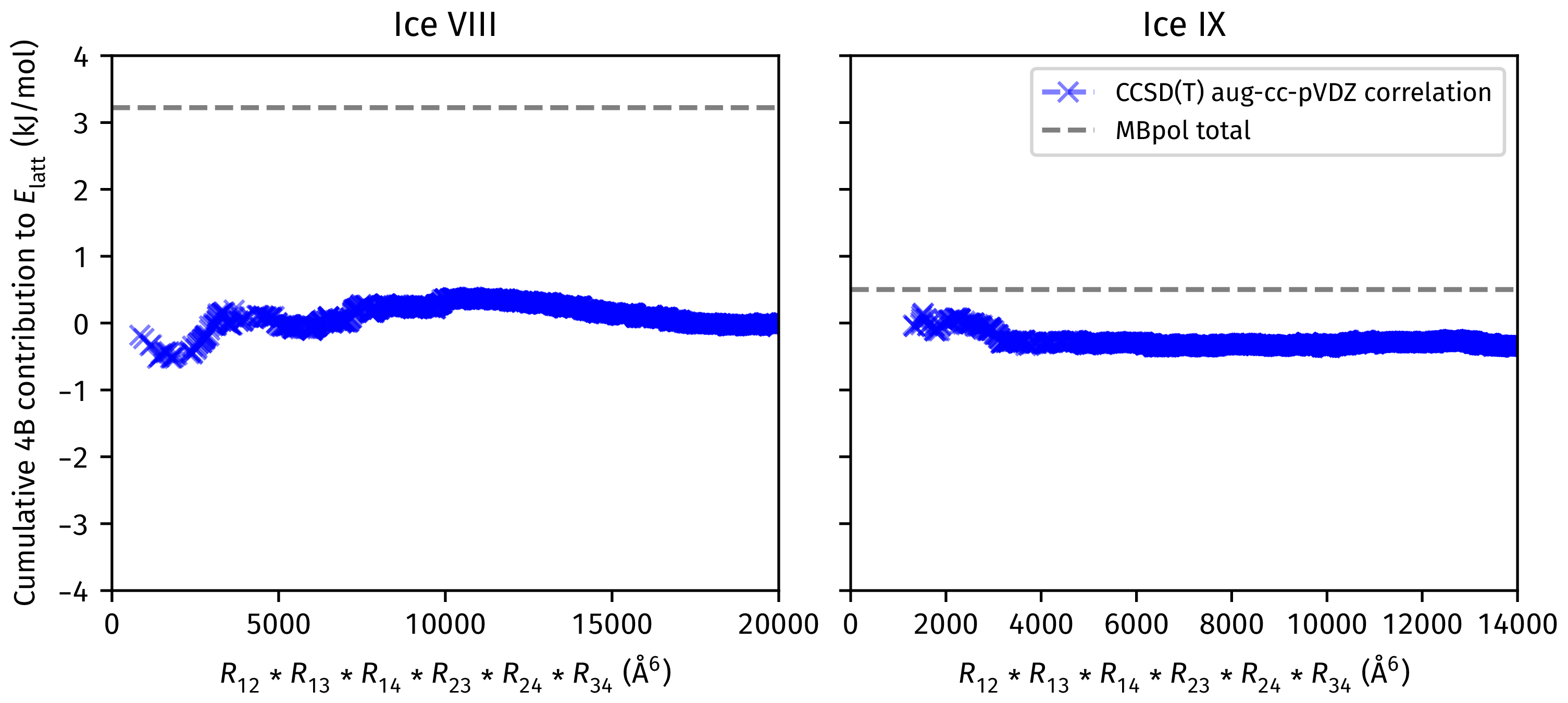}
    \caption{\label{fig:4b_ice_conv}The convergence with distance cutoff for the four-body contribution of the CCSD(T) correlation energy to $E_\text{latt}$ for ice VIII and IX. These are compared to the converged total energy contribution approximated by MB-pol estimates from Ref.~\citenum{hermanFormulationManyBodyExpansion2023b}.}
\end{figure}

We have considered only 2B and 3B contributions to the MBE so far.
Ice VIII is a system where higher-order terms in the many-body expansion, such as the four-body (4B) contribution can have a significant contribution to the lattice energy.
This contribution is significant for the \textit{total} energy (taken for MB-pol from Ref.~\citenum{hermanFormulationManyBodyExpansion2023b}), but we show here that it is small for the CCSD(T) correlation energy, meaning most of the contribution comes from the HF energy.
As shown in Figure~\ref{fig:4b_ice_conv}, the corresponding contribution from the CCSD(T) correlation energy is small, being less than $0.5\,$kJ/mol compared to the $3.2\,$kJ/mol for the total 4B contribution (approximated using MB-pol).
We also computed this contribution for ice IX, finding a similarly small contribution at ${\sim}0.4\,$kJ/mol.
Overall, these results indicate that we can neglect many-body contributions beyond the 3B term for the molecular crystals we have studied throughout this work.

\subsubsection{\label{sec:local_converge}Local approximation convergence}

We look at the convergence with local natural orbital (LNO) approximations thresholds for the (pre-defined) {\tt Tight} and {\tt vTight} settings, together with the local approximation free (LAF) extrapolation, comparing these to the canonical results for the aug-cc-pVTZ, aug-cc-pVQZ and CBS(aug-cc-pVTZ/aug-cc-pVQZ) basis sets.
This has been compared for the 12 phases of ice in Tables~\ref{tab:2b_lno_conv} and~\ref{tab:3b_lno_conv} for the 2B and 3B contributions, respectively.
The errors are generally small, being on average less than $1\,$kJ/mol for the 2B contribution.
This error gets halved, to an MAD of $0.5\,$kJ/mol for the {\tt vTight} LNO threshold and is further halved to $\sim0.2\,$kJ/mol with the LAF extrapolation.
In general we find that the errors do not change much between different basis set choices.
For the 3B terms, the dependence on the LNO thresholds is weaker, coming to only a maximum MAD of $0.3\,$kJ/mol across 12 phases of ice due to the already small contribution of this term.

\begin{table}[h]

\caption{\label{tab:2b_lno_conv}Convergence of 2B contribution for 12 ice phases with LNO thresholds: {\tt Tight}, {\tt vTight} and extrapolated locality approximation free (LAF) for aug-cc-pVTZ, aug-cc-pVQZ and CBS(aug-cc-pVTZ/aug-cc-pVQZ) basis sets}
\begin{tabular}{lrrrrrrrrrrrr}
\toprule
 & \multicolumn{4}{c}{TZ} & \multicolumn{4}{c}{QZ} & \multicolumn{4}{c}{CBS} \\ \cmidrule(lr){2-5} \cmidrule(lr){6-9} \cmidrule(lr){10-13} 
 & {\tt Tight} & {\tt vTight} & LAF & Canonical & {\tt Tight} & {\tt vTight} & LAF & Canonical & {\tt Tight} & {\tt vTight} & LAF & Canonical \\
\midrule
Ih & -19.5 & -19.9 & -20.1 & -20.3 & -21.6 & -22.1 & -22.4 & -22.5 & -23.1 & -23.7 & -24.0 & -24.0 \\
II & -21.9 & -22.5 & -22.8 & -23.1 & -24.0 & -24.7 & -25.0 & -25.2 & -25.5 & -26.2 & -26.5 & -26.7 \\
III & -21.5 & -21.7 & -21.8 & -22.2 & -23.6 & -23.9 & -24.1 & -24.4 & -25.0 & -25.5 & -25.7 & -26.0 \\
IV & -23.1 & -23.5 & -23.7 & -24.1 & -25.0 & -25.6 & -25.8 & -26.1 & -26.4 & -27.1 & -27.4 & -27.6 \\
VI & -24.0 & -24.7 & -25.1 & -25.3 & -26.1 & -26.9 & -27.2 & -27.5 & -27.6 & -28.4 & -28.8 & -29.0 \\
VIII & -26.2 & -27.0 & -27.4 & -27.6 & -28.1 & -29.0 & -29.4 & -29.6 & -29.5 & -30.4 & -30.9 & -31.0 \\
IX & -20.6 & -20.9 & -21.0 & -21.4 & -22.7 & -23.1 & -23.2 & -23.6 & -24.2 & -24.6 & -24.8 & -25.1 \\
XI & -19.9 & -20.3 & -20.5 & -20.7 & -22.1 & -22.6 & -22.9 & -23.0 & -23.6 & -24.3 & -24.6 & -24.6 \\
XIII & -22.3 & -22.7 & -23.0 & -23.3 & -24.3 & -24.9 & -25.2 & -25.5 & -25.8 & -26.4 & -26.7 & -27.0 \\
XIV & -23.2 & -23.8 & -24.0 & -24.4 & -25.3 & -25.9 & -26.2 & -26.6 & -26.8 & -27.4 & -27.7 & -28.1 \\
XV & -23.6 & -24.2 & -24.6 & -24.8 & -25.6 & -26.4 & -26.7 & -26.9 & -27.1 & -27.9 & -28.2 & -28.4 \\
XVII & -18.9 & -19.3 & -19.6 & -19.7 & -21.1 & -21.6 & -21.9 & -21.9 & -22.6 & -23.2 & -23.5 & -23.5 \\
\midrule MAD & 1.0 & 0.5 & 0.3 & 0.0 & 1.1 & 0.5 & 0.2 & 0.0 & 1.1 & 0.5 & 0.2 & 0.0 \\
\bottomrule
\end{tabular}

\end{table}

\begin{table}[h]

\caption{\label{tab:3b_lno_conv}Convergence of 3B contribution for 12 ice phases with LNO thresholds: {\tt Tight}, {\tt vTight} and extrapolated locality approximation free (LAF) for aug-cc-pVDZ, aug-cc-pVTZ and CBS(aug-cc-pVDZ/aug-cc-pVTZ) basis sets}
\begin{tabular}{lrrrrrrrrrrrr}
\toprule
 & \multicolumn{4}{c}{DZ} & \multicolumn{4}{c}{TZ} & \multicolumn{4}{c}{CBS} \\ \cmidrule(lr){2-5} \cmidrule(lr){6-9} \cmidrule(lr){10-13} 
 & {\tt Tight} & {\tt vTight} & LAF & Canonical & {\tt Tight} & {\tt vTight} & LAF & Canonical & {\tt Tight} & {\tt vTight} & LAF & Canonical \\
\midrule
Ih & 0.6 & 0.3 & 0.1 & 0.1 & -0.0 & 0.2 & 0.3 & 0.1 & -0.4 & 0.1 & 0.4 & 0.1 \\
II & 1.5 & 1.7 & 1.8 & 1.6 & 1.5 & 1.7 & 1.8 & 1.7 & 1.5 & 1.7 & 1.8 & 1.7 \\
III & 0.6 & 0.7 & 0.8 & 0.6 & 0.7 & 0.8 & 0.8 & 0.7 & 0.8 & 0.8 & 0.8 & 0.7 \\
IV & 1.2 & 1.3 & 1.3 & 1.3 & 1.6 & 1.5 & 1.4 & 1.4 & 1.8 & 1.6 & 1.5 & 1.5 \\
VI & 1.5 & 1.8 & 2.0 & 1.8 & 1.7 & 1.9 & 2.0 & 1.9 & 1.8 & 2.0 & 2.0 & 1.9 \\
VIII & 3.0 & 2.6 & 2.4 & 2.3 & 2.7 & 2.3 & 2.1 & 2.4 & 2.6 & 2.1 & 1.9 & 2.5 \\
IX & 1.0 & 1.1 & 1.1 & 1.0 & 1.4 & 1.0 & 0.8 & 1.0 & 1.6 & 0.9 & 0.6 & 1.1 \\
XI & 0.2 & 0.2 & 0.2 & 0.0 & -0.2 & -0.0 & 0.1 & 0.0 & -0.5 & -0.2 & -0.0 & 0.0 \\
XIII & 0.9 & 1.4 & 1.7 & 1.3 & 1.4 & 1.4 & 1.5 & 1.4 & 1.7 & 1.5 & 1.3 & 1.5 \\
XIV & 1.5 & 1.8 & 1.9 & 1.7 & 1.6 & 1.8 & 1.8 & 1.8 & 1.7 & 1.8 & 1.8 & 1.9 \\
XV & 1.4 & 1.8 & 2.0 & 1.7 & 1.8 & 1.8 & 1.8 & 1.8 & 2.0 & 1.9 & 1.8 & 1.9 \\
XVII & -0.2 & 0.3 & 0.6 & 0.2 & 0.0 & 0.3 & 0.5 & 0.3 & 0.1 & 0.3 & 0.4 & 0.3 \\
\midrule MAD & 0.3 & 0.1 & 0.2 & 0.0 & 0.2 & 0.1 & 0.1 & 0.0 & 0.3 & 0.1 & 0.2 & 0.0 \\
\bottomrule
\end{tabular}

\end{table}

In Table~\ref{tab:benzene_lno_conv}, we have also compared these LNO thresholds with the aug-cc-pVDZ basis set for the 2B and 3B terms for benzene.
Similarly, it can be seen that for the 2B terms, the LAF can come into near exact agreement with canonical estimates.
For the 3B terms, the results appear to already be well-converged to within $0.5\,$kJ/mol with the {\tt Tight} LNO thresholds and do not improve by going to {\tt vTight} or LAF treatments.

\begin{table}[h]

\caption{\label{tab:benzene_lno_conv}Convergence of 2B and 3B contributions, for the benzene molecular crystal in the X23 dataset, with LNO thresholds: {\tt Tight}, {\tt vTight} and extrapolated locality approximation free (LAF) for aug-cc-pVDZ basis set}
\begin{tabular}{lrrrrrrrr}
\toprule
 & \multicolumn{4}{c}{2B} & \multicolumn{4}{c}{3B} \\ \cmidrule(lr){2-5} \cmidrule(lr){6-9} 
 & {\tt Tight} & {\tt vTight} & LAF & Canonical & {\tt Tight} & {\tt vTight} & LAF & Canonical \\
\midrule
CCSD(T) aug-cc-pVDZ (kJ/mol) & -70.7 & -70.2 & -70.0 & -70.0 & 4.7 & 4.6 & 4.5 & 5.1 \\
\bottomrule
\end{tabular}

\end{table}

Overall, these tests indicate that an LAF extrapolation treatment benefits 2B terms when aiming for sub-kJ/mol accuracy while its effects are negligible or worse for 3B terms, where {\tt Tight} or {\tt vTight} settings are sufficient already.

\clearpage

\subsubsection{\label{sec:cheaper_settings}Final parameters}

Based on the observations made within the preceding sections for the MBE treatment of the lattice energy $E_\text{latt}$, we come up with several approaches for computing $E_\text{latt}$ that can retain a high level of accuracy.
The procedures are as follows:
\begin{itemize}
    \item \textbf{Canonical} --- This is the most accurate as it makes extensive use of canonical CCSD(T).
    We use for the following basis sets for the many-body terms:
    \begin{itemize}
        \item 1B terms: CBS(aug-cc-pVQZ/aug-cc-pV5Z)
        \item 2B terms: CBS(aug-cc-pVTZ/aug-cc-pVQZ)
        \item 3B terms: CBS(aug-cc-pVDZ/aug-cc-pVTZ)
    \end{itemize}
    \item \textbf{LAF} --- In this more economical approach, we use LNO-CCSD(T) and aim to reach the local approximation free limit by extrapolating using {\tt Tight} and {\tt vTight} settings. The exact same sets of AO basis functions are used as for the Canonical approach.
    
    \item \textbf{Composite} --- To further lower costs, we make use of a composite procedure whereby the canonical MP2 is used to correct for both the local (i.e., LNO) approximation errors as well as for basis set incompleteness effects.
    One way to explain how it works (using the 2B terms as an example) is that we start from a CBS(aug-cc-pVTZ/aug-cc-pVQZ) extrapolated canonical MP2 estimates, using the same basis set combinations as listed above, and add on correction between LNO-CCSD(T) and local MP2 (LMP2) at the smaller aug-cc-pVTZ basis set.
    \begin{align}
    E_\text{latt}^{\Delta\text{MP2}} &= 
        E^\text{MP2}_\text{latt}[\text{CBS(aug-cc-pVTZ/aug-cc-pVQZ)}] \notag \\
        &\quad \Bigl[ + E_\text{latt}^\text{LAF LNO-CCSD(T)}[\text{aug-cc-pVTZ}] \notag \\
        &\quad - E_\text{latt}^\text{LAF LMP2}[\text{aug-cc-pVTZ}] \Bigr]
    \end{align}
    Alternatively, this may also be thought as starting from a LAF extrapolated LNO-CCSD(T) estimate at the aug-cc-pVTZ basis set with a subsequent correction from canonical CBS(aug-cc-pVTZ/aug-cc-pVQZ) MP2 to correct for local approximation errors and basis set errors in LMP2 aug-cc-pVTZ:
    \begin{align}
    E_\text{latt}^{\Delta\text{MP2}} &= E_\text{latt}^\text{LAF LNO-CCSD(T)}[\text{aug-cc-pVTZ}]
         \notag \\
        &\quad \Bigl[ + E^\text{MP2}_\text{latt}[\text{CBS(aug-cc-pVTZ/aug-cc-pVQZ)}] \notag \\
        &\quad - E_\text{latt}^\text{LAF LMP2}[\text{aug-cc-pVTZ}] \Bigr].
    \end{align}    
    This procedure is expected to work very well for the LNO-CCSD(T) implementation within {\sc Mrcc} because the LMP2\cite{nagyIntegralDirectLinearScalingSecondOrder2016a,szaboLinearScalingOpenShellMP22021} energy is exactly the second-order part of LNO-CCSD(T)~\cite{nagyOptimizationLinearScalingLocal2018,nagyApproachingBasisSet2019,nagyStateoftheartLocalCorrelation2024}, thus their local approximations also match and consequently, both can be consitently improved by the canonical MP2 correction.
\end{itemize}

In Table~\ref{tab:canonical_cheaper}, we have compared the lattice energy for these three approaches across 12 ice phases in their small-sized unit cells.
Table~\ref{tab:canonical_cheaper_medium} gives the same estimates for the medium-sized unit cells to assess whether there might be a build-up in errors when more terms have to be computed.~\cite{altunAddressingSystemSizeDependence2021a}
We find that compared to the canonical results, the LAF approach introduces errors of only $0.2\,$kJ/mol, while Composite introduces errors of $0.3\,$kJ/mol in terms of MAD across the 12 ice phases.
We find no change in errors when using medium-sized unit cells, indicating no build up of errors, arising because the lattice energy is a per molecule quantity.

\begin{table}[h]

\caption{\label{tab:canonical_cheaper}Comparison of various approaches [Canonical (Can.), LAF and Composite (Comp.) described in the text] for computing the lattice energy of several ice phases in their small-sized unit cells.}
\begin{tabular}{lrrrrrrrrrrrr}
\toprule
 & \multicolumn{3}{c}{1B} & \multicolumn{3}{c}{2B} & \multicolumn{3}{c}{3B} & \multicolumn{3}{c}{Total} \\ \cmidrule(lr){2-4} \cmidrule(lr){5-7} \cmidrule(lr){8-10} \cmidrule(lr){11-13} 
 & Can. & LAF & Comp. & Can. & LAF & Comp. & Can. & LAF & Comp. & Can. & LAF & Comp. \\
\midrule
Ih & -8.0 & -8.0 & -8.1 & -24.0 & -24.0 & -23.7 & 0.1 & 0.4 & 0.1 & -31.9 & -31.6 & -31.7 \\
II & -7.1 & -7.1 & -7.2 & -26.7 & -26.5 & -26.4 & 1.7 & 1.8 & 1.8 & -32.1 & -31.8 & -31.8 \\
III & -8.0 & -8.0 & -8.1 & -26.0 & -25.7 & -25.6 & 0.7 & 0.8 & 0.7 & -33.3 & -32.9 & -33.0 \\
IV & -7.1 & -7.1 & -7.1 & -27.6 & -27.4 & -27.3 & 1.5 & 1.5 & 1.5 & -33.2 & -33.0 & -32.9 \\
VI & -6.8 & -6.8 & -6.8 & -29.0 & -28.8 & -28.7 & 1.9 & 2.0 & 2.0 & -33.8 & -33.5 & -33.5 \\
VIII & -5.3 & -5.3 & -5.3 & -31.0 & -30.9 & -30.7 & 2.5 & 1.9 & 2.6 & -33.8 & -34.3 & -33.4 \\
IX & -7.5 & -7.5 & -7.6 & -25.1 & -24.8 & -24.8 & 1.1 & 0.6 & 1.1 & -31.6 & -31.7 & -31.3 \\
XI & -8.6 & -8.6 & -8.6 & -24.6 & -24.6 & -24.2 & 0.0 & -0.0 & 0.0 & -33.1 & -33.2 & -32.8 \\
XIII & -6.9 & -6.9 & -6.9 & -27.0 & -26.7 & -26.7 & 1.5 & 1.3 & 1.5 & -32.4 & -32.3 & -32.1 \\
XIV & -6.6 & -6.6 & -6.7 & -28.1 & -27.7 & -27.8 & 1.9 & 1.8 & 1.9 & -32.9 & -32.6 & -32.5 \\
XV & -6.5 & -6.5 & -6.6 & -28.4 & -28.2 & -28.1 & 1.9 & 1.8 & 1.9 & -33.1 & -33.0 & -32.8 \\
XVII & -8.4 & -8.4 & -8.4 & -23.5 & -23.5 & -23.1 & 0.3 & 0.4 & 0.3 & -31.6 & -31.5 & -31.3 \\
\midrule MAD & 0.0 & 0.0 & 0.0 & 0.0 & 0.2 & 0.3 & 0.0 & 0.2 & 0.0 & 0.0 & 0.3 & 0.3 \\
\bottomrule
\end{tabular}

\end{table}

\begin{table}[h]

\caption{\label{tab:canonical_cheaper_medium}Comparison of various approaches [Canonical (Can.), LAF and Composite (Comp.) described in the text] for computing the lattice energy of several ice phases in their medium-sized unit cells.}
\begin{tabular}{lrrrrrrrrrrrr}
\toprule
 & \multicolumn{3}{c}{1B} & \multicolumn{3}{c}{2B} & \multicolumn{3}{c}{3B} & \multicolumn{3}{c}{Total} \\ \cmidrule(lr){2-4} \cmidrule(lr){5-7} \cmidrule(lr){8-10} \cmidrule(lr){11-13} 
 & Can. & LAF & Comp. & Can. & LAF & Comp. & Can. & LAF & Comp. & Can. & LAF & Comp. \\
\midrule
Ih & -8.6 & -8.6 & -8.6 & -23.8 & -23.9 & -23.4 & 0.1 & 0.4 & 0.1 & -32.2 & -32.0 & -31.9 \\
III & -7.5 & -7.5 & -7.6 & -25.5 & -25.2 & -25.1 & 0.6 & 0.8 & 0.6 & -32.4 & -31.9 & -32.1 \\
IV & -7.0 & -7.0 & -7.0 & -27.3 & -27.0 & -27.0 & 1.3 & 1.1 & 1.3 & -33.0 & -32.9 & -32.7 \\
VII & -5.6 & -5.6 & -5.6 & -31.8 & -31.7 & -31.5 & 3.2 & 3.1 & 3.2 & -34.2 & -34.2 & -33.9 \\
XVII & -8.4 & -8.4 & -8.4 & -23.4 & -23.5 & -23.0 & 0.2 & 0.4 & 0.2 & -31.5 & -31.5 & -31.2 \\
\midrule MAD & 0.0 & 0.0 & 0.0 & 0.0 & 0.2 & 0.3 & 0.0 & 0.2 & 0.0 & 0.0 & 0.2 & 0.3 \\
\bottomrule
\end{tabular}

\end{table}

We have also tested these approaches for the benzene molecular crystal in Table~\ref{tab:x23_benzene_lno_conv}.
We find that the Composite procedure is accurate to within $0.1\,$kJ/mol w.r.t. the canonical result.
This comes from a cancellation of ${\sim}0.6\,$kJ/mol errors on the 2B and 3B contributions to the total lattice energy.
The LAF approach is about $2.2\,$kJ/mol off from the canonical estimates and this arises from errors in the 3B contribution, while there is agreement to within $0.1\,$kJ/mol on the 2B terms.
It seems that the LAF extrapolation formula, while working for 2B terms, is less successful for 3B terms, as we have observed in Section~\ref{sec:local_converge}.

\begin{table}[h]

\caption{\label{tab:x23_benzene_lno_conv}Convergence of the LNO-MBE-CCSD(T) lattice energy of benzene with respect to the LNO threshold, compared to the canonical result. All values are in kJ/mol.}
\begin{tabular}{lrrrr}
\toprule
 & 1B & 2B & 3B & Total \\ 
\midrule
{\tt Tight} & -1.3 & -80.0 & 3.6 & -77.7 \\
{\tt vTight} & -1.3 & -79.3 & 5.8 & -74.8 \\
LAF & -1.3 & -78.9 & 6.8 & -73.4 \\
Composite & -1.3 & -79.6 & 5.3 & -75.7 \\
Canonical & -1.3 & -79.0 & 4.7 & -75.6 \\
\bottomrule
\end{tabular}

\end{table}

\clearpage

\subsection{\label{sec:rel_ene_details}Relative energy}
\subsubsection{Final parameters} 
Compared to the lattice energies, relative energies are expected to be easier to converge since there can be cancellation of errors (e.g., in basis set, MBE, or local approximation) between the two polymorphs being compared.
Within this section, we will demonstrate this for the relative energy between ice forms Ih and VIII as well as the experimental and vanEijck-3 forms of molecule X from the third Crystal Structure Prediction (CSP) Blind Test~\cite{dayThirdBlindTest2005}.
Its easier convergence means that it is possible to come up with cheaper procedures than what was discussed for the lattice energy that can scale up towards larger molecular crystal systems.
Specifically, we show in Table~\ref{tab:deltamp2_cheaper} that the Composite with the original basis sets (dubbed `Full') can be loosened towards using smaller basis sets (dubbed `Reduced') and less tight LNO approximations for the various terms that make up the Composite relative energy:
\begin{equation}
    E_\text{rel}^\text{final} =  E_\text{rel}^\text{LNO-CCSD(T)} + [E_\text{rel}^\text{MP2} - E_\text{rel}^\text{LMP2}].
\end{equation}
In this `Reduced' basis set, for the 2B terms, we use aug-cc-pVDZ and aug-cc-pVTZ pair as opposed to the aug-cc-pVTZ and aug-cc-pVQZ pair.
For the 3B terms, we introduce the jul-cc-pVDZ and jul-cc-pVTZ pairs, where the `jul' basis sets remove the diffuse augmentation functions on the H atoms for an overall smaller basis set.
Additionally, we have also used canonical MP2 to correct CCSD(T) with {\tt Tight} LNO thresholds as opposed to LAF extrapolated (from the {\tt Tight} and {\tt vTight} threshold pair).

\begin{sidewaystable}[p]
\caption{\label{tab:deltamp2_cheaper}The basis sets used for the original -- termed `Full' -- Composite procedure in Section~\ref{sec:local_converge} for computing the lattice energy and the cheaper procedure -- termed `Reduced' for computing relative energy.}
\centering
{\small
\begin{tabular}{@{}llrrrrrr@{}}
\toprule
\multirow{2}{*}{Approach}    & \multirow{2}{*}{Quantity} & \multicolumn{2}{c}{1B contribution}                   & \multicolumn{2}{c}{2B contribution}                   & \multicolumn{2}{c}{3B contribution}                   \\ \cmidrule(lr){3-4} \cmidrule(lr){5-6} \cmidrule(lr){7-8} 
                             &                           & \multicolumn{1}{c}{Basis} & \multicolumn{1}{c}{LNO} & \multicolumn{1}{c}{Basis} & \multicolumn{1}{c}{LNO} & \multicolumn{1}{c}{Basis} & \multicolumn{1}{c}{LNO} \\ \midrule
\multirow{3}{*}{Composite Full}   & $E_\text{rel}^\text{LNO-CCSD(T)}$                   & aug-cc-pVQZ                  & LAF                    & aug-cc-pVTZ                  & LAF                    & aug-cc-pVDZ                  & LAF                    \\
                             & $E_\text{rel}^\text{MP2}$                   & aug-cc-pVQZ                  & LAF                    & aug-cc-pVTZ                  & LAF                    & aug-cc-pVDZ                  & LAF                    \\
                             & $E_\text{rel}^\text{LMP2}$                  & CBS(aug-cc-pVQZ/5Z) & Canonical              & CBS(aug-cc-pVTZ/QZ) & Canonical              & CBS(aug-cc-pVDZ/TZ) & Canonical              \\ \midrule
\multirow{3}{*}{Composite Reduced} & $E_\text{rel}^\text{LNO-CCSD(T)}$                   & jul-cc-pVTZ                  & {\tt Tight}                  & aug-cc-pVDZ                  & {\tt Tight}                  & jul-cc-pVDZ                  & {\tt Tight}                  \\
                             & $E_\text{rel}^\text{MP2}$                   & jul-cc-pVTZ                  & {\tt Tight}                  & aug-cc-pVDZ                  & {\tt Tight}                  & jul-cc-pVDZ                  & {\tt Tight}                  \\
                             & $E_\text{rel}^\text{LMP2}$                  & CBS(jul-cc-pVTZ/QZ) & Canonical              & CBS(aug-cc-pVDZ/TZ) & Canonical              & CBS(jul-cc-pVDZ/TZ) & Canonical              \\ \bottomrule
\end{tabular}
}
\end{sidewaystable}

\clearpage

\subsubsection{\label{sec:molecule_x_conv}Experimental and vanEijck-3 forms of molecule X}

Molecule X from the third CSP Blind Test has represented a challenge for many DFAs~\cite{whittletonExchangeHoleDipoleDispersion2017,greenwellInaccurateConformationalEnergies2020} due to its conformational flexibility, which requires an accurate treatment of both the intermolecular and intramolecular interactions.
In particular, the relative energy between the experimental and vanEijck-3 form of molecule X has been challenging to predict.

\begin{figure}[h]
    \includegraphics[width=\textwidth]{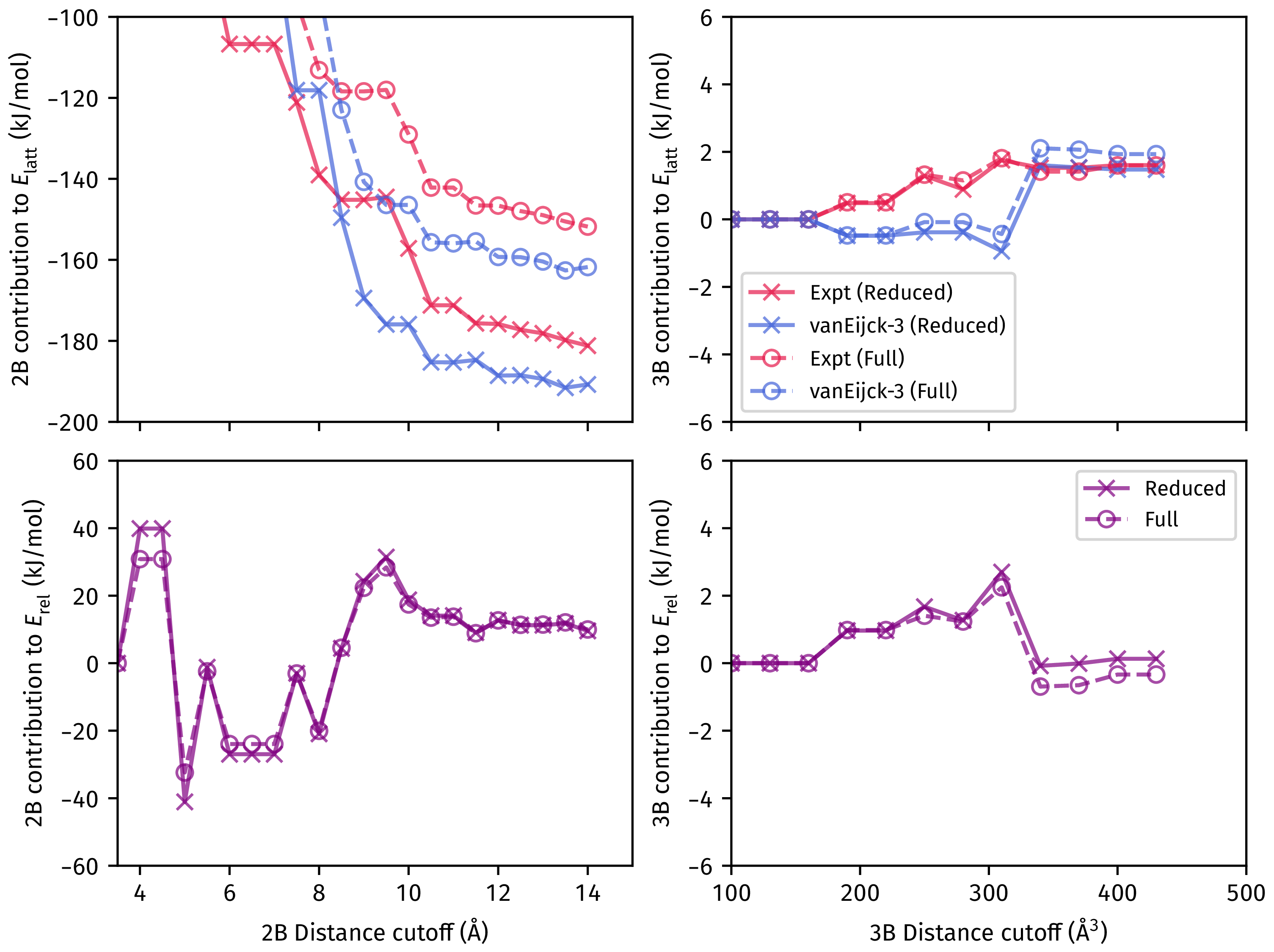}
    \caption{\label{fig:X_2b_3b_rel_conv}Demonstrating the quicker convergence of the relative energy $E_\text{rel}$ between the experimental and vanEijck-3 polymorphs of molecule X over the lattice energy $E_\text{latt}$ of these individual polymorphs for both the two-body (2B) and three-body (3B) contributions of the correlation energy to the many-body expansion (MBE).}
\end{figure}

In Figure~\ref{fig:X_2b_3b_rel_conv}, we demonstrate the convergence of the 2B and 3B contributions to the lattice energy $E_\text{latt}$ from the correlation energy of the (aforementioned) two polymorphs of molecule X as well as their relative energy $E_\text{rel}$.
It can be seen that in terms of absolute value, the use of the Reduced basis sets leads to a large ${\sim}30\,$kJ/mol difference compared to the Full basis sets for the 2B terms.
However, when we consider energy differences, it is less than $0.5\,$kJ/mol for the 2B contribution, highlighting the near-perfect cancellation of errors.
This is similarly observed for the 3B contributions, albeit to a much smaller extent.

In Table~\ref{tab:x_method_convergence}, we show that the relative energy difference between the vanEijck-3 and experiment forms are to within $0.6\,$kJ/mol of each other, regardless of the errors in the absolute $E_\text{latt}$.
Overall, this highlights that we can expect cancellation in errors between the local approximation as well as the basis set when computing relative energies.
Furthermore, it can be seen that the relative energy convergences significantly quicker with cutoff than the absolute contribution to the lattice energy.

\begin{table}[h]

\caption{\label{tab:x_method_convergence}Convergence of the relative energies between the experimental and vanEijck-3 polymorphs of X arising from CCSD(T) correlation with respect to basis set size. Both the results with the `Reduced' basis set (used in the main text) and `Full' basis sets are given. All values are in kJ/mol.}
\begin{tabular}{lrrrrr}
\toprule
 &  & 1B & 2B & 3B & Total \\ 
\midrule
\multirow[c]{3}{*}{Reduced basis} & Experimental & -0.0 & -152.4 & 1.6 & -150.8 \\
 & vanEijck-3 & 0.6 & -165.1 & 1.9 & -162.6 \\
 & $E_\text{rel}$ & -0.6 & 12.8 & -0.3 & 11.8 \\
\cline{1-6}
\multirow[c]{3}{*}{Full basis} & Experimental & -0.0 & -181.2 & 1.6 & -179.6 \\
 & vanEijck-3 & 1.2 & -190.8 & 1.5 & -188.1 \\
 & $E_\text{rel}$ & -1.2 & 9.6 & 0.1 & 8.5 \\
\cline{1-6}
\bottomrule
\end{tabular}

\end{table}

\clearpage

\subsubsection{Ice Ih and VIII}

\begin{figure}
    \includegraphics[width=\textwidth]{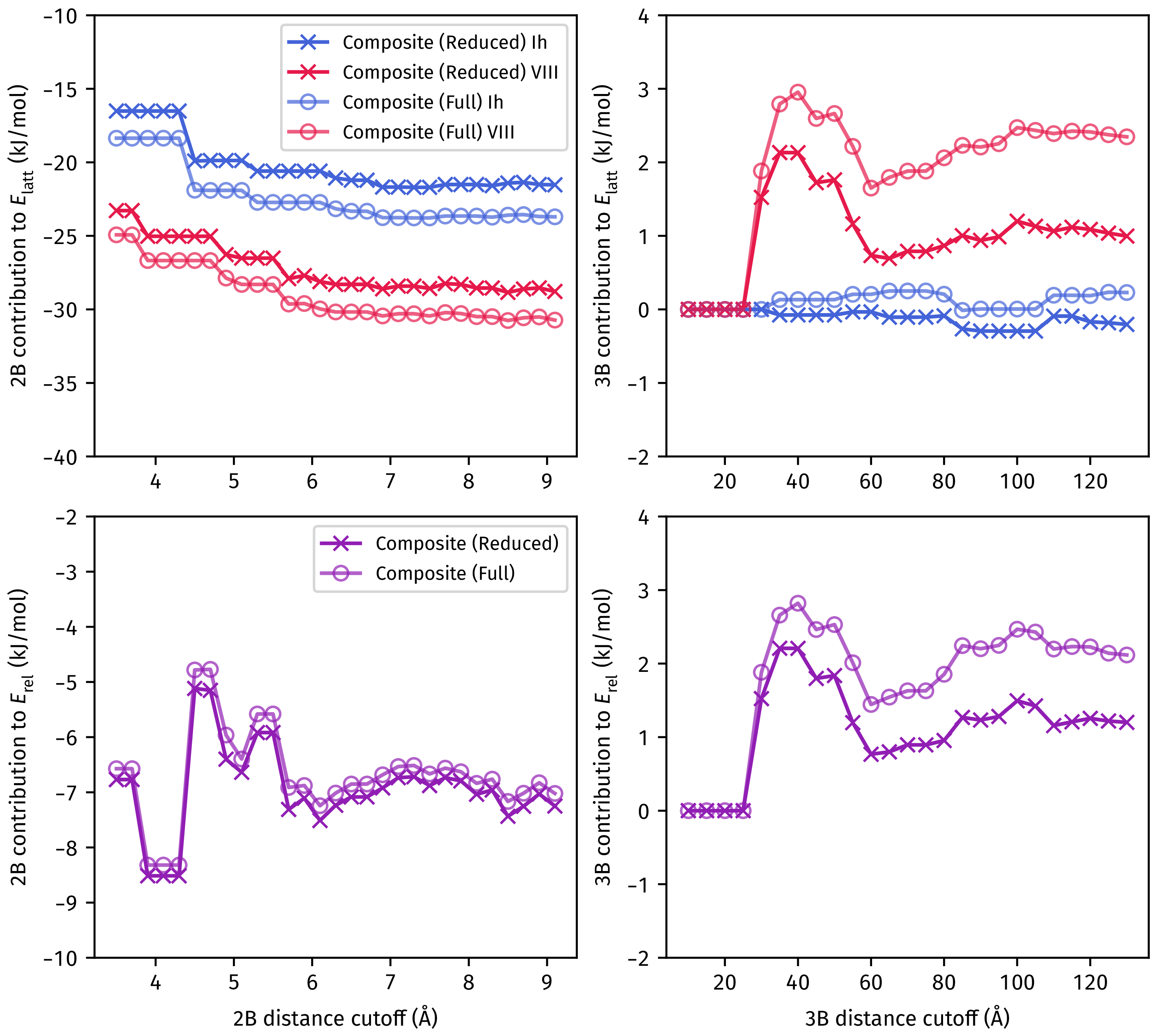}
    \caption{\label{fig:ice_2b_3b_rel_ene_conv}Convergence of the 2B and 3B terms to the MBE for ice Ih (left) and VIII (right) for different local threshold procedures to LNO-CCSD(T): {\tt Tight}, {\tt vTight} and local approximation free (LAF) extrapolation.}
\end{figure}

The relative energy between Ice Ih and VIII represents a particularly challenging task for the MBE because there is significantly different many-body behavior between the two phases, the former being a low-pressure/density phase while the latter is a high-pressure/density phase.
In Figure~\ref{fig:ice_2b_3b_rel_ene_conv}, we show the contribution from the 2B and 3B terms to the total lattice energies of Ih and VIII.
It can be seen that Ih has almost no contributions to the CCSD(T) correlation energy from three-body terms while VIII has significant contributions from three-body terms.
Furthermore, there is also a large difference in density between the two systems, whereby ice Ih has a density of $0.93\,$g/cm$^3$ at $10\,$K~\cite{rottgerLatticeConstantsThermal1994} and low pressures, while ice VIII has a density of $1.49\,$g/cm$^3$ at zero temperature and pressure~\cite{whalleyEnergiesPhasesIce1984}
Thus, there is also a different convergence behavior of the 2B and 3B terms with cutoff distance.
Given these considerations, we expect the relative energy between ice Ih and VIII represents a system that can really stress-test the MBE approach.  

In Table~\ref{tab:ice_rel_ene_mbe_conv}, we compare the relative energy between forms VIII and Ih using the `Reduced' basis set Composite approach with respect to the several treatments with the `Full' basis set used for the lattice energy in Section~\ref{sec:cheaper_settings}.
It can be seen that this cheaper treatment can obtain the relative energy to within $1.1\,$kJ/mol of the range of values from the (more expensive) approaches used for the lattice energy.
This error largely comes from the three-body term, which is underestimated by the Composite Reduced approach.
We also observe similar large errors with the LAF Full basis treatment, suggesting that it is hard to leverage cancellation of errors arising from the local approximation due to the different density behaviors.

To summarize, we find that relative energies can still be obtained to within $1\,$kJ/mol even for these challenging pair of polymorphs.
We do not expect systems with such differences in character and density to be compared under general circumstances, as they would appear under different thermodynamic conditions.
For typical systems, we expect a better control of errors, as observed for the molecule X polymorphs.

\begin{table}[h]

\caption{\label{tab:ice_rel_ene_mbe_conv}The MBE contributions to the lattice energy of ice Ih and VIII for several LNO approximations ({\tt Tight}, {\tt vTight} and LAF extrapolated) to the CCSD(T) in the Composite `Reduced' approach.}
\begin{tabular}{lrrrr}
\toprule
 &  & Ih $E_\text{latt}$ & VIII $E_\text{latt}$ & (VIII - Ih) $E_\text{rel}$ \\ 
\midrule
\multirow[c]{4}{*}{1B} & Composite Reduced & -7.9 & -5.2 & 2.7 \\
 & LAF Full & -8.0 & -5.3 & 2.8 \\
 & Canonical Full & -8.0 & -5.3 & 2.8 \\
 & Composite Full & -8.1 & -5.3 & 2.8 \\
\cline{1-5}
\multirow[c]{4}{*}{2B} & Composite Reduced & -21.5 & -28.8 & -7.2 \\
 & LAF Full & -24.0 & -30.9 & -6.9 \\
 & Canonical Full & -24.0 & -31.0 & -7.0 \\
 & Composite Full & -23.7 & -30.7 & -7.0 \\
\cline{1-5}
\multirow[c]{4}{*}{3B} & Composite Reduced & -0.3 & 1.3 & 1.6 \\
 & LAF Full & 0.4 & 1.9 & 1.5 \\
 & Canonical Full & 0.1 & 2.5 & 2.4 \\
 & Composite Full & 0.1 & 2.6 & 2.5 \\
\cline{1-5}
\multirow[c]{4}{*}{Total} & Composite Reduced & -29.8 & -32.7 & -2.9 \\
 & LAF Full & -31.6 & -34.3 & -2.7 \\
 & Canonical Full & -31.9 & -33.8 & -1.8 \\
 & Composite Full & -31.7 & -33.4 & -1.8 \\
\cline{1-5}
\bottomrule
\end{tabular}

\end{table}

\clearpage

\section{\label{sec:x23}Lattice energy of X23 Dataset}

Within this section, we will arrive at our final LNO-MBE-CCSD(T) estimates of the lattice energy for the X23 dataset, comparing it to experimental lattice energies (corrected from sublimation enthalpies) and quantum diffusion Monte Carlo (DMC) estimates from the literature.

\subsection{\label{sec:x23_exp_analysis}Experimental analysis}

Experiments do not measure the lattice energy, but instead the sublimation enthalpy $\Delta H$, which include temperature and vibrational contributions.
Thus, these sublimation enthalpies must be converted to lattice energies (calculating the temperature contribution $E_\text{temp}$ from simulations) for an apples-to-apples comparison to our LNO-MBE-CCSD(T) estimates.
Additionally, for each of 23 systems within the X23 dataset, there are often several experimental measurements~\cite{dewitThermodynamicPropertiesMolecular1983,ribeirodasilvaThermodynamicStudySublimation2001,acreePhaseTransitionEnthalpy2016,acreePhaseTransitionEnthalpy2017} made with different instruments and techniques as well as at different years throughout time.
These experiments can vary quite significantly, as discussed in Ref.~\citenum{dellapiaHowAccurateAre2024}, with ranges of over $10\,$kJ/mol on the sublimation enthalpy for some studies.
A good estimate of for the error bars on the average of the experiments is also required to enable proper comparison to the simulations.

\begin{figure}[h]
    \includegraphics[width=\textwidth]{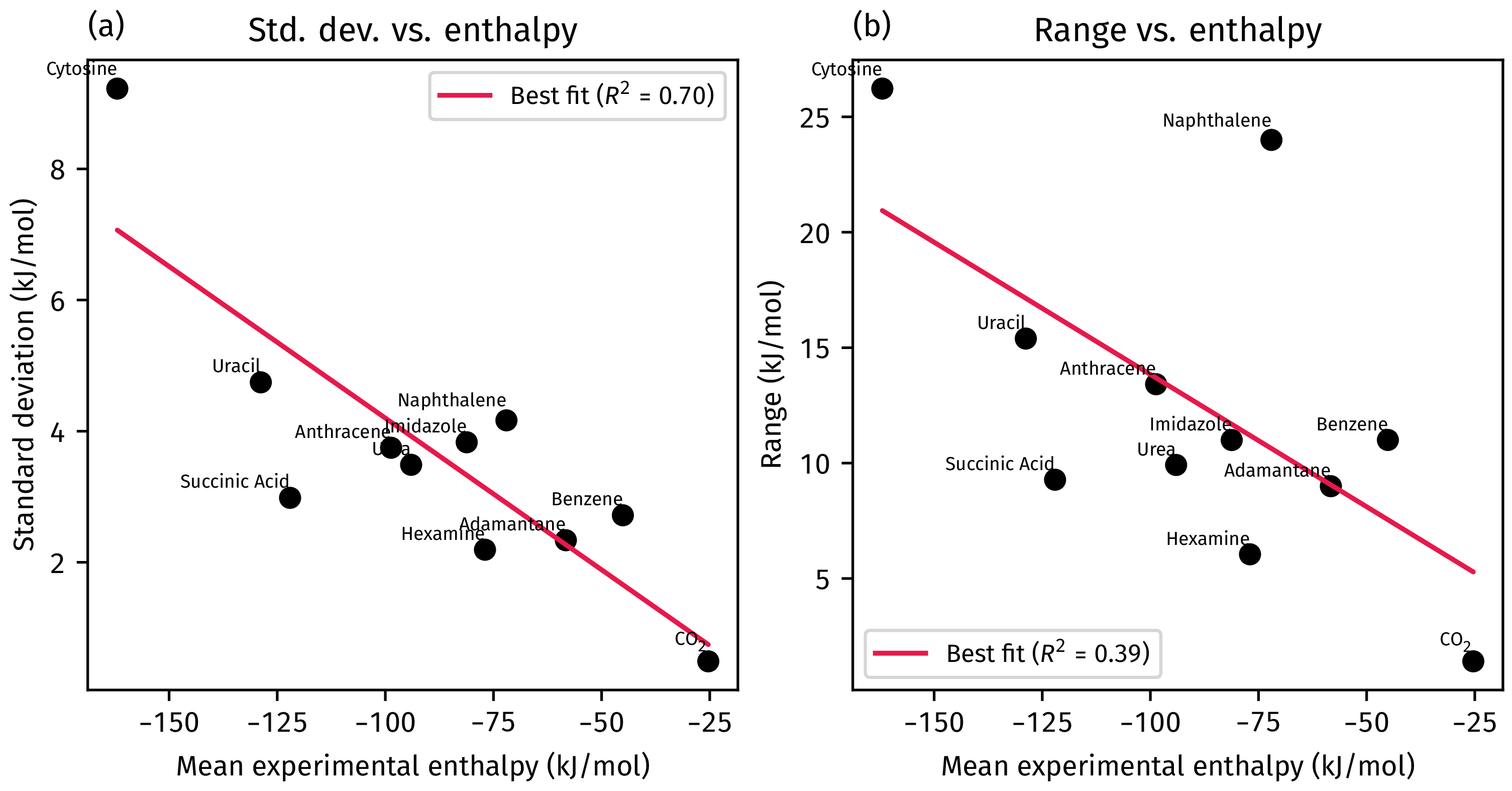}
    \caption{\label{fig:exp_enthalpy_std}Correlation between the $\Delta H$ and (a) standard deviation as well as (b) range for the set of X23 systems with 5 or more experimental measurements.}
\end{figure}

We consider two approaches to estimate the error on the mean value of the experimental sublimation enthalpy.
For the systems with 5 or more experimental measurements, we take the standard deviation as the error, quoting $2\sigma$ errors to account for 95\% confidence interval.
For the remaining systems, we make a prediction of the $2\sigma$ standard deviation.
In Figure~\ref{fig:exp_enthalpy_std}a, we show that there is a good correlation between the value of the mean experimental sublimation enthalpy as well as the standard deviation, with an $R^2$ of 0.70 for the set of systems with 5 or more sublimation enthalpy measurements.
The line of best fit we obtain has the following equation (in kJ/mol units):
\begin{equation}
    \sigma_\text{predicted} = -0.0462642\times\Delta H_\text{mean} + -0.427082.
\end{equation}
In panel b, we also consider the correlation with the range of the sublimation enthalpy measurements, finding significantly worse correlation with $R^2$ of 0.39.

To convert the sublimation enthalpies $\Delta H$ into lattice energies $E_\text{latt}$, we have used recent~\cite{dellapiaAccurateEfficientMachine2025} temperature corrections $E_\text{temp}$ calculated using machine-learned interatomic potentials (MLIPs) trained to DFT data:
The chosen DFT functional within this work is vdW-DF2~\cite{hamadaVanWaalsDensity2014}, which has been shown to demonstrate excellent agreement to DMC estimates for the X23 dataset.
Specifically, we take sublimation enthalpy values computed using path-integral molecular dynamics (PIMD) in Table S32 of the Supplementary Material for Ref.~\citenum{dellapiaAccurateEfficientMachine2025} simulations, accounting fully for anharmonicity and nuclear quantum effects.
The temperature correction is calculated as:
\begin{equation}
    E_\text{temp} = \Delta H - E_\text{latt}.
\end{equation}
As described in Ref.~\citenum{dellapiaAccurateEfficientMachine2025}, such an approach is particularly important for systems such as succinic acid due to its conformational flexibility and anharmonicity.
The final experimental estimate of the lattice energy and error bars are given in Table~\ref{tab:x23_experimental_lattice_energies}

\LTcapwidth=\textwidth

\begin{longtable}{llrrrrrrrr}
\caption{\label{tab:x23_experimental_lattice_energies}Compilation of experimental lattice energies of the X23 set, together with the mean and (analyzed) error estimate. All values are in kJ/mol.} \\

\toprule
Systems & Enthalpy measurements & \rotatebox{90}{$E_\text{temp}$~\cite{dellapiaAccurateEfficientMachine2025}} & \rotatebox{90}{Mean enthalpy} & \rotatebox{90}{2$\sigma$} & \rotatebox{90}{Predicted 2$\sigma$} & \rotatebox{90}{Range} & \rotatebox{90}{Lattice energy} & \rotatebox{90}{Error} & \rotatebox{90}{Error type} \\ 
\midrule
\endfirsthead

\caption[]{(continued)} \\
\endhead

\multicolumn{10}{r}{{Continued on next page}} \\
\endfoot

\bottomrule
\endlastfoot

\midrule
1,4-cyclohexanedione & \makecell[t]{-84.2, -75.0, -84.0} & 6.4 & -81.1 & - & 6.6 & - & -87.5 & 6.6 & Predicted \\
Acetic Acid & \makecell[t]{-66.2, -68.7} & 5.0 & -67.5 & - & 5.4 & - & -72.4 & 5.4 & Predicted \\
Adamantane & \makecell[t]{-58.5, -59.6, -56.0, -55.3\\-59.5, -60.2, -59.8, -54.2\\-59.2, -59.0, -53.0, -58.0\\-59.0, -60.0, -55.0, -59.0\\-60.0, -60.0, -62.0} & 6.7 & -58.3 & 4.7 & 4.5 & 9.0 & -64.9 & 4.7 & Real \\
Ammonia & \makecell[t]{-31.1} & 5.8 & -31.1 & - & 2.0 & - & -36.8 & 2.0 & Predicted \\
Anthracene & \makecell[t]{-98.8, -98.5, -99.1, -96.2\\-99.7, -103.5, -97.1, -103.3\\-100.7, -101.1, -99.5, -103.3\\-95.6, -91.9, -91.9, -96.1\\-100.6, -98.1, -95.9, -102.6\\-101.3, -99.2, -90.6, -104.1\\-99.3, -102.9, -102.9, -93.1\\-91.3, -94.2, -98.8, -98.0\\-100.0, -96.0, -99.0, -104.0\\-103.0} & 5.6 & -98.7 & 7.5 & 8.3 & 13.4 & -104.2 & 7.5 & Real \\
Benzene & \makecell[t]{-41.5, -45.0, -45.1, -52.5\\-48.0, -45.6, -43.8, -42.3\\-44.6, -46.6, -44.5, -42.5\\-44.3} & 7.7 & -45.1 & 5.4 & 3.3 & 11.0 & -52.8 & 5.4 & Real \\
CO$_2$ & \makecell[t]{-26.1, -24.7, -25.5, -25.5\\-25.0} & 3.8 & -25.3 & 1.0 & 1.5 & 1.4 & -29.1 & 1.0 & Real \\
Cyanamide & \makecell[t]{-75.8, -75.0} & 4.2 & -75.4 & - & 6.1 & - & -79.6 & 6.1 & Predicted \\
Cytosine & \makecell[t]{-168.8, -155.3, -149.8, -155.0\\-167.0, -176.0} & 5.3 & -162.0 & 18.5 & 14.1 & 26.2 & -167.2 & 18.5 & Real \\
Ethyl carbamate & \makecell[t]{-77.2, -72.3, -89.1, -76.0} & 6.2 & -78.6 & - & 6.4 & - & -84.9 & 6.4 & Predicted \\
Formamide & \makecell[t]{-72.2, -71.7, -72.4} & 6.5 & -72.1 & - & 5.8 & - & -78.6 & 5.8 & Predicted \\
Hexamine & \makecell[t]{-78.7, -77.0, -79.1, -74.0\\-80.0, -75.0, -75.0} & 6.7 & -77.0 & 4.4 & 6.3 & 6.0 & -83.7 & 4.4 & Real \\
Imidazole & \makecell[t]{-83.1, -80.8, -83.0, -74.0\\-85.0} & 4.6 & -81.2 & 7.7 & 6.7 & 11.0 & -85.8 & 7.7 & Real \\
Naphthalene & \makecell[t]{-87.8, -72.3, -73.0, -79.0\\-71.3, -73.7, -71.0, -72.5\\-75.9, -73.3, -72.5, -72.7\\-77.1, -72.0, -71.2, -74.2\\-73.3, -74.6, -72.6, -67.5\\-63.8, -72.8, -71.8, -66.0\\-65.7, -68.7, -72.0, -72.2\\-70.0, -72.0, -73.0, -72.0\\-72.0, -72.0, -73.0, -64.0\\-66.0} & 5.3 & -72.0 & 8.3 & 5.8 & 24.0 & -77.3 & 8.3 & Real \\
Oxalic Acid $\alpha$ & \makecell[t]{-93.7, -94.0} & 2.5 & -93.8 & - & 7.8 & - & -96.4 & 7.8 & Predicted \\
Oxalic Acid $\beta$ & \makecell[t]{-93.6} & 1.0 & -93.6 & - & 7.8 & - & -94.6 & 7.8 & Predicted \\
Pyrazine & \makecell[t]{-56.3} & 6.9 & -56.3 & - & 4.4 & - & -63.2 & 4.4 & Predicted \\
Pyrazole & \makecell[t]{-73.9, -72.2, -74.0, -69.0} & 4.1 & -72.3 & - & 5.8 & - & -76.3 & 5.8 & Predicted \\
Succinic Acid & \makecell[t]{-128.7, -119.4, -121.7, -119.6\\-123.0, -122.0, -120.0} & -5.2 & -122.0 & 6.0 & 10.4 & 9.3 & -116.8 & 6.0 & Real \\
Triazine & \makecell[t]{-56.9, -54.2} & 6.6 & -55.6 & - & 4.3 & - & -62.2 & 4.3 & Predicted \\
Trioxane & \makecell[t]{-56.7, -55.6, -56.5, -56.2} & 7.8 & -56.2 & - & 4.3 & - & -64.1 & 4.3 & Predicted \\
Uracil & \makecell[t]{-127.4, -130.8, -129.3, -134.3\\-122.2, -123.8, -137.6, -128.9\\-122.6, -131.0} & 4.5 & -128.8 & 9.5 & 11.1 & 15.4 & -133.3 & 9.5 & Real \\
Urea & \makecell[t]{-95.8, -96.0, -95.5, -92.3\\-88.7, -97.8, -96.4, -88.9\\-89.2, -95.5, -98.6} & 5.2 & -94.1 & 7.0 & 7.8 & 9.9 & -99.2 & 7.0 & Real \\
\end{longtable}

\clearpage

\subsection{\label{sec:x23_hf}Periodic HF}

The first step to our final LNO-MBE-CCSD(T) lattice energy is the calculation of the Hartree-Fock (HF) using the periodic plane-wave implemention in VASP (described in Section~\ref{sec:vasp_details}).
The lattice energy is defined as:
\begin{equation}
    E_{\text{latt}} = E_{\text{crys}} - E_{\text{gas}},
\end{equation}
where $E_{\text{crys}}$ is the total energy per molecule in the crystal phase while $E_{\text{gas}}$ is that for the molecule in the (isolated) gas phase.
Achieving accurate estimates of the lattice energy requires hard PAW potentials~\cite{brandenburgBenchmarkingDFTSemiempirical2015,yourdkhaniUsingNoncovalentInteractions2023} with a (1) large energy cutoff of $1000\,$eV together with a (2) converged $k$-point mesh.
We follow a composite scheme to reach both criteria, which goes as:
\begin{itemize}
    \item The main calculation is performed with hard PAW potentials (with $1000\,$eV energy cutoff) and a moderate $k$-point spacing of 0.25~\AA{}$^{-1}$.
    \item The $\Delta_k$ correction towards denser $k$-point spacing of 0.18~\AA{}$^{-1}$ (on top of a spacing of 0.25~\AA{}$^{-1}$) is performed using standard (Std) PAW potentials and a $520\,$eV energy cutoff.
\end{itemize}

These separate estimates are given in Table~\ref{tab:x23_hf_contributions}.
In general, it can be seen that the first calculation is sufficient for accuracy of within $0.2\,$kJ/mol for all of the systems, making the $\Delta$ correction relatively negligible.
It also highlights in the importance of hard PAW potentials as the standard PAW potentials can differ by as much as $7\,$kJ/mol for some of the X23 systems.

As described in the main text, one of the new developments we leverage is new implementations over the past decade that use graphic processing units~\cite{haceneAcceleratingVASPElectronic2012} (GPU) together with the Adaptively Compressed Exchange~\cite{linAdaptivelyCompressedExchange2016} (ACE) operator.
We report the GPU costs (in GPUh) for the individual contributions in Table~\ref{tab:x23_hf_gpu_cost}.
Overall, it can be seen that for the dominant hard PAW calculations, the cost is less than $5\,$GPUh on average, with a maximum of $25\,$GPUh for Trioxane.
In fact, the cost for the negligible $\Delta_k$ correction is more, at a sum of ${\sim}11\,$GPU on average.

\begin{table}[h]

\caption{\label{tab:x23_hf_contributions}Contributions that make up the final HF lattice energies for the X23 set. All values are in kJ/mol.}
\begin{tabular}{lrrrrr}
\toprule
 & \rotatebox{90}{Hard (KSPACING=0.25)} & \rotatebox{90}{Std (KSPACING=0.25)} & \rotatebox{90}{Std (KSPACING=0.18)} & \rotatebox{90}{$\Delta_k$} & \rotatebox{90}{Final} \\ 
\midrule
1,4-cyclohexanedione & -3.4 & -5.7 & -5.4 & 0.2 & -3.2 \\
Acetic Acid & -17.4 & -20.1 & -20.1 & 0.0 & -17.4 \\
Adamantane & 41.5 & 41.5 & 41.4 & -0.1 & 41.5 \\
Ammonia & -9.1 & -9.4 & -9.4 & -0.0 & -9.1 \\
Anthracene & 54.9 & 55.0 & 55.0 & -0.0 & 54.9 \\
Benzene & 23.0 & 23.0 & 23.0 & -0.1 & 22.9 \\
CO$_2$ & -3.4 & -4.3 & -4.3 & -0.0 & -3.5 \\
Cyanamide & -29.9 & -30.7 & -30.8 & -0.1 & -29.9 \\
Cytosine & -48.0 & -51.1 & -51.1 & 0.0 & -48.0 \\
Ethyl carbamate & -17.4 & -19.7 & -19.8 & -0.0 & -17.4 \\
Formamide & -30.3 & -33.0 & -33.1 & -0.0 & -30.4 \\
Hexamine & 18.0 & 18.2 & 18.2 & -0.1 & 18.0 \\
Imidazole & -14.3 & -14.7 & -14.7 & -0.0 & -14.3 \\
Naphthalene & 43.7 & 43.8 & 43.8 & -0.0 & 43.7 \\
Oxalic Acid $\alpha$ & -35.3 & -40.7 & -40.8 & -0.0 & -35.3 \\
Oxalic Acid $\beta$ & -28.2 & -33.4 & -33.5 & -0.0 & -28.2 \\
Pyrazine & 14.5 & 14.1 & 14.0 & -0.0 & 14.5 \\
Pyrazole & -3.8 & -4.2 & -4.2 & -0.0 & -3.9 \\
Succinic Acid & -22.6 & -26.9 & -27.1 & -0.2 & -22.7 \\
Triazine & 2.4 & 1.8 & 1.8 & -0.0 & 2.4 \\
Trioxane & 6.2 & 4.2 & 4.2 & 0.0 & 6.2 \\
Uracil & -38.5 & -42.6 & -42.7 & -0.1 & -38.6 \\
Urea & -45.4 & -49.4 & -49.7 & -0.3 & -45.6 \\
\bottomrule
\end{tabular}

\end{table}
\begin{table}[h]

\caption{\label{tab:x23_hf_gpu_cost}Estimated computational cost (in GPUh) for the composite VASP calculations used to compute the HF lattice energies of the X23 dataset on an NVIDIA A100 GPU. The three different settings correspond to different PAW potentials and plane-wave cutoffs (with $k$-point spacing given in parenthesis), as described in the text.}
\begin{tabular}{lrrr}
\toprule
 & \rotatebox{90}{Hard (0.25)} & \rotatebox{90}{Std (0.25)} & \rotatebox{90}{Std (0.18)} \\ 
\midrule
1,4-cyclohexanedione & 3.4 & 1.6 & 11.2 \\
Acetic Acid & 2.9 & 1.4 & 8.5 \\
Adamantane & 1.4 & 0.7 & 4.7 \\
Ammonia & 0.2 & 0.2 & 0.9 \\
Anthracene & 4.0 & 1.7 & 7.1 \\
Benzene & 2.4 & 0.9 & 4.5 \\
CO$_2$ & 0.5 & 0.2 & 1.2 \\
Cyanamide & 4.3 & 2.0 & 17.3 \\
Cytosine & 4.2 & 1.7 & 10.4 \\
Ethyl carbamate & 2.6 & 1.4 & 15.0 \\
Formamide & 2.3 & 1.2 & 6.8 \\
Hexamine & 1.8 & 1.0 & 4.7 \\
Imidazole & 2.2 & 1.1 & 7.8 \\
Naphthalene & 3.9 & 1.9 & 6.0 \\
Oxalic Acid $\alpha$ & 5.6 & 2.7 & 10.7 \\
Oxalic Acid $\beta$ & 3.9 & 1.9 & 9.2 \\
Pyrazine & 1.8 & 1.0 & 5.0 \\
Pyrazole & 9.8 & 3.5 & 22.2 \\
Succinic Acid & 2.5 & 1.2 & 7.9 \\
Triazine & 8.3 & 2.6 & 7.4 \\
Trioxane & 24.8 & 8.4 & 26.3 \\
Uracil & 14.3 & 5.4 & 8.9 \\
Urea & 0.7 & 0.3 & 1.6 \\
Average & 4.7 & 1.9 & 8.9 \\
\bottomrule
\end{tabular}

\end{table}

\clearpage

\subsection{\label{sec:x23_mp2_ccsd}MBE cWFT correlation}

We have used the Composite approach with `Full' basis set described in Section~\ref{sec:cheaper_settings} to calculate the 1B, 2B and 3B contributions from the correlation energy to the lattice energy, which we report in Table~\ref{tab:x23_mbe_contributions}.
We have reported several post-HF methods besides CCSD(T), such as second-order M{\o}ller-Plesset perturbation theory~\cite{mollerNoteApproximationTreatment1934} (MP2) as well as coupled cluster theory with single and double excitations.
Within the Composite approach, MP2 is devoid of local approximations, while CCSD and CCSD(T) make use of the LNO approximation.
We leave comparisons of the performance of these methods to Section~\ref{sec:dft_benchmark} and discuss general trends here.

We find that the 2B contribution makes up the bulk of the lattice energy, being an order of magnitude or more than the 3B contributions.
MP2 generally predicts less-repulsive values (by an average of $2.6\,$kJ/mol) on the 3B contributions than CCSD or CCSD(T) due to missing (repulsive) three-body dispersion effects~\cite{huangReliablePredictionThreebody2015}.
Conversely, CCSD can reproduce 3B contributions well~\cite{rezacBenchmarkCalculationsThreeBody2015} to a mean absolute deviation (MAD) of $0.6\,$kJ/mol and tends to underbind the 2B contributions significantly by an MAD of $13.8\,$kJ/mol which is equivalent to the mean-signed deviation.
MP2 has an MAD of $6.9\,$kJ/mol on the 2B terms and most of this comes from an overbinding of dispersion-bound compounds~\cite{e.rileyMP2XGeneralizedMP252011}.

\begin{table}[h]

\caption{\label{tab:x23_mbe_contributions}The MBE contributions to the lattice energy of the X23 set for several correlated wavefunction theory methods in the Composite `Full' basis approach. All values are in kJ/mol. We also report the mean signed deviation (MSD) and mean absolute deviation (MAD) of MP2 and CCSD compared to CCSD(T).}
\begin{tabular}{lrrrrrrrrrrrrrrrrrrrr}
\toprule
 & \multicolumn{4}{c}{MP2} & \multicolumn{4}{c}{CCSD} & \multicolumn{4}{c}{CCSD(T)} \\ \cmidrule(lr){2-5} \cmidrule(lr){6-9} \cmidrule(lr){10-13} 
 & 1B & 2B & 3B & Total & 1B & 2B & 3B & Total & 1B & 2B & 3B & Total \\
\midrule
1,4-cyclohexanedione & -6.4 & -86.7 & 0.7 & -92.5 & -4.5 & -71.3 & 3.0 & -72.8 & -5.8 & -86.4 & 3.6 & -88.6 \\
Acetic Acid & -4.8 & -49.1 & 0.8 & -53.1 & -3.7 & -42.1 & 2.2 & -43.6 & -4.5 & -50.5 & 2.6 & -52.4 \\
Adamantane & -0.4 & -115.3 & 1.2 & -114.4 & -0.6 & -94.5 & 4.3 & -90.9 & -1.0 & -113.1 & 4.7 & -109.3 \\
Ammonia & -0.9 & -27.6 & 0.4 & -28.1 & -0.9 & -24.2 & 1.3 & -23.8 & -1.1 & -28.7 & 1.4 & -28.4 \\
Anthracene & -2.4 & -212.0 & 1.8 & -212.5 & -1.8 & -141.0 & 4.4 & -138.4 & -2.7 & -171.7 & 5.3 & -169.1 \\
Benzene & -1.0 & -92.7 & 2.1 & -91.6 & -0.9 & -66.3 & 4.6 & -62.6 & -1.3 & -79.6 & 5.3 & -75.7 \\
CO$_2$ & 0.1 & -26.8 & 1.0 & -25.7 & 0.1 & -22.4 & 2.0 & -20.2 & 0.1 & -26.9 & 2.3 & -24.5 \\
Cyanamide & -3.4 & -59.6 & 2.2 & -60.8 & -2.1 & -43.4 & 3.8 & -41.7 & -2.5 & -53.5 & 4.1 & -51.9 \\
Cytosine & -10.7 & -113.7 & -0.1 & -124.4 & -8.3 & -88.5 & 3.3 & -93.5 & -10.3 & -107.5 & 3.8 & -114.1 \\
Ethyl carbamate & -2.9 & -64.9 & 0.3 & -67.4 & -2.3 & -56.1 & 2.1 & -56.3 & -3.0 & -67.1 & 2.5 & -67.6 \\
Formamide & -7.9 & -41.8 & 0.2 & -49.5 & -6.0 & -37.0 & 2.0 & -41.0 & -7.5 & -44.8 & 2.6 & -49.7 \\
Hexamine & -0.9 & -111.7 & 1.0 & -111.6 & -0.5 & -90.8 & 3.7 & -87.6 & -0.7 & -108.9 & 4.0 & -105.6 \\
Imidazole & -5.0 & -84.6 & 1.1 & -88.4 & -3.6 & -61.5 & 3.4 & -61.6 & -4.5 & -74.1 & 3.9 & -74.6 \\
Naphthalene & -1.4 & -159.7 & 1.2 & -159.9 & -1.2 & -109.0 & 2.8 & -107.5 & -1.8 & -132.2 & 3.2 & -130.8 \\
Oxalic Acid $\alpha$ & 1.8 & -71.1 & 4.1 & -65.3 & 1.5 & -58.4 & 4.6 & -52.2 & 2.4 & -71.6 & 5.1 & -64.1 \\
Oxalic Acid $\beta$ & -3.5 & -73.5 & 5.3 & -71.7 & -2.7 & -59.8 & 4.9 & -57.5 & -2.7 & -74.3 & 6.5 & -70.5 \\
Pyrazine & -2.4 & -96.3 & 1.8 & -96.8 & -1.8 & -67.7 & 5.4 & -64.0 & -2.6 & -81.7 & 6.4 & -77.9 \\
Pyrazole & -4.0 & -86.5 & 1.7 & -88.8 & -2.9 & -62.9 & 3.8 & -62.0 & -3.7 & -75.6 & 4.3 & -75.1 \\
Succinic Acid & -7.7 & -100.8 & 3.2 & -105.3 & -6.3 & -83.8 & 6.0 & -84.1 & -7.5 & -101.8 & 7.2 & -102.2 \\
Triazine & -2.8 & -71.0 & 2.2 & -71.6 & -2.0 & -54.2 & 4.5 & -51.8 & -2.9 & -65.8 & 5.3 & -63.4 \\
Trioxane & -4.7 & -67.7 & 1.5 & -70.9 & -3.9 & -58.4 & 3.2 & -59.0 & -5.1 & -68.8 & 3.9 & -70.1 \\
Uracil & -12.3 & -94.3 & -0.5 & -107.1 & -8.8 & -73.1 & 1.7 & -80.1 & -11.2 & -90.1 & 2.7 & -98.6 \\
Urea & -5.8 & -53.7 & -2.6 & -62.1 & -3.9 & -47.3 & 0.6 & -50.6 & -4.5 & -56.9 & 0.3 & -61.2 \\
MSD & -0.2 & -5.6 & -2.6 & -8.4 & 0.8 & 13.8 & -0.6 & 14.0 & 0.0 & 0.0 & 0.0 & 0.0 \\
MAD & 0.4 & 6.9 & 2.6 & 8.5 & 0.8 & 13.8 & 0.6 & 14.0 & 0.0 & 0.0 & 0.0 & 0.0 \\
\bottomrule
\end{tabular}

\end{table}

\clearpage

\subsection{\label{sec:ct_effects}Effects from CCSD(cT)-fit}

As discussed in the main text, there are discrepancies between DMC and LNO-MBE-CCSD(T) for $\pi-\pi$ dispersion-bound complexes such as naphthalene and anthracene of over $10\,$kJ/mol.
This is in line with observations by Al-Hamdani and Nagy {\it et al.}~\cite{al-hamdaniInteractionsLargeMolecules2021}, which found even larger discrepancies for the interaction energy between $\pi-\pi$ bound supramolecules including coronene and circumcoronene.
It is currently a matter of open debate where these errors lie~\cite{schaferUnderstandingDiscrepanciesNoncovalent2025,fishmanAnotherAngleBenchmarking2025,lambieApplicabilityCCSDTDispersion2025,laoCanonicalCoupledCluster2024}, with studies indicating either errors in the DMC nodal surface~\cite{lambieNodalErrorDiscrepancies2025} or level of convergence up to the perturbative triples (T) contribution~\cite{schaferUnderstandingDiscrepanciesNoncovalent2025} in CCSD(T).

Sch{\"a}fer \etal{}~\cite{schaferUnderstandingDiscrepanciesNoncovalent2025} have suggested that part of this discrepancy arises from missing contributions in (T) that can be accounted by the (cT) approach.
They have demonstrated that there exists an empirical relationship between the (cT) and the (T) correlation energy contributions
 for dispersion-dominated complexes.
This empirical relationship makes use of the MP2, CCSD and (T) correlation energies, with a (cT)-fit of the form:
\begin{equation}
    \dfrac{\text{(T)}}{\text{(cT)-fit}} = a + b \cdot \dfrac{\text{MP2 corr.}}{\text{CCSD corr.}},
\end{equation}
where $a$ and $b$ were parameters fitted from a set of dispersion-bound complexes.
While this means that CCSD(cT)-fit should only be applied to study dispersion-bound systems, we find that it does not affect performance for H-bonded systems both within this work and for dimers of the S66 dataset~\cite{shiSystematicDiscrepanciesReference2025}, while giving improvements in agreement to DMC for dispersion-bound systems.

\begin{figure}
    \includegraphics{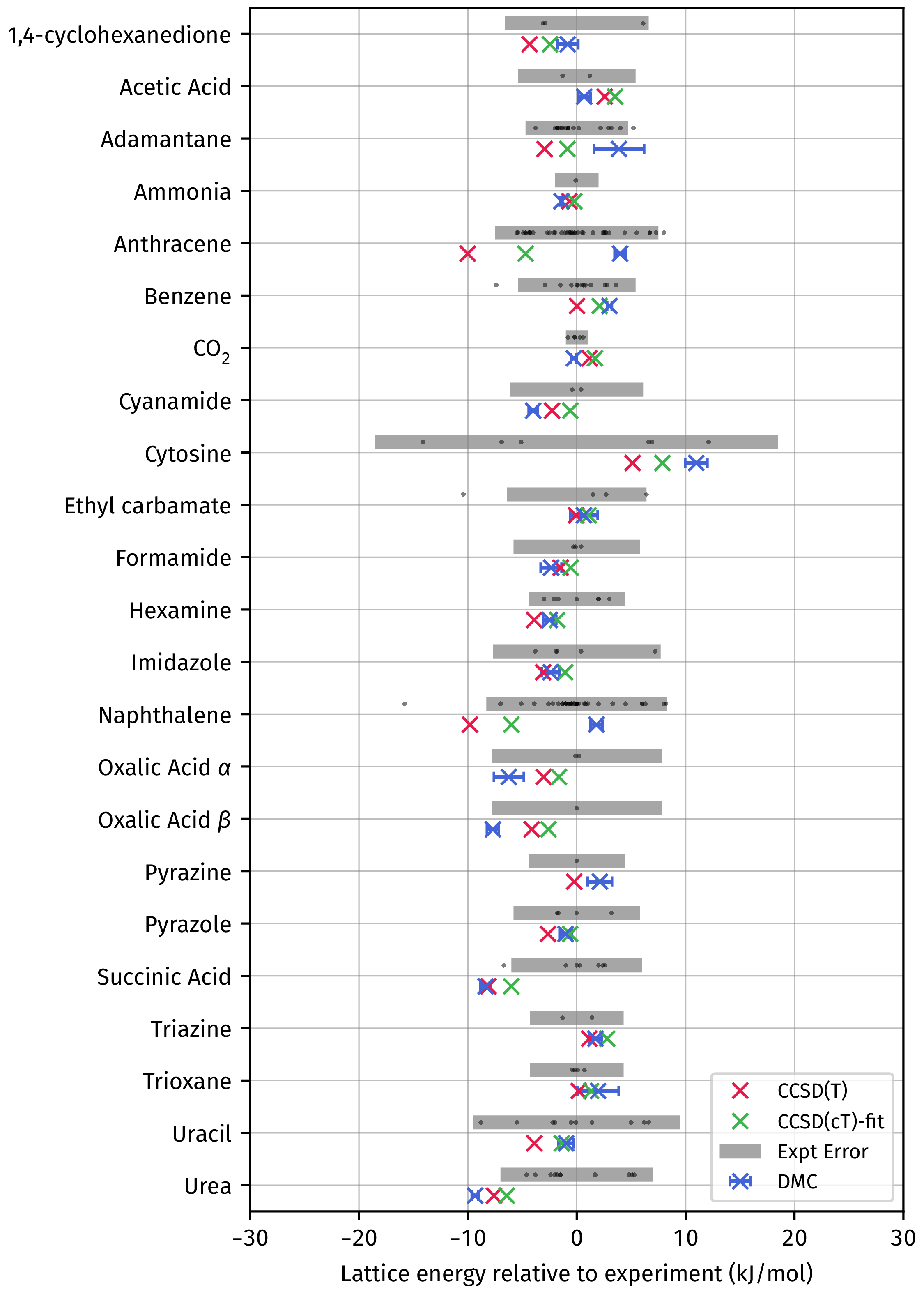}
    \caption{\label{fig:ct_effect}Comparing the relative differences of DMC, CCSD(T) and CCSD(cT)-fit w.r.t. the experimental mean and error.}
\end{figure}

In Table~\ref{tab:x23_cT_mbe_contributions}, we have computed the MBE contributions of the CCSD(cT)-fit correlation energy and compared it to CCSD(T).
Its effect is small for the 1B and 3B terms and the general trend is to weaken the 2B contributions, leading to a MSD and MAD of $1.7\,$kJ/mol w.r.t.\ the CCSD(T) estimate.
Importantly, this is most prominent for dispersion-bound systems such as anthracene and naphthalene, amounting to $5.1\,$kJ/mol and $3.7\,$kJ/mol, respectively.
As seen in Figure~\ref{fig:ct_effect}, in both cases, this brings CCSD(cT)-fit closer in the direction of the DMC estimates than CCSD(T), albeit not fully in exact agreement with DMC, commensurate with observations by~\citet{schaferUnderstandingDiscrepanciesNoncovalent2025}.
This in turn has improved the MAD from $3.1\,$kJ/mol to $2.5\,$kJ/mol for the entire X23 dataset (Table~\ref{tab:x23_final_compare}). On the other hand, other studies suggest that beyond CCSD(cT) effects from quadruple excitations move the CC results back close to CCSD(T) for $\pi$-$\pi$ dimers,\cite{fishmanAnotherAngleBenchmarking2025} so one cannot provide more conclusive results at the moment.

\begin{table}[h]

\caption{\label{tab:x23_cT_mbe_contributions}The MBE contributions to the lattice energy of the X23 set between CCSD(T) and CCSD(cT)-fit in the Composite `Full' basis approach. All values are in kJ/mol. We also report the mean signed deviation (MSD) and mean absolute deviation (MAD) compared to CCSD(T).}
\begin{tabular}{lrrrrrrrrrrrrrrrr}
\toprule
 & \multicolumn{4}{c}{CCSD(T)} & \multicolumn{4}{c}{CCSD(cT)-fit} \\ \cmidrule(lr){2-5} \cmidrule(lr){6-9} \cmidrule(lr){10-13} \cmidrule(lr){14-17} 
 & 1B & 2B & 3B & Total & 1B & 2B & 3B & Total \\
\midrule
1,4-cyclohexanedione & -5.8 & -86.4 & 3.6 & -88.6 & -5.6 & -84.8 & 3.7 & -86.7 \\
Acetic Acid & -4.5 & -50.5 & 2.6 & -52.4 & -4.4 & -49.7 & 2.7 & -51.5 \\
Adamantane & -1.0 & -113.1 & 4.7 & -109.3 & -1.0 & -111.1 & 4.8 & -107.2 \\
Ammonia & -1.1 & -28.7 & 1.4 & -28.4 & -1.1 & -28.3 & 1.5 & -27.9 \\
Anthracene & -2.7 & -171.7 & 5.3 & -169.1 & -2.6 & -166.6 & 5.4 & -163.8 \\
Benzene & -1.3 & -79.6 & 5.3 & -75.7 & -1.3 & -77.7 & 5.4 & -73.6 \\
CO$_2$ & 0.1 & -26.9 & 2.3 & -24.5 & 0.1 & -26.4 & 2.4 & -24.0 \\
Cyanamide & -2.5 & -53.5 & 4.1 & -51.9 & -2.4 & -52.0 & 4.1 & -50.2 \\
Cytosine & -10.3 & -107.5 & 3.8 & -114.1 & -10.1 & -105.2 & 3.9 & -111.3 \\
Ethyl carbamate & -3.0 & -67.1 & 2.5 & -67.6 & -2.9 & -66.1 & 2.6 & -66.4 \\
Formamide & -7.5 & -44.8 & 2.6 & -49.7 & -7.3 & -44.1 & 2.7 & -48.8 \\
Hexamine & -0.7 & -108.9 & 4.0 & -105.6 & -0.6 & -107.0 & 4.0 & -103.5 \\
Imidazole & -4.5 & -74.1 & 3.9 & -74.6 & -4.3 & -72.3 & 4.0 & -72.6 \\
Naphthalene & -1.8 & -132.2 & 3.2 & -130.8 & -1.7 & -128.5 & 3.3 & -127.0 \\
Oxalic Acid $\alpha$ & 2.4 & -71.6 & 5.1 & -64.1 & 2.3 & -70.1 & 5.1 & -62.7 \\
Oxalic Acid $\beta$ & -2.7 & -74.3 & 6.5 & -70.5 & -2.7 & -72.7 & 6.4 & -69.0 \\
Pyrazine & -2.6 & -81.7 & 6.4 & -77.9 & -2.5 & -79.6 & 6.5 & -75.6 \\
Pyrazole & -3.7 & -75.6 & 4.3 & -75.1 & -3.6 & -73.8 & 4.3 & -73.0 \\
Succinic Acid & -7.5 & -101.8 & 7.2 & -102.2 & -7.4 & -99.9 & 7.3 & -100.1 \\
Triazine & -2.9 & -65.8 & 5.3 & -63.4 & -2.8 & -64.4 & 5.4 & -61.8 \\
Trioxane & -5.1 & -68.8 & 3.9 & -70.1 & -5.0 & -67.9 & 3.9 & -69.0 \\
Uracil & -11.2 & -90.1 & 2.7 & -98.6 & -10.8 & -88.0 & 2.9 & -96.1 \\
Urea & -4.5 & -56.9 & 0.3 & -61.2 & -4.4 & -56.1 & 0.5 & -60.0 \\
MSD & 0.0 & 0.0 & 0.0 & 0.0 & 0.1 & 1.7 & 0.1 & 1.9 \\
MAD & 0.0 & 0.0 & 0.0 & 0.0 & 0.1 & 1.7 & 0.1 & 1.9 \\
\bottomrule
\end{tabular}

\end{table}

\clearpage

\subsection{\label{sec:final_estimates}Final estimates and comparison to DMC}

In Table~\ref{tab:x23_final_compare}, we compare the final (LNO-MBE-)CCSD(T) and CCSD(cT)-fit estimates to both DMC and experiments.
Within this context, it's not clear what should be used as the reference, given the large error bars on the experimental lattice energies and the open question on whether CCSD(T) or DMC has larger error bars.
Regardless, the agreement between all of the approaches is great, reaching sub-chemical accuracy of below $4\,$kJ/mol regardless of whether the experiment, LNO-MBE-CCSD(T) or DMC are used as the reference.
For experiments, both DMC and LNO-MBE-CCSD(T) have an MAD of $3.4\,$kJ/mol w.r.t.\ experiments, although the large error bars of experiments (the majority exceeding $4\,$kJ/mol) will make it difficult to assess the accuracy between these methods at this level of precision.

\begin{figure}[h]
    \includegraphics[width=\textwidth]{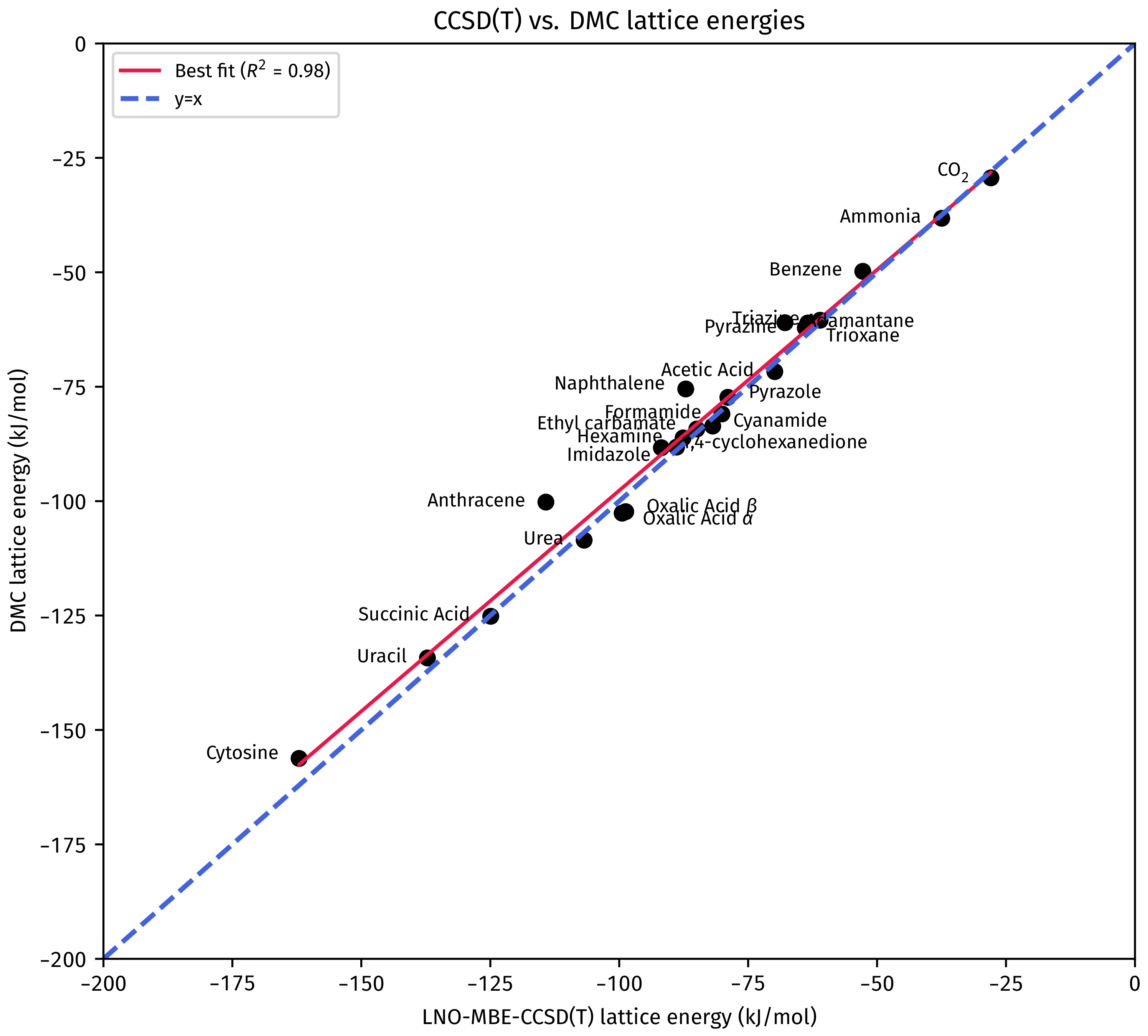}
    \caption{\label{fig:x23_cc_dmc}Parity plot of the LNO-MBE-CCSD(T) and DMC lattice energies for the X23 dataset.}
\end{figure}

When comparing the LNO-MBE-CCSD(T) estimates to DMC (compared visually in Figure~\ref{fig:x23_cc_dmc}), it can be seen that there is a tendency to predict stronger binding, with an MSD of $-1.8\,$kJ/mol.
As shown in Table~\ref{tab:x23_vdw_bonded_compare}, this stronger binding comes predominantly from the vdW dispersion-bound systems (as defined in Ref.~\citenum{reillyUnderstandingRoleVibrations2013b}).
We have compared the lattice energy between LNO-MBE-CCSD(T) and DMC for this set of systems and it can be seen that the MSD [negative indicating stronger binding of CCSD(T)] lies close in magnitude of the MAD.
This trend is in agreement with a recent comparison of dimers of different binding characters in Ref.~\citenum{shiSystematicDiscrepanciesReference2025}, which indicates stronger binding of CCSD(T) w.r.t.\ DMC.

In Table~\ref{tab:x23_h_bonded_compare}, we have also compared the lattice energies between DMC and LNO-MBE-CCSD(T) for the subset of systems with predominantly H-bonding character~\cite{reillyUnderstandingRoleVibrations2013b}.
It follows similar trends for those recently observed for dimers~\cite{shiSystematicDiscrepanciesReference2025} whereby CCSD(T) predicts weaker binding than DMC, indicated by a positive MSD that is very close in magnitude to the MAD.

\begin{table}[h]

\caption{\label{tab:x23_final_compare}Final lattice energy (in kJ/mol) for the X23 dataset computed with different methods, compared to experimental values. The DMC values are taken from Ref.~\citenum{dellapiaHowAccurateAre2024}. The mean absolute deviation (MAD) is also reported with respect to (LNO-MBE-)CCSD(T), experiment and DMC.}
\begin{tabular}{lrrrrrr}
\toprule
 & CCSD(T) & CCSD(cT)-fit & DMC & DMC Error & Expt & Expt Error \\ 
\midrule
1,4-cyclohexanedione & -91.8 & -89.9 & -88.3 & 1.0 & -87.5 & 6.6 \\
Acetic Acid & -69.8 & -68.9 & -71.7 & 0.6 & -72.4 & 5.4 \\
Adamantane & -67.9 & -65.8 & -61.0 & 2.3 & -64.9 & 4.7 \\
Ammonia & -37.5 & -37.0 & -38.2 & 0.1 & -36.8 & 2.0 \\
Anthracene & -114.2 & -108.9 & -100.2 & 0.5 & -104.2 & 7.5 \\
Benzene & -52.8 & -50.7 & -49.8 & 0.2 & -52.8 & 5.4 \\
CO$_2$ & -27.9 & -27.4 & -29.4 & 0.2 & -29.1 & 1.0 \\
Cyanamide & -81.9 & -80.2 & -83.6 & 0.4 & -79.6 & 6.1 \\
Cytosine & -162.1 & -159.3 & -156.2 & 1.0 & -167.2 & 18.5 \\
Ethyl carbamate & -85.0 & -83.8 & -84.2 & 1.3 & -84.9 & 6.4 \\
Formamide & -80.1 & -79.2 & -81.0 & 1.0 & -78.6 & 5.8 \\
Hexamine & -87.6 & -85.5 & -86.2 & 0.6 & -83.7 & 4.4 \\
Imidazole & -88.9 & -86.9 & -88.2 & 0.8 & -85.8 & 7.7 \\
Naphthalene & -87.1 & -83.3 & -75.5 & 0.5 & -77.3 & 8.3 \\
Oxalic Acid $\alpha$ & -99.4 & -98.0 & -102.6 & 1.4 & -96.4 & 7.8 \\
Oxalic Acid $\beta$ & -98.7 & -97.2 & -102.3 & 0.6 & -94.6 & 7.8 \\
Pyrazine & -63.4 & -61.1 & -61.1 & 1.1 & -63.2 & 4.4 \\
Pyrazole & -78.9 & -76.9 & -77.3 & 0.5 & -76.3 & 5.8 \\
Succinic Acid & -124.9 & -122.8 & -125.2 & 0.5 & -116.8 & 6.0 \\
Triazine & -61.1 & -59.4 & -60.5 & 0.6 & -62.2 & 4.3 \\
Trioxane & -63.9 & -62.8 & -62.1 & 1.9 & -64.1 & 4.3 \\
Uracil & -137.2 & -134.7 & -134.3 & 0.7 & -133.3 & 9.5 \\
Urea & -106.8 & -105.6 & -108.5 & 0.3 & -99.2 & 7.0 \\
MAD [CCSD(T)] & 0.0 & 1.9 & 3.1 & - & 3.4 & - \\
MAD (Expt) & 3.4 & 2.6 & 3.4 & - & 0.0 & - \\
MAD (DMC) & 3.1 & 2.5 & 0.0 & - & 3.4 & - \\
MSD (DMC) & -1.8 & 0.1 & 0.0 & - & 0.7 & - \\
\bottomrule
\end{tabular}

\end{table}

\begin{table}[h]

\caption{\label{tab:x23_vdw_bonded_compare}Lattice energy (in kJ/mol) for the vdW-bonded subset of the X23 dataset computed with different methods, compared to experimental values. The DMC values are taken from Ref.~\citenum{dellapiaHowAccurateAre2024}. The mean absolute deviation (MAD) is also reported with respect to (LNO-MBE-)CCSD(T), experiment and DMC.}
\begin{tabular}{lrrrrrr}
\toprule
 & CCSD(T) & CCSD(cT)-fit & DMC & DMC Error & Expt & Expt Error \\ 
\midrule
1,4-cyclohexanedione & -91.8 & -89.9 & -88.3 & 1.0 & -87.5 & 6.6 \\
Adamantane & -67.9 & -65.8 & -61.0 & 2.3 & -64.9 & 4.7 \\
Anthracene & -114.2 & -108.9 & -100.2 & 0.5 & -104.2 & 7.5 \\
Benzene & -52.8 & -50.7 & -49.8 & 0.2 & -52.8 & 5.4 \\
CO$_2$ & -27.9 & -27.4 & -29.4 & 0.2 & -29.1 & 1.0 \\
Hexamine & -87.6 & -85.5 & -86.2 & 0.6 & -83.7 & 4.4 \\
Naphthalene & -87.1 & -83.3 & -75.5 & 0.5 & -77.3 & 8.3 \\
Pyrazine & -63.4 & -61.1 & -61.1 & 1.1 & -63.2 & 4.4 \\
Pyrazole & -78.9 & -76.9 & -77.3 & 0.5 & -76.3 & 5.8 \\
Triazine & -61.1 & -59.4 & -60.5 & 0.6 & -62.2 & 4.3 \\
Trioxane & -63.9 & -62.8 & -62.1 & 1.9 & -64.1 & 4.3 \\
MAD [CCSD(T)] & 0.0 & 2.3 & 4.4 & - & 3.3 & - \\
MAD (Expt) & 3.3 & 2.4 & 2.1 & - & 0.0 & - \\
MAD (DMC) & 4.4 & 2.6 & 0.0 & - & 2.1 & - \\
MSD (DMC) & -4.1 & -1.8 & 0.0 & - & -1.3 & - \\
\bottomrule
\end{tabular}

\end{table}

\begin{table}[h]

\caption{\label{tab:x23_h_bonded_compare}Lattice energy (in kJ/mol) for the H-bonded subset of the X23 dataset computed with different methods, compared to experimental values. The DMC values are taken from Ref.~\citenum{dellapiaHowAccurateAre2024}. The mean absolute deviation (MAD) is also reported with respect to (LNO-MBE-)CCSD(T), experiment and DMC.}
\begin{tabular}{lrrrrrr}
\toprule
 & CCSD(T) & CCSD(cT)-fit & DMC & DMC Error & Expt & Expt Error \\ 
\midrule
Acetic Acid & -69.8 & -68.9 & -71.7 & 0.6 & -72.4 & 5.4 \\
Ammonia & -37.5 & -37.0 & -38.2 & 0.1 & -36.8 & 2.0 \\
Cyanamide & -81.9 & -80.2 & -83.6 & 0.4 & -79.6 & 6.1 \\
Ethyl carbamate & -85.0 & -83.8 & -84.2 & 1.3 & -84.9 & 6.4 \\
Formamide & -80.1 & -79.2 & -81.0 & 1.0 & -78.6 & 5.8 \\
Oxalic Acid $\alpha$ & -99.4 & -98.0 & -102.6 & 1.4 & -96.4 & 7.8 \\
Oxalic Acid $\beta$ & -98.7 & -97.2 & -102.3 & 0.6 & -94.6 & 7.8 \\
Succinic Acid & -124.9 & -122.8 & -125.2 & 0.5 & -116.8 & 6.0 \\
Urea & -106.8 & -105.6 & -108.5 & 0.3 & -99.2 & 7.0 \\
MAD [CCSD(T)] & 0.0 & 1.3 & 1.6 & - & 3.3 & - \\
MAD (Expt) & 3.3 & 2.5 & 4.5 & - & 0.0 & - \\
MAD (DMC) & 1.6 & 2.7 & 0.0 & - & 4.5 & - \\
MSD (DMC) & 1.5 & 2.7 & 0.0 & - & 4.2 & - \\
\bottomrule
\end{tabular}

\end{table}

\clearpage

\subsection{\label{sec:x23_mbe_lit}Comparison to other work}

In Table~\ref{tab:x23_references_compare}, we compare LNO-MBE-CCSD(T) (as well as experiment and DMC) to several previous works employing beyond-DFT methods:
\begin{itemize}
    \item Syty~\etal{}~\cite{sytyMultiLevelCoupledClusterDescription2025} have been able to study the entire X23 dataset through a multilevel approach, which performs LNO-CCSD(T) for 1B and short-range 2B terms within the MBE, while an energy-corrected RPA (RPA+ph) is used for longer-range dimers and short-range trimers, and periodic HF to cover remaining contributions. The {\tt vvTight} thresholds were used on LNO-CCSD(T) for most systems, but this was decreased to {\tt vTight} for anthracene.
    \item \v{C}ervinka~\etal{}~\cite{cervinkaCCSDTCBSFragmentbased2016} studied a subet of the X23 dataset (alongside several other small crystalline molecules), utilizing a MBE approach up to 3B contributions, followed by calculating electrostatic interactions using atomic partial charges. 
    \item Klime\v{s} applied RPA with singles correlation energy contributions at the GW singles excitations (GWSE) level under periodic boundary conditions to study a subset of X23.
    \item Sherrill and co-workers have performed MBE calculations with CCSD(T) for the two-body contributions~\cite{sargentBenchmarkingTwobodyContributions2023a} of the entire X23 dataset and three-body contributions for a subset~\cite{xieAssessmentThreebodyDispersion2023a,nelsonConvergenceManybodyExpansion2024}. We include estimates for three dispersion-bound systems as higher-body terms are expected to be required for H-bonded systems. To get the final lattice energy, we have added 1B contributions calculated within this work.
\end{itemize}

Our LNO-MBE-CCSD(T) approach achieves good agreement to the previous works.
Against the estimates by Syty~\etal{}, it achieves a mean absolute deviation (MAD) of $2.0\,$kJ/mol, while DMC achieves $3.1\,$kJ/mol and experimental agreement is $4.3\,$kJ/mol.
Similarly, for the other aforementioned studies, our LNO-MBE-CCSD(T) approach achieves the best agreement compared to DMC and experiment.

\begin{table}[h]

\caption{\label{tab:x23_references_compare}Comparison of reference literature lattice energies (in kJ/mol) for X23. The mean absolute deviation (MAD) is reported with respect to LNO-MBE-CCSD(T), experiment and DMC~\cite{dellapiaHowAccurateAre2024}.}
\begin{tabular}{lrrrrr}
\toprule
 & \rotatebox{90}{\shortstack{Multilevel\\Syty \etal{}~\cite{sytyMultiLevelCoupledClusterDescription2025}}} & \rotatebox{90}{\shortstack{CCSD(T)\\1B+2B+3B+ELST($>$3B) MBE\\(\v{C}ervinka~\cite{cervinkaCCSDTCBSFragmentbased2016})}} & \rotatebox{90}{\shortstack{RPA+GWSE\\(Klime\v{s}~\cite{klimesLatticeEnergiesMolecular2016})}} & \rotatebox{90}{\shortstack{HMBI CCSD(T)\\1B+2B+FF($>$2B)\\(Wen and Beran~\cite{wenAccurateMolecularCrystal2011a})}} & \rotatebox{90}{\shortstack{CCSD(T)\\1B+2B+3B MBE\\(Sherrill \etal{}~\cite{sargentBenchmarkingTwobodyContributions2023a,xieAssessmentThreebodyDispersion2023a})}} \\ 
\midrule
1,4-cyclohexanedione & -94.4 & - & - & - & - \\
Acetic Acid & -71.2 & -70.6 & - & - & - \\
Adamantane & -66.0 & - & -69.4 & - & - \\
Ammonia & -37.6 & -39.1 & -37.2 & -40.2 & - \\
Anthracene & -111.4 & - & -112.7 & - & - \\
Benzene & -51.6 & - & -55.3 & -54.0 & -53.8 \\
CO$_2$ & -29.4 & -21.2 & -28.4 & -29.5 & -28.7 \\
Cyanamide & -82.6 & - & -79.7 & - & - \\
Cytosine & -163.1 & - & - & - & - \\
Ethyl carbamate & -86.5 & - & - & - & - \\
Formamide & -84.0 & -89.9 & - & -80.4 & - \\
Hexamine & -88.0 & - & - & - & - \\
Imidazole & -88.4 & - & - & -90.8 & - \\
Naphthalene & -82.5 & - & -81.7 & - & - \\
Oxalic Acid $\alpha$ & -103.0 & - & -96.3 & - & - \\
Oxalic Acid $\beta$ & -101.6 & - & -96.1 & - & - \\
Pyrazine & -64.3 & - & - & - & - \\
Pyrazole & -79.3 & - & - & - & - \\
Succinic Acid & -128.2 & - & - & - & - \\
Triazine & -60.3 & - & - & - & -52.2 \\
Trioxane & -67.8 & - & - & - & - \\
Uracil & -139.7 & - & - & - & - \\
Urea & -111.0 & - & -102.5 & - & - \\
MAD [CCSD(T)] & 2.0 & 4.7 & 2.4 & 1.5 & 3.6 \\
MAD (Expt) & 4.3 & 5.8 & 2.6 & 2.4 & 3.8 \\
MAD (DMC) & 3.1 & 4.8 & 5.7 & 1.9 & 4.3 \\
\bottomrule
\end{tabular}

\end{table}

\clearpage
\subsection{\label{sec:cc_validation}Validating CCSD(T) estimates}

As we have briefly discussed in Section~\ref{sec:d4_corr_ene} that the standard MBE of the CCSD(T) correlation energy can be enhanced by using subtractive embedding (by calculating $\Delta_\text{D4}^\text{CCSD(T)}$ on top of the D4 dispersion).
On top of its improved convergence speed with distance cutoff, this also offers an alternative approach for validation.
In the limit of infinite 2B and 3B cutoffs (assuming negligible contribution from 4B terms), we expect both approaches to converge towards the same value.
In Table~\ref{tab:x23_d4_compare}, we have compared both approaches for the same set of 2B and 3B cutoffs.
The differences between the two are very small, with an MAD of $0.4\,$kJ/mol, reassuring the estimates the MBE estimates we have made of the correlation energy.

\begin{table}[h]

\caption{\label{tab:x23_d4_compare}Lattice energy (in kJ/mol) for the X23 dataset when LNO-MBE-CCSD(T) is corrected with (and without) the D4 dispersion correction.}
\begin{tabular}{lrrr}
\toprule
 & Without D4 & With D4 & Difference \\ 
\midrule
1,4-cyclohexanedione & -91.8 & -92.2 & 0.4 \\
Acetic Acid & -69.8 & -70.3 & 0.5 \\
Adamantane & -67.9 & -67.5 & -0.3 \\
Ammonia & -37.5 & -38.6 & 1.1 \\
Anthracene & -114.2 & -113.9 & -0.3 \\
Benzene & -52.8 & -52.8 & 0.0 \\
CO$_2$ & -27.9 & -27.9 & -0.0 \\
Cyanamide & -81.9 & -82.4 & 0.5 \\
Cytosine & -162.1 & -161.9 & -0.2 \\
Ethyl carbamate & -85.0 & -85.0 & 0.1 \\
Formamide & -80.1 & -80.8 & 0.7 \\
Hexamine & -87.6 & -88.1 & 0.5 \\
Imidazole & -88.9 & -88.8 & -0.1 \\
Naphthalene & -87.1 & -83.8 & -3.3 \\
Oxalic Acid $\alpha$ & -99.4 & -99.3 & -0.1 \\
Oxalic Acid $\beta$ & -98.7 & -98.8 & 0.1 \\
Pyrazine & -63.4 & -63.5 & 0.0 \\
Pyrazole & -78.9 & -78.5 & -0.4 \\
Succinic Acid & -124.9 & -125.0 & 0.1 \\
Triazine & -61.1 & -61.5 & 0.4 \\
Trioxane & -63.9 & -64.1 & 0.2 \\
Uracil & -137.2 & -137.2 & -0.0 \\
Urea & -106.8 & -107.0 & 0.2 \\
Mean & - & - & 0.4 \\
\bottomrule
\end{tabular}

\end{table}

\clearpage

\section{\label{sec:comp_cost}Cost comparison for X23 dataset}

In this section, we will compare the cost to perform LNO-MBE-CCSD(T) against standard approaches such as periodic GGA-DFT (with the PBE DFA) and for hybrid-DFT (with the PBE0 DFA).
Furthermore, we will compare it to fixed-node quantum diffusion Monte Carlo (DMC), which has become a relatively mature high-level theory that has been applied to many molecular crystals in the past~\cite{hongoDiffusionMonteCarlo2015,zenFastAccurateQuantum2018a,dellapiaDMCICE13AmbientHigh2022b,dellapiaHowAccurateAre2024}.
We will use the 23 molecular crystals within the X23 dataset to assess the cost to perform these calculations as they feature a diverse range of elements commonly found in molecular crystals together with varying system and unit cell sizes.

\begin{table}[h]

\caption{\label{tab:x23_method_cost}Estimated computational cost (in CPUh) for the different methods used to compute the lattice energies of the X23 dataset. The LNO-MBE-CCSD(T) costs are given for the low (2B cutoff of $6.5\,$\AA{} and 3B cutoff of $100\,$\AA{}$^3$), moderate (2B cutoff of $9.0\,$\AA{} and 3B cutoff of $200\,$\AA{}$^3$) and high (2B cutoff of $12.0\,$\AA{} and 3B cutoff of $300\,$\AA{}$^3$) models. The DMC costs are estimated for a standard deviation of 2$\sigma$ of 1, 2 and 4 kJ/mol, respectively.}
\begin{tabular}{lrrrrrrrr}
\toprule
 & \rotatebox{90}{HF} & \rotatebox{90}{DFT (GGA)} & \rotatebox{90}{CCSD(T) [low]} & \rotatebox{90}{CCSD(T) [moderate]} & \rotatebox{90}{CCSD(T) [high]} & \rotatebox{90}{DMC [4]} & \rotatebox{90}{DMC [2]} & \rotatebox{90}{DMC [1]} \\ 
\midrule
1,4-cyclohexanedione & 1297 & 3 & 800 & 2276 & 7824 & 1984 & 23818 & 317584 \\
Acetic Acid & 1684 & 4 & 139 & 878 & 2195 & 433 & 5198 & 69307 \\
Adamantane & 365 & 3 & 624 & 1795 & 5585 & 4334 & 52013 & 693512 \\
Ammonia & 93 & 1 & 50 & 193 & 411 & 18 & 219 & 2925 \\
Anthracene & 2496 & 6 & 2822 & 6586 & 21941 & 2065 & 24791 & 330552 \\
Benzene & 1201 & 4 & 241 & 1000 & 2802 & 1221 & 14657 & 195439 \\
CO$_2$ & 162 & 1 & 24 & 116 & 277 & 160 & 1926 & 25689 \\
Cyanamide & 1993 & 5 & 137 & 465 & 1149 & 349 & 4198 & 55984 \\
Cytosine & 2778 & 6 & 754 & 5566 & 10764 & 1285 & 15430 & 205738 \\
Ethyl carbamate & 1048 & 3 & 484 & 1604 & 4160 & 1044 & 12528 & 167046 \\
Formamide & 1149 & 3 & 180 & 662 & 1505 & 366 & 4395 & 58600 \\
Hexamine & 1104 & 4 & 548 & 1045 & 2637 & 2191 & 26293 & 350579 \\
Imidazole & 1308 & 4 & 274 & 1575 & 3242 & 764 & 9178 & 122386 \\
Naphthalene & 1148 & 4 & 1238 & 3758 & 11119 & 2627 & 31534 & 420464 \\
Oxalic Acid $\alpha$ & 2411 & 3 & 128 & 1154 & 2263 & 1845 & 22146 & 295291 \\
Oxalic Acid $\beta$ & 1677 & 2 & 293 & 1351 & 2495 & 1329 & 15952 & 212701 \\
Pyrazine & 798 & 3 & 212 & 991 & 1750 & 886 & 10636 & 141823 \\
Pyrazole & 6483 & 8 & 409 & 2836 & 5588 & 764 & 9178 & 122386 \\
Succinic Acid & 916 & 5 & 356 & 1050 & 4786 & 1182 & 14195 & 189276 \\
Triazine & 6337 & 5 & 238 & 640 & 1025 & 886 & 10636 & 141823 \\
Trioxane & 11773 & 10 & 194 & 1135 & 1830 & 1601 & 19214 & 256190 \\
Uracil & 7081 & 9 & 1297 & 6717 & 13454 & 1285 & 15430 & 205738 \\
Urea & 170 & 2 & 108 & 456 & 736 & 290 & 3481 & 46414 \\
Mean & 2412 & 4 & 502 & 1906 & 4762 & 1257 & 15089 & 201193 \\
\bottomrule
\end{tabular}

\end{table}

We report the cost for each of the 23 molecular crystals as well as their average in Table~\ref{tab:x23_method_cost}.
For each of the approaches, we have made our best attempts to calculate costs from a seasoned user of the corresponding approach, based on our collective expertise, which we will discuss in Secs.~\ref{sec:cost-hybrid-DFT},~\ref{sec:cost_ccsdt} and~\ref{sec:cost_dmc} for DFT, LNO-MBE-CCSD(T) and DMC, respectively.
In particular, we have reported several costs for LNO-MBE-CCSD(T), defined by `low', `moderate', which loosen both the 2B and 3B cutoffs in the MBE relative to the original `high' estimate.
The cutoffs are specified in Section~\ref{sec:cost_ccsdt}, where we show that `low' and `moderate' lead to mean absolute deviations (MADs) of $2$ and $1\,$kJ/mol against `high', respectively.
We note here that these metrics utilize the subtractive embedding of D4 dispersion described in Section~\ref{sec:d4_corr_ene} due to its improved convergence behavior.
Similarly, for DMC, we present costs for increasing precision, namely with a standard deviation $2\sigma$ of the energy estimation of 4, 2 and $1\,$kJ/mol (where a 95\% confidence interval corresponds to the interval $E\pm 2\sigma$).

The cost of the `high' LNO-MBE-CCSD(T) are already relatively cheap and comparable to periodic HF, where the former has costs ranging from 277 to 21941 CPUh while the latter has costs ranging from 93 to 11773 CPUh.
Already, this cost is highly economical, particularly within the context of crystal structure prediction (CSP), where it has been recently suggested~\cite{firahaPredictingCrystalForm2023a} that the threshold for a computational method to be economically viable is `about one day on 1,000 cores' (amounting to 24,000$\,$CPUh) within a final screening stage.
In the case of~\citet{firahaPredictingCrystalForm2023a}, such a threshold was for going beyond static electronic energies towards free energies but we expect that it could be similarly important to get LNO-MBE-CCSD(T) lattice energies.
Going to the `moderate' or `low' LNO-MBE-CCSD(T) approaches leads to significantly cheaper costs, with the `medium' coming to an average cost of $1906\,$CPUh, while `low' brings this down to an average of $502\,$CPUh, with the majority (20 out of the 23) of systems requiring fewer than $1000\,$CPUh.
Importantly, this makes the cost cheaper than HF or hybrid DFT.
The GGA-DFT remains the cheapest approach by far but that comes with an accuracy trade-off as shown in Section~\ref{sec:dft_benchmark}.

Compared to the LNO-MBE-CCSD(T) costs, the DMC costs are larger.
For a 95\% confidence interval of $1\,$kJ/mol, it would cost ${\sim}200,000\,$CPUh on average.
If less accuracy is needed, the cost is on average ${\sim}15,000$ and ${\sim}1,300\,$CPUh for a 2 and $4\,$kJ/mol error respectively.
For relative energies between some polymorphs (particularly competing polymorphs), the differences may be $1\,$kJ/mol or less, necessitating that level of statistical accuracy on the DMC calculations.

\subsection{Calculating cost of hybrid DFT}\label{sec:cost-hybrid-DFT}

As our HF calculations were actually performed on GPUs (see Section~\ref{sec:x23_hf}) and it is challenging to make a one-to-one comparison with costs on CPU, we reperformed these calculations on the CPU.
Both the HF and DFT (GGA) calculations were performed in Quantum Espresso~\cite{giannozziQUANTUMESPRESSOModular2009a}, where we have also accelerated the calculations with the Adaptively Compressed Exchange~\cite{linAdaptivelyCompressedExchange2016} (ACE) operator.
This is expected to speed up costs by ${\approx}3$.
These costs were estimated on 48-core Cascadelake nodes with $768\,$GB of RAM.

\subsection{\label{sec:cost_ccsdt}Calculating cost of LNO-MBE-CCSD(T)}

For the X23 dataset, all LNO-MBE-CCSD(T) calculations were performed on nodes involving $2{\times}18$ 2.1 GHz Intel Xeon E5-2695 (Broadwell) series processors with $256\,$GB of RAM in total.
For the X23 dataset, each calculation was performed on 8-cores with $46\,$GB of RAM allocated.
Our costs include not only the cost to perform the LNO-CCSD(T) calculations, but also the composite correction using the MP2 as well within the Composite procedure with 'Full' basis sets (see Section~\ref{sec:cheaper_settings}).
These costs exclude the underlying costs to perform the periodic HF calculations because their costs are negligible compared to the CCSD(T) calculations with the help of GPUs (see Section~\ref{sec:x23_hf}).
It also enables fairer comparison to DMC calculations, where the costs to (1) generate the DFT trial wave-function and (2) optimizing the Jastrow factor have been neglected during the cost analysis.

\begin{table}[h]

\caption{\label{tab:x23_smaller_models}Lattice energy (in kJ/mol) for the X23 dataset with the large (2B cutoff of $12.0\,$\AA{} and 3B cutoff of $300\,$\AA{}$^3$) LNO-MBE-CCSD(T) model, and the medium (2B cutoff of $6.5\,$\AA{} and 3B cutoff of $100\,$\AA{}$^3$) as well as small models (2B cutoff of $9.0\,$\AA{} and 3B cutoff of $200\,$\AA{}$^3$).}
\begin{tabular}{lrrr}
\toprule
 & Small & Medium & Large \\ 
\midrule
1,4-cyclohexanedione & -92.9 & -92.2 & -92.2 \\
Acetic Acid & -70.9 & -69.5 & -70.3 \\
Adamantane & -65.4 & -66.9 & -67.5 \\
Ammonia & -37.4 & -37.9 & -38.6 \\
Anthracene & -110.4 & -111.3 & -113.9 \\
Benzene & -53.2 & -52.3 & -52.8 \\
CO$_2$ & -28.4 & -28.3 & -27.9 \\
Cyanamide & -82.9 & -81.8 & -82.4 \\
Cytosine & -163.8 & -157.5 & -161.9 \\
Ethyl carbamate & -83.7 & -85.1 & -85.0 \\
Formamide & -76.8 & -81.1 & -80.8 \\
Hexamine & -85.1 & -85.3 & -88.1 \\
Imidazole & -87.8 & -87.9 & -88.8 \\
Naphthalene & -81.7 & -82.7 & -83.8 \\
Oxalic Acid $\alpha$ & -102.5 & -99.1 & -99.3 \\
Oxalic Acid $\beta$ & -101.2 & -98.1 & -98.8 \\
Pyrazine & -64.4 & -63.4 & -63.5 \\
Pyrazole & -78.5 & -78.1 & -78.5 \\
Succinic Acid & -128.9 & -128.5 & -125.0 \\
Triazine & -61.4 & -61.5 & -61.5 \\
Trioxane & -64.1 & -63.8 & -64.1 \\
Uracil & -141.6 & -137.2 & -137.2 \\
Urea & -103.7 & -108.8 & -107.0 \\
MAD & 1.8 & 1.0 & 0.0 \\
\bottomrule
\end{tabular}

\end{table}

As discussed in the preceding section, we have defined several presets of 2B and 3B cutoffs assuming a D4 dispersion embedding (due to its faster convergence):
\begin{itemize}
    \item low --- 2B cutoff of $6.5\,$\AA{} and 3B cutoff of $100\,$\AA{}$^3$,
    \item moderate --- 2B cutoff of $9.0\,$\AA{} and 3B cutoff of $200\,$\AA{}$^3$,
    \item high --- 2B cutoff of $12.0\,$\AA{} and 3B cutoff of $300\,$\AA{}$^3$.
\end{itemize}
The cost for each molecular crystal within the X23 dataset is shown in Table~\ref{tab:x23_smaller_models}.

\subsection{\label{sec:cost_dmc}Calculating the cost of DMC}

Diffusion Monte Carlo (DMC) estimations are affected by systematic and random errors.
Systematic errors arise from approximations in the approach; the crucial ones are the fixed-node (FN) approximation, the use of pseudopotentials (PPs), and the finite-timestep ($\tau$) approximation.
The first two (FN and PP) are difficult to investigate because they cannot be reduced systematically; they can only be assessed by performing sets of calculations with different nodal surfaces and different PPs. For this reason, we treat FN and PP errors as a feature of the method that may affect its accuracy.
By contrast, the finite-$\tau$ error can be made negligible by performing simulations at decreasing $\tau$ and either extrapolating to $\tau\!\to\!0$ or choosing a $\tau$ such that the timestep bias is negligible for the target. This comes at the price of increased computational cost, which scales approximately as $1/\tau$ for a DMC simulation at timestep $\tau$ (see below).

Random errors are due to the stochastic nature of the approach and are estimated in each simulation from the standard deviation $\sigma$ of the estimator. The smaller $\sigma$, the higher the precision.
The stochastic error can be reduced to any target by increasing sampling, with $\sigma\propto$ (sampling)$^{-1/2}$ (and thus $\sigma\propto T^{-1/2}$ in computational time; see below). Consequently, reducing $\sigma$ by a factor $x$ costs about $x^2$ more compute.
There is no point in having a tiny stochastic error but a large timestep bias (or vice versa): simulations should be planned so that both errors are comparable and sufficiently small for the desired precision and accuracy; otherwise total DMC cost will be unnecessarily large.

When the final estimator is a combination of results from multiple DMC runs, the overall stochastic error is obtained by propagating the independent errors.
This is the case for lattice energies $E_{\mathrm{latt}}$ of molecular crystals, computed as the difference
$E_{\mathrm{latt}} = E_{\mathrm{crys,1mol}} - E_{\mathrm{gas}}$
between the energy per molecule of the crystal and the gas-phase molecular energy.
If the two are estimated independently with stochastic errors $\sigma_{\mathrm{crys,1mol}}$ and $\sigma_{\mathrm{gas}}$, then
$\sigma_{\mathrm{latt}} = \sqrt{\sigma_{\mathrm{crys,1mol}}^2 + \sigma_{\mathrm{gas}}^2}$.

Despite having the same target precision, the costs of $E_{\mathrm{crys,1mol}}$ and $E_{\mathrm{gas}}$ are not comparable.
The crystal calculation is typically much more expensive, whereas the gas-phase calculation is often negligible or a small overhead.
This follows from DMC’s scaling with system size and from the fact that $E_{\mathrm{crys,1mol}}$ is obtained from a periodic simulation cell containing several molecules. The simulated cell is not necessarily the primitive cell because finite-size errors (FSEs) can bias the result. Two strategies are common: (i) brute-force extrapolation by computing a sequence of supercells and extrapolating to the thermodynamic limit, or (ii) applying finite-size corrections to each simulated cell. Several FSE corrections exist; see Refs.~\citenum{KZK:prl2008, Chiesa:size_effects:prl2006, MPC:Fraser1996, MPC:Will1997, MPC:Kent1999}.
Applying an FSE correction to a small cell is far cheaper than supercell extrapolation but is only as reliable as the correction itself; any bias propagates to $E_{\mathrm{crys,1mol}}$ and therefore to $E_{\mathrm{latt}}$. If only a single cell is used there is no internal check of the correction’s reliability. Using multiple cells reduces the correction magnitude (it vanishes in the thermodynamic limit) but increases cost toward that of an explicit extrapolation. For molecular crystals there is evidence that the model periodic Coulomb (MPC)-based correction\cite{MPC:Fraser1996, MPC:Will1997, MPC:Kent1999} is reliable whenever the inscribed-sphere radius of the Wigner–Seitz cell of the supercell is $\gtrsim 5$~\AA; see Ref.~\cite{zenFastAccurateQuantum2018a}. Our cost estimates below assume this setup.

We evaluate the computational cost for DMC lattice energies of the X23 molecular crystals as reported in Table~\ref{tab:x23_method_cost}. This evaluation is based on calculations performed on the Peta4-Skylake nodes of the University of Cambridge CSD3 HPC system (Intel Xeon Scalable “Skylake” processors), published in Refs.~\citenum{zenFastAccurateQuantum2018a, dellapiaHowAccurateAre2024}. The Cascade Lake nodes used for hybrid DFT estimates (Section~\ref{sec:cost-hybrid-DFT}) are slightly newer; we expect broadly comparable performance.

\paragraph{Stochastic cost model.}
Let $\sigma$ be the target stochastic error for the total energy of a system.
The DMC energy is the average over $N_s$ electronic configurations $\{X_i\}_{i=1}^{N_s}$ sampled by DMC of the local energy $E_L[X_i]$:
$E_{\mathrm{DMC}} = \frac{1}{N_s}\sum_{i=1}^{N_s} E_L[X_i]$.
If the $E_L[X_i]$ were independent, the error would be $\sqrt{\mathrm{Var}(E_L)/N_s}$.
Because configurations are correlated, we introduce an effective sample size $N_{\mathrm{eff}}<N_s$ and estimate the error as
$\sigma_{\mathrm{DMC}}=\sqrt{\mathrm{Var}(E_L)/N_{\mathrm{eff}}}$.
To achieve a target $\sigma$ for a given $\mathrm{Var}(E_L)$, we therefore require
$N_{\mathrm{eff}}=\mathrm{Var}(E_L)/\sigma^2$.

A DMC run begins with $n_w(t{=}0)$ “walkers”. The algorithm generates new configurations from the previous iteration, and a population-control step keeps the walker count close to a target $n_w^{\mathrm{target}}$ each iteration. After equilibration, the sampling phase consists of
$N_{\mathrm{iter}} = N_s / n_w^{\mathrm{target}}$ iterations.
Within an iteration, walkers are (approximately) independent; correlations are significant between consecutive iterations. Define the autocorrelation
$\rho(k)=\gamma(k)/\gamma(0)$, with $\gamma(k)=\mathrm{Cov}(\langle E_L\rangle_t,\langle E_L\rangle_{t+k})$ over sampling iterations (after discarding equilibration).
Let $N_{\mathrm{ac}}=1+2\sum_{k=1}^{\infty}\rho(k)$ be the integrated autocorrelation time in units of iterations. Then
\[
\frac{N_s}{N_{\mathrm{eff}}}\approx N_{\mathrm{ac}} \quad\Rightarrow\quad
N_{\mathrm{eff}}\approx \frac{\tau}{t_{\mathrm{ac}}}\,N_s,
\]
where we used the empirical observation that $N_{\mathrm{ac}}\propto 1/\tau$ and define a (system-dependent) correlation time
$t_{\mathrm{ac}}=\tau\,N_{\mathrm{ac}}$ (constant for a given system, in atomic units of time).
Hence the required sampling for target $\sigma$ is
\[
N_s = \frac{t_{\mathrm{ac}}}{\tau}\,\frac{\mathrm{Var}(E_L)}{\sigma^2},\qquad
N_{\mathrm{iter}}=\frac{1}{n_w^{\mathrm{target}}}\frac{t_{\mathrm{ac}}}{\tau}\frac{\mathrm{Var}(E_L)}{\sigma^2}.
\]
The wall time per iteration, $T_{\mathrm{iter}}$, is proportional to $n_w^{\mathrm{target}}$ and only weakly dependent on $\tau$ in common implementations (nonlocal PP treatments can add a weak $\tau$-dependence). Writing
$T_{\mathrm{iter}} = n_w^{\mathrm{target}}\,T_w$,
where $T_w$ is the time to advance one walker by one iteration, the total cost (neglecting equilibration) scales as $1/\tau$ and $1/\sigma^2$.

Let $T_{\mathrm{DMC}}$ be the computational cost in \emph{CPU hours} (core-hours) for a DMC simulation with timestep $\tau$ achieving error $\sigma$:
\[
T_{\mathrm{DMC}} = \frac{T_w}{3600}\,\frac{t_{\mathrm{ac}}}{\tau}\,\frac{\mathrm{Var}(E_L)}{\sigma^2}.
\]
(The factor $1/3600$ converts seconds to hours if $T_w$ is measured in seconds.)
The variance $\mathrm{Var}(E_L)$ is extensive (for a fixed wavefunction/PP setup), roughly proportional to the number of electrons $N_e$:
$\mathrm{Var}(E_L) \approx N_e\,V_{1e}$, with $V_{1e}$ the average variance per electron.
In many DMC implementations,
\[
T_w = C\,(N_e^2 + \epsilon N_e^3),
\]
where $C$ and $\epsilon$ are system- and architecture-dependent (typically $\epsilon\!\lesssim\!10^{-3}$).
Thus,
\begin{equation}\label{eq:DMCcost}
T_{\mathrm{DMC}}(\sigma,\tau)
= \frac{C}{3600}\,\bigl(N_e^3 + \epsilon N_e^4\bigr)\,
\frac{t_{\mathrm{ac}}}{\tau}\,\frac{V_{1e}}{\sigma^2}\,.
\end{equation}

For the X23 systems on the Peta4-Skylake cluster\cite{dellapiaAccurateEfficientMachine2025}, analysis of our runs indicates broadly similar $C$, $\epsilon$, $t_{\mathrm{ac}}$, and $V_{1e}$ across crystals. We adopt the averages:
$\tilde C = 2.4\times 10^{-6}\ \mathrm{s}$,
$\tilde\epsilon = 4\times 10^{-4}$,
$\tilde V_{1e} = 0.026$ a.u.,
and $\tilde t_{\mathrm{ac}} = 0.73$ a.u. (atomic units of time). Note that Eq.~\ref{eq:DMCcost} shows the scaling of DMC with the number of simulated electrons. Since $\epsilon\sim 10^{-4}$, DMC scales with $N^3_e$ for typically simulated systems with fewer than $\sim 10^4$ electrons. The scaling is instead $o(N^4_e)$ for $N_e \gtrsim 10^4$.

The DMC cost for a lattice-energy evaluation is the sum of the costs for $E_{\mathrm{crys,1mol}}$ and $E_{\mathrm{gas}}$ computed via Eq.~\eqref{eq:DMCcost}.
In the crystal, the periodic cell contains $N_{\mathrm{mol}}$ molecules, so $N_e = N_{\mathrm{mol}} N_{e,1\mathrm{mol}}$ and $E_{\mathrm{crys,1mol}} = E_{\mathrm{crys}}/N_{\mathrm{mol}}$.
To achieve $\sigma_\mathrm{crys,1mol}$ \emph{per molecule}, we target a stochastic error on the total cell energy of
$\sigma_\mathrm{crys} = N_{\mathrm{mol}}\,\sigma_\mathrm{crys,1mol}$.
If our target $\sigma$ and the timestep $\tau$ are the same on the gas-phase and on the periodic cell simulations,  the crystal simulation is roughly $N_{\mathrm{mol}}$ times more expensive than the gas-phase one, as implied by Eq.~\eqref{eq:DMCcost}.
Therefore, considering the overall cost of both simulations, it can be shown that the most efficient setup for the simulations (i.e., the setup minimising the stochastic error on the lattice energy given the overall computational cost) is to set 
$\sigma_\mathrm{crys,1mol}^2 \sim {\sqrt{N_\text{mol}} \over 1+\sqrt{N_\text{mol}}} \sigma_\text{latt}^2$
and
$\sigma_\mathrm{mol}^2 \sim {1 \over 1+\sqrt{N_\text{mol}}} \sigma_\text{latt}^2$,
being $\sigma_\text{latt}$ our target stochastic error on the lattice energy evaluation.

\paragraph{Choice of timestep.}
The timestep bias decreases with $\tau$.
From the analysis reported in the SI of Ref.~\citenum{dellapiaHowAccurateAre2024} for X23 molecular crystals, typical bounds are:
$\tau=0.03$ a.u. $\Rightarrow$ bias $< 2$ kJ/mol;
$\tau=0.01$ a.u. $\Rightarrow$ bias $< 1$ kJ/mol;
$\tau=0.003$ a.u. $\Rightarrow$ bias $< 0.5$ kJ/mol.
These $\tau$ values were used for the DMC[4], DMC[2], DMC[1] entries in Table~\ref{tab:x23_method_cost}.
%
%

\clearpage

\section{\label{sec:ice13_dataset}Lattice and relative energy of ICE13 Dataset}

In this section, we report the final lattice energy estimates computed by the MBE for the ICE13 dataset and compare it to both experiment and DMC.
We also compute the relative energies w.r.t.\ ice Ih for the other 12 phases.
Throughout this section, we use experimental lattice energies from~\citet{whalleyEnergiesPhasesIce1984}, which have been extrapolated to zero temperature and pressure, with zero-point energies removed.

\subsection{\label{sec:ice13_final_latt}Final lattice energies}

In Table~\ref{tab:ice13_lattice_energies}, we have collated the LNO-MBE-CCSD(T) estimates for the Small, Medium and Large unit cells.
As mentioned in the main text, there are significant errors in the lattice energy predicted for ice VII for the Small unit cell due to the large dipole moment within the unit cell.
Thus, when comparing to e.g., DMC calculations, we instead use values from the Medium cell for ice VII, while reporting the Small unit cell for all other phases.
These estimates are given under `Final'.

The lattice energies predicted between the Small, Medium or Large unit cells are generally within $1\,$kJ/mol and oftentimes much better than that.
Besides the noted difference for ice VII, the differences can arise due to differences in the ferroelectric character of the unit cells or because the small unit cell uses the Canonical approach described in Section~\ref{sec:cheaper_settings}, while the other unit cell sizes uses the LAF approach.

We have also included estimates for both MP2 and CCSD (just showing the Final set).
It can be seen that going towards CCSD(T) is absolutely necessary, with MP2 and CCSD both performing worse compared to DMC.
In fact, CCSD performs poorly, with an MAD of ${\sim}6.3\,$kJ/mol across the 13 ice polymorphs.
MP2 performs better, with an MAD of ${\sim}1\,$kJ/mol.

As shown in Table~\ref{tab:ice13_dft_elatt_mad}, we find that the differences between CCSD(T) and DMC are sufficiently small that the MAD (across the 13 ice phases) on $E_\text{latt}$ are similar to within $0.5\,$kJ/mol when used to benchmark several density functional approximations taken from Ref.~\citenum{dellapiaDMCICE13AmbientHigh2022b}.
A more detailed comparison of DFAs is given in Section~\ref{sec:dft_benchmark}.

\begin{table}[h]

\caption{\label{tab:ice13_lattice_energies}Lattice energies of the ICE13 set from DMC~\cite{dellapiaDMCICE13AmbientHigh2022b}, Experiment~\cite{whalleyEnergiesPhasesIce1984} and LNO-MBE-CCSD(T). We consider the Small, Medium and Large unit cell sizes for LNO-MBE-CCSD(T) and also a Final estimate that corrects for errors in ice VII when using a Small unit cell. We also give Final estimates at the MP2 and CCSD levels. All values are in kJ/mol.}
\begin{tabular}{lrrrrrrrrr}
\toprule
\rotatebox{90}{Ice Phase} & \rotatebox{90}{DMC (Unitcell)} & \rotatebox{90}{Experiment} & \rotatebox{90}{CCSD(T) [Small cell]} & \rotatebox{90}{CCSD(T) [Medium cell]} & \rotatebox{90}{CCSD(T) [Large cell]} & \rotatebox{90}{CCSD(T) [Final]} & \rotatebox{90}{CCSD [Final]} & \rotatebox{90}{MP2 [Final]} \\ 
\midrule
Ih & -59.5 $\pm$ 0.1 & -58.9 $\pm$ 1.0 & -59.3 & -58.4 & -58.3 & -59.3 & -53.5 & -59.2 \\
II & -59.1 $\pm$ 0.1 & -58.8 $\pm$ 1.0 & -58.3 & N/A & N/A & -58.3 & -52.5 & -57.7 \\
III & -58.2 $\pm$ 0.1 & -57.9 $\pm$ 1.0 & -57.3 & -57.0 & -56.8 & -57.3 & -51.3 & -57.1 \\
IV & -55.6 $\pm$ 0.1 & N/A & -56.0 & -55.5 & -54.8 & -56.0 & -50.0 & -55.3 \\
VI & -57.7 $\pm$ 0.1 & -57.2 $\pm$ 1.0 & -56.9 & N/A & -56.2 & -56.9 & -50.8 & -56.1 \\
VII & -54.5 $\pm$ 0.1 & -54.7 $\pm$ 1.0 & -57.2 & -54.6 & -54.4 & -54.6 & -48.5 & -53.1 \\
VIII & -55.2 $\pm$ 0.1 & -55.7 $\pm$ 1.7 & -55.6 & N/A & N/A & -55.6 & -49.6 & -54.3 \\
IX & -58.9 $\pm$ 0.1 & -58.5 $\pm$ 1.0 & -57.7 & N/A & N/A & -57.7 & -52.0 & -57.4 \\
XI & -59.3 $\pm$ 0.1 & N/A & -59.3 & N/A & N/A & -59.3 & -53.3 & -59.4 \\
XIII & -57.3 $\pm$ 0.1 & N/A & -57.4 & N/A & N/A & -57.4 & -51.6 & -56.9 \\
XIV & -57.8 $\pm$ 0.1 & N/A & -56.9 & N/A & N/A & -56.9 & -51.0 & -56.2 \\
XV & -57.7 $\pm$ 0.1 & N/A & -56.9 & N/A & N/A & -56.9 & -50.9 & -56.1 \\
XVII & -57.7 $\pm$ 0.1 & N/A & -57.7 & -57.4 & -57.1 & -57.7 & -52.0 & -57.8 \\
MAD (DMC) &  &  &  &  &  & 0.5 & 6.3 & 0.9 \\
MAD (Exp) & 0.4 &  &  &  &  & 0.4 & 6.2 & 1.1 \\
\bottomrule
\end{tabular}

\end{table}

\clearpage

\begin{table}[h]

\caption{\label{tab:ice13_dft_elatt_mad}Mean absolute deviations (MAD) with respect to LNO-MBE-CCSD(T) and DMC~\cite{dellapiaDMCICE13AmbientHigh2022b} for several DFT functionals on the lattice energies of the ICE13 set. All values are in kJ/mol.}
\begin{tabular}{lrr}
\toprule
 & MAD [DMC] & MAD [CCSD(T)] \\ 
\midrule
PBE-D3(BJ) & 7.9 & 8.3 \\
optB86b-vdW & 9.3 & 9.7 \\
SCAN+rVV10 & 9.6 & 9.9 \\
revPBE0-D3 & 1.5 & 1.2 \\
B3LYP-D4 & 2.3 & 2.6 \\
\bottomrule
\end{tabular}

\end{table}

\subsection{Final relative energies}

\begin{figure}[h]
    \includegraphics[width=\textwidth]{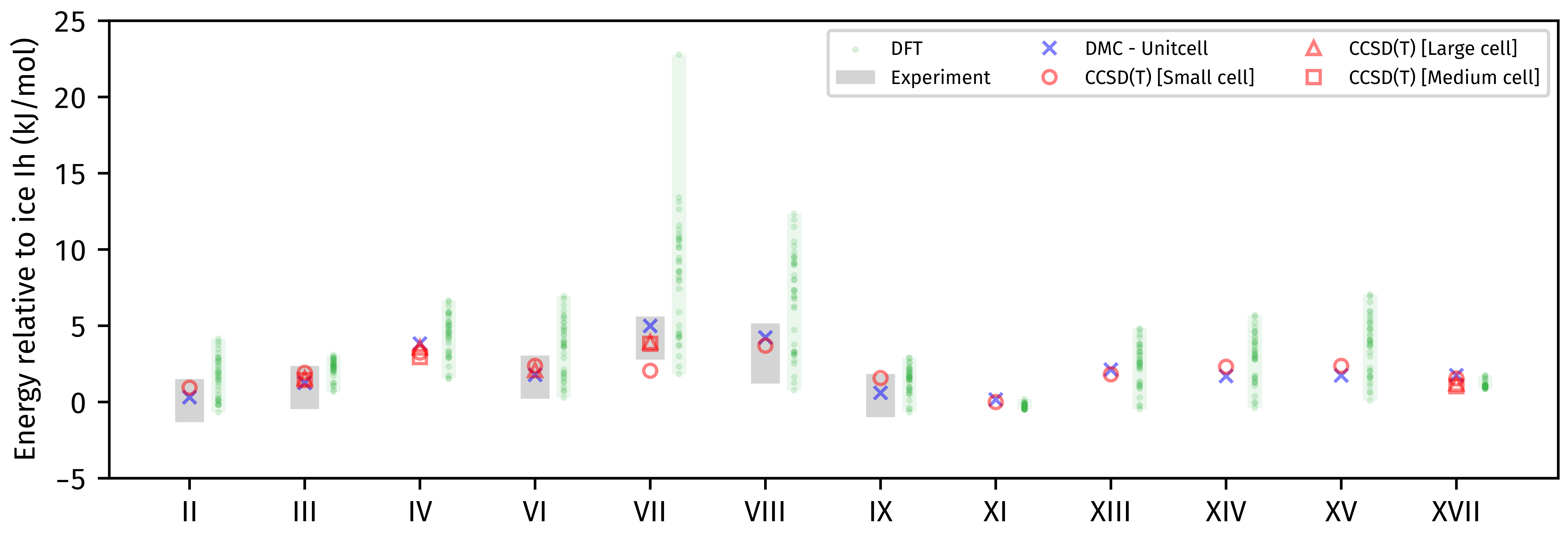}
    \caption{\label{fig:ice13_rel_ene}Comparison of the energy relative to ice Ih for the 13 phases of ice. Besides estimates from LNO-MBE-CCSD(T), we include DMC and DFT estimates from Ref.~\citenum{dellapiaDMCICE13AmbientHigh2022b} and experimental values from Ref.~\citenum{whalleyEnergiesPhasesIce1984}.}
\end{figure}

The relative energy between ice polymorphs lie within a small $5\,$kJ/mol range for both LNO-MBE-CCSD(T) and DMC as shown in Figure~\ref{fig:ice13_rel_ene}.
This makes it a stringent test system, where minute differences can have ramifications in the resulting ice phase diagram~\cite{reinhardtQuantummechanicalExplorationPhase2021,zhangPhaseDiagramDeep2021,boreRealisticPhaseDiagram2023}.
Within this figure, we also highlight a range of DFT values, which can span more than $20\,$kJ/mol on the relative energy depending on the chosen functional, taken from Ref.~\citenum{dellapiaDMCICE13AmbientHigh2022b}.

We compare the mean absolute deviation for the 12 phases of ice (excluding Ih) against DMC and experiment in Table~\ref{tab:ice13_relative_energies}.
To be consistent with DMC, we use results for the small unit cell, except for ice VII, where we used the medium cell, and dub this combined set of numbers `Final'.
The resulting MAD against DMC is better than $1\,$kJ/mol, at $0.6\,$kJ/mol.
The MAD against experiment for DMC and LNO-MBE-CCSD(T) are $0.5\,$kJ/mol and $0.8\,$kJ/mol respectively, both within the expected error range for experiment.
Here, it is interesting to note that despite its poor performance on the lattice energy, CCSD performs exceptionally well for relative energies, with similar MADs (w.r.t.\ DMC and experiment) as CCSD(T).
For this quantity, MP2 performs worse, with an MAD of $0.8\,$kJ/mol and $1.5\,$kJ/mol relative to DMC and experiment, respectively.

\begin{table}[h]

\caption{\label{tab:ice13_relative_energies}Relative lattice energies of the ICE13 set with respect to ice Ih from DMC~\cite{dellapiaDMCICE13AmbientHigh2022b}, Experiment~\cite{whalleyEnergiesPhasesIce1984} and LNO-MBE-CCSD(T). We consider the Small, Medium and Large unit cell sizes for LNO-MBE-CCSD(T) and also a Final estimate that corrects for errors in ice VII when using a Small unit cell. We also give Final estimates at the MP2 and CCSD levels. All values are in kJ/mol. All values are in kJ/mol.}
\begin{tabular}{lrrrrrrrrr}
\toprule
\rotatebox{90}{Ice Phase} & \rotatebox{90}{DMC (Unitcell)} & \rotatebox{90}{Experiment} & \rotatebox{90}{CCSD(T) [Small cell]} & \rotatebox{90}{CCSD(T) [Medium cell]} & \rotatebox{90}{CCSD(T) [Large cell]} & \rotatebox{90}{CCSD(T) [Final]} & \rotatebox{90}{CCSD [Final]} & \rotatebox{90}{MP2 [Final]} \\ 
\midrule
II & 0.3 $\pm$ 0.1 & 0.1 $\pm$ 1.4 & 0.9 & N/A & N/A & 0.9 & 1.0 & 1.5 \\
III & 1.2 $\pm$ 0.1 & 0.9 $\pm$ 1.4 & 1.9 & 1.4 & 1.6 & 1.9 & 2.1 & 2.1 \\
IV & 3.8 $\pm$ 0.1 & N/A & 3.2 & 2.9 & 3.6 & 3.2 & 3.5 & 3.9 \\
VI & 1.8 $\pm$ 0.1 & 1.6 $\pm$ 1.4 & 2.4 & N/A & 2.1 & 2.4 & 2.7 & 3.1 \\
VII & 5.0 $\pm$ 0.1 & 4.2 $\pm$ 1.4 & 2.0 & 3.8 & 3.9 & 3.8 & 4.9 & 6.1 \\
VIII & 4.2 $\pm$ 0.1 & 3.2 $\pm$ 2.0 & 3.7 & N/A & N/A & 3.7 & 3.9 & 4.9 \\
IX & 0.6 $\pm$ 0.1 & 0.4 $\pm$ 1.4 & 1.6 & N/A & N/A & 1.6 & 1.5 & 1.8 \\
XI & 0.2 $\pm$ 0.1 & N/A & -0.0 & N/A & N/A & -0.0 & 0.2 & -0.2 \\
XIII & 2.1 $\pm$ 0.1 & N/A & 1.8 & N/A & N/A & 1.8 & 1.9 & 2.3 \\
XIV & 1.7 $\pm$ 0.1 & N/A & 2.3 & N/A & N/A & 2.3 & 2.5 & 2.9 \\
XV & 1.7 $\pm$ 0.1 & N/A & 2.4 & N/A & N/A & 2.4 & 2.5 & 3.1 \\
XVII & 1.8 $\pm$ 0.1 & N/A & 1.5 & 1.0 & 1.2 & 1.5 & 1.5 & 1.4 \\
MAD (DMC) &  &  &  &  &  & 0.6 & 0.5 & 0.8 \\
MAD (Exp) & 0.5 &  &  &  &  & 0.8 & 0.9 & 1.5 \\
\bottomrule
\end{tabular}

\end{table}

We compare the MADs predicted for a subset of DFAs from Ref.~\cite{dellapiaDMCICE13AmbientHigh2022b} against both DMC and LNO-MBE-CCSD(T) for predicting relative energies for the ICE13 dataset in Table~\ref{tab:ice13_dft_erel_comparison}
This highlights that the precision in both DMC and CCSD(T) are sufficiently accurate that they will discern good-performing DFAs.
In fact, the differences between DMC and LNO-MBE-CCSD(T) are now smaller, with MADs differing on average ${\sim}0.1\,$kJ/mol to either DMC or LNO-MBE-CCSD(T) across the DFAs.
A more detailed comparison of DFAs is given in Section~\ref{sec:dft_benchmark}.

\begin{table}[h]

\caption{\label{tab:ice13_dft_rel_ene_mad}Mean absolute deviations (MAD) with respect to LNO-MBE-CCSD(T) and DMC~\cite{dellapiaDMCICE13AmbientHigh2022b} for several DFT functionals on the relative lattice energies of the ICE13 set (using ice Ih as the reference). All values are in kJ/mol. All values are in kJ/mol.}
\begin{tabular}{lrr}
\toprule
 & MAD [DMC] & MAD [CCSD(T)] \\ 
\midrule
PBE-D3(BJ) & 3.5 & 3.3 \\
optB86b-vdW & 0.5 & 0.7 \\
SCAN+rVV10 & 0.7 & 0.7 \\
revPBE0-D3 & 1.0 & 1.0 \\
B3LYP-D4 & 1.9 & 1.7 \\
\bottomrule
\end{tabular}

\end{table}

\clearpage

\section{\label{sec:opt_xdm}Towards optimized density functional approximations for molecular crystals}

The large set of LNO-MBE-CCSD(T) lattice energy estimates we have generated make for useful benchmarks.
This can be used to assess the performance of density functional approximations (DFAs), as we will do in Section~\ref{sec:dft_benchmark}.
In addition, these benchmarks can be used to parametrize new density functional approximations, as has been common using small dimer binding energies~\cite{rezacS66WellbalancedDatabase2011,rezacExtensionsApplicationsA242015} but has been absent for molecular crystals to date.
In particular, our LNO-MBE-CCSD(T) estimates provide low-errors compared to the large (error) range of possible experimental data for several systems within the X23 dataset.

In this section, we have focused efforts on optimizing the parametrization of the XDM parameters for B86bPBE50~\cite{a.priceXDMcorrectedHybridDFT2023} to our LNO-MBE-CCSD(T) X23 data, which we will dub B86bPBE50-revXDM.
This base DFA was selected because its large amount of exact exchange makes it more resilient towards delocalization errors~\cite{bryentonDelocalizationErrorGreatest2023} which may be prominent when tackling the flexible pharmaceutical-size molecular crystals that we will study later.
Furthermore, we have optimized the XDM parameters specifically for the compact \texttt{lightdense} basis sets in FHI-AIMS.
These basis sets are particularly ideal as they are less susceptible to basis set superposition error at small sizes~\cite{yangDevelopingCorrelationconsistentNumeric2024}.
Inspired by the ``-3c'' approaches~\cite{brandenburgLowCostQuantumChemical2014} and others~\cite{kellerSmallBasisSet2024}, we hope to capture the basis set incompleteness errors within the XDM term.
Overall, the development of such a cheap approach would make it useful for crystal structure prediction, where it might be used to screen thousands of polymorphs (at as low a cost as possible).
Additionally, this also makes it useful in the context of training machine-learned interatomic potentials (MLIP) to obtain thermal contributions -- particular for large pharmaceutical molecular crystals -- as we will do later.

The XDM dispersion is computed as a damped asymptotic pairwise term,
\begin{equation}
E_{\text{XDM}} = - \sum_{n=6,8,10} \sum_{i>j} \frac{C_{n,ij}}{R_{ij}^n + R_{\text{vdW},ij}^n},
\end{equation}
which is added to the base density functional energy (in our case B86bPBE50 with `lightdense' basis sets),
\begin{equation}
E = E_{\text{base}} + E_{\text{XDM}}.
\end{equation}
Here, $i$ and $j$ run over atoms, $R_{ij}$ are interatomic distances, $C_{n,ij}$ are dispersion coefficients, and the damping lengths are
\begin{equation}
R_{\text{vdW},ij} = a_1 R_{c,ij} + a_2.
\end{equation}
These $a_1$ and $a_2$ coefficients are what needs to be parameterized or optimized.

\begin{figure}[h]
    \includegraphics[]{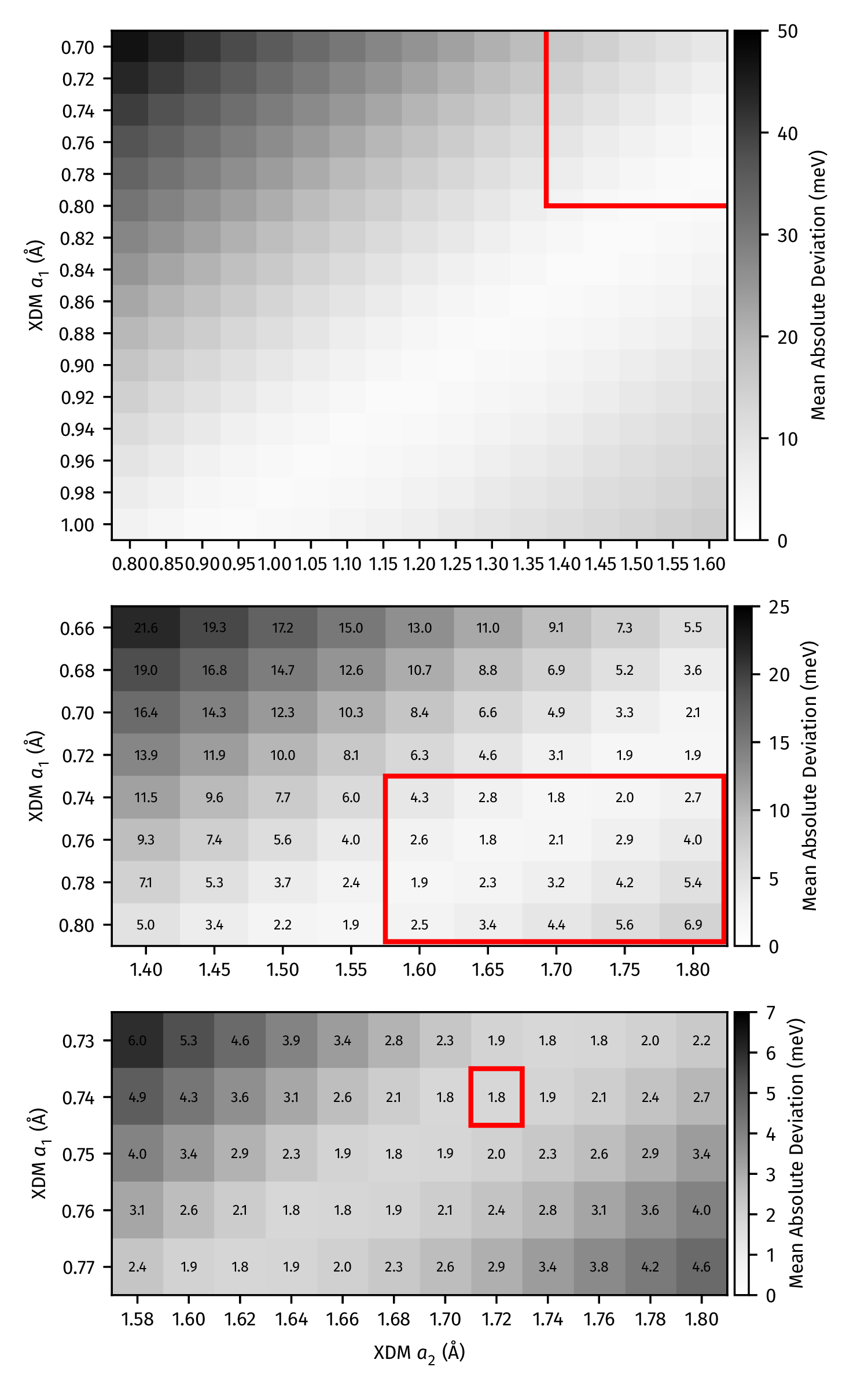}
    \caption{\label{fig:xdm_opt}Optimizing the XDM $a_1$ and $a_2$ coefficients against the X23 dataset. This was performed in a three step process starting from a large (top panel), medium (middle panel) and small (bottom panel) of values to come to the final coefficients to within one decimal place. We used meV units ($1\,$kJ/mol$=10.4\,$meV) to make differences easier to discern.}
\end{figure}

In Figure~\ref{fig:xdm_opt}, we plot the B86bPBE50-XDM mean absolute deviation (MAD) against our LNO-MBE-CCSD(T) benchmarks for the X23 dataset for a range of $a_1$ and $a_2$ parameters.
We used meV units ($1\,$kJ/mol$=10.4\,$meV) to make differences easier to discern.
We start with a wide search of the $a_1$ and $a_2$ parameters at the top panel, covering a range between $0.70 - 1.00\,$\AA{} and $0.80 - 1.60\,$\AA{}, in intervals of $0.02$ and $0.05\,$\AA{}, respectively.
We then perform a focused search between $0.66 - 0.80\,$\AA{} and $1.40 - 1.80\,$\AA{} in the middle panel, as that region was found to have the lowest MAD.
Subsequently, we perform a final detailed search in the bottom panel (involving intervals of $0.01$ and $0.02\,$\AA{} for $a_1$ and $a_2$, respectively) to identify $a_1=0.74$ and $a_2=1.72$ as the most optimal parameters, with an MAD of $18.9\,$meV and root mean square deviation of $24.7\,$meV.
As shown in Table~\ref{tab:x23_xdm_compare}, this improves over the default parameters for the \texttt{lightdense} basis set with B86bPBE50 significantly, which has an MAD of $45.3\,$meV, higher than chemical accuracy.
We note here that this MAD differs from that in Table~\ref{tab:x23_hybrid_dft_comparison} because that utilized a two step process to correct for errors in the basis set (using the cheaper B86bPBE-XDM) functional, as described in Section~\ref{sec:fhiaims_parameters}.

A recurring challenge, as we have highlighted in the main text and also illustrated in Ref.~\citenum{a.priceXDMcorrectedHybridDFT2023} is that methods which describe molecular crystals in the X23 dataset do not necessarily describe the ice polymorphs in the ICE13 dataset well.
In Table~\ref{tab:ice13_xdm_compare}, we have also calculated the MAD for the lattice energies of the ICE13 dataset and it can be seen that the agreement remains excellent, improving over the default XDM parameters, with MAD better than the best DFA observed in Ref.~\citenum{dellapiaDMCICE13AmbientHigh2022b}.

Finally, we briefly discuss the low cost of B86bPBE50-revXDM, shown for the X23 data in Table~\ref{tab:x23_xdm_cpuh}.
The cost is on average $10\,$CPU core-hours (CPUh) for the X23 dataset, in a tight range between $5{-}17\,$CPUh.
This cost is comparable to a GGA with plane-waves, as shown in Section~\ref{sec:comp_cost} while being much cheaper than the corresponding hybrid calculations.

\begin{table}[h]

\caption{\label{tab:x23_xdm_compare}Comparison of X23 lattice energies (kJ/mol) calculated with B86bPBE50-XDM using the `lightdense' basis sets and associated XDM parameters against B86bPBE50-revXDM parameters ($a_1 = 0.74$ \AA{} and $a_2 = 1.72$ \AA{}) with `lightdense' basis sets in FHI-aims. The reference (LNO-MBE-)CCSD(T) lattice energies are also shown.}
\begin{tabular}{lrrr}
\toprule
 & CCSD(T) & B86bPBE50-XDM (lightdense) & B86bPBE50-revXDM \\ 
\midrule
1,4-cyclohexanedione & -92.2 & -85.6 & -91.2 \\
Acetic Acid & -70.3 & -69.5 & -73.6 \\
Adamantane & -67.5 & -59.2 & -63.5 \\
Ammonia & -38.6 & -34.4 & -36.5 \\
Anthracene & -113.9 & -106.0 & -112.8 \\
Benzene & -52.8 & -50.2 & -53.4 \\
CO$_2$ & -27.9 & -24.6 & -27.7 \\
Cyanamide & -82.4 & -79.7 & -82.8 \\
Cytosine & -161.9 & -154.0 & -160.4 \\
Ethyl carbamate & -85.0 & -79.6 & -84.2 \\
Formamide & -80.8 & -76.2 & -79.8 \\
Hexamine & -88.1 & -86.7 & -91.2 \\
Imidazole & -88.8 & -86.1 & -89.6 \\
Naphthalene & -83.8 & -78.0 & -82.8 \\
Oxalic Acid $\alpha$ & -99.3 & -97.3 & -104.0 \\
Oxalic Acid $\beta$ & -98.8 & -94.8 & -101.4 \\
Pyrazine & -63.5 & -59.8 & -63.8 \\
Pyrazole & -78.5 & -76.0 & -79.6 \\
Succinic Acid & -125.0 & -123.8 & -131.4 \\
Triazine & -61.5 & -56.3 & -60.1 \\
Trioxane & -64.1 & -58.0 & -63.4 \\
Uracil & -137.2 & -132.4 & -138.7 \\
Urea & -107.0 & -100.4 & -104.5 \\
MAD (kJ/mol) &  & 4.4 & 1.8 \\
\bottomrule
\end{tabular}

\end{table}

\begin{table}[h]

\caption{\label{tab:ice13_xdm_compare}Comparison of ICE13 lattice energies (in kJ/mol) calculated with B86bPBE50-XDM using the `lightdense' basis sets and associated XDM parameters against B86bPBE50-revXDM parameters ($a_1 = 0.74$ \AA{} and $a_2 = 1.72$ \AA{}) obtained in this work, against the DMC-ICE13 reference values.}
\begin{tabular}{lrrr}
\toprule
 & DMC & B86bPBE50-XDM (lightdense) & B86bPBE50-revXDM \\ 
\midrule
Ih & -59.5 & -59.7 & -60.9 \\
II & -59.1 & -57.5 & -59.1 \\
III & -58.2 & -56.9 & -58.3 \\
IV & -55.6 & -54.4 & -56.1 \\
VI & -57.7 & -55.0 & -56.8 \\
VII & -54.5 & -50.9 & -53.0 \\
VIII & -55.2 & -52.0 & -54.1 \\
IX & -58.9 & -57.6 & -59.1 \\
XI & -59.3 & -59.6 & -60.8 \\
XIII & -57.3 & -56.5 & -58.2 \\
XIV & -57.7 & -55.9 & -57.6 \\
XV & -57.7 & -55.1 & -56.9 \\
XVII & -57.7 & -58.4 & -59.5 \\
MAD (kJ/mol) &  & 1.6 & 0.8 \\
\bottomrule
\end{tabular}

\end{table}

\begin{table}[h]

\caption{\label{tab:x23_xdm_cpuh}Computational cost in CPU core-hours (CPUh) required for B86bPBE50-revXDM calculations with `lightdense' basis sets in FHI-aims for the X23 set on 128 cores for the molecular calculations and 512 cores for the molecular crystal calculations. These were performed on AMD EPYC 7742 2.25 GHz processors.}
\begin{tabular}{lr}
\toprule
 & B86bPBE50-revXDM CPUh \\ 
System &  \\
\midrule
1,4-cyclohexanedione & 9 \\
Acetic Acid & 8 \\
Adamantane & 11 \\
Ammonia & 5 \\
Anthracene & 12 \\
Benzene & 9 \\
CO$_2$ & 6 \\
Cyanamide & 10 \\
Cytosine & 14 \\
Ethyl carbamate & 7 \\
Formamide & 8 \\
Hexamine & 11 \\
Imidazole & 9 \\
Naphthalene & 11 \\
Oxalic Acid $\alpha$ & 11 \\
Oxalic Acid $\beta$ & 7 \\
Pyrazine & 7 \\
Pyrazole & 15 \\
Succinic Acid & 8 \\
Triazine & 13 \\
Trioxane & 17 \\
Uracil & 14 \\
Urea & 7 \\
Mean & 10 \\
\bottomrule
\end{tabular}

\end{table}

\clearpage

\section{\label{sec:pharma_polymorphs}Relative energy calculations on pharmaceutical molecular crystals}

In this section, we tabulate the relative energy between pairs of challenging polymorphs for four different molecular crystals, with the largest (Axitinib and Rotigotine) being comparable in size to those found in pharmaceutical drugs.
The molecular crystals and their polymorphs are as follows:
\begin{enumerate}
    \item \textbf{ROY} --- There are 14 known forms of ROY~\cite{weatherstonPolymorphicROYalty14th2025}, nearly half of which were only proposed since 2019.
    Importantly, it has been a significant challenge for DFT~\cite{ranaCorrectingPdelocalisationErrors2023b} to get the difference between forms R and Y correctly, where the latter is the experimentally predicted form at ambient conditions~\cite{stephensonConformationalColorPolymorphism1995,yuThermochemistryConformationalPolymorphism2000}.

    \item \textbf{Rotigotine} --- Used as a dopamine agonist to treat Parkinson's (and restless leg) disease~\cite{rascolRotigotineTransdermalDelivery2009} under the brand name Neupro. 
    Rotigotine was thought to exist as only one polymorph form (I) since its release in 1985.
    However, in 2008, the appearance of a more stable (and less soluble) polymorph (II) led to a massive batch recall~\cite{rietveldRotigotineUnexpectedPolymorphism2015}, with implications towards public health and the economy.
    
    \item \textbf{Axitinib} --- Sold under the brand name  Inlyta and used to treat kidney cancer~\cite{riniAG013736MultitargetTyrosine2005}  with potential applications for treating breast cancer as well~\cite{wilmesAG013736NovelInhibitor2007}.
    Axitinib represents another large pharmaceutical-sized molecular crystal where the energy difference between its various polymorphs has served as a major challenge for DFAs~\cite{beranPolymorphsCocrystalsSalts2025}.

    \item \textbf{Molecule X from the third Blind Test} --- This small molecular crystal featured in an early Blind Test~\cite{dayThirdBlindTest2005}.
    For this system, the energy differences between the experimentally observed and vanEijck-3 forms~\cite{whittletonExchangeHoleDipoleDispersion2017} are small and prone to change with the chosen DFA.
    This system was mostly included as its small size (and small energy difference) makes it ideal for benchmarking out electronic structure parameters in Section~\ref{sec:molecule_x_conv}
\end{enumerate}

\clearpage

\subsection{Convergence of 2B and 3B terms as a function of cutoff}

\begin{figure}[h]
    \includegraphics[width=\textwidth]{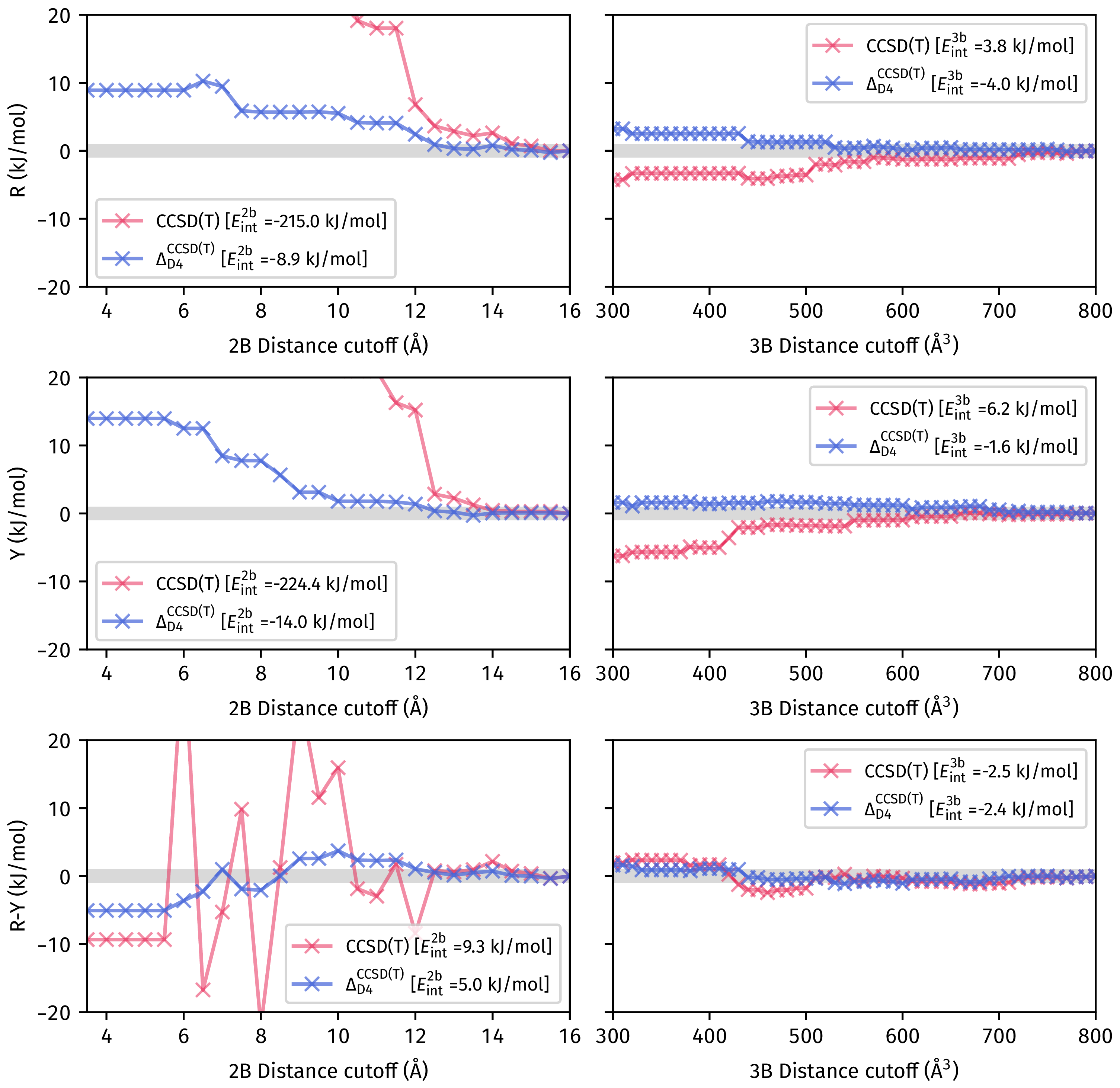}
    \caption{\label{fig:roy_2b_3b}Convergence of the 2B and 3B MBE components of ROY R and Y as well as the difference between forms R and Y.}
\end{figure}

\begin{figure}[h]
    \includegraphics[width=\textwidth]{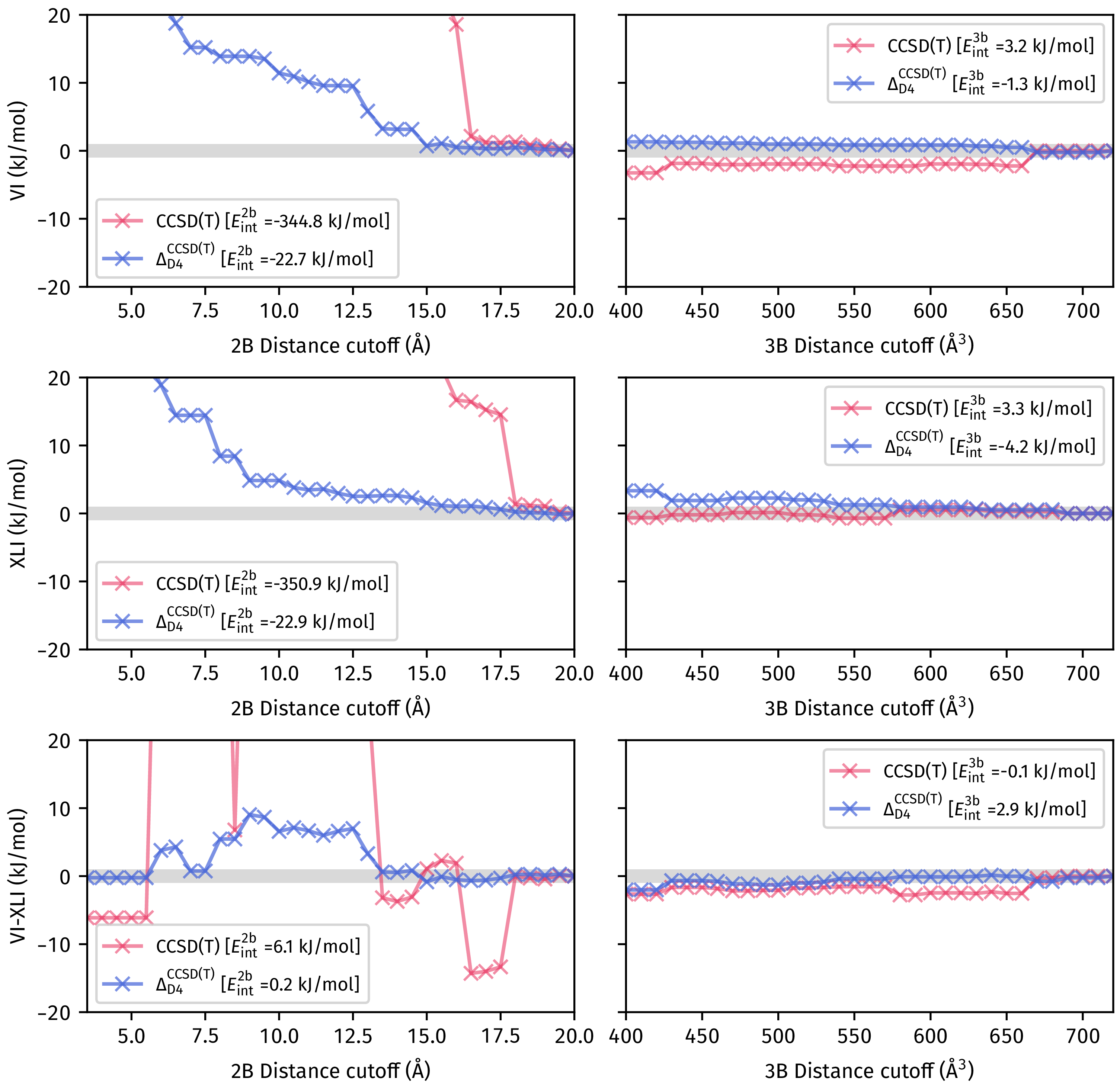}
    \caption{\label{fig:axitinib_2b_3b}Convergence of the 2B and 3B MBE components of Axitinib VI and XLI as well as the difference between forms VI and XLI.}
\end{figure}

\begin{figure}[h]
    \includegraphics[width=\textwidth]{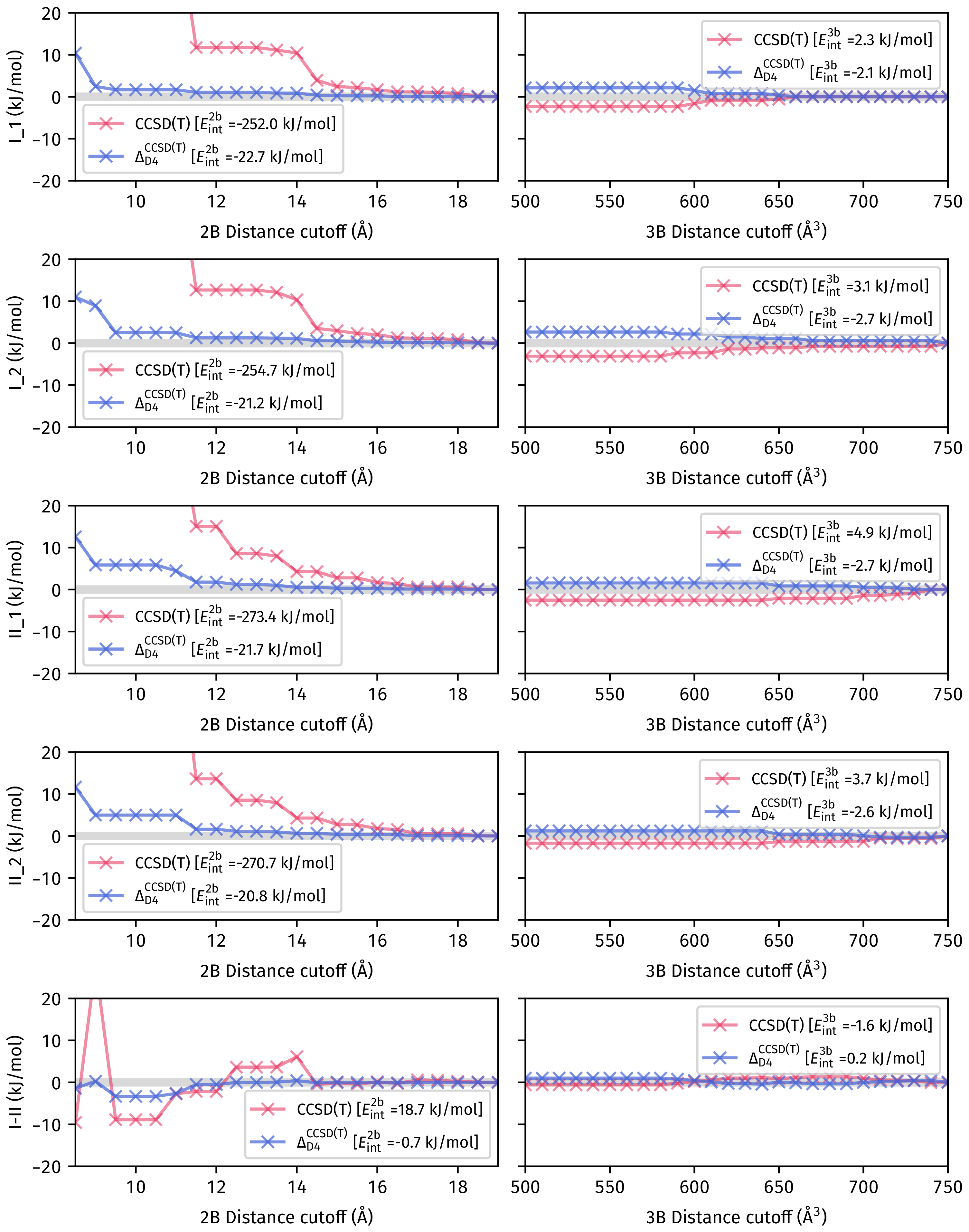}
    \caption{\label{fig:rotigotine_2b_3b}Convergence of the 2B and 3B MBE components of Rotigotine I\textsubscript{1}, I\textsubscript{2}, II\textsubscript{1}, II\textsubscript{2} as well as the difference between forms I and II.}
\end{figure}

\begin{figure}[h]
    \includegraphics[width=\textwidth]{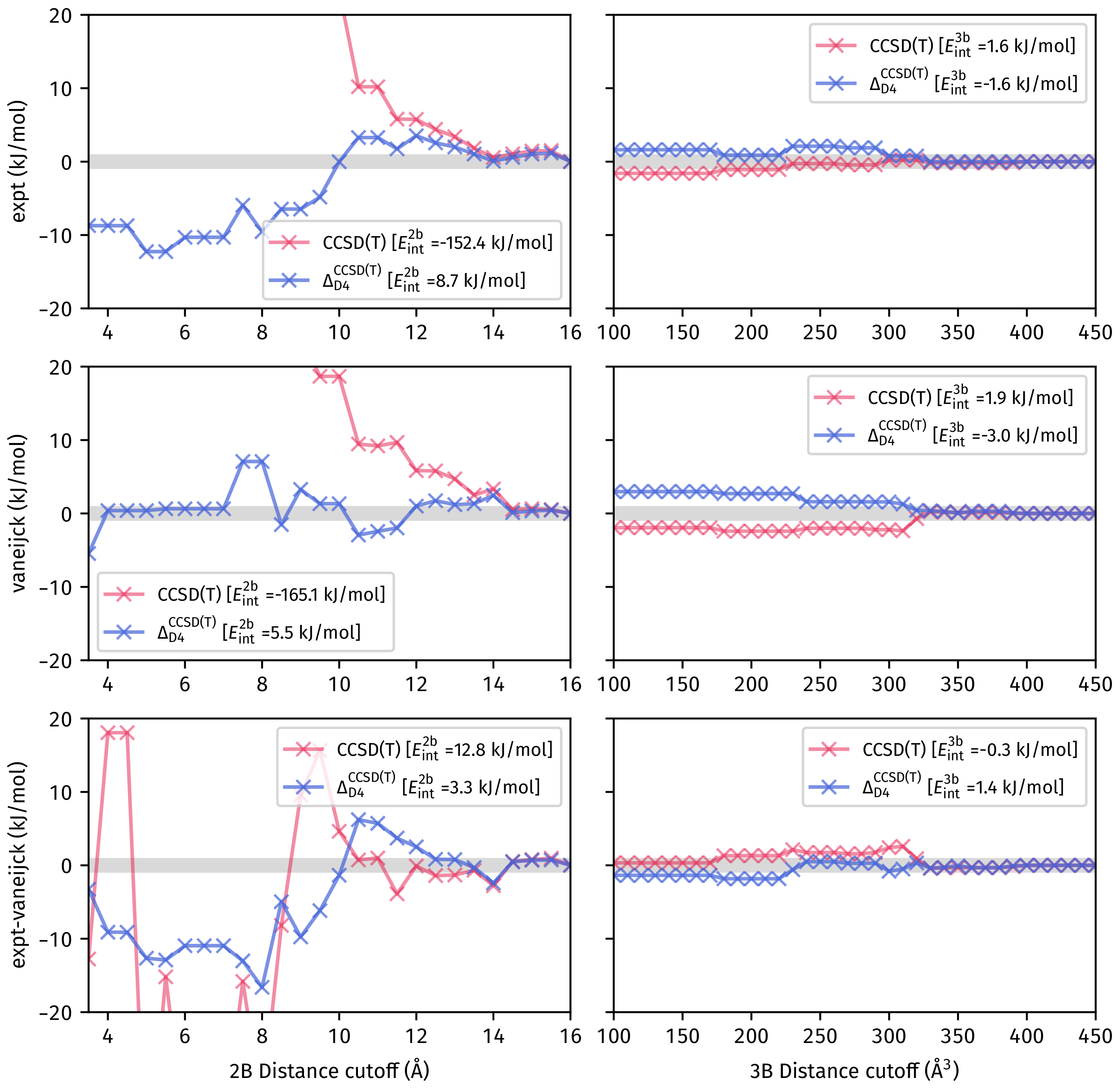}
    \caption{\label{fig:x_2b_3b}Convergence of the 2B and 3B MBE components for the experiment and vanEijck-3 forms of molecule X from the third Blind test, as well as the difference between these two forms.}
\end{figure}

In Figs.~\ref{fig:roy_2b_3b} to~\ref{fig:x_2b_3b}, we demonstrate the convergence of the 2B and 3B terms for molecular crystal polymorphs (cutoffs specified in Section~\ref{sec:mbe_details}).
We also plot the convergence of the difference between the polymorph pairs.
In general, we find that the relative energy converges faster with distance cutoff for both the 2B and 3B terms.
For example, for ROY, a 2B (center-of-mass) cutoff of up to $16\,$\AA{} is required to converge the 2B cutoff of polymorph to within $1\,$kJ/mol, while the relative energy between R and Y converges by a $12\,$\AA{} cutoff.
Similarly, the 3B cutoff also converges faster, requiring around $600\,$\AA$^3$ to converge for polymorph Y, while it is converged by $450\,$\AA$^3$ for the relative energy.
This faster convergence is observed for the other molecular crystals as well.

For all molecular crystal polymorphs, we find that performing a subtractive embedding to the CCSD(T) correlation energy with D4 improves the convergence of the 2B terms with cutoff, mirroring the observations found in Section~\ref{sec:d4_corr_ene} for the X23 dataset.
Importantly, this also further decreases the required distance cutoff to converge the relative energy.
For example, for Rotigotine, the relative energy between forms I and II is converged to within $1\,$kJ/mol by $12\,$\AA{} while $14\,$\AA{} is required without the D4 subtractive embedding.

\clearpage

\clearpage

\subsection{\label{sec:pharma_dist_conv}MBE contributions to the relative energies}

We summarize the final 1B, 2B, 3B contributions to the LNO-MBE-CCSD(T) (correlation energy) MBE as well as the periodic HF estimates in Tables~\ref{tab:roy_mbe_contributions} to~\ref{tab:x_mbe_contributions}.
We include results both with and without incorporating subtractive embedding using D4.

\begin{table}[h]

\caption{\label{tab:roy_mbe_contributions}Contributions to the CCSD(T) relative energies between the R and Y polymorphs of ROY from the many-body expansion (MBE) up to 3-body terms. Both the results without and with (corrections from HF with) D4 dispersion corrections are given. 1B terms are computed relative to the monomer taken from polymorph Y. All values are in kJ/mol.}
\begin{tabular}{lrrrrrrrrrrrr}
\toprule
D4 Correction & \multicolumn{6}{c}{without D4} & \multicolumn{6}{c}{with D4} \\ \cmidrule(lr){2-7} \cmidrule(lr){8-13} 
Method & \multicolumn{1}{c}{HF} & \multicolumn{4}{c}{CCSD(T) corr} & \multicolumn{1}{c}{Final} & \multicolumn{1}{c}{HF} & \multicolumn{4}{c}{CCSD(T) corr - D4} & \multicolumn{1}{c}{Final} \\ \cmidrule(lr){2-2} \cmidrule(lr){3-6} \cmidrule(lr){7-7} \cmidrule(lr){8-8} \cmidrule(lr){9-12} \cmidrule(lr){13-13}
MBE Level & Periodic & 1B & 2B & 3B & Sum & Total & Periodic & 1B & 2B & 3B & Sum & Total \\
\midrule
R & -31437.1 & -4.6 & -215.0 & 3.8 & -215.8 & -31653.0 & -31978.3 & -4.2 & -8.9 & -4.0 & -17.1 & -31995.4 \\
Y & -31435.7 & -0.0 & -224.3 & 6.2 & -218.1 & -31653.9 & -31980.6 & -0.0 & -14.0 & -1.6 & -15.6 & -31996.1 \\
\bottomrule
\end{tabular}

\end{table}

\begin{table}[h]

\caption{\label{tab:rotigotine_mbe_contributions}Contributions to the CCSD(T) relative energies between the I and II polymorphs of Rotigotine from the many-body expansion (MBE) up to 3-body terms. Both the results without and with (corrections from HF with) D4 dispersion corrections are given. 1B terms are computed relative to the monomer taken from polymorph II\textsubscript{1}. All values are in kJ/mol.}
\begin{tabular}{lrrrrrrrrrrrr}
\toprule
D4 Correction & \multicolumn{6}{c}{without D4} & \multicolumn{6}{c}{with D4} \\ \cmidrule(lr){2-7} \cmidrule(lr){8-13} 
Method & \multicolumn{1}{c}{HF} & \multicolumn{4}{c}{CCSD(T) corr} & \multicolumn{1}{c}{Final} & \multicolumn{1}{c}{HF} & \multicolumn{4}{c}{CCSD(T) corr - D4} & \multicolumn{1}{c}{Final} \\ \cmidrule(lr){2-2} \cmidrule(lr){3-6} \cmidrule(lr){7-7} \cmidrule(lr){8-8} \cmidrule(lr){9-12} \cmidrule(lr){13-13}
MBE Level & Periodic & 1B & 2B & 3B & Sum & Total & Periodic & 1B & 2B & 3B & Sum & Total \\
\midrule
I\textsubscript{1} & -46115.9 & -1.4 & -252.0 & 2.3 & -251.1 & -46367.0 & -46865.8 & -0.6 & -22.7 & -2.1 & -25.5 & -46891.3 \\
I\textsubscript{2} & -46114.5 & 0.1 & -254.7 & 3.1 & -251.5 & -46366.0 & -46866.9 & -0.7 & -21.2 & -2.6 & -24.6 & -46891.4 \\
II\textsubscript{1} & -46105.4 & -0.0 & -273.4 & 4.9 & -268.6 & -46373.9 & -46874.7 & 0.0 & -21.7 & -2.7 & -24.4 & -46899.1 \\
II\textsubscript{2} & -46104.8 & -1.3 & -270.7 & 3.8 & -268.2 & -46373.0 & -46874.3 & 0.5 & -20.9 & -2.6 & -22.9 & -46897.2 \\
\bottomrule
\end{tabular}

\end{table}

\begin{table}[h]

\caption{\label{tab:axitinib_mbe_contributions}Contributions to the CCSD(T) relative energies between the VI and XLI polymorphs of Axitinib from the many-body expansion (MBE) up to 3-body terms. Both the results without and with (corrections from HF with) D4 dispersion corrections are given. 1B terms are computed relative to the monomer taken from polymorph XLI. All values are in kJ/mol.}
\begin{tabular}{lrrrrrrrrrrrr}
\toprule
D4 Correction & \multicolumn{6}{c}{without D4} & \multicolumn{6}{c}{with D4} \\ \cmidrule(lr){2-7} \cmidrule(lr){8-13} 
Method & \multicolumn{1}{c}{HF} & \multicolumn{4}{c}{CCSD(T) corr} & \multicolumn{1}{c}{Final} & \multicolumn{1}{c}{HF} & \multicolumn{4}{c}{CCSD(T) corr - D4} & \multicolumn{1}{c}{Final} \\ \cmidrule(lr){2-2} \cmidrule(lr){3-6} \cmidrule(lr){7-7} \cmidrule(lr){8-8} \cmidrule(lr){9-12} \cmidrule(lr){13-13}
MBE Level & Periodic & 1B & 2B & 3B & Sum & Total & Periodic & 1B & 2B & 3B & Sum & Total \\
\midrule
VI & -51601.7 & 1.9 & -344.9 & 3.2 & -339.7 & -51941.4 & -52506.3 & -6.3 & -22.7 & -1.3 & -30.3 & -52536.6 \\
XLI & -51600.3 & 0.0 & -350.9 & 3.3 & -347.6 & -51948.0 & -52517.0 & 0.0 & -22.9 & -4.2 & -27.1 & -52544.1 \\
\bottomrule
\end{tabular}

\end{table}

\begin{table}[h]

\caption{\label{tab:x_mbe_contributions}Contributions to the CCSD(T) relative energies between the experimental and vanEijck-3 polymorphs of X from the many-body expansion (MBE) up to 3-body terms. Both the results without and with (corrections from HF with) D4 dispersion corrections are given. 1B terms are computed relative to the monomer taken from the vanEijck-3 polymorph. All values are in kJ/mol.}
\begin{tabular}{lrrrrrrrrrrrr}
\toprule
D4 Correction & \multicolumn{6}{c}{without D4} & \multicolumn{6}{c}{with D4} \\ \cmidrule(lr){2-7} \cmidrule(lr){8-13} 
Method & \multicolumn{1}{c}{HF} & \multicolumn{4}{c}{CCSD(T) corr} & \multicolumn{1}{c}{Final} & \multicolumn{1}{c}{HF} & \multicolumn{4}{c}{CCSD(T) corr - D4} & \multicolumn{1}{c}{Final} \\ \cmidrule(lr){2-2} \cmidrule(lr){3-6} \cmidrule(lr){7-7} \cmidrule(lr){8-8} \cmidrule(lr){9-12} \cmidrule(lr){13-13}
MBE Level & Periodic & 1B & 2B & 3B & Sum & Total & Periodic & 1B & 2B & 3B & Sum & Total \\
\midrule
expt & -30679.5 & -0.0 & -152.4 & 1.6 & -150.8 & -30830.3 & -31124.8 & -0.0 & 8.7 & -1.6 & 7.1 & -31117.8 \\
vaneijck & -30669.8 & 0.6 & -165.1 & 1.9 & -162.6 & -30832.4 & -31123.0 & 0.4 & 5.5 & -3.0 & 2.9 & -31120.1 \\
\bottomrule
\end{tabular}

\end{table}

\clearpage

\subsection{\label{sec:big_pharma_final_rel}Final relative energies}

We report the final relative energies between the polymorph pairs in Table~\ref{tab:final_relative_energies} for all four molecular crystals.
These are compared to experimental values, where we have performed the conversion from relative sublimation enthalpies to relative energies using molecular dynamics simulations with machine-learning interatomic potentials in Section~\ref{sec:mlip_thermal}.
For LNO-MBE-CCSD(T), the agreement with experiments is excellent, being within $0.5\,$kJ/mol for all systems (barring X, where there isn't any experimental values).
We are able to obtain two values of the relative energy, given both with and without subtracting D4 embedding.
These values are close to one another, being within $1\,$kJ/mol (and often much less) for all of the systems, giving confidence in the estimates provided.

We also report relative energies predicted by CCSD and MP2.
It can be seen that both MP2 and CCSD are unable to consistently agree with experiments.
For example, there is a ${\sim}5\,$kJ/mol error on MP2 for the VI--XLI relative energy of Axitinib and ${\sim}3\,$kJ/mol error on CCSD for Rotigotine (I--II).

We also briefly discuss our results for molecule X.
Our results are the highest level of accuracy [CCSD(T)] for this system so far.
It is in agreement with previous studies that the vanEijck-3 and experimental forms are very competitive in stability, with CCSD predicting them to nearly degenerate.
However, both CCSD(T) and MP2 suggests that the vanEijck-3 form is more stable by $2\,$kJ/mol.
This, combined with the fact that free energy contributions~\cite{whittletonExchangeHoleDipoleDispersion2017} further favors the vanEijck-3 form suggest that it may indeed be more stable than the experimental crystal form.

\begin{table}[h]

\caption{\label{tab:final_relative_energies}Final relative energies between the most stable polymorphs of each pharmaceutical molecule studied here with different levels of theory, both without and with (corrections from HF with) D4 dispersion corrections. All values are in kJ/mol.}
\begin{tabular}{lrrrrrrr}
\toprule
 & Experiment & \multicolumn{3}{c}{without D4} & \multicolumn{3}{c}{with D4} \\ \cmidrule(lr){3-5} \cmidrule(lr){6-8} 
 & (kJ/mol) & MP2 & CCSD & CCSD(T) & MP2 & CCSD & CCSD(T) \\
\midrule
ROY R - Y & 1.2$\pm$0.5~\cite{yuPolymorphismMolecularSolids2010b} & 1.6 & 2.0 & 0.9 & 1.5 & 1.8 & 0.8 \\
Axitinib VI - XLI & 7.5$\pm$0.8~\cite{campetaDevelopmentTargetedPolymorph2010a} & 11.4 & 5.8 & 6.6 & 12.4 & 6.8 & 7.5 \\
Rotigotine I - II & 7.0$\pm$0.5~\cite{mortazaviComputationalPolymorphScreening2019} & 7.8 & 4.4 & 7.0 & 7.6 & 4.1 & 6.8 \\
X vanEijck-3 - Expt & N/A & -2.5 & -0.0 & -2.1 & -2.8 & -0.3 & -2.4 \\
\bottomrule
\end{tabular}

\end{table}

\subsection{\label{sec:pharma_dfa}Density functional approximation comparison}
In Table~\ref{tab:pharma_dft_comparison}, we have compared the relative energies predicted by the polymorph pairs between the 4 molecular crystals, reporting their difference w.r.t. the LNO-MBE-CCSD(T) values in Table~\ref{tab:final_relative_energies} (the average of when `Without D4' and `With D4' subtractive embedding).

Overall, we find that there is significant discrepancy between CCSD(T) and the DFAs.
For example, for ROY, only one DFA [B86bPBE50-revXDM] can correctly predict the stability of form Y over R.
Similarly, this is also a challenging property to predict for Axitinib, with some DFAs predicting errors in excess of $8\,$kJ/mol; these are DFAs such as B86bPBE-XDM as well as r$^2$SCAN-D4, which perform well on absolute lattice energies for the X23 dataset.
Rotigotine represents a relatively simple system to get right, with many DFAs lying within $1\,$kJ/mol.

\begin{table}[h]

\caption{\label{tab:pharma_dft_comparison}Relative lattice energies (in kJ/mol) for polymorph pairs of the four molecular crystals with various density functional approximation, together with their error with respect to the CCSD(T) estimates.}
\begin{tabular}{lrrrrrrrr}
\toprule
 & \multicolumn{2}{c}{ROY} & \multicolumn{2}{c}{Axitinib} & \multicolumn{2}{c}{Rotigotine} & \multicolumn{2}{c}{X} \\ \cmidrule(lr){2-3} \cmidrule(lr){4-5} \cmidrule(lr){6-7} \cmidrule(lr){8-9} 
 & $E_\text{rel}$ & Error & $E_\text{rel}$ & Error & $E_\text{rel}$ & Error & $E_\text{rel}$ & Error \\
\midrule
PBE-D3(BJ) & -6.2 & -7.0 & -2.4 & -9.5 & 5.1 & -1.8 & 0.8 & 3.1 \\
PBE-TS & 0.0 & -0.8 & 0.0 & -7.0 & 0.0 & -6.9 & 0.0 & 2.2 \\
revPBE-D30 & -5.7 & -6.5 & 7.8 & 0.8 & 7.6 & 0.8 & 1.1 & 3.3 \\
B86bPBE-XDM & -5.9 & -6.7 & -2.0 & -9.0 & 7.4 & 0.5 & 1.6 & 3.8 \\
r$^2$SCAN-D4 & -6.3 & -7.1 & -1.6 & -8.6 & 7.3 & 0.4 & 0.2 & 2.5 \\
optB86b-vdW & -5.8 & -6.6 & 0.6 & -6.5 & 11.9 & 5.0 & 2.4 & 4.6 \\
vdW-DF2 & -6.3 & -7.1 & 0.6 & -6.4 & 10.0 & 3.1 & 0.5 & 2.7 \\
B86bPBE25-XDM & -2.5 & -3.3 & 1.3 & -5.7 & 7.9 & 1.0 & 1.4 & 3.6 \\
B86bPBE50-XDM & -0.1 & -0.9 & 3.4 & -3.7 & 8.1 & 1.2 & 0.6 & 2.8 \\
B86bPBE50-revXDM & 1.4 & 0.6 & 3.3 & -3.8 & 7.5 & 0.6 & 0.3 & 2.5 \\
PBE0-MBD & -2.7 & -3.6 & 1.4 & -5.7 & 7.3 & 0.4 & 0.7 & 2.9 \\
\bottomrule
\end{tabular}

\end{table}

\subsection{Estimation of geometrical errors}

As discussed in Section~\ref{sec:geom_details}, we utilized DFT-optimized geometries for the pharmaceutical polymorphs [ROY (R and Y), Axitinib (VI and XLI), Rotigotine (I and II) and Molecule X (van-Eijck and experimental)] for our LNO-MBE-CCSD(T) calculations.
Obtaining CCSD(T)-optimized geometries would be highly challenging as energy gradients are generally unavailable in most codes.
As such, we are limited towards using geometries generated by a lower level of theory such as DFT.
There is thus an error in the resulting $E_\text{rel}$ arising from the use of a DFT geometry.

\begin{figure}
    \includegraphics[width=\textwidth]{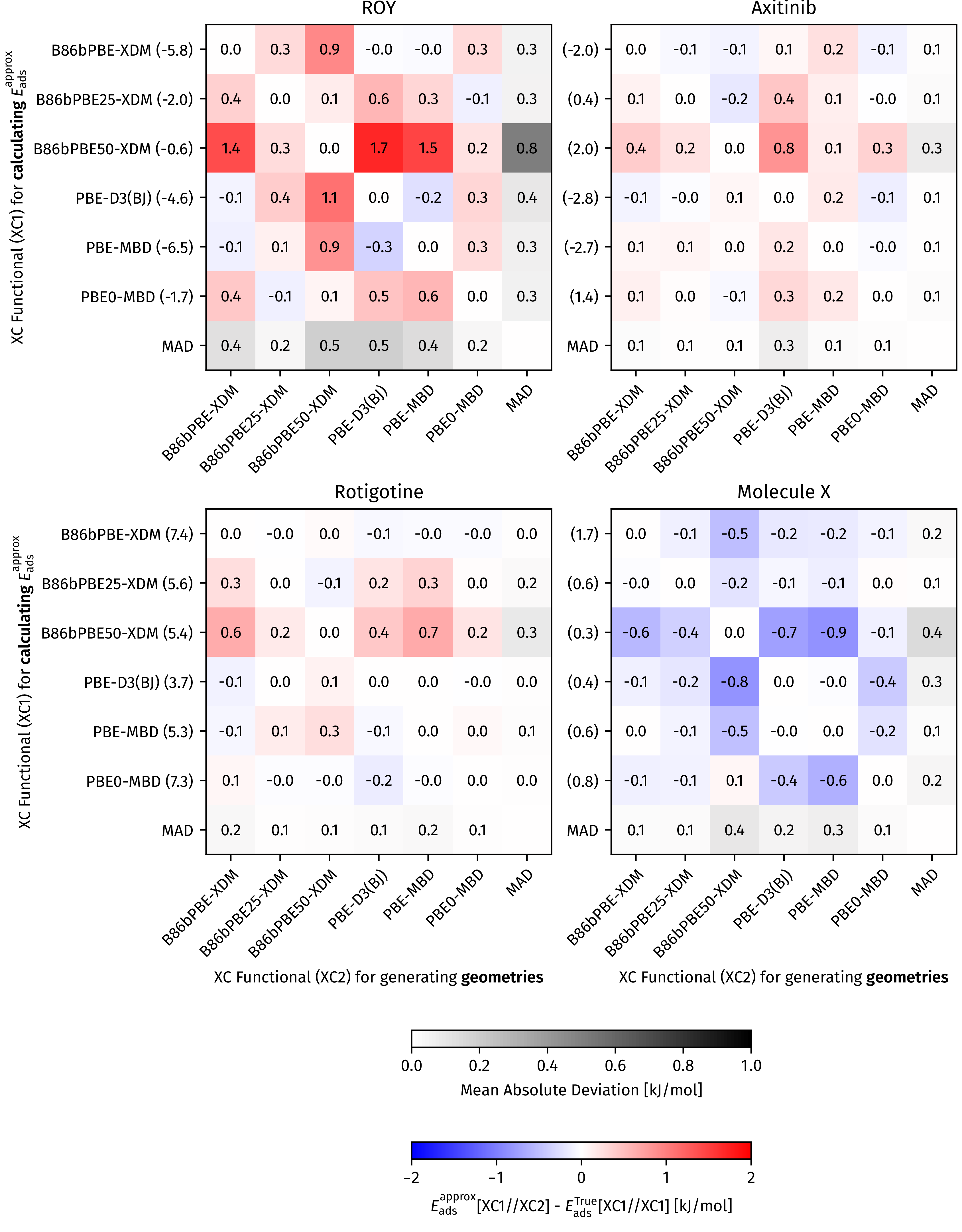}
    \caption{\label{fig:xc_geom_error} Estimating the errors in $E_\text{rel}$ for using inconsistent geometries for the polymorph pairs of ROY (R-Y), Axitinib (VI-XLI), Rotigotine (I-II)and  Molecule X (van-Eijck-experimental). We considered 6 DFAs. For the geometry generated by each DFA (on the x-axis), an approximate $E_\text{rel}$ is computed for the other DFAs and compared to the true estimate with the appropriate geometry. The difference in $E_\text{rel}$ is plotted, with a corresponding mean absolute deviation given in the bottom row.}
\end{figure}

Here, we aim to find the typical errors expected from utilizing an inconsistent level of theory by considering 6 DFT functionals (selected from DFAs discussed in the next section): B86bPBE-XDM, B86bPBE25-XDM, B86bPBE50-XDM, PBE-D3(BJ), PBE-MBD and PBE0-MBD.
We relaxed each of the polymorph geometries with these 6 different DFAs, to obtain a ``true'' $E_\text{rel}$ with the consistent DFA for the geometry.
We then obtained 5 other ``approximate'' $E_\text{rel}$ for each geometry on the other DFAs.
The difference between the approximate and true $E_\text{rel}$ is plotted in Figure~\ref{fig:xc_geom_error}, with the mean absolute deviation from the true $E_\text{rel}$ given.
The MAD given along the column to the right is a measure of how sensitive a DFA is to the geometry, while the MAD given along the row at the bottom is a measure of the error expected when applying an inconsistent level of theory to a DFA's geometry.

We find that the geometrical errors are small, on the order of $0.5\,$kJ/mol or less.
This is most significant for ROY, while the other systems have MADs mostly in the $0.1$ to $0.2\,$kJ/mol range.
B86bPBE50-XDM appears to be the most sensitive DFA to the geometry, probably arising from its higher exact exchange fraction, as evidenced by the fact that it is strongly affected in ROY --- a system where delocalization error is prominent.

\clearpage

\section{\label{sec:mlip_thermal}Thermal contributions from machine-learning interatomic potentials}

Experimental techniques typically measure the sublimation enthalpy, where the thermal and vibrational contributions must be removed if we want to make a direct `apples-to-apples' comparison to the lattice energies (or relative energies) computed directly from our simulation methods.
However, computing these contributions rigorously may not be trivial.
Even for several systems~\cite{dellapiaAccurateEfficientMachine2025} within the X23 dataset (notably succinic acid), the standard quasi-harmonic approximation (QHA) may lead to significant errors (${\sim}10\,$kJ/mol).
Incorporating anharmonicity requires performing classical molecular dynamics (MD) or path-integral MD (PIMD) to further incorporate quantum nuclear effects.
However, these methods are computationally prohibitive, requiring hundreds of thousands of force evaluations.

We leverage recent developments in machine learning interatomic potentials (MLIPs) by~\citet{dellapiaAccurateEfficientMachine2025} to overcome the cost to perform (PI)MD.
We have modified this framework to enable tackling multiple polymorphs simultaneously.
Furthermore, its accuracy is improved further through the use of B86bPBE50-revXDM discussed in Section~\ref{sec:opt_xdm}.
We will use this approach to calculate the thermal contributions to the relative energy for the ROY (R and Y) and Axitinib (XLI and VI) polymorphs.
For Rotigotine Forms I and II, we have taken the thermal contributions calculated by~\citet{mortazaviComputationalPolymorphScreening2019}.

\subsection{The framework}

Here, we describe our extensions to the previous framework of~\citet{dellapiaAccurateEfficientMachine2025} to compute relative energies between molecular crystal polymorphs -- commonly the energy of interest for crystal structure prediction.
This previous work focused on relatively rigid molecules, which does not represent the interplay of intra- and intermolecular interactions commonly found in many organic crystal polymorphs.
Our extensions to the framework enable us to study the relative enthalpies between polymorph pairs of Axitinib and ROY, both systems which exemplify not only a strong interplay of intra- and intermolecular interactions but are also challenges to density functional approximations.
We observe that no significant increases in data are needed to tackle these more flexible systems.

\begin{figure}
    \includegraphics[width=0.85\textwidth]{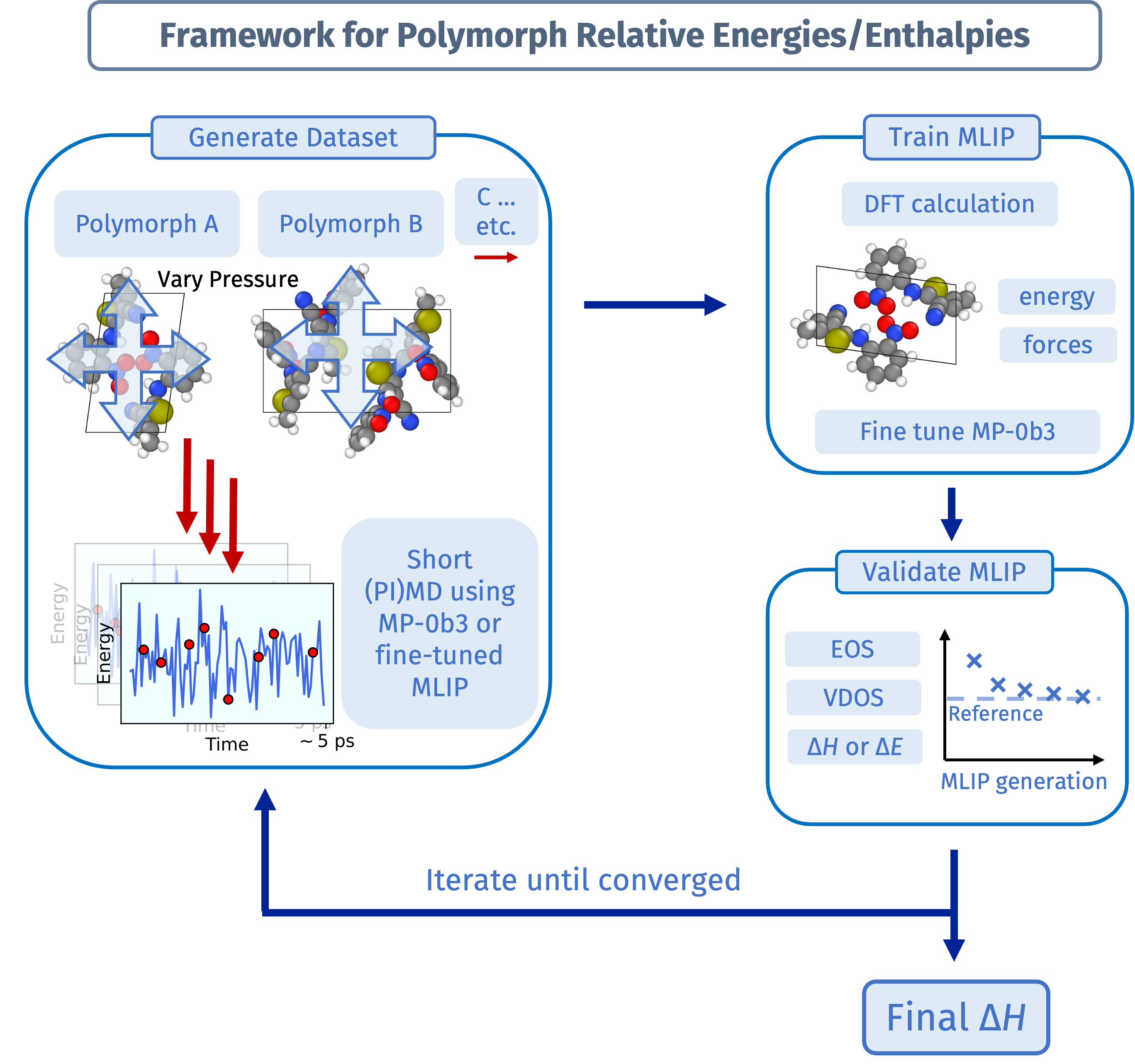}
    \caption{\label{fig:mlip_framework}Illustrating the framework used to train DFT-level MLIPs for flexible polymorphs.}
\end{figure}

The framework we use is illustrated in Figure~\ref{fig:mlip_framework}.
The framework goes over several generations, with the aim to converge the target property of interest.
In our work, it will be the relative enthalpy from molecular dynamics simulations.
We start by generating structures across a range of 5 densities, from relaxations at pressures of -4, -2, -1, 0, 1, 2 and 4 kbar.
For each of these pressures, we will perform short NVT (path-integral) molecular dynamics simulations.
We will start with the MACE-MP-0b3 MLIP~\cite{NEURIPS2022_4a36c3c5} -- a foundational model trained to PBE~\cite{perdewGeneralizedGradientApproximation1996d} data -- which we finetune at each subsequent generation with additionally generated data (including the data from previous generations).

While we have focused on a pair of polymorphs within this work, it is straightforward to tackle a larger number of polymorphs by incorporating data from those additional polymorphs.
In the future, we will look into generalizing this approach to be able to study unseen polymorphs of molecular crystals.

\subsection{Computational and dataset details}

We refer to the reader to the Methods section in Ref.~\citenum{dellapiaAccurateEfficientMachine2025} for all the computational details on the MACE~\cite{NEURIPS2022_4a36c3c5} MLIP architecture (and corresponding hyperparameters) and how we calculate the sublimation enthalpy from (path-integral) molecular dynamics as well as the quasi-harmonic approximation (QHA).
The final dataset we used consisted of 5 generations, whereby the first generation was generated by MD simulations using MACE-MP-0b3 foundation model with the other generations using the subsequently trained models.
This resulted in 888 structures for both Axitinib and ROY, amounting to $37{,}320$ and $21{,}725\,$CPUh in total, respectively, on Intel Xeon Gold 6248 2.5GHz processors.
We used the primitive unit cells of each polymorph, corresponding to 4 (184 atoms) and 2 (92 atoms) molecules for forms XLI and VI of Axitnib, respectively, and also 4 (108 atoms) and 2 (54 atoms) molecules for forms Y and R of ROY.

\subsection{Convergence of the relative sublimation enthalpies}

We use several metrics to assess the convergence of our MLIPs, including the vibrational density of states (VDOS), equation of state (EOS) as well as directly computing the relative enthalpy $\Delta H$.

\begin{figure}[h]
    \includegraphics[width=\textwidth]{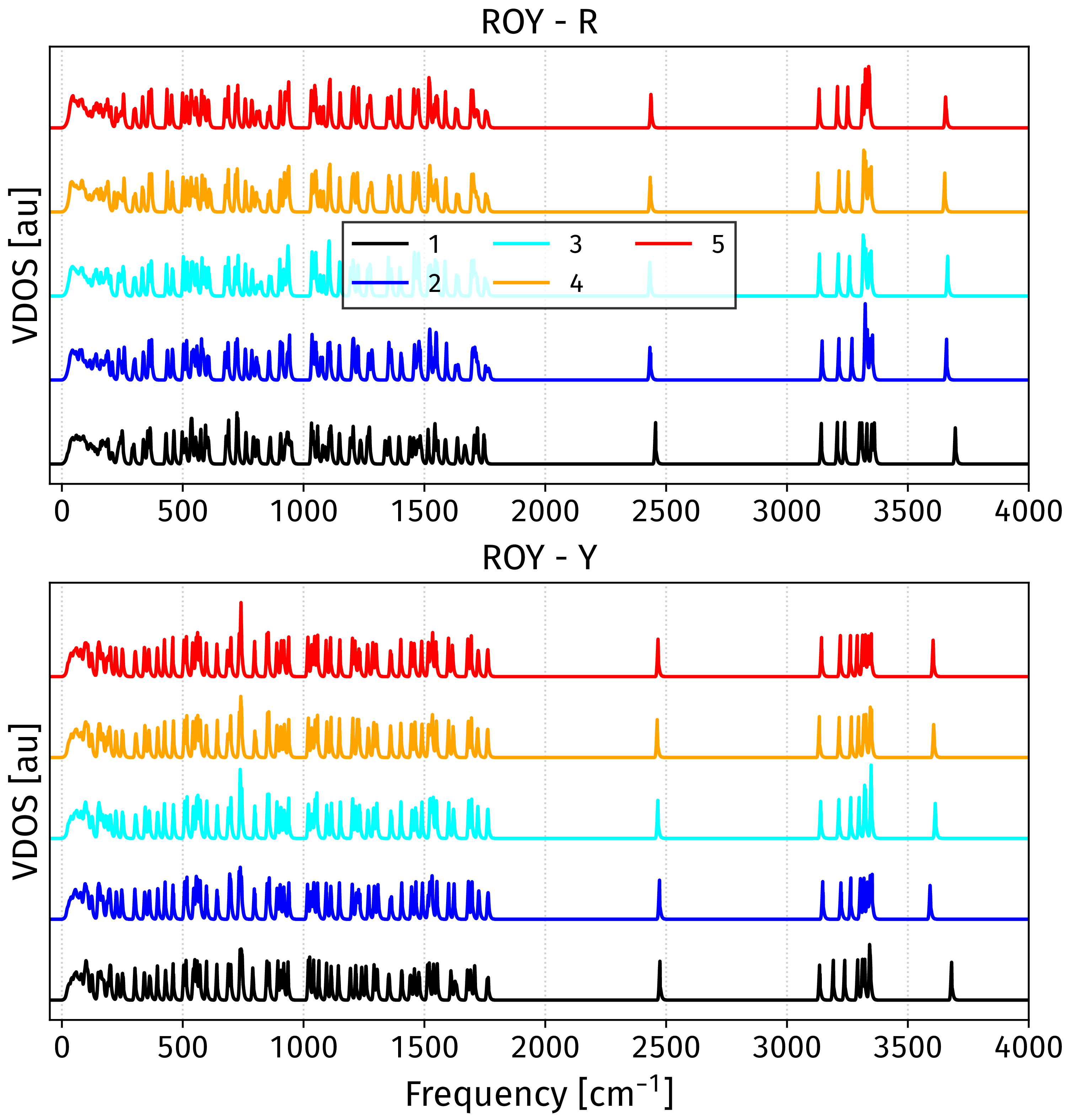}
    \caption{\label{fig:ROY_vdos}Convergence of the vibrational density of states for ROY forms R (upper panel) and XLI (lower panel) with MLIP generations.}
\end{figure}

\begin{figure}[h]
    \includegraphics[width=\textwidth]{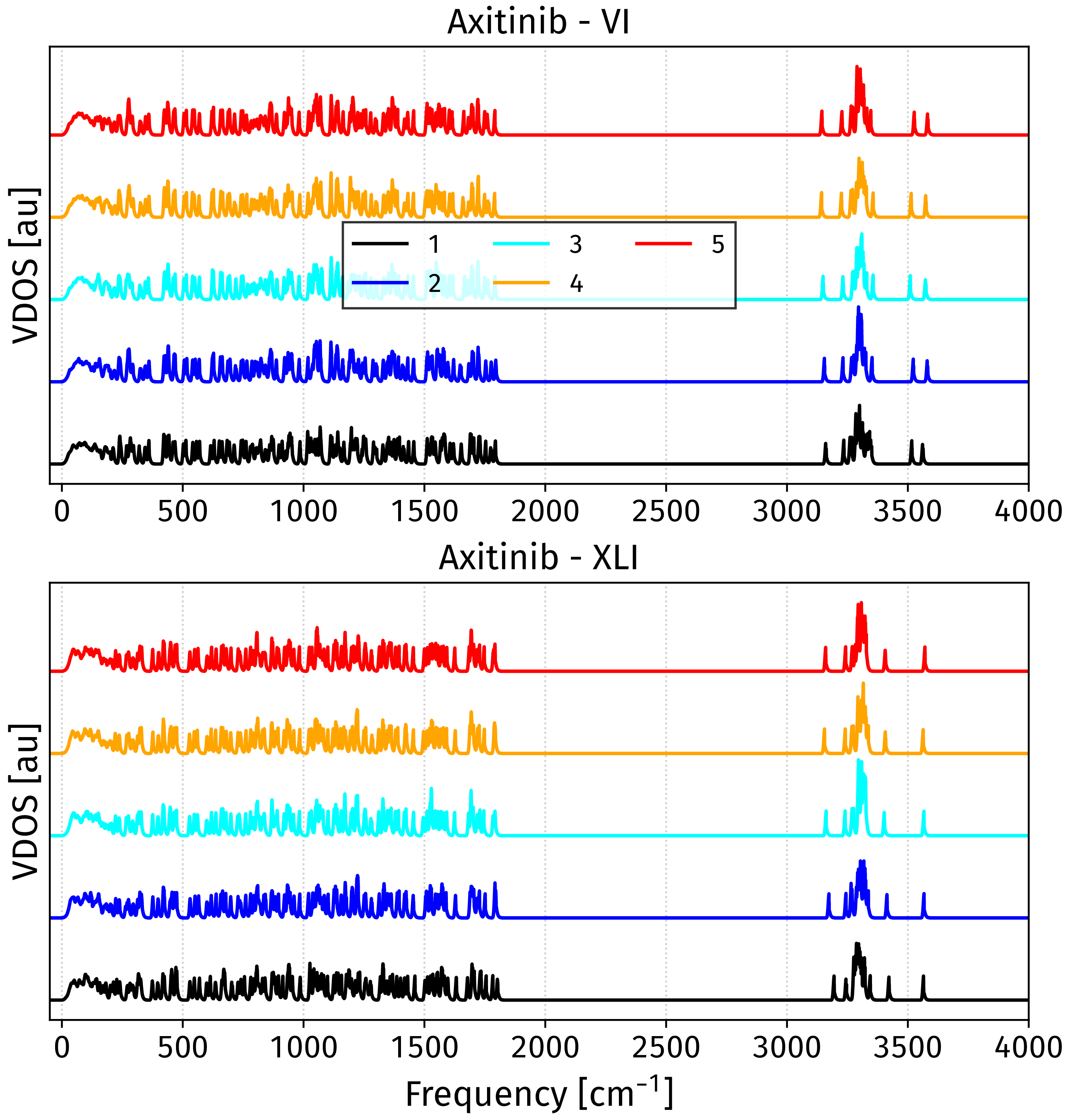}
    \caption{\label{fig:Axitinib_vdos}Convergence of the vibrational density of states for Axitinib forms VI (upper panel) and XLI (lower panel) with MLIP generations.}
\end{figure}

We have computed the VDOS for the 5 generations of MLIPs for the polymorphs of ROY and Axitinib in Figs.~\ref{fig:ROY_vdos} and~\ref{fig:Axitinib_vdos}, respectively.
By visual inspection, it can be seen (particularly at the higher frequency peaks) that from generation 3 onwards, there is little noticeable change in the VDOS for the polymorphs of both ROY and Axitinib.

\begin{figure}[h]
    \includegraphics[width=\textwidth]{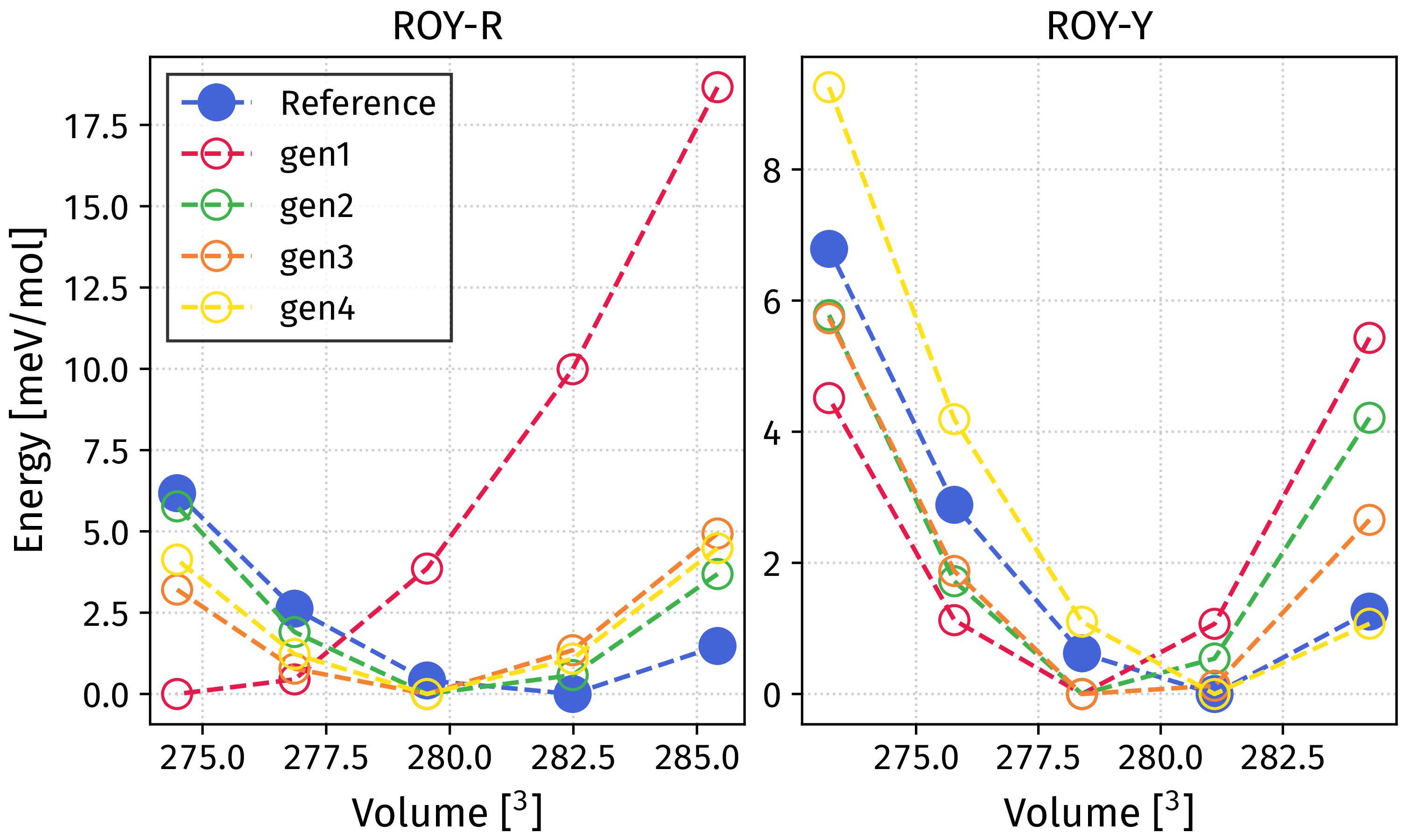}
    \caption{\label{fig:ROY_eos}Convergence of the equation of states for ROY forms R (left panel) and Y (right panel) with MLIP generations.}
\end{figure}

\begin{figure}[h]
    \includegraphics[width=\textwidth]{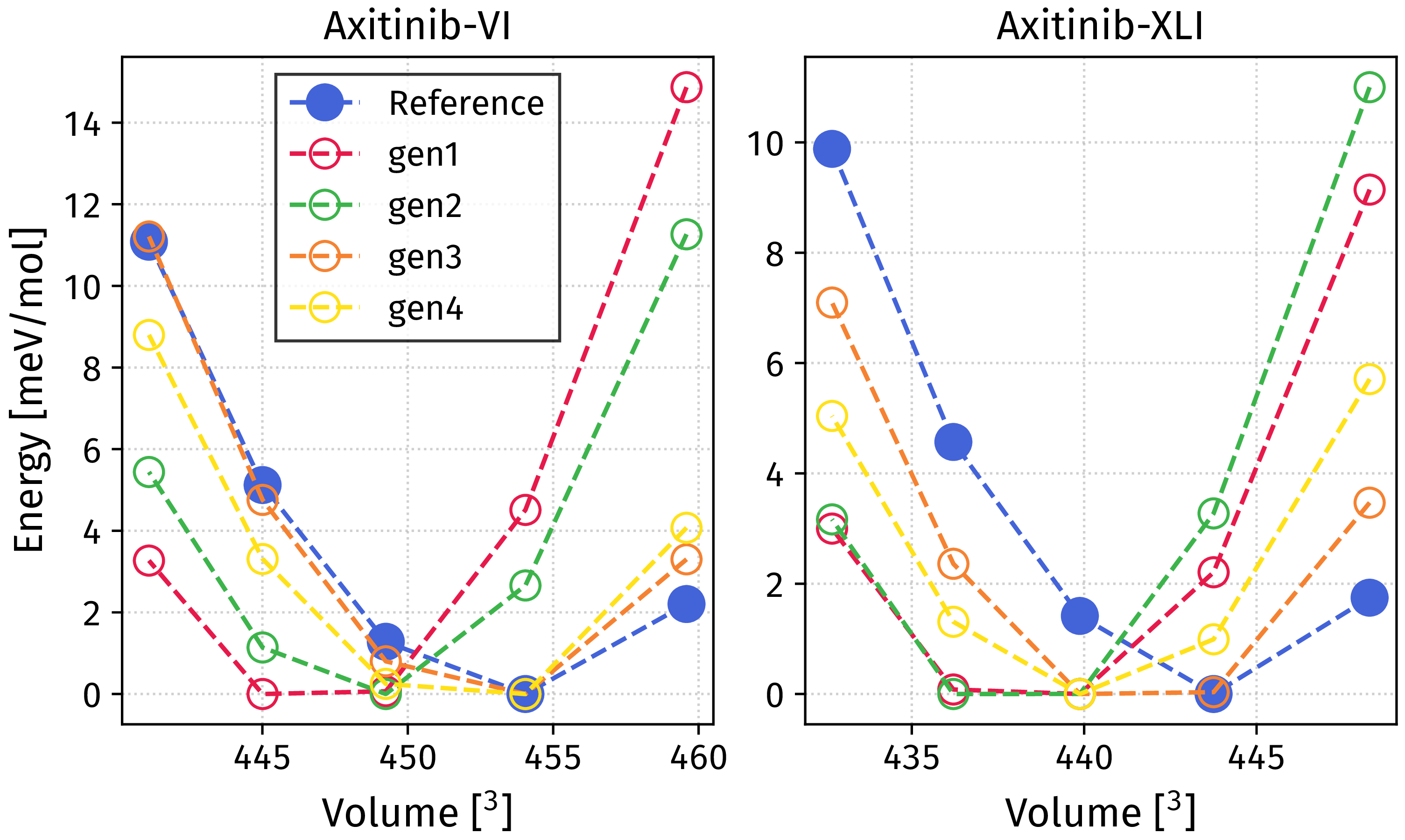}
    \caption{\label{fig:Axitinib_eos}Convergence of the equation of states for Axitinib forms VI (left panel) and XLI (right panel) with MLIP generations.}
\end{figure}

In Figs.~\ref{fig:ROY_eos} and ~\ref{fig:Axitinib_eos}, we show the convergence of EOS for the polymorphs of ROY and Axitinib.
Besides the generation 1 MLIP, which were trained only to PBE-level structures generated from the original MACE-MP-0b3, the other generations are able to reach close to $3\,$meV per molecule (${\sim}0.3\,$kJ/mol) on the EOS relative to the B86bPBE50-revXDM DFA.

\clearpage

\begin{figure}[h]
    \includegraphics[width=\textwidth]{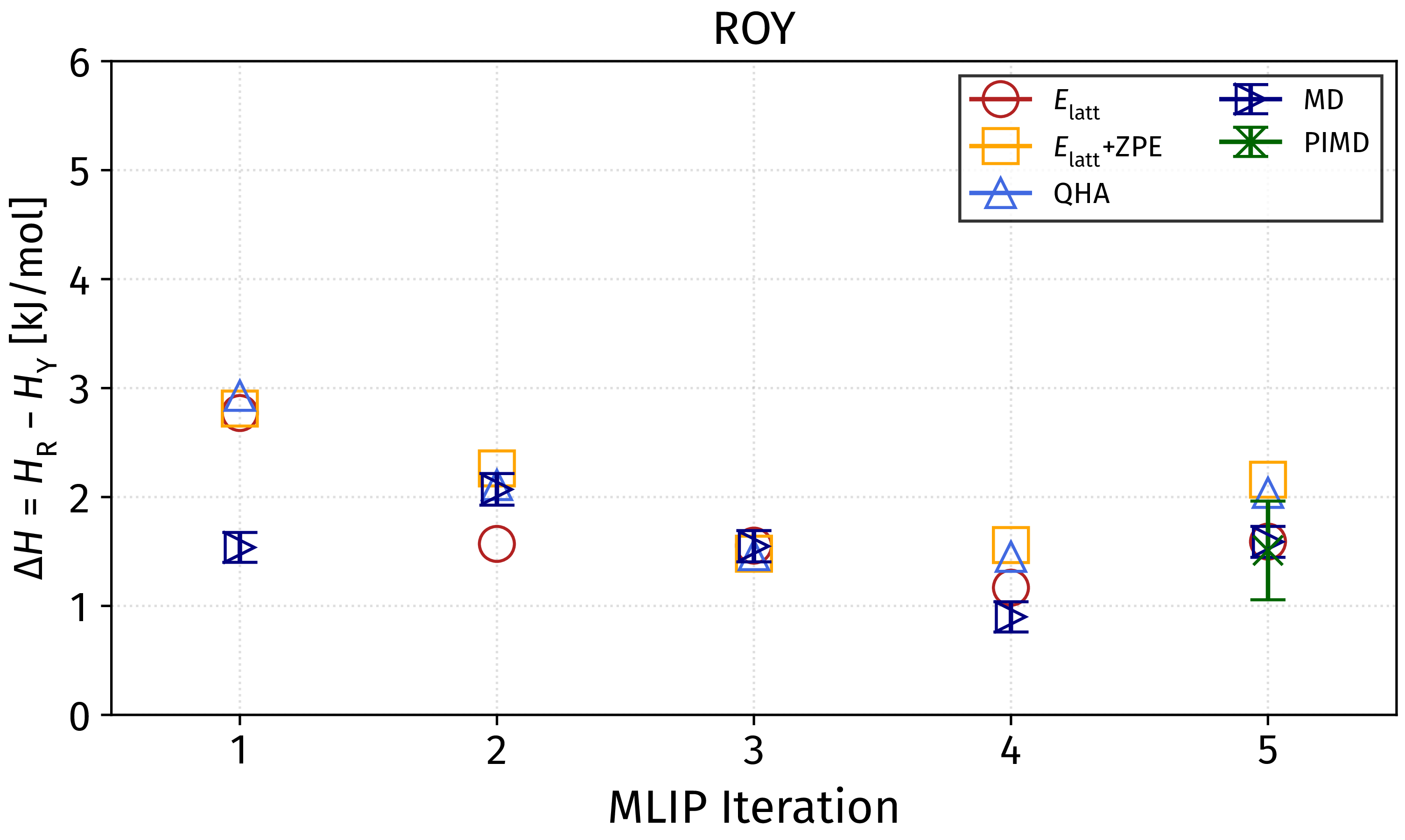}
    \caption{\label{fig:ROY_deltah}Convergence of the relative energy between ROY forms R and Y with MLIP generations.}
\end{figure}

\begin{figure}[h]
    \includegraphics[width=\textwidth]{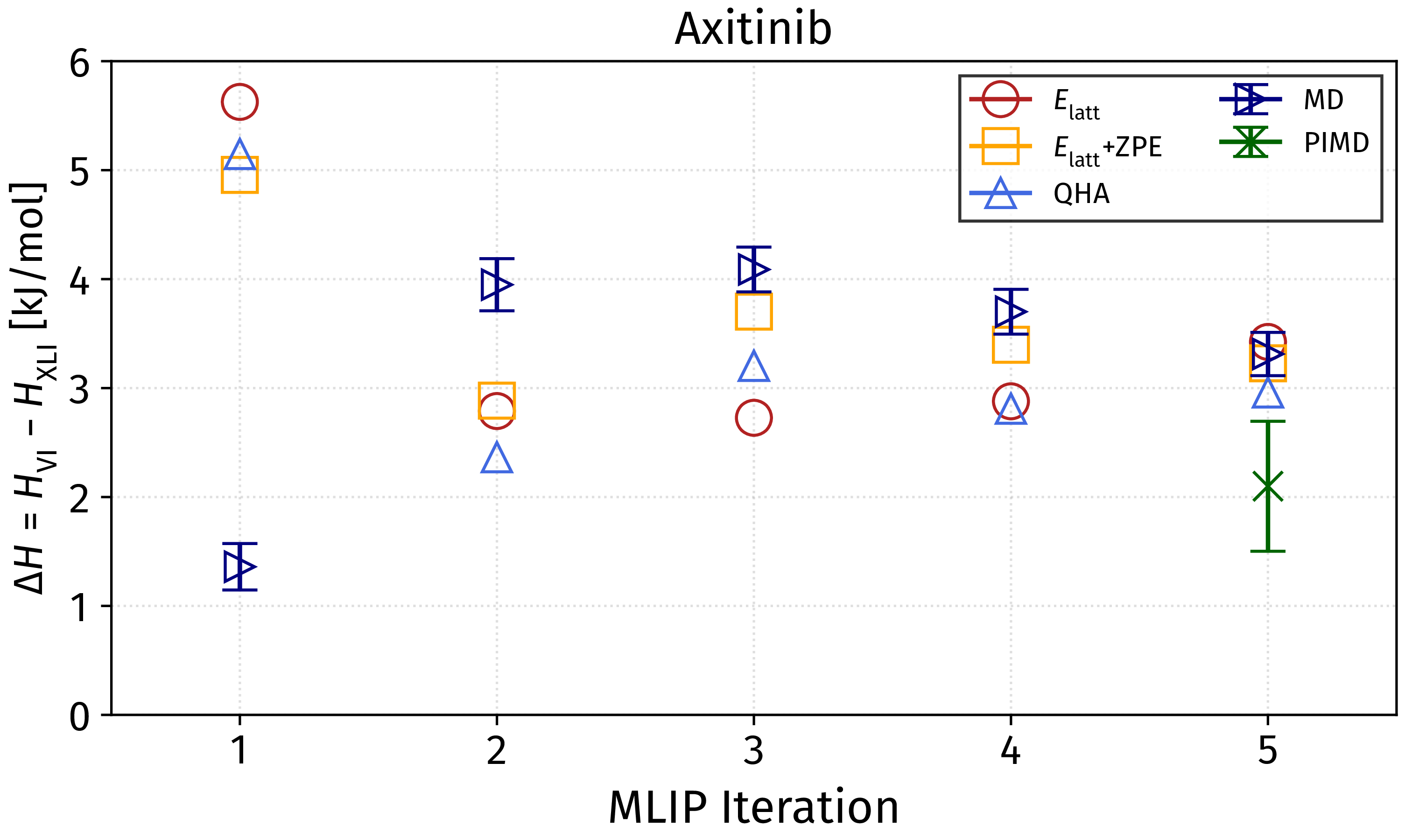}
    \caption{\label{fig:Axitinib_deltah}Convergence of the relative energy between Axitinib forms VI and XLI with MLIP generations.}
\end{figure}

We finally look at relative energy in Figs.~\ref{fig:ROY_deltah} and ~\ref{fig:Axitinib_deltah} for ROY and Axitinib.
We consider various related quantities:
\begin{itemize}
    \item $\Delta E$ --- the relative energy at 0K.
    \item $\Delta E$ + ZPE --- the relative energy at 0K with zero-point energy contributions calculated using the harmonic approximation.
    \item QHA --- the relative enthalpy calculated using the quasi-harmonic approximation.
    \item MD --- the relative enthalpy calculated using classical molecular dynamics.
    \item PIMD --- the relative enthalpy calculated using path-integral molecular dynamics.
\end{itemize}
These estimates are tabulated in Table~\ref{tab:pharma_deltaH}.
It can be seen that for all of these relative energy quantities, there is a convergence to within $0.5\,$kJ/mol by the generation 3 MLIP onwards.
For both systems, the finite-temperature effects, going up to $\Delta H^\text{MD}$ are small compared to $\Delta E$, being less than $0.5\,$kJ/mol for both ROY and Axitinib.

For the the final (fifth) generation, we have also computed PIMD estimates to $\Delta$H. 
The error bars are generally larger due to the significantly higher computational costs compared to MD.
For ROY, we see that the MD and PIMD $\Delta$H estimates are within their respective error bars.
For Axitinib, there is some effect (of the order of $1\,$kJ/mol) due to nuclear quantum effects but overall, these differences are not significant given the experimental value of $\Delta$H ($7.5\,$kJ/mol).

\clearpage

\begin{table}[h]

\caption{\label{tab:pharma_deltaH}Relative between ROY (forms R and Y) and Axitinib (forms VI and XLI)as predicted by MLIPs of different generations. All values are in kJ/mol. Statistical uncertainties ($1\sigma$) are reported for MD and PIMD results.}
\begin{tabular}{lrrrrrr}
\toprule
 &  & \multicolumn{5}{c}{MLIP Generation} \\ 
 &  & 1 & 2 & 3 & 4 & 5 \\
\midrule
\multirow[c]{5}{*}{ROY} & Elatt & 2.8 & 1.6 & 1.6 & 1.2 & 1.6 \\
 & Elatt + ZPE & 2.8 & 2.3 & 1.5 & 1.6 & 2.2 \\
 & QHA & 2.9 & 2.1 & 1.5 & 1.5 & 2.0 \\
 & MD DeltaH & 1.5$\pm$0.1 & 2.1$\pm$0.1 & 1.5$\pm$0.1 & 0.9$\pm$0.1 & 1.6$\pm$0.1 \\
 & PIMD DeltaH & - & - & - & - & 1.5$\pm$0.5 \\
\cline{1-7}
\multirow[c]{5}{*}{Axitinib} & Elatt & 5.6 & 2.8 & 2.7 & 2.9 & 3.4 \\
 & Elatt + ZPE & 5.0 & 2.9 & 3.7 & 3.4 & 3.2 \\
 & QHA & 5.1 & 2.4 & 3.2 & 2.8 & 3.0 \\
 & MD DeltaH & 1.4$\pm$0.2 & 3.9$\pm$0.2 & 4.1$\pm$0.2 & 3.7$\pm$0.2 & 3.3$\pm$0.2 \\
 & PIMD DeltaH & - & - & - & - & 2.1$\pm$0.6 \\
\cline{1-7}
\bottomrule
\end{tabular}

\end{table}

\subsection{Final experimental relative energy estimates}
We report the final thermal contributions (calculated with MD) in Table~\ref{tab:pharma_expt_deltaH}, which we used to correct the experimental $\Delta$H measurements.
In general, we set $0.5\,$kJ/mol error bars on the experimental $\Delta$H values.
This is based upon the study by \citet{campetaDevelopmentTargetedPolymorph2010a}, which shows that experimental differential scanning calorimetry (DSC) can give values in the range of ${\sim}1\,$kJ/mol for across several measurements for a single polymorph.
This is in line with previous works that highlight errors up to $5\,$\% on DSC measurements~\cite{rudtschUncertaintyHeatCapacity2002}.
For axitinib specifically, we set the error bar to $0.8\,$kJ/mol due to the increased errors found specifically for form VI by \citet{campetaDevelopmentTargetedPolymorph2010a}.
We use the (classical) MD estimates of $\Delta H$ for ROY and Rotigotine, which has been shown to agree with the PIMD estimates.
For rotigotine, we report $\Delta E$ and $\Delta H$ values taken from Mortazavi \etal{}~\cite{mortazaviComputationalPolymorphScreening2019}, where the enthalpy is estimated from vibrational free energies within the harmonic approximation.

\begin{table}[h]

\caption{\label{tab:pharma_expt_deltaH}DFT static relative energy and relative enthalpy from MD simulations. Their difference is used to correct experimental enthalpies for the polymorph pairs of rotigotine (I and II), axitinib (VI and XLI) and ROY (R and Y).}
\begin{tabular}{lrrrr}
\toprule
 & DFT $\Delta E$ & DFT $\Delta H^\text{MD}$ & Expt. $\Delta H$ & Expt. $\Delta E$ \\ 
\midrule
Rotigotine & 7.1 & 7.6 & 7.5$\pm$0.5~\cite{mortazaviComputationalPolymorphScreening2019} & 7.0$\pm$0.5 \\
Axitinib & 3.4 & 3.3 & 7.3$\pm$0.8~\cite{campetaDevelopmentTargetedPolymorph2010a} & 7.5$\pm$0.8 \\
ROY & 1.6 & 1.6 & 1.2$\pm$0.5~\cite{yuPolymorphismMolecularSolids2010b} & 1.2$\pm$0.5 \\
\bottomrule
\end{tabular}

\end{table}

\clearpage

\section{\label{sec:dft_benchmark}Benchmarking density functional and wave function methods}
We have benchmarked the performance of a selection of density functional approximations (DFAs) in this section for the X23 and ICE13 datasets.
While not aiming to cover all possible DFAs, the DFAs we have selected encompass a broad set of exchange-correlation functionals (across Jacob's ladder~\cite{perdewJacobLadderDensity2001b}) and dispersion corrections~\cite{grimmeDispersionCorrectedMeanFieldElectronic2016}.
These are as follows:
\begin{itemize}
    \item PBE-D3(BJ) -- PBE~\cite{perdewGeneralizedGradientApproximation1996d} with Grimme’s D3 dispersion using the Becke–Johnson damping.
    \item PBE-TS -- PBE with the pairwise~\citet{tkatchenkoAccurateMolecularVan2009}, formally equivalent to Grimme’s D2 dispersion but with charge-density dependent dispersion coefficients and damping function.
    \item PBE-MBD -- PBE together with a many-body dispersion by~\citet{tkatchenkoAccurateEfficientMethod2012}. This uses the ``rsSCS'' formulation.
    \item revPBE-D3(0) -- revPBE~\cite{zhangCommentGeneralizedGradient1998} together with Grimme's D3 dispersion using zero damping.
    \item B86bPBE-XDM -- B86b exchange and PBE correlation with exchange–hole dipole moment (XDM) dispersion correction~\citep{beckeExchangeholeDipoleMoment2007}.
    \item r$^2$SCAN-D4 -- Regularized SCAN meta-GGA functional~\cite{furnessAccurateNumericallyEfficient2020a,ningWorkhorseMinimallyEmpirical2022} with Grimme’s D4 dispersion correction~\citep{caldeweyherGenerallyApplicableAtomiccharge2019}, including environment-dependent atomic charges.
    \item optB86b-vdW -- Optimized B86b exchange combined with nonlocal van der Waals correlation (vdW-DF) as developed by~\citet{klimesChemicalAccuracyVan2010}.
    \item vdW-DF2 -- Second version of the nonlocal van der Waals density functional (vdW-DF2) of~\citet{leeHigheraccuracyVanWaals2010}
    \item B86bPBE25-XDM -- Hybrid B86bPBE functional with 25\% Hartree-Fock exchange, combined with XDM dispersion correction.
    \item B86bPBE50-XDM -- Hybrid B86bPBE functional with 50\% Hartree-Fock exchange, combined with XDM dispersion correction.
    \item PBE0-MBD -- Hybrid PBE0 functional combined with many-body dispersion (MBD) correction within the rsSCS formulation.
\end{itemize}

\subsection{X23 lattice energy}

We tabulate the lattice energies of the DFAs in Tables~\ref{tab:x23_dft_comparison} and~\ref{tab:x23_hybrid_dft_comparison} for the semilocal and hybrid DFAs, respectively.
We also give the mean absolute deviation (MAD) against experimental, DMC and LNO-MBE-CCSD(T) references for each DFA.
In general, the trends are well-captured irrespective of the reference methods used.
For example, the DFA with the largest error (optB86b-vdW) is the same for all three methods.
However, for the DFAs with small errors, with MAD below $4\,$kJ/mol, it may be more difficult to identify the best DFA.
For example, B86bPBE25-XDM gives the best performance w.r.t.\ LNO-MBE-CCSD(T), but r$^2$SCAN-D4 is the best according to DMC, while B86bPBE25-XDM is the best when compared to experiments.

\begin{table}[h]

\caption{\label{tab:x23_dft_comparison}Comparison of several semilocal density functional approximations to experiments, LNO-MBE-CCSD(T) and DMC (in kJ/mol) for the X23 dataset}
\begin{tabular}{lrrrrrrrr}
\toprule
 & \rotatebox{90}{PBE-D3(BJ)} & \rotatebox{90}{PBE-TS} & \rotatebox{90}{PBE-MBD} & \rotatebox{90}{revPBE-D3(0)} & \rotatebox{90}{B86bPBE-XDM} & \rotatebox{90}{r$^2$SCAN-D4} & \rotatebox{90}{optB86b-vdW} & \rotatebox{90}{vdW-DF2} \\ 
\midrule
1,4-cyclohexanedione & -88.6 & -105.6 & -91.9 & -90.6 & -88.4 & -90.6 & -115.1 & -99.5 \\
Acetic Acid & -74.3 & -81.9 & -76.4 & -67.7 & -73.5 & -74.9 & -87.3 & -73.4 \\
Adamantane & -71.4 & -105.5 & -77.7 & -73.9 & -70.5 & -63.5 & -98.9 & -82.0 \\
Ammonia & -42.6 & -43.7 & -42.3 & -38.4 & -40.5 & -40.5 & -44.1 & -39.7 \\
Anthracene & -106.7 & -135.9 & -107.6 & -115.1 & -102.6 & -103.1 & -135.3 & -103.7 \\
Benzene & -54.9 & -66.7 & -54.5 & -58.4 & -51.6 & -51.2 & -67.3 & -54.4 \\
CO$_2$ & -24.2 & -24.6 & -23.1 & -22.6 & -24.6 & -28.9 & -34.5 & -32.8 \\
Cyanamide & -92.7 & -93.4 & -91.4 & -84.5 & -89.2 & -91.0 & -99.3 & -86.9 \\
Cytosine & -160.4 & -172.5 & -161.0 & -153.7 & -156.8 & -159.9 & -179.2 & -152.6 \\
Ethyl carbamate & -87.8 & -98.5 & -90.9 & -82.6 & -86.3 & -85.8 & -104.6 & -91.1 \\
Formamide & -82.3 & -85.9 & -82.4 & -75.9 & -80.5 & -81.8 & -90.0 & -79.7 \\
Hexamine & -86.3 & -115.5 & -90.7 & -87.2 & -87.9 & -84.6 & -112.5 & -92.7 \\
Imidazole & -92.1 & -101.7 & -93.4 & -87.4 & -90.9 & -89.6 & -103.0 & -87.2 \\
Naphthalene & -80.1 & -101.2 & -80.7 & -86.8 & -76.6 & -77.4 & -101.3 & -77.9 \\
Oxalic Acid $\alpha$ & -91.0 & -98.4 & -94.1 & -83.2 & -93.6 & -103.8 & -116.7 & -104.1 \\
Oxalic Acid $\beta$ & -94.8 & -102.3 & -97.4 & -87.4 & -96.7 & -105.5 & -119.3 & -102.6 \\
Pyrazine & -65.3 & -74.8 & -62.5 & -63.1 & -62.7 & -60.4 & -78.2 & -66.0 \\
Pyrazole & -81.0 & -88.7 & -82.2 & -78.7 & -79.1 & -78.4 & -91.6 & -76.0 \\
Succinic Acid & -130.5 & -146.8 & -135.8 & -120.4 & -130.6 & -136.4 & -156.7 & -124.8 \\
Triazine & -60.2 & -67.3 & -57.1 & -58.4 & -57.8 & -57.2 & -73.7 & -64.6 \\
Trioxane & -58.2 & -75.3 & -63.8 & -59.0 & -62.1 & -58.3 & -80.2 & -72.9 \\
Uracil & -137.2 & -148.2 & -137.7 & -128.7 & -134.5 & -138.5 & -156.5 & -133.8 \\
Urea & -108.0 & -111.9 & -109.8 & -100.0 & -105.1 & -110.5 & -117.7 & -104.0 \\
MAD [CCSD(T)] & 3.9 & 12.4 & 4.2 & 4.2 & 3.5 & 4.1 & 17.1 & 5.0 \\
MAD [DMC] & 4.4 & 14.6 & 5.4 & 5.9 & 2.9 & 2.8 & 18.9 & 4.4 \\
MAD [Expt] & 4.8 & 15.0 & 5.9 & 5.2 & 3.8 & 4.8 & 19.7 & 5.3 \\
\bottomrule
\end{tabular}

\end{table}

\begin{table}[h]

\caption{\label{tab:x23_hybrid_dft_comparison}Comparison of several hybrid density functional approximations to experiments, LNO-MBE-CCSD(T) and DMC (in kJ/mol) for the X23 dataset}
\begin{tabular}{lrrr}
\toprule
 & \rotatebox{90}{B86bPBE25-XDM} & \rotatebox{90}{B86bPBE50-XDM} & \rotatebox{90}{PBE0-MBD} \\ 
\midrule
1,4-cyclohexanedione & -88.5 & -89.7 & -94.2 \\
Acetic Acid & -71.9 & -71.3 & -76.1 \\
Adamantane & -68.2 & -64.7 & -77.3 \\
Ammonia & -36.7 & -33.9 & -39.0 \\
Anthracene & -106.5 & -107.9 & -115.2 \\
Benzene & -52.1 & -51.6 & -56.4 \\
CO$_2$ & -24.1 & -23.9 & -23.7 \\
Cyanamide & -85.2 & -82.2 & -88.3 \\
Cytosine & -155.9 & -155.5 & -162.8 \\
Ethyl carbamate & -83.7 & -81.8 & -89.9 \\
Formamide & -78.2 & -76.7 & -81.4 \\
Hexamine & -89.3 & -90.5 & -95.3 \\
Imidazole & -88.7 & -86.3 & -92.6 \\
Naphthalene & -79.2 & -80.1 & -85.9 \\
Oxalic Acid $\alpha$ & -94.7 & -97.4 & -97.9 \\
Oxalic Acid $\beta$ & -94.9 & -95.0 & -98.2 \\
Pyrazine & -61.3 & -59.6 & -62.9 \\
Pyrazole & -77.5 & -76.0 & -82.0 \\
Succinic Acid & -127.8 & -127.1 & -136.0 \\
Triazine & -57.2 & -56.9 & -58.0 \\
Trioxane & -60.4 & -59.4 & -64.5 \\
Uracil & -134.0 & -133.8 & -139.6 \\
Urea & -103.5 & -103.1 & -109.4 \\
MAD [CCSD(T)] & 3.0 & 3.4 & 3.5 \\
MAD [DMC] & 2.8 & 3.2 & 5.8 \\
MAD [Expt] & 3.1 & 3.3 & 6.3 \\
\bottomrule
\end{tabular}

\end{table}

\clearpage

\subsection{ICE13 lattice energy and relative energy}

We compare the lattice energy and relative energy predicted by the DFAs for the ICE13 dataset in Tables~\ref{tab:ice13_dft_comparison} and~\ref{tab:ice13_dft_erel_comparison}, respectively.
Their MAD against both DMC and LNO-MBE-CCSD(T) are also tabulated.
The conclusions drawn from comparisons to LNO-MBE-CCSD(T) are similar to DMC, highlighting the high precision in both the DMC and LNO-MBE-CCSD(T) estimates.
In general, the MAD for the deviation between a DFA and LNO-MBE-CCSD(T) is within $0.5\,$kJ/mol the deivation against DMC as well.
For example, in agreement with Ref.~\citenum{dellapiaDMCICE13AmbientHigh2022b}, we find that revPBE-D3 with zero-damping performs really well at predicting both the absolute lattice energy and the relative energy.

\begin{turnpage}
\begin{table}[h]

\caption{\label{tab:ice13_dft_comparison}Comparison of several density functional approximations to LNO-MBE-CCSD(T) and DMC (in kJ/mol) for the ICE13 dataset}
\begin{tabular}{lrrrrrrrrrrr}
\toprule
 & \rotatebox{90}{PBE-D3(BJ)} & \rotatebox{90}{PBE-TS} & \rotatebox{90}{PBE-MBD} & \rotatebox{90}{revPBE-D3(0)} & \rotatebox{90}{B86bPBE-XDM} & \rotatebox{90}{r$^2$SCAN-D4} & \rotatebox{90}{optB86b-vdW} & \rotatebox{90}{vdW-DF2} & \rotatebox{90}{B86bPBE25-XDM} & \rotatebox{90}{B86bPBE50-XDM} & \rotatebox{90}{PBE0-MBD} \\ 
\midrule
Ih & -70.4 & -69.5 & -70.5 & -59.1 & -68.2 & -66.5 & -68.7 & -59.5 & -60.7 & -55.3 & -63.0 \\
II & -66.6 & -67.8 & -67.4 & -57.8 & -65.3 & -65.8 & -67.9 & -60.2 & -58.7 & -54.0 & -61.3 \\
III & -67.5 & -67.4 & -67.8 & -56.7 & -65.6 & -64.6 & -67.5 & -58.4 & -58.1 & -52.7 & -60.6 \\
IV & -64.1 & -65.3 & -64.8 & -55.1 & -62.7 & -63.2 & -65.8 & -58.0 & -55.9 & -51.0 & -58.4 \\
VI & -64.0 & -65.9 & -64.9 & -56.2 & -63.0 & -64.4 & -66.9 & -59.2 & -56.5 & -51.9 & -59.1 \\
VII & -57.9 & -59.1 & -59.2 & -55.0 & -57.6 & -61.6 & -63.0 & -56.6 & -52.2 & -48.7 & -54.6 \\
VIII & -58.9 & -60.4 & -60.2 & -55.8 & -58.7 & -62.7 & -64.0 & -57.8 & -53.3 & -49.9 & -55.8 \\
IX & -67.7 & -68.0 & -68.2 & -57.5 & -66.0 & -65.4 & -68.3 & -59.9 & -58.9 & -53.7 & -61.4 \\
XI & -70.9 & -70.0 & -71.0 & -59.3 & -68.6 & -66.9 & -69.2 & -59.5 & -60.8 & -55.1 & -63.2 \\
XIII & -66.0 & -67.3 & -66.7 & -56.7 & -64.6 & -65.1 & -67.7 & -60.0 & -57.9 & -53.0 & -60.5 \\
XIV & -65.2 & -66.8 & -65.9 & -56.3 & -64.0 & -64.9 & -67.5 & -59.9 & -57.3 & -52.5 & -59.9 \\
XV & -63.8 & -65.7 & -64.7 & -56.1 & -62.8 & -64.3 & -66.7 & -59.3 & -56.5 & -52.0 & -59.0 \\
XVII & -69.3 & -68.5 & -69.5 & -58.0 & -67.2 & -65.5 & -67.3 & -58.1 & -59.5 & -54.0 & -61.9 \\
MAD [CCSD(T)] & 8.3 & 9.1 & 9.0 & 0.5 & 7.0 & 7.5 & 9.7 & 1.7 & 1.0 & 4.6 & 2.7 \\
MAD [DMC] & 8.0 & 8.7 & 8.6 & 0.9 & 6.6 & 7.1 & 9.4 & 1.4 & 1.0 & 5.0 & 2.3 \\
\bottomrule
\end{tabular}

\end{table}
\end{turnpage}

\begin{turnpage}
\begin{table}[h]

\caption{\label{tab:ice13_dft_erel_comparison}Comparison of several density functional approximations to LNO-MBE-CCSD(T) and DMC~\cite{dellapiaDMCICE13AmbientHigh2022b} (in kJ/mol) for the relative energy of the ICE13 dataset}
\begin{tabular}{lrrrrrrrrrrr}
\toprule
 & \rotatebox{90}{PBE-D3(BJ)} & \rotatebox{90}{PBE-TS} & \rotatebox{90}{PBE-MBD} & \rotatebox{90}{revPBE-D3(0)} & \rotatebox{90}{B86bPBE-XDM} & \rotatebox{90}{r$^2$SCAN-D4} & \rotatebox{90}{optB86b-vdW} & \rotatebox{90}{vdW-DF2} & \rotatebox{90}{B86bPBE25-XDM} & \rotatebox{90}{B86bPBE50-XDM} & \rotatebox{90}{PBE0-MBD} \\ 
\midrule
II & 3.8 & 1.8 & 3.1 & 1.3 & 2.9 & 0.7 & 0.8 & -0.7 & 2.0 & 1.3 & 1.7 \\
III & 2.9 & 2.1 & 2.7 & 2.3 & 2.6 & 1.9 & 1.2 & 1.0 & 2.5 & 2.6 & 2.4 \\
IV & 6.3 & 4.2 & 5.7 & 4.0 & 5.5 & 3.3 & 3.0 & 1.5 & 4.8 & 4.3 & 4.6 \\
VI & 6.4 & 3.7 & 5.6 & 2.8 & 5.2 & 2.1 & 1.9 & 0.3 & 4.2 & 3.4 & 3.9 \\
VII & 12.5 & 10.5 & 11.3 & 4.1 & 10.6 & 4.9 & 5.8 & 2.9 & 8.5 & 6.6 & 8.4 \\
VIII & 11.5 & 9.1 & 10.3 & 3.3 & 9.5 & 3.8 & 4.8 & 1.6 & 7.3 & 5.3 & 7.2 \\
IX & 2.7 & 1.5 & 2.3 & 1.6 & 2.2 & 1.2 & 0.5 & -0.4 & 1.8 & 1.6 & 1.6 \\
XI & -0.5 & -0.5 & -0.5 & -0.2 & -0.4 & -0.3 & -0.5 & -0.0 & -0.1 & 0.2 & -0.2 \\
XIII & 4.4 & 2.2 & 3.8 & 2.3 & 3.6 & 1.5 & 1.0 & -0.5 & 2.8 & 2.3 & 2.6 \\
XIV & 5.2 & 2.7 & 4.5 & 2.7 & 4.2 & 1.7 & 1.3 & -0.4 & 3.4 & 2.7 & 3.2 \\
XV & 6.6 & 3.8 & 5.8 & 2.9 & 5.4 & 2.3 & 2.1 & 0.1 & 4.2 & 3.3 & 4.0 \\
XVII & 1.1 & 1.1 & 1.0 & 1.0 & 1.0 & 1.0 & 1.5 & 1.4 & 1.2 & 1.3 & 1.2 \\
MAD [CCSD(T)] & 3.3 & 1.6 & 2.7 & 0.4 & 2.4 & 0.3 & 0.7 & 1.6 & 1.5 & 0.8 & 1.4 \\
MAD [DMC] & 3.4 & 1.7 & 2.9 & 0.8 & 2.6 & 0.5 & 0.5 & 1.5 & 1.7 & 1.0 & 1.5 \\
\bottomrule
\end{tabular}

\end{table}
\end{turnpage}

\bibliography{references.bib}